\documentclass[11pt,a4paper]{book}

\usepackage{setspace}
\usepackage{amssymb}
\usepackage[polutonikogreek, english]{babel}
\usepackage{fancyhdr}
\usepackage{booktabs}
\usepackage{cite}
\usepackage{amsmath}
\usepackage{dsfont}
\usepackage{amsfonts}
\usepackage{epsfig}
\usepackage{latexsym}
\usepackage{slashed}
\usepackage{xcolor}
\usepackage{cancel}
\usepackage{verbatim}
\usepackage{nicefrac}
\usepackage{amsbsy}
\usepackage{graphicx}
\usepackage[heightrounded]{geometry}
\usepackage{comment}
\allowdisplaybreaks[1]

\usepackage{titlesec}
\titleformat{\part}[display]{\bf\LARGE\filcenter}{\titlerule[1.5pt] \vspace{1pc} \Huge\MakeUppercase{\partname} \thepart}{1pc}{\titlerule[1.5pt] \vspace{1pc} \Huge}
\titleformat{\chapter}[display]{\bf\Large\filcenter}{\titlerule[1pt] \vspace{1pc} \LARGE\MakeUppercase{\chaptertitlename} \thechapter}{1pc}{\titlerule[1pt] \vspace{1pc} \LARGE}

\usepackage[small,bf]{caption}
\def\nb{n_B}
\def\nbb{n_{\bar{B}}}
\def\ng{n_\gamma}

\newcommand{\rmeq}{{\rm eq}}
\newcommand{\barphi}{\bar{\phi}}
\newcommand{\barell}{\bar{\ell}}
\newcommand{\ml}{\mathcal{L}}

\newcommand{\bof}[1]{{\bf #1}}
\newcommand{\bfk}{{\bf k}}
\newcommand{\bh}[1]{\hat{\bf #1}}

\newcommand{\rmd}{{\rm d}}
\newcommand{\rmi}{{\rm i}}
\newcommand{\rme}{{\rm e}}
\newcommand{\tr}{{\rm tr}}
\newcommand{\ew}{{\rm EW}}
\newcommand{\CP}{{C\!P}}
\newcommand{\hc}{{\rm h.c.}}
\newcommand{\gb}{\boldsymbol{\gamma}}

\newcommand{\rmIm}{{\rm Im}}
\newcommand{\D}{\Delta}
\newcommand{\Dt}{\tilde{\Delta}}

\def\openone{\leavevmode\hbox{\small1\kern-3.8pt\normalsize1}}

\def\a{\alpha}
\def\b{\beta}
\def\c{\chi}
\def\d{\delta}
\def\e{\epsilon}
\def\f{\phi}
\def\g{\gamma}
\def\h{\eta}

\def\j{\psi}

\def\l{\lambda}
\def\m{\mu}
\def\n{\nu}
\def\o{\omega}
\def\p{\pi}
\def\q{\theta}
\def\r{\rho}
\def\s{\sigma}
\def\t{\tau}

\def\D{\Delta}

\def\G{\Gamma}

\def\O{\Omega}
\def\P{\Pi}

\def\S{\Sigma}

\def\cl{{\cal L}}

\def\bo{{\raise-.3ex\hbox{\large$\Box$}}}               
\def\face{{\raise.2ex\hbox{$\displaystyle \bigodot$}\mskip-2.2mu \llap {$\ddot
        \smile$}}}                                      

\def\leftrightarrowfill{$\mathsurround=0pt \mathord\leftarrow \mkern-6mu
        \cleaders\hbox{$\mkern-2mu \mathord- \mkern-2mu$}\hfill
        \mkern-6mu \mathord\rightarrow$}       
\def\dvec#1{\vbox{\ialign{##\crcr
        \leftrightarrowfill\crcr\noalign{\kern-1pt\nointerlineskip}
        $\hfil\displaystyle{#1}\hfil$\crcr}}}           

\def\beq{\begin{equation}}
\def\eeq{\end{equation}}

\def\beqx{\begin{displaymath}}
\def\eeqx{\end{displaymath}}

\def\bea{\begin{eqnarray}}
\def\eea{\end{eqnarray}}

\def\bs{\begin{split}}
\def\ensp{\end{split}}

\def\ba{\begin{align}}
\def\ea{\end{align}}

\def\bef{\begin{figure}}
\def\eef{\end{figure}}

\def\bec{\begin{center}}
\def\eec{\end{center}}

\catcode`@=11
\def\@citex[#1]#2{\if@filesw\immediate\write\@auxout{\string\citation{#2}}\fi
  \def\@citea{}\@cite{\@for\@citeb:=#2\do
    {\@citea\def\@citea{,\penalty\@m}\@ifundefined
      {b@\@citeb}{{\bf ?}\@warning
       {Citation `\@citeb' on page \thepage \space undefined}}%
\hbox{\csname b@\@citeb\endcsname}}}{#1}}

\def\citer{\@ifnextchar [{\@tempswatrue\@citexr}{\@tempswafalse\@citexr[]}}
\def\@citexr[#1]#2{\if@filesw\immediate\write\@auxout{\string\citation{#2}}\fi
  \def\@citea{}\@cite{\@for\@citeb:=#2\do
    {\@citea\def\@citea{--\penalty\@m}\@ifundefined
       {b@\@citeb}{{\bf ?}\@warning
       {Citation `\@citeb' on page \thepage \space undefined}}%
\hbox{\csname b@\@citeb\endcsname}}}{#1}}
\catcode`@=12

\renewcommand{\chaptermark}[1]%
         {\markboth{\thechapter.\ #1}{}}
\renewcommand{\sectionmark}[1]%
         {\markright{\thesection\ #1}}
\makeatletter
\def\cleardoublepage{\clearpage\if@twoside \ifodd\c@page\else
    \hbox{}
    \thispagestyle{plain}
    \newpage
    \if@twocolumn\hbox{}\newpage\fi\fi\fi}
\makeatother \clearpage{\pagestyle{plain}\cleardoublepage}

\makeatletter
\renewcommand\tableofcontents{%
    \if@twocolumn
      \@restonecoltrue\onecolumn
    \else
      \@restonecolfalse
    \fi
    \chapter*{\contentsname
        \@mkboth{%
           \contentsname}{\contentsname}}%
    \@starttoc{toc}%
    \if@restonecol\twocolumn\fi
    }
\makeatother

\newcommand{\LMUTitle}[9]{
  \thispagestyle{empty}
  \vspace*{\stretch{1}}
  {\parindent0cm
   \rule{\linewidth}{.7ex}}
  \begin{flushright}

    \vspace*{\stretch{1}}
    \sffamily\bfseries\Huge
    #1\\
    \vspace*{\stretch{1}}
    \sffamily\bfseries\large
    #2
    \vspace*{\stretch{1}}
  \end{flushright}
  \rule{\linewidth}{.7ex}
  \vspace*{\stretch{5}}
  \begin{center}
    \includegraphics[width=2in]{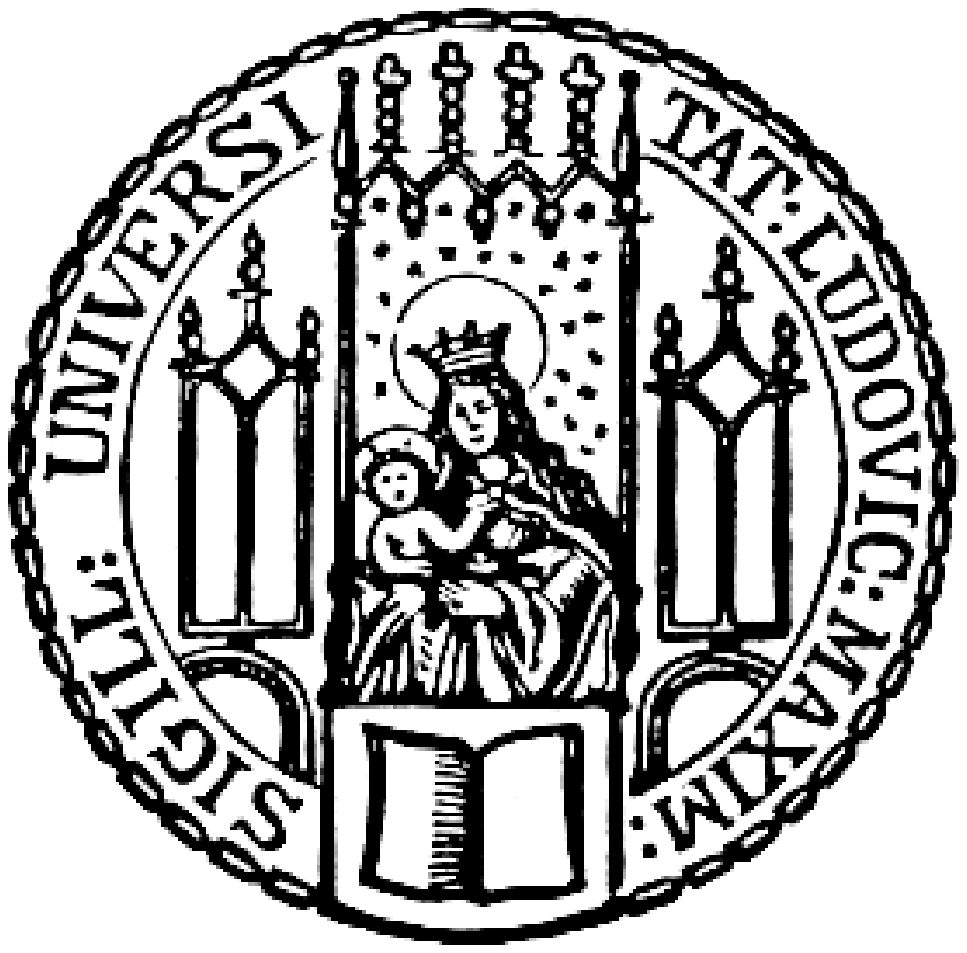} \hspace{2cm}
  \end{center}
  \vspace*{\stretch{1}}
  \begin{center}\sffamily\LARGE{#5}\end{center}
  \newpage
  \thispagestyle{empty}

  \cleardoublepage
  \thispagestyle{empty}

  \vspace*{\stretch{1}}
  {\parindent0cm
  \rule{\linewidth}{.7ex}}
  \begin{flushright}
    \vspace*{\stretch{1}}
    \sffamily\bfseries\Huge
    #1\\
    \vspace*{\stretch{1}}
    \sffamily\bfseries\large
    #2
    \vspace*{\stretch{1}}
  \end{flushright}
  \rule{\linewidth}{.7ex}

  \vspace*{\stretch{3}}
  \begin{center}
    \Large Dissertation\\
    \Large an der #4\\
    \Large der Ludwig--Maximilians--Universit\"at\\
    \Large M\"unchen\\
    \vspace*{\stretch{1}}
    \Large vorgelegt von\\
    \Large #2\\
    \Large aus #3\\
    \vspace*{\stretch{2}}
    \Large M\"unchen, den #6
  \end{center}

  \newpage
  \thispagestyle{empty}

\begin{large}
  \noindent This thesis is based on the author's work partly published
  in \cite{Kiessig:2009cm, Kiessig:2010pr, Kiessig:2010ev,
    Kiessig:2010zz} conducted from November 2007 until March 2011 at
  the Max--Planck--Institut f\"ur Physik
  (Werner--Heisenberg--Institut), M\"unchen, under the supervision of
  Dr.\ Michael Pl\"umacher.
\end{large} 

  \vspace*{\stretch{1}}

  \begin{flushleft}
    \large Erstgutachter:  #7 \\[1mm]
    \large Zweitgutachter: #8 \\[1mm]
  \end{flushleft}

  \cleardoublepage
}

\begin{document}

   \frontmatter

   \LMUTitle
      {\Huge Quasiparticles in Leptogenesis\\ \LARGE A hard-thermal-loop study}  
      {Clemens Paul Kie\ss ig}                       
      {Starnberg}                             
      {Fakult\"at f\"ur Physik}                         
      {M\"unchen 2011}                          
      {1.~April 2011}                            
      {PD Dr. Georg Raffelt}                          
      {Prof. Dr. Gerhard Buchalla}                         
      {}                         

\renewcommand\contentsname{Table of Contents} 
  \tableofcontents
   \markboth{Table of Contents}{Table of Contents}



   \addcontentsline{toc}{chapter}{\protect Abstract}

\chapter*{Abstract}

We analyse the effects of thermal quasiparticles in leptogenesis using
hard-thermal-loop-resummed propagators in the imaginary time formalism
of thermal field theory. We perform our analysis in a leptogenesis toy
model with three right-handed heavy neutrinos $N_1$, $N_2$ and
$N_3$. We consider decays and inverse decays and work in the
hierarchical limit where the mass of $N_2$ is assumed to be much
larger than the mass of $N_1$, that is $M_2 \gg M_1$. We neglect
flavour effects and assume that the temperatures are much smaller than
$M_2$ and $M_3$. We pay special attention to the influence of
fermionic quasiparticles. We allow for the leptons to be either
decoupled from each other, except for the interactions with neutrinos,
or to be in chemical equilibrium by some strong interaction, for
example via gauge bosons. In two additional cases, we approximate the
full hard-thermal-loop lepton propagators with zero-temperature
propagators, where we replace the zero-temperature mass by the thermal
mass of the leptons $m_\ell(T)$ in one case and the asymptotic mass of
the positive-helicity mode $\sqrt{2} \, m_\ell(T)$ in the other
case. We calculate all relevant decay rates and $\CP$-asymmetries and
solve the corresponding Boltzmann equations we derived. We compare the
final lepton asymmetry of the four thermal cases and the vacuum case
for three different initial neutrino abundances; zero, thermal and
dominant abundance. The final asymmetries of the thermal cases differ
considerably from the vacuum case and from each other in the weak
washout regime for zero abundance and in the intermediate regime for
dominant abundance. In the strong washout regime, where no influences
from thermal corrections are commonly expected, the final lepton
asymmetry can be enhanced by a factor of two by hiding part of the
lepton asymmetry in the quasi-sterile minus-mode in the case of
strongly interacting lepton modes.

  \markboth{Abstract}{Abstract}

   \addcontentsline{toc}{chapter}{\protect Zusammenfassung}

\chapter*{Zusammenfassung}

Wir analysieren die Effekte von thermischen Quasiteilchen in
Leptogenese, wobei wir Propagatoren verwenden, die durch harte
thermische Schleifen resummiert sind. Wir arbeiten im
Imagin\"ar\-zeit-Formalismus der ther\-mischen Feld\-theorie. Unsere
Analyse wird in einem Beispielmodell von Leptogenese mit drei
rechtsh\"andigen Neutrinos $N_1$, $N_2$ und $N_3$ durchgef\"uhrt. Wir
betrachten Zerf\"alle und inverse Zerf\"alle und nehmen den
hierarchischen Grenzfall an, in dem die $N_2$-Masse wesentlich
gr\"o\ss er als die $N_1$-Masse ist, das hei\ss t $M_2 \gg M_1$. Wir
vernachl\"assigen Flavoureffekte und nehmen an, dass die Temperaturen
sehr viel kleiner sind als $M_2$ und $M_3$. Wir legen besonderes
Augenmerk auf den Einfluss der fermionischen Quasiteilchen und lassen
sowohl eine v\"ollige Entkopplung der Leptonmoden bis auf die
Neutrinowechselwirkung zu, als auch eine starke Kopplung der Moden,
beispiels\-weise durch Eichbosonen. In zwei zus\"atzlichen F\"allen
n\"ahern wir den vollst\"andigen, durch harte thermische Schleifen
resummierten Leptonpropagator durch normale Vakuumpropagatoren an, die
eine Masse erhalten, die in einem Fall der thermischen Leptonmasse
$m_\ell(T)$ und in einem anderen Fall der asymptotischen Masse der
positiven Helizit\"atsmode, $\sqrt{2} \, m_\ell (T)$ entspricht. Wir
berechnen alle relevanten Zerfallsraten und $\CP$-Asymmetrien und
l\"osen die hergeleiteten Boltzmann-Gleichungen.  Wir vergleichen die
resultierende Leptonasymmetrie der vier ther\-mischen F\"alle und des
Vakuumfalls f\"ur drei verschiedene Anfangswerte der
Neutrino\-verteilung; verschwindende, thermische und dominante
Anfangsverteilung. Die Lepton\-asymmetrien der thermischen F\"alle
unter\-scheiden sich in gewissen Parameterregionen stark vom
Vakuumszenario als auch unter\-einander; n\"amlich im schwachen
Washout-Regime f\"ur verschwindende Anfangsverteilung und im
intermedi\"aren Regime f\"ur dominante Anfangsverteilung. Au{\ss}erdem
vergr\"o{\ss}ert sich die \mbox{finale} Leptonasymmetrie im starken
Washout-Regime, wo typischerweise keine thermischen Effekte erwartet
werden, um einen Faktor von etwa zwei, wenn ein Teil der
Leptonasymmetrie auf die quasisterile Minus-Mode im Falle stark
wechselwirkender Leptonmoden \"ubertragen wird.

  \markboth{Zusammenfassung}{Zusammenfassung}

  \mainmatter\setcounter{page}{1}


\clearpage
\pagenumbering{arabic}
\setcounter{page}{1}
\vspace{2cm}
\noindent
\chapter*{Introduction}
\addcontentsline{toc}{chapter}{Introduction}
\markboth{Introduction}{Introduction}

\hfill
\parbox{0.4 \textwidth}{\vspace{-5ex} \raggedright \small And the angel of the presence
  spake~[\dots]: Write the complete history of the creation, how in
  six days
  the Lord God finished all His works and all that He created~[\dots].\cite{Jubilees:1917aa}}
\begin{flushright}
\small \it Leptogenesis 2, 1\\[4ex]
\end{flushright}

The question of the origin of all things that we observe and that are
present in our life and surroundings, which in its last consequence is
nothing else than the question of the origin of mankind itself, has
always fascinated us and driven us to search for answers in
science, religion, philosophy and the arts. On the scientific side,
physics as the study of the laws of nature (\greektext <h fusik'h
\latintext ``nature'') and within physics, cosmology (\greektext <o k`osmoc
\latintext ``order'') as the science of the order and the evolution of the
universe, address this question and have their own formulation of
it. What is the origin of the matter that is the building block of all
things we observe, including ourselves?

The matter in nature consists of electrons, which belong to the
leptons (\greektext lept'os \latintext ``small''), and the much heavier
protons and neutrons, which belong to the baryons (\greektext bar'uc
\latintext ``heavy'') and are in turn made up of quarks. The theory that
describes how the smallest ingredients of matter, the elementary
particles, interact, is the standard model of particle physics (SM),
which has been tested to great accuracy by experiment. According to
the SM, matter particles, quarks or leptons, can only be created in
pairs together with their antiparticles, that is, antiquarks and
antileptons. If we assume that the early universe was indeed without
form and void\footnote{In fact, according to standard cosmology, the
  early universe was by no means void, but rather a vibrant soup of
  all particles that we know and possibly many more species
  interacting rapidly with each other.}~\cite{KingJames:1611aa} , that
is in the language of particle physics, there was no excess of one
particle species over the other, there would have to be an equal
amount of particles and antiparticles today. More specifically, since
annihilation of particles and antiparticles proceeds at fast rates, no
structures like atoms, molecules, galaxies, stars, planets, DNA, cells
and finally living organisms could have formed and we would
observe\footnote{Or rather, not observe.} a universe populated almost
exclusively by photons and the slowly interacting neutrinos. There are
only two possible ways out of this obviously wrong scenario: Either we
assume an excess of matter over antimatter as an initial condition of
the big bang or we find some mechanism which does not strictly obey
the conservation of baryon and lepton number and can create such an
asymmetry dynamically in some early phase of the evolution of the
universe.

It might seem tempting to assume an excess of particles over
antiparticles as an initial condition of the universe, or more
specifically the excess of baryons over antibaryons, since this is the
asymmetry we measure on cosmological scales. The fact that this
approach has to be refused is not even mainly because it would be a
highly unsatisfactory approach from a scientific point of view, but
the assumption itself clashes with another important theory in early
universe cosmology, inflation. There is broad consensus that, in order
to cure serious problems of cosmology, the early universe must have
undergone such a phase of rapid expansion, which is so fast that it
would dilute any baryon asymmetry we could realistically impose as an
initial condition of the young universe. We are thus bound to find a
mechanism that creates a baryon asymmetry dynamically, we need a
theory for baryogenesis.

Among the many baryogenesis theories that solve the problem of the
matter-antimatter asymmetry, we focus on a variant called
leptogenesis~\cite{Fukugita:1986hr}, which is a particularly
attractive model since it simultaneously solves two problems: The
creation of a baryon asymmetry via the detour of a lepton asymmetry on
the one hand, and the explanation of why the neutrinos have such a
small mass compared to all the other particles of the SM via the
seesaw
mechanism~\cite{Minkowski:1977sc,Yanagida:1979as,GellMann:1980vs,
  Mohapatra:1980yp} on the other hand. In short, one adds heavy,
right-handed neutrinos to the SM, which interact with the SM neutrinos
and suppress their mass. In the early universe, these heavy neutrinos
decay into leptons and Higgs bosons and create a lepton asymmetry,
which is lateron converted to a baryon asymmetry by anomalous SM
processes~\cite{Klinkhamer:1984di,Kuzmin:1985mm}.

Ever since the development of the theory 25 years ago, the
calculations of leptogenesis dynamics have become more refined and
many effects and scenarios that have initially been neglected have
been considered\footnote{For an excellent review of the development in
  this field, we refer to reference\cite{Davidson:2008bu}.}. Notably
the question how the hot and dense medium of SM particles influences
leptogenesis dynamics has received increasing attention over the last
years~\cite{Covi:1997dr,Giudice:2003jh,Anisimov:2010dk,Anisimov:2010gy,
  Garny:2010nz,Beneke:2010dz,Beneke:2010wd,Garbrecht:2010sz}. At high
temperature, particles show a different behaviour than in vacuum due
to their interaction with the medium: they acquire thermal masses,
modified dispersion relations and modified helicity properties. All
these properties can be summed up by viewing the particles as thermal
quasiparticles with different behaviour than their zero-temperature
counterparts, much like the large zoo of single-particle and
collective excitations that are known in high density situations in
solid-state physics. At high temperature, notably fermions can occur in two
distinct states with a positive or negative ratio of helicity over
chirality and different dispersion relations than at zero temperature.

Thermal effects have been considered by
references~\cite{Covi:1997dr,Giudice:2003jh,Anisimov:2010dk,Anisimov:2010gy,
  Garny:2010nz,Beneke:2010dz,Beneke:2010wd,Garbrecht:2010sz}. Notably
reference~\cite{Giudice:2003jh} performs an extensive analysis of the
effects of thermal masses that arise by resumming propagators using
the hard thermal loop (HTL) resummation within thermal field theory
(TFT). However, the authors approximated the two fermionic helicity
modes with one simplified mode that behaves like a vacuum 
particle
with its zero-temperature mass replaced by a thermal
mass\footnote{Moreover, an incorrect thermal factor for the
  $\CP$-asymmetry was obtained, as has been pointed out in
  reference~\cite{Garny:2010nj}.}. Due to their chiral nature, there
are serious consequences to assigning a chirality breaking mass to
fermions, hence the effects of abandoning this property should be
examined. Moreover, it seems questionable to completely neglect the
negative-helicity fermionic state which, according to TFT, will be
populated at high temperature. We argue in this study that one should
include the effect of the fermionic quasiparticles in leptogenesis
calculations and possibly in other early universe dynamics, since they
behave differently from zero-temperature states with thermal masses,
both conceptually and regarding their numerical influence on the final
lepton asymmetry. We do this by analysing the dynamics of a
leptogenesis toy model that includes only decays and inverse decays
of neutrinos and Higgs bosons, but takes into account all HTL
corrections to the leptons and Higgs bosons, paying special attention
to the two fermionic quasiparticles. In a slightly different scenario,
we assume chemical equilibrium among the two leptonic modes, thereby
simulating a scenario where the modes interact very fast. As a
comparison, we calculate the dynamics for two models where we
approximate the lepton modes with ordinary zero-temperature states and
modified masses, the thermal mass $m_\ell(T)$ and the asymptotic mass
of the positive-helicity mode, $\sqrt{2} \, m_\ell(T)$.

The thesis is structured as follows: In
chapter~\ref{cha:leptogenesis}, we present a short overview over
leptogenesis and explain the standard dynamics. We also discuss the
limitations of our approach, where we do not include flavour and
resonant effects or effects from a possible abundance of $N_2$ or
$N_3$. Chapter~\ref{cha:tft} is devoted to a brief and comprehensive
introduction into TFT in the framework of the imaginary time formalism
(ITF), where we present the necessary ingredients for the further
analysis, in particular frequency sums for fermions and bosons. We pay
special attention to the resummation of hard thermal loops, which form
the basis for the description of thermal quasiparticles. In
chapter~\ref{decayrate}, we present the toy model of leptogenesis and
calculate decays and inverse decays. At high temperature, the thermal
mass of the Higgs bosons becomes larger than the neutrino mass, such
that the neutrino decay is no longer possible and is replaced by the
decay of Higgs bosons into neutrinos and leptons. We discuss in detail
the conceptual and numerical differences of the full two-mode approach
to the one-mode approach and the vacuum result. The $\CP$-asymmetry
for the different approaches is the main topic of
chapter~\ref{cpas}. The $\CP$-asymmetry in the two-mode approach
consists of four different contributions due to the two possibilities
for the leptons in the loops. We present some useful rules for
performing calculations with the fermionic modes and compare the
analytical expressions for the $\CP$-asymmetries in different
cases. We restrict ourselves to the hierarchical limit where the mass
of $N_1$ is much smaller than the mass of $N_2$, that is $M_2 \gg
M_1$. The temperature dependence of the $\CP$-asymmetry is discussed
in detail for the one-mode approach, the two-mode approach and the
vacuum case. The differences between the asymmetries and the physical
interpretation of certain features of the asymmetries are explained
in detail. Chapter~\ref{boltzmanneq} deals with the evaluation of the
Boltzmann equations. We derive the equations and explicitly perform
the subtraction of on-shell propagators for our cases in
appendix~\ref{sec:subtr-shell-prop}. We compare our four thermal
scenarios, wich are the decoupled and strongly coupled two-mode
approach and the one-mode approach with thermal and asymptotic mass,
to the vacuum case. We show the evolution of the abundances for three
different initial conditions for the neutrinos, that is zero, thermal
and dominant abundance. We explain the dynamics of the different cases
in detail and find considerable differences both of the thermal
approaches to the vacuum case and of the two-mode cases to the
one-mode cases. We summarise the main insights of this work in the
Conclusions and give an outlook on future work and prospects.

In appendix~\ref{sec:green0}, we present Green's functions at zero
temperature, while in appendix~\ref{cha:analyt-solut-htl}, we derive
the analytical solution for the lepton dispersion relations. Some
quantities relating to leptogenesis in the vacuum case are derived in
appendix~\ref{cha:quantities-at-zero}. In
appendix~\ref{cha:other-cuts}, we present analytical expressions for
the $\CP$-asymmetry contributions of the two cuts through
$\{N',\ell'\}$ and $\{N',\phi'\}$\footnote{We shamelessly stole our
  notation for the cuts in the vertex contribution from reference
  \cite{Garbrecht:2010sz}.}, which we did not consider in
chapter~\ref{cpas}, since we are working in the hierarchical limit.
Appendix~\ref{sec:subtr-shell-prop} is devoted to the detailed
description of a correct subtraction of on-shell propagators in our
scenario.





\clearpage
\chapter{Leptogenesis}
\label{cha:leptogenesis} 

\section{The Matter-Antimatter Asymmetry}

The matter-antimatter asymmetry of the universe is usually expressed
as
\begin{equation}
\h \equiv \left. \frac{\nb - \nbb}{\ng} \right |_0\, , 
\end{equation}
where $\nb$, $\nbb$, and $\ng$ are the number densities of baryons,
antibaryons, and photons, respectively, and the subscript 0 implies
that the value is measured at present cosmic time. There might be an
excess of leptons over antileptons as well, but its contribution to
the energy density of the universe is small compared to the
contribution of the baryons.

\begin{figure}
\begin{center}
\includegraphics[width=0.60\textwidth]{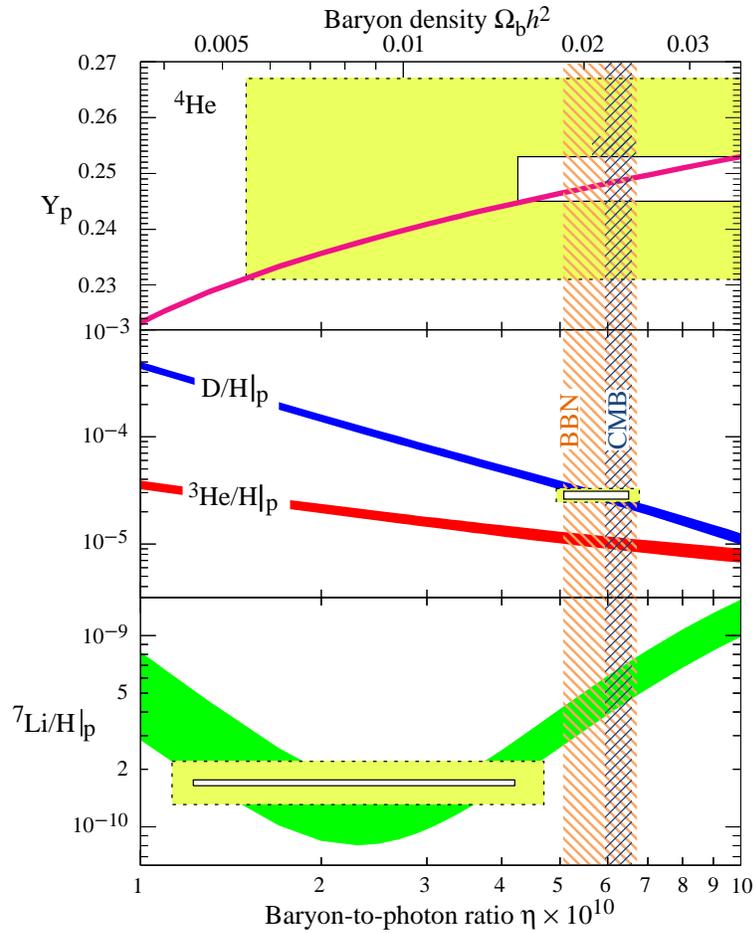}
\caption[BBN abundances]{The observed abundances of light elements
  compared to the standard BBN
  predictions~\protect\cite{Nakamura:2010zzi}. The smaller boxes
  indicate $2\s$ statistical errors only, the larger ones the combined
  $2\s$ statistical and systematic errors.}
\label{fig:bbn}
\end{center}
\end{figure}

\begin{figure}
\begin{center}
\includegraphics[width=0.6\textwidth]{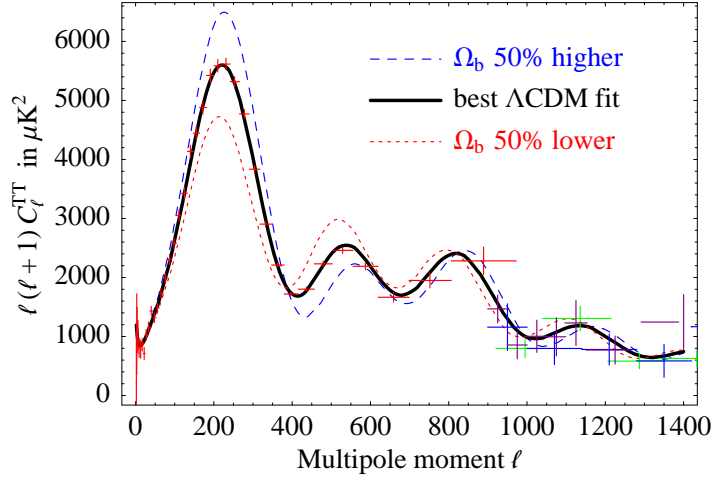}
\caption[CMB anisotropies]{The dependence of CMB temperature anisotropies on the 
baryon abundance $\O_b \equiv \r_B/\ \r_{\rm crit}$ compared with
data~\protect\cite{Strumia:2006qk}, where $\r_{\rm crit}$ is the critical
energy density of the universe. The relation to $\h$ is given by
$\h=2.74 \times 10^{-8} \; \O_b \; h^2$, where $h \equiv H_0/({\rm 100 \;
 km \; s^{-1} \; Mpc^{-1})=0.72 \pm 0.08}$ is the present Hubble
parameter.}
\label{fig:cmb}
\end{center}
\end{figure}

The photon density is proportional to $T^3$ and the temperature of the
universe is inferred via the cosmic microwave background radiation
(CMB), which shows an almost perfect blackbody spectrum. The baryon
density can be inferred in two ways: First, from the abundances of
the light elements D, $^3$He, $^4$He and $^7$Li, which are a direct
probe of the primordial abundances formed during the big bang
nucleosynthesis (BBN) phase at redshifts $z \sim 10^{10}$. Of these,
the deuterium abundance depends most sensitively on the
baryon-to-photon ratio $\h$, while the other elements exhibit a weaker
dependence. A measurement of the abundances
gives~\cite{Nakamura:2010zzi}
\begin{equation} 
\h = (5.8 \pm 0.7) \times 10^{-10}
\end{equation} 
at 95\% confidence level, see figure~\ref{fig:bbn}.
Second, there exist even more stringent constraints on $\h$, which
originate in the observation of the CMB anisotropies. The anisotropies
reflect the acoustic oscillations of the photon-baryon fluid at the
time of decoupling, which in turn depend on the baryon-to-photon ratio
at this time, at a redshift $z \sim 1000$. A high baryon density
enhances the odd peaks in the angular power spectrum relative to the
even peaks, as can be seen in figure~\ref{fig:cmb}. The measurement of
the CMB anisotropies from the 7-year Wilkinson Microwave Anisotropy
Probe (WMAP) data gives~\cite{Komatsu:2010fb}
\begin{equation}
\h=(6.15 \pm 0.25) \times 10^{-10}\, .
\end{equation}
The agreement of these two indicators at extremely different redshifts
is an important success of standard big-bang cosmology.
One might be tempted to assume this asymmetry as an initial condition
of the universe, but as we hinted at in the introduction, there are at
least two grave arguments against such an assumption. The first
argument is that such an initial condition would be a highly
fine-tuned one, since it would imply that for every 100 million
quark-antiquark pairs, there would have been one additional quark.
The second argument comes from the broad agreement that the universe
has undergone an inflationary expansion in its very early phase that
solves some otherwise unresolvable problems from cosmological
observations. The inflationary phase would wash out any asymmetry that
might have existed after the big bang. Thus, we need a theory that
creates the baryon asymmetry dynamically after inflation. Such
theories are commonly referred to as baryogenesis theories.

\section{Sakharov's Legacy}

Sakharov has formulated three necessary
conditions~\cite{Sakharov:1967dj} that baryogenesis theories have to
fulfil\footnote{There are exotic models that do not fulfill all of
  these conditions, but still produce a baryon asymmetry, such as
  Dirac leptogenesis for example, which does not require a violation
  of lepton number~\cite{Dick:1999je}.}:
\begin{itemize}
\item {\sl\noindent Baryon number $B$ non-conservation:}\\
It is immediately clear that processes which create a baryon asymmetry
do not conserve baryon number $B$.

\item {\sl\noindent $C$ and $C \! P$ symmetry violation:}\\
  Processes that are invariant under the discrete transformations of
  charge conjugation $C$ or the product of charge conjugation and
  parity reversal, $\CP$, will not create a baryon asymmetry. This is
  because in these cases, the $B$ violating processes that create
  baryons proceed at the same rate as their $C$- and $\CP$-conjugated
  processes which create an equal amount of antibaryons. Thus we need
  $C$- {\bf and} $\CP$-symmetry violation.

\item {\sl\noindent Deviation from thermal equilibrium:}\\
In chemical equilibrium, the entropy is maximal if the chemical
potentials associated with non-conserved quantum numbers, such as $B$
in our case, vanish. Since also the masses of quarks and antiquarks
are equal by $C\!PT$-invariance, the phase space densities 
\begin{equation}
  \label{eq:5}
f_i(p)=\frac{1}{1+\rme^{\sqrt{p^2+m_i^2}/T}}  
\end{equation}
are equal in thermal equilibrium and thus also 
the number densities.
\end{itemize}

\section{Why Does the Standard Model Fail?}
In principle, all three Sakharov conditions are met in the standard
model of particle physics (SM).

\subsection[$B$ non-conservation]{$\boldsymbol{B}$ non-conservation}
Baryon and lepton number, $B$ and $L$, are accidental symmetries in the SM
and are conserved at tree level. However, at the one-loop level, $B$
and $L$ are not conserved in anomalous diagrams where the $B$ and $L$
currents couple to two electroweak gauge bosons through a fermion
triangle~\cite{Adler:1969gk, Bell:1969ts, Dimopoulos:1978kv,
  Manton:1983nd}. In the vacuum structure of the electroweak theory,
there are non-perturbative transitions from one vacuum to a different
vacuum with differing $B$ and $L$ numbers~\cite{'tHooft:1976up}. These
instanton transitions change $B$ and $L$ by three each and conserve
$B-L$. At zero temperature, the instantons are suppressed by the
instanton action, i.e.~$\exp(-8 \p^2/g^2)$, where $g$ is the $SU(2)$
coupling constant, so that their rate is negligible. However, there
are static, but unstable field configurations which correspond to
saddle points of the field energy of the gauge-Higgs system, so-called
sphaleron configurations~\cite{Klinkhamer:1984di}. These saddle points
can be reached thermally and mediate a thermal transition from one
vacuum to another vacuum. At temperatures above the electroweak
scale, $T > T_\ew \sim 100 \, {\rm GeV}$, the sphaleron processes are
fast and in equilibrium~\cite{Kuzmin:1985mm}. Thus, they will also
wash out any previously existing $B$ or $L$ asymmetry that is not
linked to a $B-L$ asymmetry.

\subsection[$C$ and $\CP$ violation]{$\boldsymbol{C}$ and
  $\boldsymbol{C\!P}$ violation}
The weak interactions violate $C$ maximally since only left-handed
particles and right-handed antiparticles couple to the gauge bosons,
whereas their charge conjugated states, left-handed antiparticles and
right-handed particles do not couple via $SU(2)_L$ gauge
interactions. Only a charge conjugation together with a parity
transformation ($\CP$) converts left-handed particles into their
right-handed antiparticles. However, even $\CP$-symmetry is broken in
the SM in $B$- and $K$-meson systems through a $\CP$-violating phase
in the quark mixing matrix~\cite{Kobayashi:1973fv}. If appropriately
normalised~\cite{Jarlskog:1985ht}, this $\CP$ violation is still
several orders of magnitude too small as to produce an asymmetry of
the order $\h \sim 10^{-10}$. This is one reason why baryogenesis does
not work in the SM without any additional assumptions. Therefore, any
reasonable baryogenesis theory needs additional sources for $\CP$
violation.

\subsection{Deviation from thermal equilibrium}
In the early universe, there may be a departure from thermal
equilibrium during the electroweak phase
transition~\cite{Rubakov:1996vz, Trodden:1998ym}, which is in
principle suitable to create a baryon asymmetry. The deviation from
equilibrium occurs in particle interactions through the bubble walls
between the broken and the unbroken phase. In order to obtain an
irreversible asymmetry, the potential barrier between the two phases
has to be large enough, that is the phase transition has to be
strongly first order. This imposes constraints on the Higgs potential,
which in turn relate to an upper bound on the Higgs mass. However, the
experimental lower bound on the Higgs mass is too high as to allow for
this kind of phase transition and we arrive at the second reason why
baryogenesis does not work in the SM. The theory requires an
additional mechanism to obtain a sufficient departure from thermal
equilibrium.

\subsection{Ways out: baryogenesis theories}
We see that successful baryogenesis needs two new ingredients: First,
new sources of $\CP$ violation and second, a different mechanism for a
departure from thermal equilibrium\footnote{Or additional degrees of
  freedom that allow for a first order phase transition, like in the
  minimal supersymmetric standard model (MSSM) with a light stop
  particle \cite{Riotto:1999yt, Cline:2006ts}.}. Moreover, any
mechanism that creates a $B$ or $L$ asymmetry at higher temperatures
also has to violate $B-L$, otherwise this asymmetry will be washed out
by the sphaleron interactions. Several possibilities to meet these
requirements have been investigated, such as
leptogenesis~\cite{Fukugita:1986hr}, grand unified theory (GUT)
baryogenesis~\cite{Ignatiev:1978uf, Yoshimura:1978ex,
  Toussaint:1978br, Weinberg:1979bt, Yoshimura:1979gy, Barr:1979ye,
  Nanopoulos:1979gx, Yildiz:1979gx}, electroweak
baryogenesis~\cite{Riotto:1999yt, Cline:2006ts}, the Affleck-Dine
mechanism~\cite{Affleck:1984fy, Dine:1995kz} and other, more exotic
variants (see, e.g.~reference~\cite{Dolgov:1991fr}).  Leptogenesis is
the model that we focus on in this work.

\section{Leptogenesis}
\subsection[The unbearable lightness of neutrino masses and the
seesaw]{The unbearable lightness of neutrino masses and the seesaw as
  a way out}

The attractiveness of leptogenesis~\cite{Fukugita:1986hr} arises from
the feature that in addition to creating a baryon asymmetry, it
simultaneously solves a seemingly unrelated puzzle, which is the
smallness of neutrino masses. If one turns the argument around, by
employing a seesaw
mechanism~\cite{Minkowski:1977sc,Yanagida:1979as,GellMann:1980vs,
  Mohapatra:1980yp} in order to explain why neutrinos have a non-zero,
but extremely small mass, we naturally arrive at a mechanism for the
generation of the baryon asymmetry without additional effort.

There is experimental evidence that at least two neutrinos have a
non-zero mass which is several orders of magnitude smaller than the
masses of the charged fermions. Oscillation experiments have
established two mass-squared differences between the neutrino mass
eigenstates and global fits give~\cite{Nakamura:2010zzi}
\begin{equation}
\begin{split}
\D m_{21}^2&=m_2^2 -m_1^2 =(7.59 \pm 0.20) \times 10^{-5} \; {\rm eV^2} 
\equiv m^2_{\rm sol}, \\
| \D m_{32}^2 | &= | m_3^2 - m_2^2 |= (2.43 \pm 0.13) 
\times 10^{-3} \; 
{\rm eV^2} \equiv m^2_{\rm atm}\, .
\end{split}
\end{equation}
Neutrinos are the only SM fermions for which a Majorana mass term is
in principle allowed since they do not carry a $U(1)$ charge,
\begin{equation}
\label{eq:majmass}
\cl_{m_\n} = \frac{1}{2} \overline{\n^c}_\a [m]_{\a \b} \n_\b + \hc \,
,
\end{equation}
where $\nu_\alpha$ are the neutrino fields and the superscript $c$
denotes charge conjugation. The subscripts $\a$ and $\b$ denote the
neutrino flavour and $[m]_{\a \b}$ is the Majorana mass mixing
matrix. This operator, however, is not invariant under the $SU(2)_L$
gauge group, so the simplest operator which respects the symmetry of
this gauge group and reduces to equation~\eqref{eq:majmass} upon
spontaneous symmetry breaking is the dimension five operator
$(\bar{\ell^c}_\a \f) (\ell_\b \f)$, where
\begin{equation}
\ell_\alpha =
\begin{pmatrix}
\nu_\alpha \\ \ell_\alpha^-
\end{pmatrix}
\end{equation}
is the lepton doublet with flavour $\alpha$ and
\begin{equation}
\phi=
\begin{pmatrix}
\phi^+\\ \phi_0
\end{pmatrix}
\end{equation}
the Higgs doublet. This operator in turn is not renormalisable, so it
must be the effective remnant of new physics that is realised at a
higher energy scale, in order not to spoil the renormalisability of
the theory. Thus, if there is new physics above the electroweak scale,
it will induce this dimension five operator at lower energies unless
some symmetry prevents it. This observation is a strong argument in
favour of Majorana mass terms. Moreover, Dirac mass terms of the form
$m \, \overline{\n}_R \, \n_L$ require the addition of right-handed
neutrinos at low energy that are singlets under the SM gauge group and
whose existence could only be inferred via the exclusion of a low
energy ``Majorana'' type mass. If the dimension five operator is
induced via tree-level interactions with heavy particles at a mass
scale $M$, which can be much higher than the electroweak breaking
scale $M_\ew$, this will automatically lead to a light neutrino mass
scale of $M_\ew^2/M$ for Yukawa couplings of order one, which explains
the smallness of neutrino masses. Since these heavy particles can be
viewed as suppressing the mass of the neutrinos, the mechanism is
called seesaw mechanism.

Since the heavy particles have to couple to a Higgs doublet and a
lepton doublet, there are three prominent possibilities, which are called
seesaw type I--III:
\begin{itemize}
\item {\bf Type I}: The heavy particles are $SU(2)_L$-singlet
  fermions~\cite{Minkowski:1977sc, Yanagida:1979as, GellMann:1980vs,
    Mohapatra:1980yp};
\item {\bf Type II}: The heavy particles are $SU(2)$-triplet
  scalars~\cite{Mohapatra:1980yp,Magg:1980ut,Schechter:1980gr,Wetterich:1981bx,Lazarides:1980nt};
\item {\bf Type III}: The heavy particles are $SU(2)$-triplet
  fermions~\cite{Foot:1988aq,Ma:1998dn,Elgaroy:2004rc}.
\end{itemize}
The type I seesaw is the simplest and the framework in which
leptogenesis is usually implemented, so we concentrate on this type.

We add two or three singlet fermions $N_i$ to the SM, sometimes
referred to as ``right-handed neutrinos'', which are assumed to have
rather large Majorana masses $M_i$, close to the scale of some
possibly underlying grand unified theory (GUT), $E_{\rm GUT} \sim
10^{15},10^{16} \, {\rm GeV}$. The additional terms of the Lagrangian
can be written in the mass basis of the charged leptons and of the
singlet fermions as
\begin{equation}
\label{eq:langrange}
\cl={\rm i} \; \bar{N}_i \partial_\mu \g^\mu N_i - \l_{i \alpha} 
\bar{N_i} (\phi^a \e_{ab} \ell_\alpha^b)- 
\frac{1}{2} \sum_i M_i \bar{N}_i N_i^c + \hc \, ,
\end{equation}
where the Higgs doublet $\phi$ is normalised such that its vacuum
expetation value (vev) in
\begin{equation}
\langle \phi \rangle =
\begin{pmatrix}
0 \\ v
\end{pmatrix}
\end{equation}
is $v \simeq 174 \, {\rm GeV}$ and $\l_{i \a}$ is the Yukawa
coupling connecting the Higgs doublet, lepton doublet and heavy
neutrino singlet. The indices $a$ and $b$ denote doublet indices and
$\e_{ab}$ is the two-dimensional total antisymmetric tensor which
ensures antisymmetric $SU(2)$-contraction.

The effective mass matrix $[m]_{\a \b}$ of the light neutrinos as defined 
in equation~\eqref{eq:majmass} can be written as
\begin{equation}
\label{eq:lightmatrix}
[m]_{\a \b} = \sum_k \l_{\a k} M_k^{-1} \l_{\b k} v^2\, .
\end{equation}
It can be diagonalised as
\begin{equation}
D_m= U^T [m] U\, ,
\end{equation}
where $D_m={\rm diag}(m_1,m_2,m_3)$ and $U$ is the leptonic mixing
matrix, also called Pontecorvo-Maki-Nakagawa-Sakata (PMNS) matrix. It
is a $3 \times 3$ unitary matrix and therefore depends, in general, on
six phases and three mixing angles. Three of the phases can be removed
by redefining the phases of the charged lepton doublet fields. Doing
this, the matrix $U$ can be conveniently parameterised as
\begin{equation}
  U=\hat{U} \cdot {\rm diag} (1,e^{\rm i \a}, e^{\rm i \b})\, ,
\end{equation}
where $\a$ and $\b$ are called Majorana phases. If the neutrinos had
Dirac mass terms, the Majorana phases could be removed by redefining
the phases of the neutrino fields. The matrix $\hat{U}$ can be parameterised
in a way similar to the Cabbibo-Kobayashi-Maskawa (CKM) matrix,
\begin{equation}
\hat{U}= \left(
\begin{array}{ccc}
c_{13} c_{12} & c_{13} s_{12} & s_{13} e^{\rm - i \d} \\
- c_{23} s_{12} - s_{23} s_{13} c_{12} e^{\rm i \d} &
c_{23} c_{12} - s_{23} s_{13} s_{12} e^{\rm i \d} & s_{23} c_{13} \\
s_{23} c_{12} - c_{23} s_{13} c_{12} e^{\rm i \d} & 
- s_{23} c_{12} - c_{23} s_{13} s_{12} e^{\rm i \d} & c_{23} c_{13}
\end{array}
\right)\, ,
\end{equation}
where $c_{ij}=\cos \theta_{ij}$ and $s_{ij} = \sin \theta_{ij}$ and
$\theta_{ij}$ are the angles of rotations in flavour space, which
connect the flavour basis with the mass basis.

Thus there are twelve physical parameters at low energies in the leptonic sector of
the SM if we add the mass matrix of equation~\eqref{eq:lightmatrix}: the three charged
lepton masses $m_e$, $m_\m$, $m_\t$, the three neutrino masses $m_1$, $m_2$,
$m_3$, and the three angles and three phases of the PMNS matrix $U$. Of
these parameters, seven have been measured, $m_e$, $m_\m$, $m_\t$, $\D
m_{21}^2$, $| \D m_{32}^2 |$, $\q_{12}$ and $\q_{23}$. There exists an
upper bound on the angle $\q_{13}$, however, the mass of the lightest
neutrino and the three phases of $U$ are experimentally not accessible at the moment. 

In the case of three heavy neutrinos, there are nine additional
parameters in the high-energy theory, amounting to 21 parameters in
total. These high-energy parameters cannot be measured at
experimentally accessible scales, but are nevertheless important for
leptogenesis. Moreover, if one assumes only two right-handed
neutrinos~\cite{Frampton:2002qc}, there are fourteen parameters in
total and one of the light neutrinos is massless such that its
corresponding phase also vanishes. This amounts to ten physical
parameters at low energy and four additional parameters at high
energy. Such ``two-right-handed-neutrino'' (2RHN) models are
attractive since they have strong predictive power. We will
concentrate on the case of three heavy neutrinos in this work, since
it allows for the possibility that all light neutrinos have mass and
the number of generations is the same as at low energy.

\subsection{Sakharov and leptogenesis}

We examine whether and how the Sakharov conditions are fulfilled in
leptogenesis. The heavy neutrinos only couple to the lepton and Higgs
doublets, thus the processes $N_j \rightarrow \ell_i \phi$ and $N_j
\rightarrow \bar{\ell_i} \barphi$ are the only possible tree-level
decays. Since the Lagrangian in equation~\eqref{eq:langrange} violates
$L$, there is more than one possibility to assign a lepton number to
the heavy neutrinos. One usually assigns to them a lepton number of
zero, then the decays violate lepton number and also, very
importantly, $B-L$ by one unit. A lepton and $B-L$ asymmetry can thus
be generated at high temperatures, which will be converted to a baryon
asymmetry by the sphaleron processes at the electroweak scale.

Charge conjugation $C$ is violated maximally in the SM and the Yukawa
couplings $\l_{\a k}$ can have $\CP$-violating phases. However,
$\CP$ violation can only arise via an interference of the tree-level
decay and its higher order corrections, most importantly the one-loop
contribution.

Concerning the third Sakharov condition, the heavy neutrinos will
decouple from thermal equilibrium when the expansion rate of the
universe is faster than the interaction rate, i.e.~if the decay rate
$\G$ of the neutrinos is smaller than the Hubble rate $H$. They will
in this case nevertheless decay into lepton and Higgs doublet, but the
decay will be out of equilibrium.

\subsection{A simple model}

We see that the Sakharov conditions can in principle be fulfilled in
leptogenesis, it remains to be determined, which lepton and baryon
asymmetry will be generated for which parameters of the high energy
theory.

\subsubsection{Creating a lepton asymmetry}

In order to introduce the calculations one has to perform in
leptogenesis, we will make some simplifying assumptions and present a
model which serves as an example for determining the leptogenesis
dynamics. Firstly, we assume that the masses of the heavy neutrinos
are strongly hierarchical, i.e.~$M_1 \ll M_2 \ll M_3$. This
corresponds to the hierarchical masses of the standard model fermions
and simplifies the calculation. If the reheating temperature is larger
than the masses of the two heavier neutrinos, one cannot neglect the
effect of producing the $N_2$ and $N_3$. Thus, in order to simplify
calculations, we assume a reheating temperature which is lower than
these masses but larger than $M_1$ in order to allow for a thermal
production of the leptogenesis protagonists. Furthermore, we will,
again for the sake of simplicity, also neglect flavour effects, even
though they play an important role. We assume that the lepton states
$\ell_{N_1}$, into which the $N_1$ decay, will keep coherence of their
flavour mixing until there are no $N_1$ processes. This is only true
if leptogenesis happens at a temperature larger than about $10^{12} \,
{\rm GeV}$, below which processes mediated by the $\tau$-Yukawa
coupling become fast~\cite{DeSimone:2006dd, Blanchet:2006ch}. However,
flavour effects depend on the Yukawa couplings and do not necessarily
have to be important below these temperatures. Moreover, the thermal
effects which are examined in this thesis can be extended to include
flavour effects if appropriately modified. This work aims at giving an
insight into the effect of quasiparticles, where adding flavour might
be an important second step in the future. We refer the reader who
wishes to learn about flavour effects and the possible influence of
$N_2$ and $N_3$ to the review by Davidson et
al.~\cite{Davidson:2008bu}. He or she will find an extensive list of
references therein.

\subsubsection[The $\CP$-asymmetry]{The $C\!P$-asymmetry}
\label{sec:cp-asymmetry}

The $\CP$-asymmetry in a lepton flavour $i$ that arises in the decay
of a heavy neutrino of generation $j$ is defined as
\begin{equation}
\label{eq:51}
\epsilon_i \equiv \frac{\Gamma(N_j \rightarrow \phi \ell_i) - 
\Gamma(N_j \rightarrow \bar{\phi} \bar{\ell_i})}{\Gamma(N_j \rightarrow \phi \ell_i) + 
\Gamma(N_j \rightarrow \bar{\phi} \bar{\ell_i})}\, .
\end{equation}
The tree-level amplitude cannot be different for the $\CP$-conjugated
process, but if one considers the one-loop level, the interference
terms in the squared sum of the matrix elements for tree level,
$\mathcal{M}_0$, and one-loop level, $\mathcal{M}_1$, can be
$\CP$-asymmetric, i.e.~$\mathcal{M}_0 \; \mathcal{M}_1^* \neq
\widetilde{\mathcal{M}}_0 \; \widetilde{\mathcal{M}}_1^*$, where
$\widetilde{\mathcal{M}_i}$ denotes the $\CP$-conjugated matrix
elements.

\begin{figure}
\begin{center}
\includegraphics[width=\textwidth]{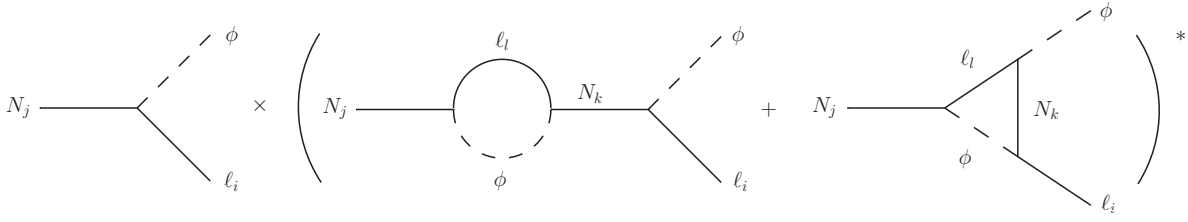}
\caption[$\CP$-asymmetry in neutrino decays]{The $\CP$-asymmetry
  $\mathcal{M}_0 \mathcal{M}_1^*$. The graph in the middle 
is the self-energy contribution, the graph on the right the
vertex contribution. In the self-energy diagram, the Higgs boson and lepton
can be particles or antiparticles.}
\label{fig:cpasfigure}
\end{center}
\end{figure}

There are two contributions to the one-loop amplitude $\mathcal{M}_1$
as displayed in figure~\ref{fig:cpasfigure}. One is the self-energy
contribution where a virtual $N_k$ is exchanged in the s-channel. The
second one is the vertex-correction graph. There is only a
$\CP$-asymmetry if the virtual $N_k$ is a different generation than
$N_j$, i.e.~$N_2$ or $N_3$ for the decay of $N_1$. The $\CP$-asymmetry
is proportional to the imaginary part of the one-loop diagram which in
turn corresponds to assuming an on-shell condition for the loop
propagators. In the self-energy diagram, one can only put the Higgs
boson and lepton on the mass shell and $N_k$ is necessarily off-shell
since $M_k \neq M_j$. In the vertex correction, it is kinematically
not possible to put the neutrino in the loop on its mass shell since
it is heavier than the Higgs boson and the lepton. Thus, again, the
Higgs boson and the lepton are on-shell. We will see in
chapter~\ref{cpas} that at finite temperature, also the neutrino can
be on its mass shell.

The calculation of the $\CP$-asymmetry is worked out in
Appendix~\ref{sec:boldsymb-asymm-vacu}. Including all contributions,
it is written as
\begin{align}
\label{eq:epsilon}
\epsilon_i = & \frac{1}{8 \pi} \frac{1}{(\l^\dagger \l)_{jj}} \; 
{\rm Im} \left[ \l_{ij}^*
(\l^\dagger \l)_{jk} \l_{ik} \right]  g(x_{kj}) + \frac{1}{8 \pi} \frac{1}{(\l^\dagger \l)_{jj}} \; 
{\rm Im} \left[ \l_{ij}^*
(\l^\dagger \l)_{kj} \l_{ik} \right]
\frac{1}{1-x_{kj}} \, ,
\end{align}
where
\begin{equation}
x_{kj} \equiv M_k^2/M_j^2 \,
\end{equation}
and
\begin{equation}
g(x)= \sqrt{x} \left[ \frac{1}{1-x} + 1 -
(1+x) \ln \left( \frac{1+x}{x}
\right) \right] \stackrel{x_{kj} \gg 1}{\longrightarrow} - \frac{3}{2 \, \sqrt{x}}
\end{equation}
and we have summed over the lepton generations in the loop.  The
second term in equation~\eqref{eq:epsilon} corresponds to the
self-energy diagram with particles in the loop as opposed to
antiparticles. It is lepton flavour changing but lepton number
conserving. Moreover, it does not exhibit an imaginary part if we sum
over the final state lepton flavors $i$, so that
\begin{equation}
\epsilon \equiv \sum_i \epsilon_i
= \frac{1}{8 \pi} \frac{1}{(\l^\dagger \l)_{jj}} 
\; {\rm Im} \left\{ \left[
(\l^\dagger \l)_{jk} \right]^2 \right\}
g(x_{kj})\, .
\end{equation}
The dominant contribution to the $\CP$-asymmetry in $N_1$ decays,
which we consider, comes from $N_2$ in the loop since $M_3 \gg
M_2$. In our further analysis, we will neglect the effect of $M_3$ and
set $j=1$ and $k=2$.  In the limit of hierarchical right-handed
neutrinos, it is possible to derive an upper bound on the
$\CP$-asymmetry in $N_1$ decays, the so called
Davidson--Ibarra-bound~\cite{Davidson:2002qv},
\begin{equation}
\epsilon \lesssim \frac{3}{8 \pi} \frac{M_1}{v^2}
(m_3-m_1) \approx \frac{3}{8 \pi} \frac{M_1}{v^2} \sqrt{\Delta
m^2_{\rm atm}}\, ,
\end{equation}
since in this limit, the light neutrino spectrum is expected to be
hierarchical as well.

Anticipating the discussion of the next two sections, it is possible
to parameterise the resulting baryon asymmetry by three numbers as
\begin{equation}
\eta = d \, \epsilon \, \kappa \, ,
\end{equation}
where $d$ is a factor which accounts for the dilution of the asymmetry
through the expansion of the universe and the distribution into
different particle species. It is of the order $\sim 0.01$. The
efficiency factor $\kappa$ parameterises the dynamics of leptogenesis
and how much of the asymmetry is washed out by other processes and is
for a vanishing initial lepton asymmetry given by
\begin{align}
  \label{eq:211}
  \kappa = \frac{n_\ell -n_{\barell}}{\epsilon \, n_\ell^\rmeq} \, .
\end{align}
It is calculated by solving the Boltzmann equations that govern the
evolution of the particle species and usually of the order $\sim
0.1$. Since we have to explain an asymmetry of $\eta \sim 5 \times
10^{-10}$, we can, for typical efficiency factors, require a
$\CP$-asymmetry of $\epsilon \sim 10^{-6}$. Together with the
Davidson--Ibarra-bound, this translates to a lower bound on the mass
of the lightest heavy neutrino of about
\begin{equation}
M_1 \gtrsim \frac{8 \pi}{3} \frac{v^2}{\sqrt{\Delta
m^2_{\rm atm}}} \, \epsilon \sim 10^9 \, {\rm GeV} \, .
\end{equation}

\subsubsection{The gravitino problem}
\label{sec:gravitino-problem}

If one assumes that the heavy neutrinos are produced thermally, a
reheating temperature of $T_{\rm RH} > M_1$ is needed in order to
produce a sufficient amount of neutrinos and this can lead to the
so-called gravitino problem, which we discuss by closely following
reference~\cite{Davidson:2008bu}. In certain models of supersymmetry
(SUSY), a high reheating temperature leads to an overproduction of
gravitinos. The gravitinos are long-lived and will decay into lighter
particles if they are not the lightest supersymmetric partner (LSP). If too many
gravitinos decay during or after BBN, the decays will destroy the
agreement between predicted and observed light element
abundances~\cite{Khlopov:1984pf, Ellis:1984eq, Kohri:2005wn,
  Rychkov:2007uq}. There are several possibilities to avoid such a
scenario, both on the SUSY side and on the leptogenesis side.  The
possibilities on the SUSY side include the following:
\begin{enumerate}
\item The gravitinos decay before BBN. Heavy enough gravitinos arise
  in anomaly-mediated scenarios~\cite{Ibe:2004tg}, but not in standard
  gravity mediated scenarios.
\item Late-time entropy production can dilute the gravitino abundance,
  but it also dilutes the $B-L$ asymmetry~\cite{Buchmuller:2006tt}
\item The gravitino is the LSP (and the dark matter particle), in
  which case the bound on $T_{\rm RH}$ is less
  restrictive~\cite{Feng:2004mt,Kanzaki:2006hm}. The gravitino is indeed the LSP in
  gauge-mediated scenarios.
\end{enumerate}
We assume that either SUSY is not realised in nature, which does not
pose a problem to our leptogenesis model, or that one of the above
scenarios solves the gravitino problem. This way, it is possible to
consider temperature regimes $T > M_1$.

\subsubsection{Converting a lepton to a baryon asymmetry}

The $L$ asymmetry, which is generated at high temperature, is partially
converted into a $B$ asymmetry by the electroweak sphaleron processes,
which are in equilibrium above $\sim T_{\rm EW}$. Taking into account
the chemical potentials of all particles, it can be worked out
(e.g. reference~\cite{Davidson:2008bu}) that
\begin{equation}
Y_B = \frac{c}{c-1} Y_L = c \, Y_{B-L} \, ,
\end{equation}
where
\begin{equation}
Y_X = \frac{n_X}{s}
\end{equation}
is the number density $n_X$ of species $X$ over the entropy density
$s$ and $c=Y_B/Y_{B-L}$ accounts for the sphaleron conversion of a
$B-L$ asymmetry into a $B$ asymmetry. In the SM, $c=12/37$ for a
smooth electroweak phase transition.

\subsubsection{Solving Boltzmann equations}
\label{sec:solv-boltzm-equat}

The intuitive requirement for the neutrino decays to be out of
equilibrium is that they decay slower than the universe expands,
that is the decay rate
\begin{equation}
\Gamma=\frac{(\lambda^\dagger \lambda)_{11} M_1}{8 \pi}
= \frac{\widetilde{m}_1 M_1^2}{8 \pi v^2}
\end{equation}
is smaller than the Hubble rate $H$, 
where
\begin{equation}
\label{eq:196}
\widetilde{m}_1 = \frac{(\lambda^\dagger \lambda)_{11} v^2}{M_1}
\end{equation}
is the conveniently defined effective neutrino mass, which is of the
order of the light neutrino mass scale. In the language of this mass,
the out-of-equilibrium condition reads
\begin{equation}
\label{eq:197}
\widetilde{m}_1 <  \left. 8 \pi \frac{v^2}{M_1^2} H \right |_{T=M_1} = m^* \simeq 1.1 \times
10^{-3} \,{\rm eV}\, ,
\end{equation}
where $m^*$ is called equilibrium neutrino mass. The parameter region
where this condition is satisfied is called strong washout
regime. However, leptogenesis is also possible when $\widetilde{m}_1>
m^*$. This can be seen if one solves the set of Boltzmann
equations~\cite{Buchmuller:2002rq, Barbieri:1999ma}
\begin{eqnarray}
\label{eq:49}
\frac{{\rm d} N_{N_1}}{{\rm d} z}&=& -(D+S) (N_{N_1}-N_{N_1}^{\rm eq}),\\
\label{eq:50}
\frac{{\rm d}N_{B-L}}{{\rm d} z}&=& - \epsilon D (N_{N_1} - N_{N_1}^{\rm
eq}) -W N_{B-L}\, ,
\end{eqnarray}
where $N_X$ is the number density of the species $X$ per comoving
volume which contained one photon when $T \gg M_1$. The temperature is
contained in $z=M_1/T$ and $(D,S,W)= (\Gamma_D,\Gamma_S,\Gamma_W)/(H
z)$, where $\Gamma_D$ is the decay rate, $\Gamma_S$ the sum of all
$N_1$ scattering rates, and $\Gamma_W$ the sum of all processes that
wash out the generated $L$ asymmetry, such as inverse decays for
example. We derive a simple version of equations~\eqref{eq:49}
and~\eqref{eq:50} in appendix~\ref{sec:boltzm-equat-at}.

Solving these equations typically gives efficiency factors of $\kappa
\sim 0.1$. The baryon asymmetry can be approximated as
\begin{equation}
Y_B \simeq \frac{135 \, \zeta(3)}{4 \pi^4 g_*} \, c \, \epsilon
\, \kappa\, ,
\end{equation}
where the first factor is the equilibrium $N_1$ number density divided
by the entropy density at $T \gg M_1$. The number of relativistic
degrees of freedom $g_*$ is given by
\begin{align}
  \label{eq:212}
  g_* = \sum_{i={\rm bosons}} g_i \left( \frac{T_i}{T} \right)^4 +
  \frac{7}{8} \sum_{i={\rm fermions}} g_i \left( \frac{T_i}{T}
  \right)^4 \, ,
\end{align}
where $i$ denotes species with mass $m_i \ll T$ and the factor $7/8$
arises from the difference in Fermi and Bose
statistics~\cite{Kolb:EarlyU}. At the temperature of leptogenesis, all
SM particles have negligible masses, so $g_*=106.75$. The expressions
$\eta$ and $Y_B$ are related via photon and entropy density today as
\begin{equation}
Y_B = \frac{n_{\gamma0}}{s_0}\, \eta \simeq \frac{\eta}{7.1}\, .
\end{equation}








\clearpage
\chapter{Thermal Field Theory}
\label{cha:tft}
The topic of this work is the role of the equilibrium quantum effects
that are implied by the presence of a hot, dense medium for
leptogenesis models. The influence of the medium is studied by means
of an effective, statistical quantum field theory, which takes into
account the temperature of the surrounding medium, hence called
thermal field theory (TFT). In this chapter, we give an introduction
into TFT and explain the methods and formalisms we use later on. The
reader who wishes to learn more about TFT is referred to the books by
Kapusta~\cite{Kapusta:1989}, LeBellac~\cite{LeBellac:1996} and
Das~\cite{Das:1997}. Our presentation follows the more intuitive
approach of Thoma's lecture notes~\cite{Thoma:2000dc}.
\section{Green's Functions at Finite Temperature}
\label{sec:green-functions-at}
Thermal field theory comprises all three basic branches of modern
physics, namely quantum mechanics, relativity and statistical physics.
We want to derive Feynman rules and diagrams at finite temperature. To
this end, we consider the two-point Green's function or propagator. We
would like to find an expression for this quantity at finite
temperature. For simplicity, we consider the case of a scalar field
$\phi$. The propagator at $T=0$ is worked out in
Appendix~\ref{sec:green0} and is the vacuum expectation value of the
time-ordered product of two fields at spacetime points $x$ and $y$,
\begin{align}
{\rm i} \, \Delta_F = \langle 0 | T \phi_x \phi_y| 0 \rangle \, .
\end{align}
At finite temperature, the vacuum expectation value of an operator $A$
has to be replaced by the quantum statistic expectation value of the
corresponding statistical ensemble,
\begin{align}
\langle 0 | A |0 \rangle \rightarrow \langle A \rangle_\rho \equiv 
{\tr}(\rho A)
 \, ,
\end{align}
where $\rho$ is the density operator of the statistical ensemble. For
the canonical ensemble, which we will consider, it is given by
\begin{align}
\label{eq:210}
\rho = \frac{1}{Z} {\rm e}^{- \beta H} \, ,
\end{align}
where
\begin{align}
\beta \equiv \frac{1}{T}
\end{align}
and $H$ is the Hamiltonian operator of the system with the discrete eigenvalues
and eigenstates
\begin{align}
H |n \rangle = E_n | n \rangle \, .
\end{align}
The partition function is
\begin{align}
Z={\tr} ({\rm e}^{- \beta H}) \, ,
\end{align}
so we can write
\begin{align}
\langle A \rangle_\rho = \frac{1}{Z} {\tr}(A {\rm e}^{-\beta H})
= \frac {1}{Z} \sum_n \langle n | A | n \rangle {\rm e}^{-\beta E_n} \, ,
\end{align}
where the sum is over all thermally excited states $|n \rangle$ of
the system, which are eigenstates of the Hamiltonian $H$. Since the
heat bath is a distinguished reference frame for our situation, our
calculations are not Lorentz-invariant. We will always work in the
rest frame of the heat bath, which is the preferred rest frame at
finite temperature.

Calculating the statistical expectation value for the scalar
propagator, we have
\begin{align}
{\rm i} \Delta_F^{T>0}(x-y) = \frac{1}{Z} \sum_n \langle n | T \{ \phi(x) 
\phi(y) \} 
| n \rangle {\rm e}^{- \beta E_n} \, .
\end{align}
Using the Fourier representation of $\phi$ in
equation~\eqref{eq:fourier} in the propagator for the case $x_0> y_0$
gives
\begin{align}
\begin{split}    
{\rm i} \Delta_F^{T>0}(x-y) = \; & \frac{1}{Z} \int \frac{\rmd^3
 k}{(2 \pi)^{3/2}} \frac{\rmd^3
 k'}{(2 \pi)^{3/2}} \frac{1}{(2 \omega_k)^{1/2}} \frac{1}{(2 \omega_k')^{1/2}}\\ 
&\times \sum_n 
\rme^{-\beta E_n} \langle n |  [
a({\bf k}) {\rm e}^{-{\rm i}\, K \cdot x} + a^\dagger({\bf k}) {\rm
e}^{\, {\rm i}\, K \cdot x} ]  [
a({\bf k'}) {\rm e}^{-{\rm i}\, K' \cdot x} + a^\dagger({\bf k'}) {\rm
e}^{\, {\rm i}\, K' \cdot x} ] | n \rangle
\end{split}
\end{align}
The states 
\begin{align}
  \label{eq:2}
| n \rangle = | n_1(\bof{k}_1), n_2(\bof{k}_2), \dots, n_i(\bof{k}_i) \rangle 
\end{align}
are orthonormalised states with $n_i$ bosons of momentum
$\bof{k}_i$. Acting with the creation and destruction operators on $|
n \rangle$ according to equation~\eqref{eq:ladder} results in
\begin{align}
  \label{eq:1}
\begin{split}
  \rmi \Delta_F^{T>0}(x-y) = \; & \frac{1}{Z} \int \frac{\rmd^3 k}{(2
    \pi)^{3/2}} \frac{\rmd^3 k'}{(2 \pi)^{3/2}} \frac{1}{(2
    \omega_k)^{1/2}} \frac{1}{(2 \omega_k')^{1/2}}\\
  &\times \sum_n \rme^{-\beta E_n} \{ [n(\bof{k})+1]
  \delta^3(\bfk-\bfk') \rme^{- \rmi K \cdot x + \rmi K' \cdot y} +
  n(\bfk) \d^3(\bfk-\bfk') \rme^{\rmi K \cdot x-\rmi K' \cdot y} \}\\
  =\; & \left. \frac{1}{Z} \int \frac{\rmd^3 k}{(2 \pi)^3} \frac{1}{2 \o_k}
  \sum_n \rme^{-\beta E_n} \{ [n(\bfk)+1] \rme^{- \rmi K \cdot(x-y)}+
  n(\bfk)\rme^{\rmi K\cdot(x-y))}\} \right |_{k_0=\o_k} \, .
\end{split}
\end{align}

We use
\begin{align}
  \label{eq:3}
  \frac{1}{Z} \sum_n n(\bfk) \rme^{-\b E_n}= \frac{1}{\rme^{\b \o_k}-1} \equiv 
  f_B(\o_k) \, ,
\end{align}
where $E_n = \sum_k \o_k n(\bfk)$ and $f_B(\o_k)$ is the Bose-Einstein
distribution. Using this relation, we find
\begin{align}
  \label{eq:propagator}
  \rmi \D^{T>0}_F(x-y)=\int \frac{\rmd^3 k}{(2 \pi)^3} \frac{1}{2 \o_k} \{
[1+f_B(\o_k)] \rme^{-\rmi K(x-y)} +f_B(\o_k) \rme^{\rmi K (x-y)} \} \, .
\end{align}
We see that for $T=0$, which implies that $f_B=0$, we arrive at the
vacuum result of equation~\eqref{eq:prop0}.  We can interpret this
result as follows: The zero-temperature part describes the usual
propagation of a particle from $y$ to $x$. However, on top of
spontaneous creation at $y$, there is also induced creation at $x$
(proportional to $n_B$) and absorption at $x$ ($\propto n_B$) due to
the presence of the thermal particles in the bath. The propagator in
equation~\eqref{eq:propagator} is the propagator of a free field
without interactions, and forms the starting point for perturbation
theory when we add interactions.

\section{Imaginary Time Formalism}
The propagator~\eqref{eq:propagator} is a useful quantity, but not
very helpful in constructing Feynman rules. We need a representation
which consists of a 4-dimensional $K$-integration, so that we can
formulate Feynman rules in momentum space. To achieve this, we
continue the propagator analytically to imaginary times $t$ with $0
\leq \tau \equiv \rmi t < \b$ and sum over the discrete energies
\begin{align}
  \label{eq:4}
  k_0=2 \pi \rmi n T \, ,
\end{align}
the so-called Matsubara frequencies instead of integrating, so that
\begin{align}
  \label{eq:6}
  \int \frac{\rmd k}{2 \pi} \to \rmi T \sum_{n=-\infty}^\infty \, .
\end{align}
The propagator can be written as
\begin{align}
  \label{eq:7}
  \rmi \Delta_F^{T>0}(x) = \rmi T \sum_n \int \frac{\rmd^3 k}{(2 \pi)^3}
  \frac{\rmi}{K^2-m^2} \rme^{- \rmi K \cdot x}\, .
\end{align}

One can also motivate the introduction of imaginary time in the
following way: for $\t=\rmi t \rightarrow \b$, the Boltzmann factor
$\exp(-\b H)$ has the form of the time evolution operator $\exp(- \rmi
H t)$. As a consequence, thermal propagators become periodic in $\b$,
\begin{align}
  \label{eq:8}
  \Delta^{T>0}_F(x-y)=\D^{T>0}_F
(\bof{x},\bof{y},\t,0) = \D^{T>0}_F(\bof{x},\bof{y},\t,\b) \, ,
\end{align}
where we have taken $\t_x=\t$ and $\t_y=0$.
In general
\begin{align}
  \label{eq:9}
  \D_F^{T>0}(\t)=\D_F^{T>0}(\t+n\b)
\end{align}
holds for any integer $n$.

The two consequences of this relation are:
\begin{enumerate}
\item The time $\t$ is restricted to the interval $[0,\b]$, the
  so-called  Kubo-Martin-Schwinger- or  KMS-condition.
\item The Fourier integral over $k_0$ in vacuum quantum field theory
  (QFT) becomes a Fourier series over the Matsubara frequencies $k_0 =
  2 \p \rmi n T$. (For fermions we have ${k_0=(2n+1) \rmi \p T}$, as
  described in section~\ref{sec:dirac-field}.
\end{enumerate}

In real life, it is usually necessary to integrate over more than one
propagator. In these cases, it is difficult to perform the summation
over the zeroth component of the loop momentum $k_0$. A convenient way
out of this problem is to use the so-called Saclay representation,
which is a mixed representation, performing the Fourier transformation
in time only. In the following, we use $\D \equiv \D_F^{T>0}$ and
write the propagator in the Saclay representation, leads to
\begin{align}
  \label{eq:10}
  \D(\t,\o_k)=-T\sum_{k_0}\rme^{- k_0 \t} \D(K)\, ,
\end{align}
where the Fourier coefficients are given by
\begin{align}
  \label{eq:11}
  \D(K)= - \int_0^\b
\rmd \t \rme^{k_0 \t} \D(\t,\o_k)\, .
\end{align}
We can perform the sum~\eqref{eq:10},
\begin{align}
  \label{eq:12}
  \D(\t,\o_k)=\frac{1}{2 \o_k} \left \{ [1+f_B(\o_k)] \rme^{-\o_k \t}
    + f_B (\o_k) \rme^{\o_k \t} \right \} \, ,
\end{align}
which agrees with equation~\eqref{eq:propagator}. 

It is often convenient to
write the propagator as the sum
\begin{align}
  \label{eq:16}
  \D(\t,\o_k)= \sum_{s=\pm 1} \frac{s}{2 \o_k} [1+f_B(s \o_k)]
  \rme^{-s \o_k \t} =\sum_{s=\pm 1} \D_s(\t,\o_k) ,
\end{align}
where we allow for negative energies $s \o_k$ in $f_B(s \o_k)$. In frequency
space, the two parts also decompose,
\begin{align}
  \label{eq:18}
  \D(K)=\sum_{s=\pm 1} \D_s(K) = \sum_{s=\pm 1} \frac{s}{2 \o_k}
  \frac{1}{k_0-s \o_k} \, ,
\end{align}
where the relations
\begin{align}
  \label{eq:19}
  \D_s(\t,\o_k)&=-T\sum_{k_0}\rme^{- k_0 \t} \D_s(K) \, , \nonumber \\
  \D_s(K)&= - \int_0^\b
  \rmd \t \rme^{k_0 \t} \D_s(\t,\o_k)
\end{align}
hold for the propagator parts as well.

\section{The Scalar Field}
\label{sec:scalar-field}
We write down the Feynman rules for a simple interaction theory, the
neutral scalar field with a $\f^4$ interaction, given by the
Lagrangian
\begin{align}
  \label{eq:13}
  \mathcal{L}=\frac{1}{2} (\partial_\m \f)(\partial_\m \f)
  -\frac{1}{2} m^2 \f^2 - g^2 \phi^4 \, ,
\end{align}
where $\f$ is a real field and $m$ the mass of the corresponding
particles. We can derive the Feynman rules for the interaction of
these fields. One can follow the operator formalism with which we
started in the derivation of the finite temperature Green's functions in
section~\ref{sec:green-functions-at} and which relies on the
interaction picture and a thermal equivalent of Wick's theorem at zero
temperature. As usual in QFT, there is a second approach, which is to
derive the Feynman rules from the path integral formalism. Both
derivations can be found in the
literature~\cite{Kapusta:1989,LeBellac:1996,Das:1997}.

The Feynman rules for this theory read as follows:
\begin{enumerate}
\item The propagator is given by
  \begin{align}
    \label{eq:14}
    \rmi \, \D_F^{T>0}(K) = \frac{\rmi}{K^2-m^2}
  \end{align}
with $k_0=2 \p \rmi \, n T$.
\item In loop integrals we make the replacement
  \begin{align}
    \label{eq:15}
    \int \frac{\rmd^4 K}{(2 \p)^4} \to \rmi \, T \sum_{k_0} \int
    \frac{\rmd^3 K}{(2 \p)^3} \, .
  \end{align}
\item The vertex reads as in vacuum $- \rmi \, 4! g^2$.
\item Symmetry factors, e.g.~$1/2$ for the tadpole, are the same as
  in vacuum.
\end{enumerate}

\begin{figure}
  \centering
  \includegraphics[width=0.20\textwidth]{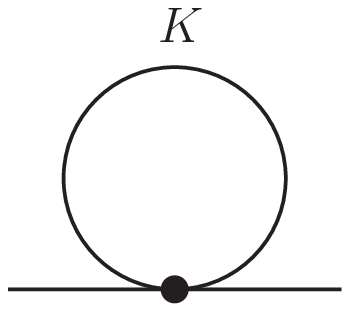}
  \caption{The $\f^4$-tadpole.}
  \label{fig:tadpole}
\end{figure}

As a simple example of a loop diagram, we compute the tadpole of the
$\f^4$-theory shown in Fig.~\ref{fig:tadpole}.
According to the above Feynman rules, we get
\begin{align}
  \label{eq:21}
  \P = \rmi \, \frac{1}{2} \left (-\rmi \, 4! g^2\right ) \rmi \, T \sum_{k_0}
  \int \frac{\rmd^3 k}{(2 \p)^3} \rmi \, \D(K) \, .
\end{align}
The Saclay representation~\eqref{eq:11} proves useful in this
calculation,
\begin{align}
  \label{eq:22}
  \P= 12 g^2 \int \frac{\rmd^3 k}{(2 \p)^3} \int_0^\b \rmd \t
  \D(\t,\o_k) T \sum_{k_0} \rme^{k_0 \t} \, ,
\end{align}
since the Matsubara frequency $k_0$, over which we have to sum,
appears in the propagator only in the exponent, which makes the
summation simple at the expense of introducing another integral over
$\t$. The sum reduces to a $\d$-function,
\begin{align}
  \label{eq:23}
  T \sum_{n=-\infty}^\infty \rme^{k_0 (\t-\t')} = T
  \sum_{n=-\infty}^\infty  \rme^{2 \p \rmi \, n T (\t-\t')} = T
  \d(T(\t-\t'))= \d(\t-\t') \, .
\end{align}
Thus we get
\begin{align}
  \label{eq:24}
  \P&=12 g^2 \int \frac{\rmd^3 k}{(2 \p)^3} \D(0,\o_k)\\ \nonumber
&= 6 g^2 \int \frac{\rmd^3 k}{(2 \p)^3} \frac{1}{\o_k} [1+2
f_B(\o_k)] \, .
\end{align}

The first term in the sum $[1+ 2 f_B(\o_k)]$ leads to an ultraviolet
(UV) divergence, but is identical to the corresponding
zero-temperature term. The second term results in a finite integral
since $f_B$ falls off exponentially fast for large momenta. We see
that the tadpole can be decomposed into a vacuum and a finite
temperature part,
\begin{align}
\P=\P^{T=0}+\P^{T>0} \, ,
\end{align}
and it is sufficient to renormalise the divergence of the vacuum term.
The finite temperature part can be integrated analytically if $m=0$
and in this case yields the simple result
\begin{align}
\label{eq:Pi}
\P^{T>0}=g^2 T^2 \, .
\end{align}

In fact, one can show that generally, renormalisation at zero
temperature is sufficient to remove all UV divergences of the theory
at finite temperature. We do not prove this statement here, but the
interested reader can find a proof in chapter~3.5 of Bellac's book for
example~\cite{LeBellac:1996}. However, the property can be understood
in a more intuitive way: temperature does not modify the theory at
distances much smaller than $1/T$ and thus, the ultraviolet
divergences are the same as at zero temperature. Infrared (IR)
divergences, on the other hand, are a different story and we will deal
with them in chapter~\ref{sec:hard-thermal-loop}.

In more complex calculations, we often have to perform the Matsubara
sum over two or more propagators, so it is useful to write down these
frequency sums for further reference: One such sum is given by
\begin{align}
  \label{eq:17}
  T \sum_{k_0} \D(K) \D(P-K) = T \sum_{k_0} \sum_{s_1,s_2} \D_{s_1}(K) 
\D_{s_2}(P-K) \, ,
\end{align}
where we can in principle allow the two bosons corresponding to the
two propagators ${\D(P-K)}$ to have different masses $m_1$ and
$m_2$. We can perform this sum by using the Saclay
representation~\eqref{eq:11}, transforming the Matsubara sum into a
$\d$-function and executing the imaginary time integrals. We arrive
at\footnote{The frequency-space propagators in Bellac's
  book~\cite{LeBellac:1996} are defined with a minus sign relative to
  our convention ${\Delta_s^{\rm this work}(K)=- \Delta_s^{\rm
      Bellac}}(K)$. He defines the Fourier transformation for the
  mixed representation with a minus sign, so the mixed-representation
  propagators agree with this work. Moverover, the frequency sums
  involve exactly two propagators, so they agree with this work as
  well.}
\begin{align}
  \label{eq:25}
 T \sum_{k_0} \D_{s_1}(K) \D_{s_2}(P-K) = - \frac{s_1 s_2}{4 \o_k \o_q} 
\frac{1+f_B(s_1 \o_1)+f_B(s_2 \o_2)}{p_0 - s_1 \o_1 - s_2 \o_2} \, , 
\end{align}
where $\o_1=\sqrt{\bof{k}^2+m_1^2}$ and $\o_2=\sqrt{(\bof{p}-\bof{k})^2+m_2^2}$ 
are the
energies of the fields.

Another frequency sum,
\begin{align}
  \label{eq:26}
  T \sum_{k_0} k_0 \D_{s_1}(K) \D_{s_2} (P-K) \, ,
\end{align}
can be evaluated by integrating the Fourier representation of $\D(\t)$
by parts~\cite{LeBellac:1996}. We arrive at
\begin{align}
  \label{eq:27}
  T \sum_{k_0} k_0 \D_{s_1}(K) \D_{s_2}(P-K) =  - \frac{s_2}{4 \o_2} 
  \frac{1+f_B(s_1 \o_1)+f_B(s_2 \o_2)}{p_0 - s_1 \o_1 - s_2 \o_2} \, , 
\end{align}
so we see that the calculation amounts to replacing $k_0$ by the
respective propagator pole $s \o_k$.

\section{The Dirac Field}
\label{sec:dirac-field}

We are considering spin 1/2 particles in a spinor representation
$\psi(x)$ with the free Lagrangian density
\begin{align}
  \label{eq:28}
  \mathcal{L}_{\rm Dirac} = \overline{\j} (\rmi \, \gamma^\m \partial_\m - m) \j \, ,
\end{align}
where we work in the  chiral or Weyl representation of
gamma matrices. 

We define a fermion Matsubara propagator through 
\begin{align}
S_{\g \d} (x-y) = \langle T \{ \j_\g(x) \overline{\j}_\d (y) \} \rangle_\r \, ,
\end{align}
where $\g$ and $\d$ denote spinor indices and $\r$ indicates that we
are taking the statistical expectation value for a density operator
$\r$, for which we take the canonical ensemble~\eqref{eq:210}.

Similar to the scalar field, the fermion propagator obeys a KMS
condition when going to imaginary time $t$ with $\t= \rmi \, t$,
however, the fermion propagator is antiperiodic in $\b$,
\begin{align}
S(x-y) = S(\bof{x},\bof{y},\t,0)= -S(\bof{x},\bof{y},\t,\b) \, ,
\end{align}
where $\t_x=\t$ and $\t_y=0$. The minus sign arises when changing the time order 
of $\j$ and $\overline{\j}$ because the spinors anticommute. In general
\begin{align}
S(\t) = (-1)^n S(\t+n\b)
\end{align}
holds for any integer $n$.

Because of the antiperiodicity, the Matsubara frequencies are given by
\begin{align}
k_0 = (2 n +1) \rmi \, \p T \, ,
\end{align}
where $n$ is an integer. Analogous to equation~\eqref{eq:7}, one can
derive the Fourier representation of the Matsubara propagator,
\begin{align}
\rmi \, S(x) = \rmi \, T \sum_{k_0} \int \frac{\rmd^3 k}{(2 \p)^3} \frac{\rmi \, 
(\slashed{K}-m)}{K^2-m^2} 
\rme^{- \rmi \, K \cdot x} \, .
\end{align}

If we write the propagator as\footnote{Note that
  $\tilde{\D}(K)=1/(K^2-m^2)$ is not the same as the scalar propagator
  $\D(K)$ because the Matsubara frequencies $k_0=(2 n+1) \rmi \, \p T$
  are different for fermions.}
\begin{align}
S(K)=\frac{\slashed{K}-m}{K^2-m^2}=(\slashed{K}-m) \cdot \tilde{\D}(K) \, ,
\end{align}
then the mixed representation is given by 
\begin{align}
\label{eq:fermionmix}
\tilde{\D} (\t,\o_k) = -T \sum_{k_0} \rme^{-k_0 \t} \tilde{\D}(K) \, ,
\end{align}
where
\begin{align}
\tilde{\D}(K)=-\int_0^\b \rmd \tau \rme^{k_0 \t} \tilde{\D}(\t,\o_k) \, .
\end{align}
As in the scalar case, one can perform the sum in equation~\eqref{eq:fermionmix},
\begin{align}
\tilde{\D}(\t,\o_k) &= \frac{1}{2 \o_k} \left \{ \left [ 1-f_F(\o_k) \right ] 
\rme^{-\o_k \t} - f_F(\o_k) \rme^{\o_k \t} 
\right \} \\ \nonumber
& =\sum_{s=\pm 1} \frac{s}{2 \o_k} [1 - f_F( s \o_k)] \rme^{-s \o_k \t}\\ \nonumber
& = \sum_{s=\pm 1} \tilde{\D}_s(\t,\o_k) \, ,
\end{align}
where
\begin{align}
f_F(\o_k)=\frac{1}{\rme^{\b \o_k}+1}
\end{align}
is the Fermi-Dirac distribution and we allow for negative energies $s
\o_k$. As in the scalar case, the two parts decompose in frequency
space as well,
\begin{align}
\label{eq:201}
\tilde{\D}=\sum_{s=\pm 1} \tilde{\D}_s(K)= \sum_{s=\pm 1} \frac{s}{2 \o_k} 
\frac{1}{k_0-s \o_k} \, ,
\end{align}
where again the relations
\begin{align}
\tilde{\D}_s(\t,\o_k)=-T \sum_{k_0} \rme^{-k_0 \t} \tilde{\D}_s(K)
\end{align}
and
\begin{align}
\tilde{\D}_s(K)= - \int_0^\b \rmd \t \rme^{k_0 \t} \tilde{\D}_s(\t,\o_k)
\end{align}
hold.

It is straightforward to calculate the following four basic frequency
sums as in Eqs.~\eqref{eq:25} and \eqref{eq:27}:
\begin{enumerate}
\item Fermion-boson case:
\begin{align}
\label{eq:fermionboson}
T \sum_{k_0} \D_{s_1}(K) \tilde{\D}_{s_2}(P-K) &= - \frac{s_1 s_2}{4
  \o_k \o_q} \frac{1+f_B(s_1 \o_1)-f_F(s_2 \o_2)}{p_0 - s_1 \o_1 - s_2
  \o_2} \, ,
\nonumber \\
T \sum_{k_0} k_0 \D_{s_1}(K) \tilde{\D}_{s_2}(P-K) &= - \frac{s_2}{4
  \o_2} \frac{1+f_B(s_1 \o_1)-f_F(s_2 \o_2)}{p_0 - s_1 \o_1 - s_2
  \o_2} \, .
\end{align}
\item Fermion-antifermion case:
\begin{align}
  \label{eq:fermionantifermion}
  T \sum_{k_0} \tilde{\D}_{s_1}(K) \tilde{\D}_{s_2}(P-K) &= -
  \frac{s_1 s_2}{4 \o_k \o_q} \frac{1-f_F(s_1 \o_1)-f_F(s_2 \o_2)}{p_0
    - s_1 \o_1 - s_2 \o_2} \, ,
  \nonumber \\
  T \sum_{k_0} k_0 \tilde{\D}_{s_1}(K) \tilde{\D}_{s_2}(P-K) &= -
  \frac{s_2}{4 \o_2} \frac{1-f_F(s_1 \o_1)-f_F(s_2 \o_2)}{p_0 - s_1
    \o_1 - s_2 \o_2} \, .
\end{align}
\end{enumerate}
We can obtain Eqs.~\eqref{eq:fermionboson} --
\eqref{eq:fermionantifermion} by replacing $f_B(s \o)$ for a bosonic
line by $-f_F(s \o)$ for a fermionic line. The substitution $f_B(k_0)
\to -f_F(k_0)$ is in fact a systematic rule for substituing a
fermionic line for a bosonic line in an arbitrary Feynman graph. We
will not prove this rule, but the interested reader is referred to the
treatment by, e.g.~Bellac~\cite{LeBellac:1996}.

For gauge fields, it can be shown~\cite{LeBellac:1996} that the
propagator reads
\begin{align}
  \label{eq:29}
  D_{\m\n}(K)=\frac{g_{\m\n}}{K^2-m^2}
\end{align}
in Feynman gauge. We will not explicitly employ the gauge field
propagator in the future discussion, even though it is used in
deriving the thermal masses of the Higgs bosons and leptons in
references~\cite{Weldon:1982bn,Klimov:1981ka,Cline:1993bd,
  Elmfors:1993re}, to which we refer in
section~\ref{sec:appl-lept}. We quote the Feynman gauge, other
gauges can be found in reference~\cite{LeBellac:1996}.





\section{Hard Thermal Loop Resummation}
\label{sec:hard-thermal-loop}

One might expect that with the formalism and techniques of the
preceding sections, it is possible to calculate all diagrams in all
finite temperature field theories. However, using the perturbation
theory as described, one encounters the following serious problems:
\begin{enumerate}
\item IR divergences: A famous example is the energy loss of a
  heavy fermion in a plasma~\cite{Braaten:1991jj,Braaten:1991we}.
\item Gauge dependent results: An example of this is the gluon damping
  rate in a hot QCD plasma, which turns out to be different in
  different gauges~\cite{Kalashnikov:1979cy, Gross:1980br}
\item Power counting: It turns out that the resummation of infinitely
  many higher order diagrams can contribute at a lower order in
  perturbation theory than expected.
\end{enumerate}

In order to cure or at least alleviate these problems, the so called
hard thermal loop (HTL) resummation technique has been invented in the
late 80s and in the beginning of the 90s by Braaten and
Pisarski~\cite{Braaten:1989mz, Braaten:1990az}. Instead of using bare
propagators and vertices, they suggested to use effective vertices and
effective propagators, constructed by resumming certain diagrams, the
so-called HTL self energies. In this way an improved perturbation
theory has been established, which we briefly present.

\subsection{HTL self energies}
\label{sec:htl-self-energies}

Our first step is to isolate the diagrams that should be resummed into
effective propagators. To this end, we distinguish between two
scales. In a hot plasma, there are two momentum scales, the hard scale
$T$ and the soft scale $g T$, where we assume for now that the
coupling constant $g$ is much smaller than one. One could add more
scales, such as $g^2 T$ for example, but two scales will be sufficient
for our discussion. The diagrams to be resummed are the HTL self
energies, one-loop diagrams in which the external momenta are soft
($\propto g T$) and all internal momenta hard ($\propto T$).

For the scalar self energy with the $\phi^4$ interaction in
Fig.~\ref{fig:tadpole}, the result in the HTL limit is the same as the
full result of equation~\eqref{eq:Pi}. However, in the case of a
fermion, the bare self energy is gauge dependent~\cite{Peshier:1998dy}
like the gluon self energy, whereas in the HTL limit we obtain a
different, gauge independent result, which we present in the
following.
\begin{figure}
  \centering
  \includegraphics[scale=0.85]{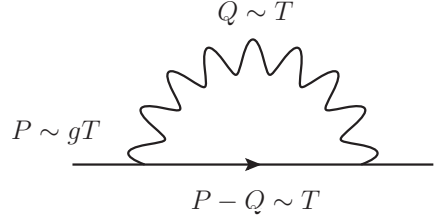}
  \caption{The HTL fermion self-energy}
  \label{fig:sigma}
\end{figure}

In all practical calculations of this thesis, the bare mass of
fermions will be negligible compared to the temperature, so we study
only the case of massless fermions. The plasma introduces the rest
frame of the heat bath as a special Lorentz frame. In a general frame,
the heat bath has four-velocity $u^\a$ with $u^\a u_\a=1$. In the rest
frame of the plasma, we can write $u^\a=(1,0,0,0)$ and $\slashed{u} =
\g_0$. The general expression for the self-energy in the rest frame of
the thermal bath is given by~\cite{Weldon:1982bn}
\begin{align}
  \label{eq:30}
  \Sigma(P)= -a(P) \slashed{P}-b(P) \slashed{u} \, ,
\end{align}
where the factors $a$ and $b$ are given by
\begin{align}
  \label{eq:31}
  a(P) &= \frac{1}{4 p^2} \left[ \tr\left(\slashed{P} \Sigma\right) -
    p_0 \tr \left( \g_0 \S \right)\right] \, , \\ \nonumber
  b(P) &= \frac{1}{4 p^2} \left[P^2 \tr\left(\g_0 \Sigma\right) -
    p_0 \tr \left( \slashed{P} \S \right) \right] \, ,
\end{align}
where the traces are evaluated in the HTL
approximation~\cite{LeBellac:1996}, and one finds
\begin{align}
  \label{eq:32}
  T_1 \equiv \tr \left(\slashed{P} \S\right) &= 4 \, m_F^2 \,
  \nonumber \\
  T_2 \equiv \tr\left(\g_0\S\right) &= 2 \, m_F^2 \frac{1}{p} \ln
  \frac{p_0+p+\rmi \, \e}{p_0-p+\rmi \, \e}
\end{align}
with the effective fermion mass
\begin{align}
  \label{eq:33}
  m_F^2 =
\begin{cases}
  e^2 T^2/8 & \text{for QED} \\
  g^2 T^2/16 & \text{for a Yukawa interaction } \mathcal{L}_Y = - g
  \overline{\j} \j \f \, .
\end{cases}
\end{align}

We will not need a resummed gauge boson propagator, therefore we will
not quote the gauge boson self energy. Again, the interested reader
will find information in the literature \cite{LeBellac:1996}.

\subsection{Effective propagators and dispersion relations}
\label{sec:effect-prop-disp}

\begin{figure}
  \centering
  \includegraphics[scale=0.7]{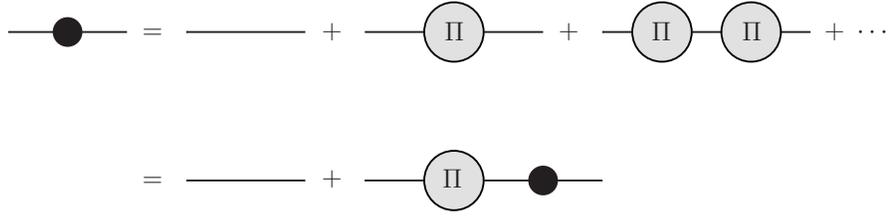}
  \caption{The resummed scalar propagator.}
  \label{fig:resummedpi}
\end{figure}

We can construct effective propagators which lead to an improved
perturbation theory by resumming the HTL self energy diagrams. For a
scalar field with a self energy $\P$, the resummed propagator is given
by the Dyson-Schwinger equation in Fig.~\ref{fig:resummedpi}.
This diagrammatic equation reads
\begin{align}
  \label{eq:34}
  \rmi \, \D^* & = \rmi \,\D + \rmi \, \D \left( - \rmi \, \P \right) \rmi \, \D +
  \dots \nonumber \\
  & = \frac{\rmi}{\D^{-1}-\P} = \frac{\rmi}{K^2-m^2-\P} \, ,
\end{align}
where $K$ is the momentum of the propagator and $m$ the zero
temperature mass.
The dispersion relation for a particle is given by the pole of its
propagator, and for the bare propagator we get
\begin{align}
  \label{eq:36}
K^2-m^2=k_0^2-\o^2(k)=0 \, ,  
\end{align}
that is
\begin{align}
  \label{eq:37}
  k_0=\o(k)=\sqrt{k^2+m^2} \, .
\end{align}
However, at finite temperature we get a different effective propagator
for a collective mode with an effective mass $m_{\rm
  eff}=\sqrt{m^2+\P}$, where it is often possible to neglect the zero
temperature mass $m$ with respect to the self energy such that $m_{\rm
  eff}=\sqrt{\P}$. The dispersion relation of the collective scalar
particle is then given by 
\begin{align}
  \label{eq:38}
 \o_{\rm eff}=\sqrt{k^2+m_{\rm eff}^2} \, .
\end{align}
Effective masses and dispersion relations as above, generated by the
interaction with a medium, have been introduced in various areas in
physics, such as the effective mass of an electron in a crystal or the
reduced velocity of a photon in a medium.

Considering fermions, we restrict ourselves to the case where the bare
mass is negligible and, in a similar way as in equation~\eqref{eq:34},
we get for the effective propagator
\begin{align}
  \label{eq:39}
  S^*(K)=\frac{1}{\slashed{K}-\Sigma_{\rm HTL}(K)}\, ,
\end{align}
where $\Sigma_{\rm HTL}$ is given by
Eqs.~\eqref{eq:30}--\eqref{eq:33}. It is very convenient to rewrite
this propagator in the helicity-eigenstate
representation~\cite{Braaten:1990wp,Braaten:1992gd},
\begin{align}
  \label{eq:35}
  S^*(K)=\frac{1}{2} \Delta_+(K) (\gamma_0-\hat{\bf k} \cdot
\boldsymbol{\gamma}) +\frac{1}{2} \Delta_-(K) (\gamma_0+\hat{\bf k} \cdot
\boldsymbol{\gamma}),
\end{align}
where $\hat{\bf k}={\bf k}/k$,
\begin{align}
\label{eq:42}
\Delta_\pm(K)=\left [ -k_0 \pm  k + \frac{m_F^2}{k} \left ( \pm 1 - 
\frac{\pm k_0 - k}{2k} \ln \frac{k_0+k}{k_0-k}  \right ) \right ]^{-1}
\end{align}
and $m_F$ is the effective fermion mass defined in
equation~\eqref{eq:33}.

\begin{figure}
  \centering
  \includegraphics[scale=0.8]{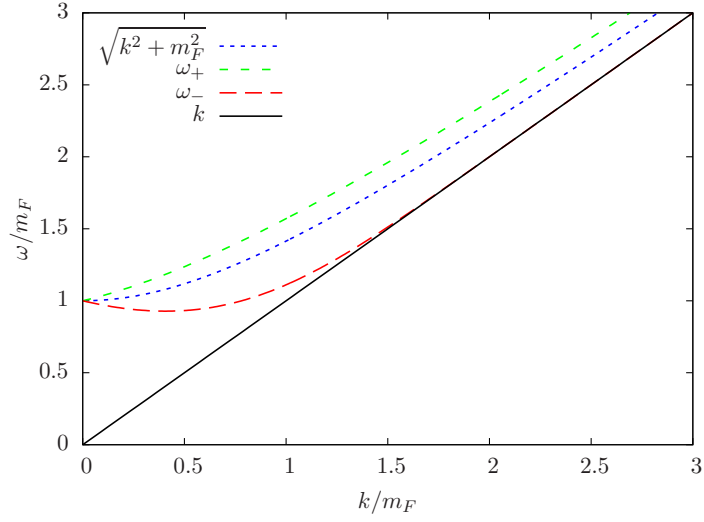}
  \caption[Fermionic dispersion relations.]{The two dispersion laws
    for fermionic excitations compared to the standard dispersion
    relation $\o^2=k^2+m_F^2$.}
  \label{fig:omega}
\end{figure}

This propagator has two poles, the zeros of the two denominators
$\D_\pm$. The poles can be seen as the dispersion relations of
collective excitations of the fermions that interact with the hot
plasma,
\begin{align}
  \label{eq:40}
  k_0=\o_\pm(k) \, .
\end{align}
We have found an analytical expression for the two dispersion
relations making use of the Lambert $W$ function
\cite{Kiessig:2010pr}, which is calculated in
appendix~\ref{cha:analyt-solut-htl}. The dispersion relations are
shown in Fig.~\ref{fig:omega}.

Note that even though the dispersion relations resemble the behaviour
of massive particles and $\o =m_F$ for zero momentum $k$, the
propagator $S^*(K)$ \eqref{eq:35} does not break chiral
invariance. Both the self energy $\S(K)$ \eqref{eq:30} and the
propagator $S^*(K)$ anticommute with $\g_5$. The Dirac spinors that
are associated with the pole at $k_0=\o_+$ are eigenstates of the
operator $(\g_0-\bh{k} \cdot \gb)$ and they have a positive ratio of
helicity over chirality, $\c=+1$. The spinors associated with
$k_0=\o_-$, on the other hand, are eigenstates of $(\g_0+\bh{k} \cdot
\gb)$ and have a negative helicity-over-chirality ratio, $\c=-1$. At
zero temperature, fermions have $\c=+1$. The introduction of a thermal
bath gives rise to collective fermionic modes which have
$\c=-1$. These modes have been called plasminos since they are new
fermionic excitations of the plasma and have first been noted in
references~\cite{Weldon:1982bn,Klimov:1981ka}.


We can introduce a spectral representation for the two parts of the
fermion propagator \eqref{eq:42} \cite{Pisarski:1989cs},
\begin{align}
  \label{eq:41}
  \D_\pm(K)=\int_{-\infty}^\infty \rmd
   \o \frac{\rho_\pm(\o,k)}{\o-k_0-\rmi \, \e} \, ,
\end{align}
where the spectral density $\rho_\pm(\o,k)$
\cite{Braaten:1990wp, Kapusta:1991qp} has two contributions, one from
the poles,
\begin{align}
  \label{eq:43}
  \rho^{\rm pole}_\pm(\o,k)= Z_\pm(\o,k) \d(\o-\o_\pm(k)) +Z_\mp(\o,k) 
  \d(\o+\o_\mp(k)) \, ,
\end{align}
and one discontinuous part,
\begin{align}
  \label{eq:44}
  \rho^{\rm disc}_\pm(\o,k)= \frac{\frac{1}{2}\, m_F^2(k\mp \o)}
  {\left\{k(\o\mp k)-m_F^2\left[Q_0(x) \mp Q_1(x)\right]\right\}^2 +
    \left[\frac{1}{2}\, \p\, m_F^2 (1 \mp x) \right]^2} \times
  \q(k^2-\o^2) \, ,
\end{align}
where $x=\o/k$, $\q(x)$ is the heaviside function and $Q_0$ and $Q_1$
are Legendre functions of the second kind,
\begin{align}
  \label{eq:45}
  Q_0(x)=\frac{1}{2} \ln \frac{x+1}{x-1} \, , \hspace{1cm} Q_1(x) = x
  Q_0(x) -1 \, .
\end{align}
The residues of the quasi-particle poles are given by
\begin{align}
  \label{eq:46}
  Z_\pm(\o,k)=\frac{\o_\pm^2(k)-k^2}{2 m_F^2} \, , \quad {\rm where}
  \quad Z_+ + Z_- = 1 \, .
\end{align}
One can describe the non-standard dispersion relations $\o_\pm$ by
momentum-dependent effective masses $m_\pm(k)$ which are given by
\begin{align}
  \label{eq:47}
  m_\pm(k)=\sqrt{\o_\pm^2(k)-k^2}=\sqrt{2 Z(\o,k)} m_F \, .
\end{align}
\begin{figure}
  \centering
  \includegraphics[scale=0.8]{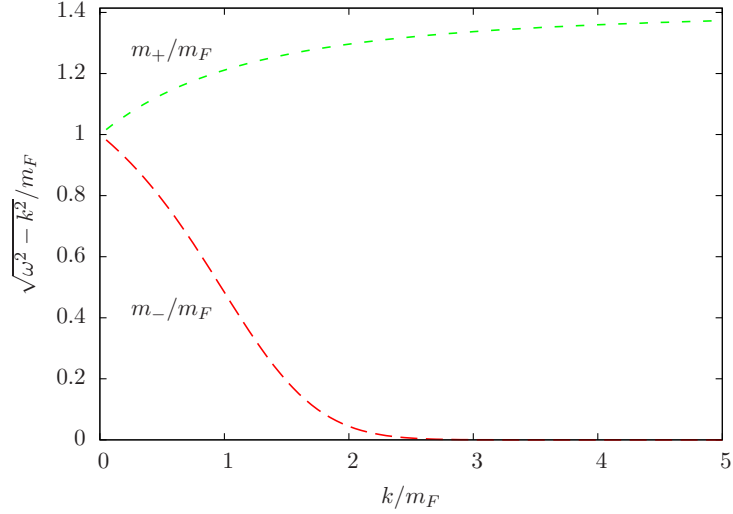}
  \caption{The momentum-dependent effective masses $m_\pm$.}
  \label{fig:mpm}
\end{figure}
These masses are shown in Fig.~\ref{fig:mpm}.

\subsection{HTL resummation technique}
\label{sec:htl-resumm-techn}

We have collected the necessary ingredients to build an improved
perturbation theory at finite temperature. We noted that naive
perturbation theory suffers from the problems of IR divergences and
gauge dependent results. The reason for this is that the naive
perturbative expansion is incomplete at $T>0$. Infinitely many higher
order diagrams can contribute to lower order in the coupling
constant. These diagrams can be taken into account by resummation.

\begin{figure}
  \centering
  \includegraphics[scale=0.62]{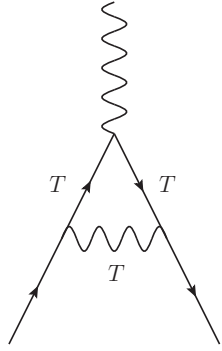}
  \caption{The HTL vertex correction}
  \label{fig:htlvertex}
\end{figure}

\begin{figure}
  \centering
  \includegraphics[scale=0.82]{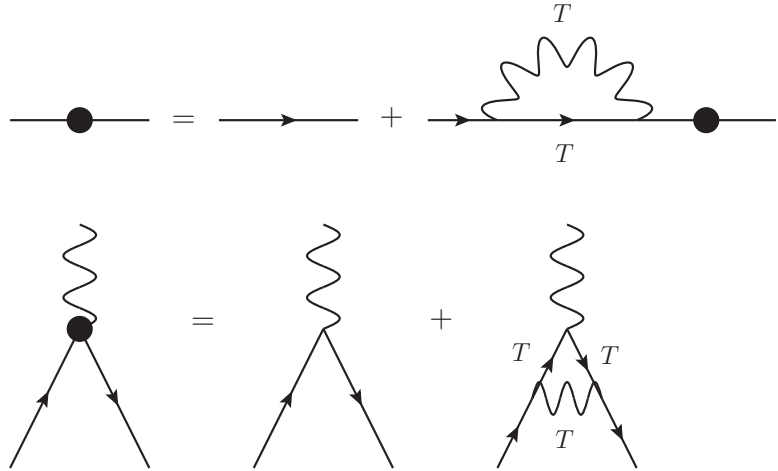}
  \caption[The effective HTL propagator and vertex.]{Examples of an
    effective HTL propagator and an effective vertex.}
  \label{fig:htlrules}
\end{figure}

We will not discuss the HTL resummation technique for gauge theories
in detail (see, e.g.~\cite{LeBellac:1996}), but we will present rules
for perturbative calculations. One has to consider the self energies
in the HTL approximation, such as $\P$, $\S^*$ and the gauge boson
polarisation tensor $\P^*_{\m\n}$\footnote{As mention above, we will
  not need the gauge boson propagator in this work, but we mention it
  in order to sketch the HTL resummation technique.}. Due to Ward
identities, fermion self energies are related to vertices, e.g.
\begin{align}
  \label{eq:48}
  \rmi \, e \left[\S(P_1) -\S(P_2) \right] = (P_1+P_2)_\m \G^\m(P_1,P_2)
  \, .
\end{align}
Therefore one also has to consider the HTL correction to vertex
contributions\footnote{Even though we will not encounter effective
  vertices in the course of this work, we add them in order to present
  a full picture of the HTL theory.}, as shown in
figure~\ref{fig:htlvertex}, where all internal lines are hard $\sim
T$.
The HTL propagators are constructed by resumming the self energies via
the Dyson-Schwinger equation as explained above. The HTL vertices are
given by adding the HTL correction to the bare vertex. Examples are
shown in figure~\ref{fig:htlrules}.
When calculating diagrams, we have to use effective propagators and
vertices if all external legs are soft; otherwies bare
propagators and vertices are sufficient. In this way, contributions of
the same order in $g$ are included, gauge independent results are
obtained and the IR behaviour of the theory is improved. After all,
the HTL improved perturbation theory has been successfully applied to
thermal QCD for the description of the quark-gluon plasma (see
e.g.~\cite{Thoma:1995ju})

Having outlined the construction of the HTL resummation technique as a
perturbation theory that cures many serious shortcomings of naive
perturbation theory at finite temperature, some remarks concerning our
use of HTL propagators seem necessary. First, even though it is
sufficient to use bare propagators and vertices if one external
momentum is hard, it is always possible to resum self-energies and
thus capture effects which arise from higher-order loop diagrams and
thus take into account the appearance of thermal masses and modified
dispersion relations in a medium. In fact, since the effective masses
we will encounter do typically not satisfy the condition $m_{\rm eff}
\ll T$ but are rather in the range $m_{\rm eff}/T \sim 0.1$ -- $1$,
the effect of resummed propagators is noticeable even when some or all
external momenta are hard $\sim T$. In summary, we always resum the
propagators of particles that are in equilibrium with the thermal
bath, that is in our case the Higgs bosons and the leptons, in order
to capture the effects of thermal masses, modified dispersion relation
and modified helicity structures. This approach is justified a
posteriori by the sizeable corrections it reveals, similar to the
treatment of meson correlation fuctions in
reference~\cite{Karsch:2000gi}.





    \clearpage
\chapter{Decays and Inverse Decays\label{decayrate}}

\section{The Quest of This Thesis}
\label{sec:quest-this-thesis}

In order to make any statement about the amount of baryon asymmetry
that is produced by a baryogenesis or, in our case, leptogenesis
model, one has to adopt a quantitative description of the dynamics
that take place at this phase and result in generating the
asymmetry. There are two ways to do this: One can either adopt the
consistent quantum mechanical view of the system and calculate how
quantum systems evolve with time when they are not in equilibrium. Or
one can view the particle distributions as classical distributions and
adopt the equations which govern them, that is, a set of Boltzmann
equations. The set of equations that govern the dynamics of the first
non-equilibrium approach are the so called Kadanoff-Baym
equations~\cite{KadanoffBaym:1962} and this approach has received some
attention in recent publications~\cite{Anisimov:2008dz, Garny:2009rv,
  Garny:2009qn, Anisimov:2010aq, Garny:2010nj, Beneke:2010wd,
  Garny:2010nz, Anisimov:2010dk, Beneke:2010dz, Drewes:2010pf,
  Drewes:2010zz}. The easier and more traditional Boltzmann approach
is possible when interactions are not fast and particles are
sufficiently close to their equilibrium distribution, such that before
and after each interaction, they can be seen as classical
particles. The interaction itself is calculated by means of quantum
field theory and appears in the collision term of the equation.

Whichever viewpoint we adopt, we have to calculate the quantum
mechanical amplitudes that enter the equations. Traditionally,
amplitudes are calculated in vacuum and inserted into Boltzmann
equations. However, as densities and temperatures are high, it is
important to calculate amplitudes at finite temperature and compare
the result to the zero-temperature case. In this work, we adopt the
Boltzmann view of particle evolution and calculate the corrections due
to finite densities and temperatures. However, the amplitudes are
related to the two-point functions one has to use when adopting the
Kadanoff-Baym approach.

The Boltzmann equations, which we stated for zero temperature in
equations~\eqref{eq:49} and \eqref{eq:50}, contain two main
quantities, which are calculated from amplitudes: First, the rates
with which interactions take place, that is, the decay and inverse
decay rates $\G_{D}$ and $\G_{ID}$, the scattering rates $\G_S$ and
the washout rates $\G_W$. Second, the difference between the $L$
violating rates and their $\CP$-conjugates, the $\CP$-asymmetry, which
is defined in equation~\eqref{eq:51} for zero temperature. Of these,
the leading order amplitudes in the interaction rates are tree level
amplitudes. However, as explained in section~\ref{sec:cp-asymmetry},
the $\CP$-asymmetry arises as an interference between tree level and
one-loop amplitudes and is proportional to the imaginary part of one
loop diagrams. Thus, for the $\CP$-asymmetry, the leading order is the
one-loop level.

The combination of SM couplings that we write as $g$ in the
HTL-corrections $gT$ is typically of the order $g \sim
\mathcal{O}(10^{-1})$. Since the creation of the lepton asymmetry
takes places at temperatures $T \sim M_1$, corrections of the order of
$gT$ are important to consider if the couplings are as large as they
are in our case. We present a consistent way of calculating these
HTL-corrections of order $gT$ and analyse the effects they imply.
Thus, we restrict ourselves to a very basic leptogenesis toy model
which is self-consistent and contains all important features of
leptogenesis even if it does not describe the full scenario in a
quantitatively accurate way. More specifically, we look at the leading
order interactions between the protagonists of leptogenesis, the heavy
neutrinos $N_1$, as well as the lepton and Higgs doublets $\ell$ and
$\phi$. These interactions are decays $N_1 \rightarrow L H$, inverse
decays $L H \rightarrow N_1$ and the $\Delta L=2$ scatterings $LH \to
LH$, where by $L$ and $H$, we denote both the doublets $\phi, \ell$
and their charge conjugated states $\overline{\phi},
\overline{\ell}$. Naturally, we have to include a calculation of the
$\CP$-asymmetry, which is a dominant quantity in the sense that there
would be no asymmetry production without it. We include
HTL-corrections for the Higgs bosons and the leptons, but not for the
neutrinos, since the Yukawa couplings are much smaller than the SM
couplings that give rise to the HTL effects for Higgs bosons and
leptons.

As temperatures are high, particles in the bath acquire different
dispersion relations as explained in
section~\ref{sec:effect-prop-disp}, which can be translated into
effective thermal masses. Due to these masses, it may happen that some
interactions are kinematically forbidden, while other processes that
would not be possible in vacuum are allowed. In leptogenesis, there is
a temperature where the processes $N_1 \leftrightarrow LH$ are
forbidden, but the processes $H \leftrightarrow LN_1$ are
possible. The leptons always have a lower thermal mass than the Higgs
bosons, so $L \leftrightarrow N_1 H$ is never allowed. These Higgs
decays and inverse decays are new processes that govern the neutrino
and lepton evolution at high temperature, so they need to be
calculated and likewise the $\CP$-asymmetry in these decays since it
leads to the production of a lepton asymmetry.

Thermal corrections to leptogenesis of various kind have been studied
before, both in the
Boltzmann-picture~\cite{Giudice:2003jh,Covi:1997dr, Basboll:2006yx,
  Garayoa:2009my, HahnWoernle:2009en, Besak:2010fb, Anisimov:2010gy,
  Garbrecht:2010sz,HahnWoernle:2009qn}, as well as in the
Kadanoff-Baym-picture~\cite{Anisimov:2008dz, Garny:2009rv,
  Garny:2009qn, Anisimov:2010aq, Garny:2010nj, Beneke:2010wd,
  Garny:2010nz, Anisimov:2010dk, Beneke:2010dz, Drewes:2010pf,
  Drewes:2010zz}. Our self-consistent calculation is a novel approach
and we discuss the differences to the aforementioned works, with a
special focus on the extensive and important work by Giudice et
al.~\cite{Giudice:2003jh}.
In this chapter, we calculate the leading order HTL corrections
to decays and inverse decays, in chapter~\ref{cpas}, we do the same
for the $\CP$-asymmetries and in chapter~\ref{boltzmanneq}, we put our
results into the appropriate Boltzmann equations and evaluate them.




 \section[Discontinuity of the Fermion Self-Energy in Yukawa
Theory]{Discontinuity of the Fermion Self-Energy in Yukawa Theory at
  Finite Temperature}
\label{sec:disc-yukawa-ferm}

We consider a leptogenesis-inspired model with a massive Majorana
fermion {\sl N} coupling to a massless Dirac fermion $\ell$ and a
massless scalar $\phi$. The interaction and mass part of the
Lagrangian reads
\begin{align} 
\label{L}
\mathcal{L}_{\rm int,mass}=
g \bar{N} \phi \ell - \frac{1}{2} M \bar{N} N^c+h.c.\: , 
\end{align} 
The HTL resummation technique has been considered in \cite{Thoma:1994yw}
for the case of a Dirac fermion with Yukawa coupling, from which the
HTL resummed propagators for the Lagrangian in equation~\eqref{L} follow directly.
We calculate the interaction rate $\Gamma$ of $N
\leftrightarrow \ell \phi$.

\begin{figure}
\begin{center}
\includegraphics[scale=0.5]{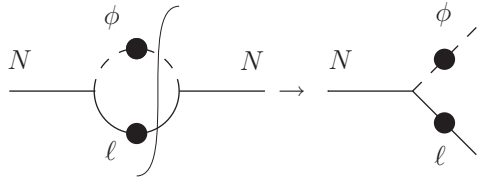}
\caption[The optical theorem in neutrino decays]{\label{cut} $N$ decay
  via the optical theorem with dressed propagators denoted by a blob}
\end{center}
\end{figure}

We cut the $N$ self-energy and use the 
HTL resummation for the fermion and scalar propagators (figure~\ref{cut}).
According to finite-temperature cutting rules
\cite{Weldon:1983jn,Kobes:1986za}, the interaction rate reads
\begin{align}
\Gamma(P) = - \frac{1}{2 p_0} \; {\rm tr} [ (\slashed{P}+M) \; {\rm Im} \; 
\Sigma(p_0 + \rmi \, \e,{\bf p})].
\end{align}
At finite temperature, the self-energy reads
\begin{align}
  \Sigma(P)=-g^2 T \sum_{k_0=\rmi \, (2 n+1) \pi T} \int \frac{{\rm d}^3 k}{(2
    \pi)^3} \: P_L \: S^*(K) \: P_R \: D^*(Q),
\end{align} 
where $P_L$ and $P_R$ are the projection operators on left- and
right-handed states and $Q=P-K$.

The HTL-resummed scalar propagator is
\begin{align}
D^*(Q)=\frac{1}{Q^2-m_\phi^2},
\end{align}
where $m_\phi^2=g^2 T^2 / 12$ is the thermal mass of the scalar,
created by the interaction with fermions, and can be calculated
analogously to the $\phi^4$ self-energy in
section~\ref{sec:scalar-field}. Due to the reduced Majorana degrees of
freedom, $m_\phi$ differs from the
 Dirac-Dirac
case by a factor 1/2 \cite{Thoma:1994yw}.

The effective fermion propagator in the helicity-eigenstate
representation is given by
equation~\eqref{eq:35}~\cite{Braaten:1990wp,Braaten:1992gd},
\begin{align}
\label{fermprop}
S^*(K)&= - \frac{1}{2} \Delta_+(K) (\gamma_0-\hat{\bf k} \cdot
\boldsymbol{\gamma}) -\frac{1}{2} \Delta_-(K) (\gamma_0+\hat{\bf k}
\cdot
\boldsymbol{\gamma}) \, , \nonumber \\
\Delta_\pm(K)&=\left [ -k_0 \pm k + \frac{m_\ell^2}{k} \left ( \pm1 -
    \frac{\pm k_0 - k}{2k} \ln \frac{k_0+k}{k_0-k} \right ) \right
]^{-1} 
\end{align}
and
\begin{align}
m_\ell^2=\frac{1}{32} g^2 T^2.
\end{align}
This again differs from the Dirac case in equation~\eqref{eq:33} by a
factor 1/2~\cite{Thoma:1994yw}.

The trace can be evaluated as
\begin{align}
{\rm tr} [(\slashed{P} +M) P_L S^*(K) P_R]= -\Delta_+ (p_0-p \eta) - \Delta_-
(p_0 +p \eta),
\end{align}
where $\eta=\cos\theta$ is the angle between {\bf p} and {\bf k}. We
evaluate the sum over Matsubara frequencies by using the Saclay method
\cite{Pisarski:1987wc}. For the scalar propagator, the Saclay
representation from equation~\eqref{eq:11} reads
\begin{align}
D^*(Q)=-\int_0^\beta {\rm d} \tau e^{q_0 \tau} \frac{1}{2 \omega_q}
\{ [ 1+f_B(\omega_q)] e^{-\omega_q \tau} + f_B(\omega_q) e^{\omega_q \tau}\},
\end{align}
where $\omega_q^2=q^2+m_\phi^2$. For the fermion propagator, it is
convenient to use the spectral representation as explained in
section~\ref{sec:effect-prop-disp}~\cite{Pisarski:1989cs},
\begin{align}
\label{eq:186}
\Delta_\pm(K)=-\int_0^\beta {\rm d \tau'} e^{k_0 \tau'} 
\int_{-\infty}^{\infty}
{\rm d} \omega \: \rho_\pm(\omega,k) [1-f_F(\omega)] e^{-\omega
\tau'},
\end{align} 
where $\rho_\pm$ is the spectral density~\cite{Braaten:1990wp}.

Since the quasi-particles are our final states, we will set $K$ such
that $1/\Delta_\pm(K)=0$. Thus, we are only interested in the pole
contribution
\begin{align}
\label{rho}
\rho_\pm^{\rm pole}(\omega,k)= - \frac{\omega^2-k^2}{2 m_\ell^2} (\delta(\omega-
\omega_\pm)+ \delta (\omega+ \omega_\mp)),
\end{align}
where $\omega_\pm$ are the dispersion relations for the two
quasiparticles, i.e.~the solutions for $k_0$ such that
$1/\Delta\pm(\omega_\pm,{\bf k})=0$, shown in
figure~\ref{fig:omega}. The analytic solutions for $\o_\pm$ are
explained in appendix~\ref{cha:analyt-solut-htl}. One assigns a
momentum-dependent thermal mass
\mbox{$m_\pm(k)^2=\omega_\pm(k)^2-k^2$} to the two modes as explained
in section~\ref{sec:effect-prop-disp} and shown in
figure~\ref{fig:mpm} and for large momenta the heavy mode $m_+$
approaches $\sqrt{2} \: m_\ell$, while the light mode becomes
massless.

In order to execute the sum over Matsubara frequencies, we write
$k_0=\rmi \, \omega_n$ with $\omega_n=(2 n+1) \pi T$ and remember
that, when evaluating frequency sums, also $p_0=\rmi \, \omega_m=\rmi \, (2
m + 1) \pi T$ can be written as a Matsubara frequency and later on be
continued analytically to real values of $p_0$
\cite{LeBellac:1996,Baym:1961aa,Dolan:1973qd}. In particular $e^{p_0
  \beta}=e^{\rmi \, \omega_m \beta}=-1$. We write
\begin{align}
  T \sum_n e^{\rmi \, \omega_n \tau}&= \sum_{n'=-\infty}^{\infty}
  \delta(\tau-n' \beta) \, , \nonumber \\
  T \sum_n e^{(p_0-k_0) \tau} e^{k_0 \tau'}&=e^{p_0 \tau}
  \delta(\tau'-\tau),
\end{align}
since $-\beta \leq \tau'-\tau \leq \beta$.  After evaluating the sum
over $k_0$ and carrying out the integrations over $\tau$ and $\tau'$,
we get
\begin{align}
\begin{split}
T \sum_{k_0} D^*(Q) \Delta_\pm(K)=
 -\int_{-\infty}^{\infty} {\rm d} \omega \,
\rho_\pm(\omega,k) \frac{1}{2 \omega_q} & \left [ 
\frac{1+f_B(\omega_q)-f_F(\omega)}{p_0 - \omega -\omega_q} \right. \\
& \left. \quad +\frac{f_B(\omega_q)+f_F(\omega)}{p_0 - \omega
    +\omega_q} \right ] \, .
\end{split}
\end{align}
Integrating $\omega$ over the pole part of $\rho_\pm$ in
equation~\eqref{rho}, we get
\begin{align}
\begin{split}
T \sum_{k_0} D^* \Delta_\pm= \frac{1}{2 \omega_q} 
& \left \{ 
\frac{\omega_\pm^2-k^2}
{2 m_\ell^2} \left [ 
\frac{1+f_B-f_F}{p_0-\omega_\pm-\omega_q}+
 \frac{f_B+f_F}{p_0-\omega_\pm+\omega_q} 
\right ]
\right. \\
& 
\label{frequencysum}
\left. +  \frac{\omega_\mp^2-k^2}{2 m_\ell^2} \left [ 
\frac{f_B+f_F}{p_0+\omega_\mp-\omega_q}+
 \frac{1+f_B-f_F}{p_0+\omega_\mp+\omega_q} \right ] 
\right \} \, ,
\end{split}
\end{align} 
where $f_B=f_B(\omega_q)$ and $f_F=f_F(\omega_\pm)$ or
$f_F(\omega_\mp)$, respectively.

The four terms in equation~\eqref{frequencysum} correspond to the processes
with the energy relations indicated in the denominator, i.e.~the decay
$N \rightarrow \phi \ell$, the production $N \phi \rightarrow \ell$,
the production $N \ell \rightarrow \phi$ and the production of $N \ell
\phi$ from the vacuum, as well as the four inverse reactions
\cite{Weldon:1983jn}. We are only interested in the process $N \leftrightarrow 
\phi \ell$, where the decay and inverse decay are illustrated by the
statistical factors
\begin{align}
1+f_B-f_F=(1+f_B)(1-f_F)+f_B f_F \, .
\end{align}
The decay is weighted by the factor $(1+f_B)(1-f_F)$ for induced
emission of a Higgs boson and a lepton, while the inverse decay is
weighted by the factor $f_B f_F$ for absorption of a Higgs boson and
a lepton from the thermal bath.  Our term reads
\begin{align}
\label{eq:123}
T \sum_{k_0} D^* \Delta_\pm \Bigg|_{N \leftrightarrow \phi \ell}=  \frac{1}{2 \omega_q} \;
\frac{\omega_\pm^2-k^2}{2 m_\ell^2} \; 
\frac{1+f_B-f_F}{p_0-\omega_\pm-\omega_q}.
\end{align}

For carrying out the integration over the angle $\eta$, we use
\begin{align}
{\rm Im} \frac{1}{p_0-\omega_\pm-\omega_q+ \rmi \, \e}= - \pi
\delta(p_0-\omega_\pm-\omega_q) = - \pi \frac{\omega_q}{k p}
\delta(\eta-\eta_\pm), 
\end{align}
where
\begin{align}
\eta_\pm=\frac{1}{2 k p} \left [ 2 p_0 \omega_\pm - M^2 - (\omega_\pm^2-k^2)
+ m_\phi^2 \right ]
\end{align}
denotes the angle for which the energy conservation
$p_0=\omega+\omega_q$ holds. The integration over $\eta$ then yields
\begin{align} 
\int_{-1}^{1} {\rm d} \eta \; {\rm Im}(T \sum_{k_0} D^*
\Delta_\pm)= - \frac{\pi}{2 k p} \; \frac{\omega_\pm^2-k^2}{2 m_\ell^2} \; 
[1+f_B(\omega_{q \pm})-f_F(\omega_\pm)],
\end{align}
where $\omega_{q \pm}=p_0-\omega_\pm$.
It follows that
\begin{align}
\begin{split}
\label{eq:20}
\Gamma(P)=&-\frac{1}{2 p_0}\; {\rm tr} [(\slashed{P} +M) {\rm Im} \; \Sigma(P)] 
\\
=& \frac{1}{2 p_0} \; {\rm Im} \left \{ g^2 T \sum_{k_0} \int
\frac{{\rm d^3} k}{(2 \pi)^3} \; {\rm tr} [(\slashed{P} +M) P_L S^* P_R] D^*
\right \} \\
=& - \frac{g^2}{8 \pi^2 p_0} \; {\rm Im} \left \{ T \sum_{k_0} \int {\rm d} k \,
{\rm d} \eta \: k^2 D^* [\Delta_+ (p_0-p \eta) +\Delta_-
(p_0 +p \eta)] \right \} \\
=& \frac{g^2}{32 \pi p_0 p} \sum_\pm \int_{-1 \leq \eta_\pm \leq 1} {\rm d} k 
\: \frac{\omega_\pm^2-k^2}{2 m_\ell^2}
[1+f_B(\omega_{q \pm}) -f_F(\omega_\pm)] \\ 
&  \hspace{3.4cm} \times [2 p_0
(k \mp \omega_\pm) \pm M^2 \pm (\omega_\pm^2-k^2) \mp m_\phi^2],
\end{split}
\end{align}
where we only integrate over regions with $-1 \leq \eta \leq 1$.

Using finite temperature cutting rules, one can also write the
interaction rates for the two modes in a way that resembles the
zero-temperature case \cite{Weldon:1983jn}
\begin{align}
\begin{split}
\Gamma_\pm(P)=\frac{1}{2 p_0}\int \, {\rm d} \tilde{k} \, {\rm d} \tilde{q} \; 
& (2 \pi)^4 \delta^4(P-K-Q) \; |\mathcal{M}_\pm(P,K)|^2 \\
 \times &  [1+f_B(\omega_q)-f_F(\omega_\pm)],
\end{split}
\end{align}
where 
\begin{align}
{\rm d} \tilde{k}=\frac{{\rm d}^3 k}{(2 \pi)^3 2 \,k_0}
\end{align}
and ${\rm d}\tilde{q}$ analogously and the matrix elements are
\begin{align}
\label{eq:130}
|\mathcal{M}_\pm(P,K)|^2=g^2 \frac{\omega_\pm^2-k^2}{2 m_\ell^2} \omega_\pm 
\left (p_0 \mp p \eta_\pm \right ). 
\end{align}

Now that we have arrived at an expression for the full HTL decay rate
of a Yukawa fermion, we would like to compare it to the conventional
approximation adopted for example by reference~\cite{Giudice:2003jh},
which we refer to as one-mode approximation. To this end, we do the
same calculation for an approximated fermion propagator
\begin{align}
\label{approx}
S^*_{\rm 
approx}(K)=\frac{1}{\slashed{K}-m_\ell} \, ,
\end{align}
This yields the following interaction rate:
\begin{align}
\begin{split}
\Gamma_{\rm approx}(P)=& \frac{g^2}{32 \pi p_0 p} \int_{k_1}^{k_2} {\rm d} k 
\: 
\frac{k}{\omega} [1+f_B(\omega_q) -f_F(\omega)] [M^2 +m_\ell^2 -m_\phi^2] 
\\
=& \frac{1}{2 p_0}\int \, {\rm d} \tilde{k} \: {\rm d} \tilde{q} \: 
  (2
\pi)^4 \delta^4(P-K-Q) |\mathcal{M}|^2 \\ 
& \hspace{1.6cm} \times 
[1+f_B(\omega_q)-f_F(\omega)],
\end{split}
\end{align}
where $\omega^2=k^2+m_\ell^2$, $\omega_q=p_0-\omega$ and the
integration boundaries
\begin{align}
k_{1,2}= \frac{1}{2 M^2} \left |p_0 \sqrt{(M^2+m_\ell^2-m_\phi^2)^2-
(2 M m_\ell)^2} \mp p (M^2+m_\ell^2-m_\phi^2) \right |
\end{align}
ensure $-1 \leq \eta \leq 1$, where 
\begin{align}
\eta=\frac{1}{2 k p} \left [ 2 p_0 \omega - M^2 - m_\ell^2
+ m_\phi^2 \right ].
\end{align}
We see that the matrix element is
\begin{align}
|\mathcal{M}|^2=\frac{g^2}{2} (M^2+m_\ell^2-m_\phi^2).
\end{align}
In addition to the dispersion relations and the phase space boundaries
for $k$, there are two major differences to the two-mode matrix
element in equation~\eqref{eq:130}. In the two-mode approach, we
integrate over the residue $Z_\pm(k)=(\omega_\pm^2-k^2)/(2 \,
m_\ell^2)$. For the plus-mode, this residue is mostly close to
unity, but can be as low as $1/2$ for low momenta. For the negative
mode, the residue is close to zero for most momenta and only up to
$1/2$ for low momenta. The rate for the plus-mode is slightly
suppressed compared to the one-mode approach, while the rate for the
minus-mode is considerably suppressed.

Another difference is the momentum product
\begin{align}
  \label{eq:111}
  P \cdot K = p_0 \omega - p k \eta = \frac{1}{2} \left( M^2
  +m_\ell^2-m_\phi^2 \right) \equiv \frac{\Sigma}{2} \, .
\end{align}
For the two-mode approach, we can introduce a chirally invariant
four-momentum for the lepton
\begin{align}
  \label{eq:121}
  K_h^\mu \equiv \omega_h (1, h \, {\bf \hat{k}}) \, , 
\end{align}
where $h=\pm 1$ denotes the helicity-over-chirality ratio. Then
\begin{align}
  \label{eq:120}
  P \cdot K_h = p_0 \omega -h p \omega \eta_h = h \frac{\Sigma_h}{2} -
  p_0 (h \omega - k) \, , 
\end{align}
where
\begin{align}
  \label{eq:154}
  \Sigma_h \equiv M^2+m_h^2(k) -m_\phi^2 \, .
\end{align}
We see that for the plus-mode, the term $p_0 (\omega-k)$ has to be subtracted
from the right-hand side of equation \eqref{eq:111}, which also
suppresses the rate compared to the one-mode calculation. For the
minus-mode, the momentum product is also different from the
one-mode approach. This difference in the momentum products is closely
linked to the fact that the modified dispersion relation of the
two-mode calculation leaves the chiral symmetry unbroken.

Continuing our discussion of the one-mode rate, we note that it
resembles the zero temperature result
\begin{align}
\Gamma_{T=0}(P)= \frac{g^2}{32 \pi p_0 p} \int_{k_1}^{k_2} {\rm d} k \: 
\frac{k}{\omega} 
\left [ M^2 +m_\ell^2 -m_\phi^2 \right ]
\end{align}
with zero temperature masses $m_\ell$, $m_\phi$.
The missing factor
\begin{align}
1+f_B-f_F=(1+f_B)(1-f_F)+f_B f_F
\end{align}
accounts for the statistical distribution of the initial or final
particles.  As pointed out in more detail in
reference~\cite{Kiessig:2009cm}, we see that the approach to treat
thermal masses like zero temperature masses in the final state
\cite{Giudice:2003jh} is justified for the decay rates, since it
equals the HTL treatment with an approximate fermion
propagator. However this approach does not equal the full HTL result.

Concluding this calculation, a caveat has to be added: In this
general calculation, the external Majorana fermion will also acquire
a thermal mass of order $gT$. Thus, if its zero temperature mass is
smaller than its thermal mass, the external fermion also needs to be
described by leptonic quasiparticles to be consistent. However, in the
leptogenesis study, the Yukawa coupling giving rise to the Majorana
neutrino decay is much smaller than the couplings giving rise to the
thermal masses of the Higgs boson (scalar) and the lepton (Dirac
fermion) and thus the thermal mass of the heavy neutrino can be
neglected, as pointed out in the previous section.

We have calculated the decay rate assuming a Majorana particle, but
the result can be very easily generalized to the case of two Dirac
fermions by inserting the appropriate factors of two in the decay rate
and the thermal masses.






\section{Decays at High Temperature}
\label{sec:decays-at-high}

When the temperature is so high that $m_\phi(T) > M$, the scalar can
decay into the Majorana fermion and the Dirac fermion\footnote{Note
  that in our model calculation, $m_\phi(T) > m_\ell(T)$}. The calculation
can be done in the same way as for the Majorana fermion decay. The
only difference is that in equation~\eqref{frequencysum}, we take the
imaginary part of the factor $1/(p_0+\omega_\mp-\omega_q)$, which
corresponds to the scalar decay. The frequency sum corresponding to
equation~\eqref{eq:123} reads
\begin{align}
  \label{eq:124}
  T \sum_{k_0} D^* \Delta_h \Bigg|_{\phi \leftrightarrow N \ell} =  \frac{1}{2 \omega_q} \;
  \frac{\omega_{-h}^2-k^2}{2 m_\ell^2} \;
  \frac{f_B(\omega_q)+f_F(\omega_{-h})}{p_0+\omega_{-h}-\omega_q}.
\end{align}
where $h=\pm 1$ denotes the helicity-over-chirality ratio.
In this case, the angle $\eta$ is given by\footnote{Note that
  in this notation, $- \bf q$ and $- \bf p$ are the three-momenta of
  the initial-state scalar and the final-state Majorana fermion, since
  their roles have been inverted.}
\begin{align}
  \label{eq:113}
  \eta_h^0 = \frac{1}{2 k p} \left[ -2 p_0 \omega_h + \Sigma_\phi
  \right] \, ,
\end{align}
where 
\begin{align}
  \label{eq:122}
  \Sigma_\phi = m_\phi^2-M^2-(\omega_h^2- k^2) \, .
\end{align}
In order to clarify the momentum relations, we revert the direction of
the three-momenta ${\bf q}$ and $\bf p$ so that they correspond to the
physical momenta of the incoming scalar and outgoing Majorana
fermion. The matrix element can be derived as
\begin{align}
  \label{eq:125}
  \left| \mathcal{M}_h(P,K) \right|^2 = g^2 Z_h \omega_h (p_0 -h \, {\bf p} \cdot \hat{\bf k}) \, ,
\end{align}
where the momentum flip ${\bf p} \to {\bf-p}$ compensates the helicity
flip $h \to -h$ and
\begin{align}
  \label{eq:126}
  Z_h = \frac{\omega_h^2-k^2}{2 m_\ell^2} \, 
\end{align}
is the residue of the modes and the angle for the reverted physical
momenta reads
\begin{align}
  \label{eq:127}
  \eta_h^0 = \frac{1}{2 k p} \left[ 2 p_0 \omega_h - \Sigma_\phi
  \right] \, .
\end{align}

\section{Application to Leptogenesis}
\label{sec:appl-lept}

When turning to leptogenesis with
\begin{align}
\delta {\mathcal L} = i \bar{N}_i \partial_\mu \gamma^\mu N_i - \lambda_{\nu,i
  \alpha} \bar{N}_i \phi^\dagger \ell_\alpha - \frac{1}{2} M_i
\bar{N}_i N_i^c + h.c.,
\end{align}
we sum over the two components of the doublets, particles and
antiparticles and the three lepton flavors. Thus we need to replace
$g^2$ by $4 (\lambda_\nu^\dagger
\lambda_\nu)_{11}$. Integrating over all neutrino momenta, the decay
density in equilibrium is
\begin{align}
\label{eq:69}
\gamma_D^{\rm eq}= \int \frac{{\rm d^3} p}{(2 \pi)^3} f_{N}^{\rm eq}(E) \, 
\Gamma_D= \frac{1}{2 \pi^2} \int_M^\infty {\rm d} E \: E \, p \, f_N^{\rm eq} 
\, 
\Gamma_D,
\end{align} 
where $E=p_0$, $f_N^{\rm eq}(E)=[\exp(E \beta)-1]^{-1}$ is the
equilibrium distribution of the neutrinos and $\Gamma_D=[1-f_N^{\rm
  eq}(E)] \, \Gamma$. We drop the subscript $1$ for the neutrino in
this section since it is the only occuring neutrino. The decay density
is also present in the Boltzmann equations,
\begin{align}
  \label{eq:128}
  \gamma(N \to H L) = \int \rmd \tilde{p}_N \rmd \tilde{p}_H
  \rmd \tilde{p}_L (2 \pi)^4 \delta^4(p_N - p_L - p_H) \left|
    \mathcal{M}_h \right|^2 f_N^\rmeq (1-f_L^\rmeq)
  (1+f_H^\rmeq) \, .
\end{align}
For the Higgs boson decay, the density reads
\begin{align}
  \label{eq:129}
    \gamma(H \to N L) = \int \rmd \tilde{p}_N \rmd \tilde{p}_H
  \rmd \tilde{p}_L (2 \pi)^4 \delta^4(p_N + p_L - p_H) \left|
    \mathcal{M}_h \right|^2 (1-f_N^\rmeq) (1-f_L^\rmeq)
  f_H^\rmeq \, ,
\end{align}
where the matrix elements for both decays are related and given by
equations \eqref{eq:130} and \eqref{eq:125}.

The thermal masses are given
by~\cite{Weldon:1982bn,Klimov:1981ka,Cline:1993bd, Elmfors:1993re}
\begin{align}
m_\phi^2(T)&=\left (\frac{3}{16} g_2^2+ \frac{1}{16} g_Y^2 +
\frac{1}{4} y_t^2 + \frac{1}{2} \lambda \right) T^2 \, , \nonumber \\
m_\ell^2(T)&=\left (\frac{3}{32} g_2^2+ \frac{1}{32} g_Y^2 \right ) T^2.
\end{align}
The couplings denote the SU(2) coupling $g_2$, the U(1) coupling
$g_Y$, the top Yukawa coupling $y_t$ and the Higgs self-coupling
$\lambda$, where we assume a Higgs mass of about $115$ GeV. The other Yukawa
couplings can be neglected since they are much smaller than unity and
the remaining couplings are renormalised at the first Matsubara mode, $2
\pi T$, as explained in reference~\cite{Giudice:2003jh} and, in more
detail, in reference~\cite{Kajantie:1995dw}.

\begin{figure}
\centering
\includegraphics[width=0.7 \textwidth]{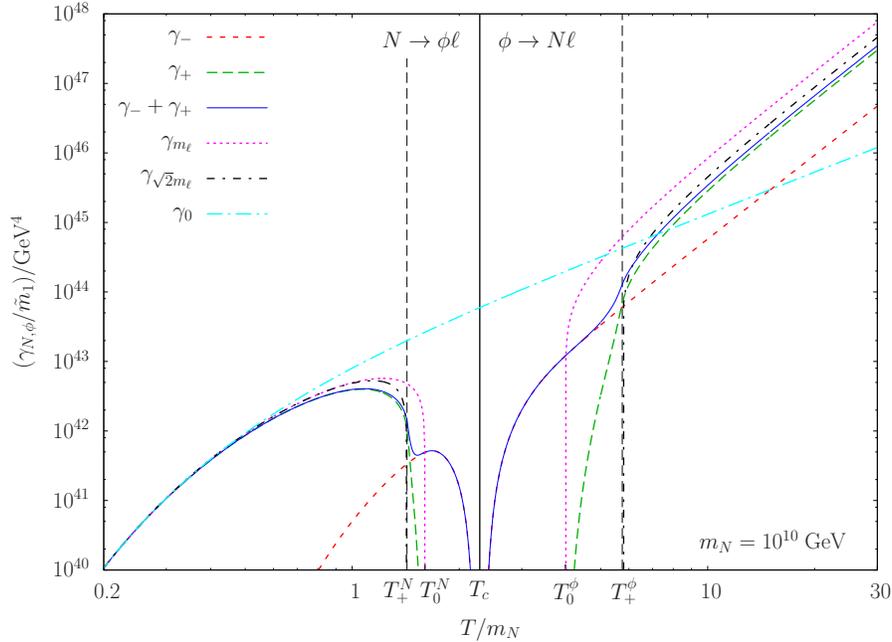}
\caption[Neutrino and Higgs boson decay densities]{The decay densities
  for the neutrino and the Higgs boson decay. We show the one-mode
  approach with the thermal mass as $\gamma_{m_{\ell}}$ and with
  the asymptotic mass as $\gamma_{\sqrt{2} m_{\ell}}$; Also the $T=0$ rate
  $\gamma_0$ and our two modes $\gamma_\pm$. The temperature
  thresholds are explained in the text.}
\label{comp} 
\end{figure}

In figure~\ref{comp}, we compare our consistent HTL calculation to the
one-mode approximation adopted by reference~\cite{Giudice:2003jh},
while we add quantum-statistical distribution functions to their
calculation, which equals the approach of using an approximated lepton
propagator $1/(\slashed{K}-m_\ell)$ as in equation~\eqref{approx}
\cite{Kiessig:2009cm}. In addition, we show the one-mode approach for
the asymptotic mass $\sqrt{2} \, m_\ell$. We evaluate the decay rates
for the $M_1=10^{10}$ GeV and normalise the rates by the effective
neutrino mass $\widetilde{m}_1$, which is often taken as
$\widetilde{m}_1 = 0.06 \, {\rm eV}$, inspired by the mass scale of
the atmospheric mass splitting.

In the one-mode approach, the decay is forbidden when the thermal
masses of Higgs boson and lepton become larger than the neutrino mass,
$M_1<m_\ell+m_\phi$ or $M_1< \sqrt{2} m_\ell +m_\phi$. Considering two
modes, the kinematics exhibit a more interesting behavior. For the
plus-mode, the phase space is reduced due to the larger
quasi-mass, and at $M_1=m_+(\infty)+m_\phi$, the decay is only
possible into leptons with small momenta, thus the rate drops
dramatically. The decay into the negative, quasi-massless mode is
suppressed since its residue is much smaller than the one of the
plus-mode. However, the decay is possible up to $M_1=m_\phi$. Due
to the various effects, the two-mode rate differs from the one-mode
approach by more than one order of magnitude in the interesting
temperature regime of $z=T/M_1 \gtrsim 1$. The $\sqrt{2} \,
m_\ell$-calculation is a better approximation to the plus-mode, but
still overestimates the rate, which is mainly due to the difference
between the momentum products in equations~\eqref{eq:111} and
\eqref{eq:120}, that is the helicity structure of the
quasiparticles. The residue also reduces the plus-rate, but the effect
is smaller since $Z_+$ is usually close to one.

At higher temperatures, when $m_\phi > M_1+m_\pm(k)$, the Higgs can
decay into neutrino and lepton modes and this process acts as a
production mechanism for neutrinos \cite{Giudice:2003jh}. The decay
$\phi \to N \ell_-$ is possible when $m_\phi > M_1$, while the decay
into $\ell_+$ is possible when $m_\phi > M_1 + m_\ell$. As for low
temperature, the rate $\gamma_+$ is unsuppressed only when $m_\phi >
M_1 + \sqrt{2} \, m_\ell$. Our decay density approaches the decay
density of reference~\cite{Giudice:2003jh} at high temperatures, but
is about a factor two below. This can be explained by the fact that
the phase space is smaller due to the larger mass of the lepton,
$m_\ell < m_\ell(k) < \sqrt{2} \, m_\ell$. Again, the asymptotic mass
calculation is a better approximation but still gives a larger rate
due to the momentum product and, to less extent, the residue. We see
that the decay rate rises as $\sim T^4$, instead of $\sim T^2$ as for
the vacuum rate $\gamma_0$. In the vacuum calculation, the squared
matrix element is proportional to $M_1^2$. In the finite temperature
calculation, it is proportional to $\Sigma_\phi = m_\phi^2-m_\ell^2(k)
- M_1^2$, so the dominant contribution is proportional to $m_\phi^2
\sim T^2$ and the rate rises by a factor $T^2$ faster than the vacuum
rate $\gamma_0$.

Summarising, we can distinguish five different thresholds for the
thermal decay rates we discussed. Going from low temperature to high
temperature, these are given by the following conditions:
\begin{align}
T_+^N: \quad M_1&=\sqrt{2} \, m_{\ell}+m_\phi \, , \nonumber \\
T_0^N: \quad M_1&=m_{\ell}+m_\phi \, , \nonumber \\
 T_c: \quad M_1&=m_\phi \, , \nonumber \\
 T_0^\phi: \quad m_\phi&=m_{\ell}+M_1 \, , \nonumber \\
 T_+^\phi: \quad m_\phi&=\sqrt{2} \, m_{\ell}+M_1 \, .
\end{align}
We will refer to these thresholds in the following chapters.

As discussed in detail in \cite{Kiessig:2009cm}, we confirm by
employing HTL resummation and finite temperature cutting rules that
treating thermal masses as kinematic masses as
in~\cite{Giudice:2003jh} is a reasonable approximation. However,
quantum statistical functions need to be included as they always
appear in thermal field theory. Moreover, the non-trivial two-mode
behaviour of the full HTL lepton propagator is not accounted for by
the conventional one-mode approach. We have calculated the effect of
the two modes in a general way, which is applicable to any decay and
inverse decay rates involving fermions at high temperature. Thus, this
calculation is a valuable tool for other particle interaction rates in
the early universe, as other leptogenesis processes, the thermal
production of gravitinos or the like.

The behaviour of the decay density of the two lepton modes can be
explained by considering the dispersion relations $\omega_\pm$ of the
modes and assigning momentum-dependent quasi-masses to them. The
thresholds for neutrino decay reported in
reference~\cite{Giudice:2003jh} are shifted and the decay density
shows deviations of more than one order of magnitude in the
interesting temperature regime $T/M_1 \sim 1$, which has implications
for the dynamics of leptogenesis as we will see in chapter
\ref{boltzmanneq}.


    \clearpage
\chapter{$\boldsymbol{\CP}$-Asymmetries \label{cpas}}

While the decay rates calculated in the last chapter are the first
important ingredient of the Boltzmann equations, the $\CP$-asymmetry
is the second quantity that enters the Boltzmann equations. This
chapter is devoted to the calculation of the $\CP$-asymmetry at finite
temperature for our different scenarios. At low temperature, we
calculate the $\CP$-asymmetry in neutrino decays, at high temperature
the $\CP$-asymmetry in Higgs boson decays

\section{Preliminaries}
\label{sec:preliminaries}

We are calculating the $\CP$-asymmetry in $N_1$ decays. We denote the
decaying $N_1$ by $N$ and the $N_2$ in the loop by $N'$. At $T=0$, the
$\CP$-asymmetry is defined in equation~\eqref{eq:51},
\begin{align}
\label{epsilonGamma}
\epsilon_0=\frac{\Gamma(N \rightarrow \phi \ell) - 
\Gamma(N \rightarrow \bar{\phi} \bar{\ell})}
{\Gamma(N \rightarrow \phi \ell) + 
\Gamma(N \rightarrow \bar{\phi} \bar{\ell})},
\end{align}
where $\Gamma$ are the decay rates of the heavy $N$s into Higgs boson
and lepton doublet and their $\CP$-conjugated processes. As we will see
in chapter \ref{boltzmanneq}, we have to calculate the $\CP$-asymmetry
via the integrated decay rates at finite temperature,
\begin{align}
\label{epsilongamma}
\epsilon_h(T)=\frac{\gamma^{T>0}(N \rightarrow \phi \ell_h) - 
\gamma^{T>0}(N \rightarrow \bar{\phi} \bar{\ell_h})}
{\gamma^{T>0}(N \rightarrow \phi \ell_h) + 
\gamma^{T>0}(N \rightarrow \bar{\phi} \bar{\ell_h})},
\end{align}
where we define the $\CP$-asymmetry for each lepton mode, denoted by
$h$, as introduced in equation~\eqref{eq:121}. We have
\begin{align}
\gamma^{T>0}=\int \frac{{\rm d^3}p_N}{(2 \pi)^3} f_N(p_N)
\Gamma^{T>0}(P_N^\mu) \, ,
\end{align}
where $f_N$ is the distribution function of the neutrinos and
$P_N^\mu$ the neutrino momentum. At $T=0$, we write
\begin{align}
\label{eq:52}
\Gamma(P^\mu)=\frac{M_1}{p_0} \Gamma_{\rm rf},
\end{align}
where $M_1$ and $p_0$ are the mass and the energy of the neutrino and
$\Gamma_{\rm rf}$ is the decay rate in the rest frame of the
neutrino. The integration over the momentum cancels out and the $\CP$
asymmetry via $\gamma$ is the same as via $\Gamma$.  At finite
temperature, however, the thermal bath breaks Lorentz invariance and
the preferred frame of reference for calculations is the rest frame of
the thermal bath. The momentum dependence of the decay rate cannot be
formulated as in equation~\eqref{eq:52} and the $\CP$-asymmetry as
defined in equation~\eqref{epsilonGamma} is momentum dependent. Since
we want to decouple the $\CP$-asymmetry from the momentum integration
in the Boltzmann-equations, the definition in
equation~\eqref{epsilongamma} is the appropriate one.

The CP asymmetry in equilibrium can be written as
\begin{align}
\epsilon_{\gamma h}^{\rm eq}(T)=\frac{ \int \frac{{\rm d^3} p}{(2 \pi)^3}
  f_N^{\rm eq} (\Gamma_{Dh}-\tilde{\Gamma}_{Dh})}
{\int \frac{{\rm d^3} p}{(2 \pi)^3}
  f_N^{\rm eq} (\Gamma_{D h}+\tilde{\Gamma}_{D h})},
\end{align}
where $\Gamma_D=\Gamma(N \rightarrow \ell \phi)$ and
$\tilde{\Gamma}_D= \Gamma(N \rightarrow \bar\ell \bar\phi)$ are the
decay rate and the $\CP$-conjugated decay rate.

We see from equations~\eqref{eq:20}, \eqref{eq:130} and \eqref{eq:69},
that the decay density is written as
\begin{align}
  \label{eq:132}
  \gamma_{Dh} = \frac{1}{2 \pi^2} \int \rmd E E p f_N^\rmeq \Gamma_{D
    h} = \frac{1}{4 (2 \pi)^3} \int \rmd E \rmd k \frac{k}{\o_h} f_N Z_D
  \left| \mathcal{M}_h \right|^2 \, .
\end{align}
where 
\begin{align}
\label{eq:167}
Z_D = (1-f_N^{\rm eq}) (1+f_\phi^\rmeq-f_\ell^\rmeq)= (1+f_\phi^\rmeq) (1-f_\ell^\rmeq)
\end{align}
is the statistical factor for the decay, with Bose-enhancement and
Fermi-blocking.  In the denominator of the $\CP$-asymmetry, it is
sufficient to take the tree-level matrix element, $\Big|
\mathcal{M}_{\rm tree} \Big|^2=\Big| \widetilde{\mathcal{M}}_{\rm
  tree} \Big|^2$. The $\CP$-asymmetry reads
\begin{align}
\epsilon_{\gamma h}(T) &= \frac{\int {\rm d}E \; {\rm d} k \;
  \frac{k}{\omega_h} \; f_N \; Z_D \;
  (|\mathcal{M}_h|^2-|\widetilde{\mathcal{M}}_h|^2)} 
{2 \; \int {\rm d}E \;
  {\rm d} k \; \frac{k}{\omega_h} \; f_N \; Z_D \; |\mathcal{M}_h|^2}
\nonumber \\
&=  \frac{1}{\gamma_h(N \to LH)} \frac{1}{4 (2 \pi)^3} \int \rmd E \rmd
k \frac{k}{\o_h} Z_D \left( \Big| \mathcal{M}_h \Big|^2 - \Big|
    \widetilde{\mathcal{M}}_h \Big|^2 \right) \, .
\end{align}

The $\CP$-asymmetry arises as the interference between tree-level and
one-loop diagrams in the decay, so we write $\mathcal{M} =
\mathcal{M}_0 + \mathcal{M}_1$, where $\mathcal{M}_0$ is the
tree-level amplitude and $\mathcal{M}_1$ the sum of all one-loop
amplitudes.  The matrix elements can be decomposed as
$\mathcal{M}_i=\lambda_i I_i$ such that the $\CP$-conjugated matrix
element is $\widetilde{\mathcal{M}_i}= \lambda_i^* I_i$. Here,
$\lambda_i$ includes the couplings and $I_i$ accounts for the
kinematics.  Thus,
\begin{align}
|\mathcal{M}|^2-|\widetilde{\mathcal{M}}|^2= - 4 \; {\rm Im} \lambda_{CP} \;
{\rm Im} I_{CP},
\end{align}
where $\lambda_{CP}=\lambda_0 \lambda_1^*$ and $I_{CP}= I_0 I_1^*$.

\section{The Vertex Contribution}
\label{sec:vertex-contribution}

\subsection{Going to finite temperature}
\label{sec:going-finite-temp}

\begin{figure}
\begin{center}
\includegraphics{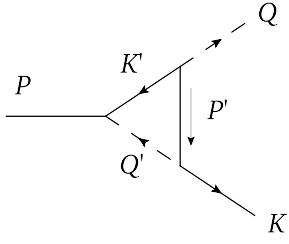}
\caption[The momentum assignments for the vertex contribution]{The
  momentum assignments for the vertex contribution to the $\CP$
  asymmetry. The solid lines without arrows are neutrinos, the ones
  with arrows the leptons and the dashed lines the Higgs bosons. All
  momenta are flowing from left to right and $P'$ as indicated.}
\label{momenta}
\end{center}
\end{figure}
Calculating the imaginary part of the kinematic term $\rmIm I_{CP}$
amounts to calculating the imaginary part of the one-loop diagram
since the tree-level diagram is real. As explained in
section~\ref{sec:cp-asymmetry}, there are two one-loop diagrams for
the neutrino decay, the vertex diagram and the self-energy
diagram. The vertex diagram is shown in figure~\ref{momenta}, along
with the momentum assignments. The coupling is
\begin{align}
  \lambda_{CP}=\lambda_0 \lambda_1^* = [(\lambda^\dagger \lambda)_{jk}]^2
  g_{SU(2)},
\end{align}
where $g_{SU(2)}=2$ denotes the sum over the Higgs and lepton
doublets, $\lambda$ is the Yukawa coupling between neutrino and Higgs
and lepton doublet, $j=1$ is the decaying neutrino family, $k=2$ is the
family of the neutrino in the loop and we have summed over all fermion
spins and the lepton families, both external and in the loop.
Moreover,
\begin{align}
\label{eq:217}
I_V= - {\rm i} \int \frac{{\rm d}^4 k'}{(2 \pi)^4} \left[ M_k
  \Delta_{N'} \Delta_{\phi'} (\overline{u}_\ell P_R u_N) (\overline{u}_N
  P_R S_{\ell'} P_L u_\ell) \right]^*,
\end{align}
where $p'$, $q'$ and $k'$ are the neutrino, Higgs boson and lepton
momentum in the loop, $\Delta_{N'}=(P'^2-M_k^2)^{-1}$ is the
denominator of the loop neutrino propagator, $\Delta_{\phi'}$
accordingly for the loop Higgs, $P_{R,L}$ are projection operators,
$S_{\ell'}$ is the loop lepton propagator and $u_N$ and $u_\ell$ are
the external neutrino and lepton spinors.

The external fermions are thermal quasiparticles and can be written as
spinors $u_\ell^\pm$~\cite{LeBellac:1996} which are eigenstates of
$(\gamma_0 \mp \hat{\bf k} \cdot \boldsymbol{\gamma})$ and have
modified dispersion relations as explained in
section~\ref{sec:effect-prop-disp}. From chapter~\ref{decayrate}, we know
that
\begin{align}
\label{eq:53}
\frac{1}{2} \sum_{s}|\mathcal{M}^s_\pm(P,K)|^2=g^2 \frac{\omega_\pm^2-k^2}{2
  m_\ell^2} \omega_\pm \left (p_0 \mp p \eta_\pm \right ),
\end{align}
where $s$ denotes the spin of the neutrino. 
We can also write the matrix element as
\begin{align}
\label{eq:54}
\frac{1}{2} \sum_{s}|\mathcal{M}^s_\pm(P,K)|^2=
\frac{1}{2} \sum_{s} g^2 (\overline{u}_\ell^\pm P_R u_N^s) (\overline{u}_N^s P_L
u_\ell^\pm) \, .
\end{align}
From equations~\eqref{eq:53} and~\eqref{eq:54} we derive a rule for
multiplying the spinors of the lepton states,\footnote{For the
  antiparticle spinors $v$, we replace $K$ by $-K$ and get
  \begin{align}
    \label{eq:202}
    v_\ell^\pm(K) \overline{v}_\ell^\pm(K) = - Z_\pm
\omega_\pm (\gamma_0 \pm \hat{\bf k} \cdot
\boldsymbol{\gamma}) \, .
  \end{align}
}
\begin{align}
\label{eq:208}
u_\ell^\pm(K) \overline{u}_\ell^\pm(K) = Z_\pm
\omega_\pm (\gamma_0 \mp \hat{\bf k} \cdot
\boldsymbol{\gamma}),
\end{align}
where
\begin{align}
  Z_\pm=\frac{\omega_\pm^2 - k^2}{2 m_\ell^2}
\end{align}
is the quasiparticle residuum.

The HTL lepton propagator is given in equations~\eqref{eq:35}
and~\eqref{fermprop}
\begin{align}
  S^*(K)&= - \frac{1}{2} \Delta_+(K) (\gamma_0-\hat{\bf k} \cdot
  \boldsymbol{\gamma}) - \frac{1}{2} \Delta_-(K) (\gamma_0+\hat{\bf k}
  \cdot
  \boldsymbol{\gamma})\, , \nonumber \\
  \Delta_\pm(K)&=\left [ -k_0 \pm k + \frac{m_\ell^2}{k} \left ( \pm1 -
      \frac{\pm k_0 - k}{2k} \ln \frac{k_0+k}{k_0-k} \right ) \right
  ]^{-1}
\end{align}
and $m_\ell$ is the thermal lepton mass. The Higgs boson propagator is 
\begin{align}
\D_{\phi'}= \frac{1}{Q'^2-m_\phi^2},
\end{align}
where $m_\phi$ is the thermal Higgs boson mass. At finite temperature,
we sum over the Matsubara modes,
\begin{align}
\int \frac{{\rm d}k'_0}{2 \pi} \rightarrow {\rm i} T \sum_{k'_0} \, ,
\end{align}
where
\begin{align}
k'_0 = (2 n + 1) \pi {\rm i} T,
\end{align}
since we are integrating over a fermion momentum.

The spin and helicity sum are evaluated as
\begin{align}
\label{eq:155}
\sum_{s,h'} (\overline{u}_\ell^h P_R u_N^s) (\overline{u}_N^s P_R S_{\ell'}^{h'} P_L
u_\ell^h) =- \sum_{h'} Z_h \omega_h M_j \Delta_{h'} (1-h h' {\bf
  \hat{k} \cdot \hat{k'}}),
\end{align}
where $h$ and $h'$ are the ratios of helicity over chirality for the
external and the loop lepton.  The integral reads
\begin{align}
  \label{eq:131}
  I_V= - T \sum_{k_0', h'} \int \frac{{\rm d}^3 k'}{(2 \pi)^3} M_k M_j Z_h \o_h
  \left[
  \D_{N'} \D_{\phi'} \Delta'_{h'} \right]^* H_-^{h h'} \, ,
\end{align}
where $H_- = 1 - h h' {\bf \hat{k} \hat{k}'}$.

In order to carry out the Matsubara sum, we use the
Saclay-representation for the propagators. For the Higgs propagator it
is given in equation~\eqref{eq:11},
\begin{align}
\Delta_{\phi}'=- \int_0^\beta {\rm d} \tau \; {\rm e}^{q_0' \tau}
\frac{1}{2 \omega_{q'}} \{ [1+f_{\phi}'(\omega_{q'})] {\rm e}^{- \omega_{q'}
  \tau} +n'_{\phi}(\omega_{q'}) {\rm e}^{\omega_{q'} \tau} \},
\label{saclayhiggs}
\end{align}
where $\omega_{q'}= \sqrt{q'^2+m_\phi^2}$ is the on-shell Higgs energy
with the thermal Higgs mass $m_\phi$ and $f_{\phi'}$ is the
Bose-Einstein distribution for the Higgs bosons with energy
$\omega_{q'}$.  For the lepton propagator the Saclay representation is
given in equation~\eqref{eq:186},
\begin{align}
  \Delta_\pm'=\Delta'(h')=- \int_0^\beta {\rm d} \tau' \; {\rm
    e}^{k_0' \tau'} \int_{-\infty}^\infty {\rm d} \omega'
  \rho_{h'}(\omega',k') [1-n'_\ell(\omega')] {\rm e}^{-\omega'
    \tau'},
\label{saclaylepton}
\end{align}
where
\begin{align}
\rho_{h'}(\omega',k')= - \frac{\omega'^2-k'^2}{2 m_\ell^2} [
  \delta(\omega'-\omega'_{h'}) +\delta(\omega'+\omega'_{-h'})]
\end{align}
is the pole part of the lepton spectral density with the two solutions
$\omega'_{h'}=\omega'_\pm$ for
$\Delta_\pm(\omega'_\pm,k')^{-1}=0$. We are only interested in the
pole part since it corresponds to the two quasiparticle
modes. Moreover, $m_\ell$ is the thermal lepton mass and
$f_{\ell'}$ the lepton Fermi-Dirac distribution. The neutrino propagator
reads
\begin{align}
\D_{N'}=-\int_0^\beta {\rm d} \tau'' \; {\rm e}^{p_0' \tau''}
\frac{1}{2 \omega_{p'}} \{ [1-f_{N'}(\omega_{p'})] {\rm e}^{-\omega_{p'}
  \tau''}-f_{N'}(\omega_{p'}) {\rm e}^{\omega_{p'} \tau''} \},
\end{align}
where 
\begin{align}
\omega_{p'}=\sqrt{p'^2+M_k^2}
\end{align}
is the neutrino on-shell energy, which is unaffected by thermal
corrections since the coupling to the bath is negligible and $f_{N'}$
is the Fermi-Dirac distribution for the neutrinos.  As usual, we can
write $p_0= {\rm i} \, (2m+1) \pi T$ as Matsubara frequency and later on
continue it analytically to real values of $p_0$. In particular ${\rm
  e}^{p_0 \beta}=-1$.

\subsection{Frequency sums for HTL fermion propagators}
\label{sec:frequency-sums-ii}


In order to deal with the HTL lepton propagator, we derive frequency
sums for the propagator parts $\D_\pm(K)$ of a fermion propagator. We
write the propagator in the Saclay representation as
\begin{align}
\label{eq:187}
\tilde{\Delta}_h(K)&=- \int_0^\beta \rmd \tau \, \rme^{k_0 \tau}
\Dt_h(\tau,{\bf k})\, , \nonumber \\
\Dt_h(\tau,{\bf k})&= \int_{-\infty}^\infty \rmd \omega \, \rho_h
f_F(-\omega) \rme^{-\omega \tau} \, .
\end{align}
Since we are only interested in the pole contribution, we write the
corresponding spectral density as
\begin{align}
  \label{eq:190}
  \rho_h^{\rm pole} &= - Z_h
  [\delta(\omega-\omega_h)+\delta(\omega+\omega_{-h})]=- \sum_s Z_{s
    h} \delta(\omega-s \omega_{s h}) \, , 
\end{align}
where $sh$ in $Z_{sh}$ and $\o_{sh}$ denotes the product of $s$ and
$h$, that is, $Z_{sh}=Z_+$ for $s=h=-1$ for example. We have for the
propagator
\begin{align}
  \label{eq:209}
  \Dt_h^{\rm pole}(\tau,{\bf k})&=- \int_{-\infty}^{\infty} \rmd \omega \, \sum_s
Z_{s h} \delta(\omega-s \omega_{s h}) f_F(-\omega) \rme^{-\omega \tau}
\nonumber \\
&= - \sum_s Z_{sh} f_F(-s \omega_{sh}) \rme^{-s \omega_{sh}\tau} 
= \sum_s \Dt_{h,s}^{\rm pole}(\tau,{\bf k}), \nonumber \\
\Dt_h^{\rm pole}(K)&= \sum_s Z_{sh} f_F(-s \omega_{sh}) \int_0^\beta
\rmd \tau \, \rme^{(k_0-s \omega_{sh})\tau} = \sum_s \Dt^{\rm
  pole}_{h,s}(K)\, , \nonumber \\
\Dt_{h,s}^{\rm pole}(K)&= Z_{sh} f_F(-s \omega_{sh}) \int_0^\beta
\rmd \tau \, \rme^{(k_0-s \omega_{sh})\tau} \nonumber \\
&= - Z_{sh} \frac{1}{k_0-s\omega_{sh}}\, ,
\end{align}
where $Z_{s h} = (\omega^2-k^2)/(2 m_\ell^2)$ is the quasiparticle
residuum.

In dealing with frequency sums of bare thermal propagators, it is very
convenient to write
\begin{align}
\Delta_s(K)= \Delta_{-s}(-K).
\end{align}
As explained in section~\ref{sec:dirac-field}, replacing a boson by a fermion
amounts to replacing $f_B(\omega)$ by $-f_F(\omega)$. Moreover,
calculating a frequency sum of $k_0$ times the propagators amounts to
replacing $k_0$ with $s \omega$ as in
\begin{align}
T \sum_{k_0} k_0 \Delta_{s_1}(K) \Delta_{s_2}(P-K)= s_1 \omega T \sum_{k_0}
\Delta_{s_1}(K) \Delta_{s_2}(P-K),
\end{align}
where $\omega=\sqrt{k^2+m^2}$ and $m$ is the mass of the first
boson. The same holds for fermions.

It is straightforward to work out the frequency sums for the resummed
lepton propagator,
\begin{align}
\label{eq:160}
T \sum_{k_0} \Dt_{h,s_1}^{\rm pole}(k_0,\omega)
\Delta_{s_2}(p_0-k_0,\omega) = Z_{s_1 h}
\frac{s_2}{2 \omega} \frac{1-f_F(s_1
  \omega_{s_1 h})+f_B(s_2 \omega)}{p_0-s_1 \omega_{s_1 h}-s_2 \omega},
\end{align}
where the other necessary frequency sums can be derived from this by
making the appropriate substitutions.

\subsection{The frequency sum for the vertex contribution}
\label{sec:freq-sums-vert}

We calculate the frequency sum of the three propagators in the vertex
loop by partial fractioning
\begin{align}
\Dt_{s,h'}^{\rm pole} \D_{s_{\phi'}} \Dt_{s_{N'}}=C_{s\phi N'} \left[
  \frac{s_{\phi'}}{2\omega_{\phi'}}\Dt_{s,h'}^{\rm
    pole}\Dt_{s_{N'}}-\frac{s_{N'}}{2\omega_{N'}}\Dt_{s,h'}^{\rm pole}\D_{s_{\phi'}}
\right] \, .
\end{align}
We are using $\Delta_{N'}(P') = \Delta_{N'}(-P')$ and
\begin{align}
C_{s\phi N'}=\frac{1}{k_0-s_{\phi'}
  \omega_{\phi'}+s_{N'}\omega_{N'}}.
\end{align}
The frequency sum is given by
\begin{align}
\label{eq:157}
T\sum_{k_0'} \Dt_{s,h'}^{\rm pole}\D_{s\phi'}\Dt_{sN'}=Z_{sh'}\frac{s_{\phi'}
s_{N'}}{4\omega_{\phi'}\omega_{N'}} C_{s\phi N'}
\left[\frac{Z_{shN'}}{N_{shN'}}-
  \frac{Z_{sh\phi'}}{N_{sh\phi'}}\right],
\end{align}
where
\begin{align}
Z_{shN'}&=1-f_F(s\omega_{sh'})-f_F(s_{N'}\omega_{N'})\, , \nonumber \\
Z_{sh\phi'}&=1-f_F(s\omega_{sh'})+f(s_{\phi'}\omega_{\phi'}) \, ,
\end{align}
and
\begin{align}
N_{shN'}&=q_0-s \omega_{sh'}-s_{N'}\omega_{N'}\, , \nonumber \\
N_{sh\phi'}&=p_0-s \omega_{sh'}-s_{\phi'}\omega_{\phi'} \, .
\end{align}
Summing over all propagator parts and the helicity-over-chirality ratios, we get
\begin{align}
T \sum_{k_0'}\sum_{h'} \D_{N'}\D_{\phi'}\Delta'H_-= \sum_{h'}
\frac{Z_{h'}}{4 \omega_{q'} \omega_{p'}} \left\{E_-H_-+E_+H_+\right\},
\end{align}
where
\begin{align}
  \label{eq:117}
 H_\pm &=1 \pm h h' \xi \, , \nonumber \\
\xi &= \bf{\hat{k} \cdot \hat{k'}} \, .
\end{align}
The coefficients $E_\pm$ are given by
\begin{align}
E_-&=F^{\phi N}A_\ell^{\phi'}-F^{\ell'
  N}A_\ell^\ell -F^{\phi\ell'}A_\ell^0+F^{\ell'\ell'}A_\ell^{N'},\\
E_+&=F^{N'\phi'}A_\ell^{\phi'}-F^{0\phi'}A_\ell^\ell -F^{N'0}A_\ell^0+F^{00}A_\ell^{N'},
\end{align}
and the coefficients $F^{ij}$ read
\begin{align}
\label{eq:114}
F^{\phi N} & = B_\phi^\phi-B_N^N, & F^{N'\phi'} & = B_\phi^{N'}-B_N^{\phi'}, \nonumber \\
F^{\ell' N} & = B_\phi^{\ell'}-B_N^N, & F^{0\phi'} & = B_\phi^0-B_N^{\phi'}, \nonumber \\
F^{\phi\ell'} & = B_\phi^\phi- B_N^{\ell'}, & F^{N'0} & = B_\phi^{N'}- B_N^0, \nonumber \\
F^{\ell'\ell'} & = B_\phi^{\ell'}- B_N^{\ell'}, & F^{00} & = B_\phi^0- B_N^0.
\end{align}
The factors $B_{N/\phi}$ and $A_\ell$ are given by
\begin{align}
\label{eq:115}
B_{N/\phi}^\psi =& \frac{Z_{N/\phi}^\psi}{N_{N/\phi}^\psi} \, , &
A_\ell^\psi =& \frac{1}{N_\ell^\psi} \, ,
\end{align}
where the numerators and denominators read
\begin{align}
\label{eq:106}
  N_N^N & = p_0-\omega'-\omega_{q'}, & N_\ell^\ell & =
  k_0-\omega_{q'}-\omega_{p'}, & N_\phi^\phi 
& = q_0-\omega'-\omega_{p'}, \nonumber \\
  N_N^0 & = p_0+\omega'+\omega_{q'}, & N_\ell^0 & =
  k_0+\omega_{q'}+\omega_{p'}, & N_\phi^0 
& = q_0+\omega'+\omega_{p'}, \nonumber \\
  N_N^{\ell'} & = p_0-\omega'+\omega_{q'}, & N_\ell^{\phi'} & =
  k_0-\omega_{q'}+\omega_{p'}, 
& N_\phi^{\ell'} & = q_0-\omega'+\omega_{p'}, \nonumber \\
  N_N^{\phi'} & = p_0+\omega'-\omega_{q'}, & N_\ell^{N'} & =
  k_0+\omega_{q'}-\omega_{p'}, & N_\phi^{N'} & = 
q_0+\omega'-\omega_{p'} \, ,
\end{align}
and
\begin{align}
\label{eq:116}
Z_N^N& = 1-f_{\ell'}+f_{\phi'} \, ,
& Z_\phi^\phi &  =  1-f_{\ell'}-f_{N'} \, , \nonumber \\
Z_N^0 & = - (1 - f_{\ell'} + f_{\phi'}) \, , 
& Z_\phi^0 &  = - (1 - f_{\ell'} - f_{N'}) \, ,  \nonumber \\
Z_N^{\ell'} & = - (f_{\ell'} + f_{\phi'}) \, , 
& Z_\phi^{\ell'} & = - (f_{\ell'} - f_{N'}) \, , \nonumber \\
Z_N^{\phi'} & = f_{\ell'} + f_{\phi'} \, , 
& Z_\phi^{N'} & = f_{\ell'} - f_{N'} \, .
\end{align}
We can write
\begin{align}
\begin{split}
T \sum_{k_0'}\sum_{h'} \D_{N'}\D_{\phi'}\Delta_{h'}H_-= \sum_{h'}
\frac{Z_{h'}}{4 \omega_{q'} \omega_{p'}} & \left\{ 
\left[ F^{\phi N}A_\ell^{\phi'}-F^{\ell'
    N}A_\ell^\ell -F^{\phi\ell'}A_\ell^0+F^{\ell'\ell'}A_\ell^{N'} \right] H_- \right. \\
& \hspace{-6pt} + \left. \left[
  F^{N'\phi'}A_\ell^{\phi'}-F^{0\phi'}A_\ell^\ell -F^{N'0}A_\ell^0+F^{00}A_\ell^{N'}
  \right] H_+
\right\}
\end{split}
\end{align}
or, more explicitly,
\begin{align}
\label{eq:102}
\begin{split}
T \sum_{k_0'}\sum_{h'} &
\D_{N'}\D_{\phi'}\Delta_{h'}H_-=\\
= \sum_{h'}
\frac{Z_{h'}}{4 \omega_{q'} \omega_{p'}} 
& \left\{ 
\left[ \left(B_\phi^\phi-B_N^N\right)A_\ell^{\phi'}-\left( B_\phi^{\ell'}-B_N^N \right)
  A_\ell^\ell - \left(B_\phi^\phi- B_N^{\ell'}\right)A_\ell^0+
  \left(B_\phi^{\ell'}- B_N^{\ell'} \right) A_\ell^{N'}
  \right] H_- \right.\\
& \hspace{-6pt} + \left. \left[
  \left(B_\phi^{N'}-B_N^{\phi'} \right) A_\ell^{\phi'}- \left( B_\phi^0-B_N^{\phi'}
  \right) A_\ell^\ell -
  \left( B_\phi^{N'}- B_N^0 \right) A_\ell^0+ \left( B_\phi^0- B_N^0 \right) A_\ell^{N'}
  \right] H_+
\right\} \, .
\end{split}
\end{align}

\section{The Self-Energy Contribution}
\label{sec:frequency-sums-self}

\begin{figure}
\begin{center}
\includegraphics{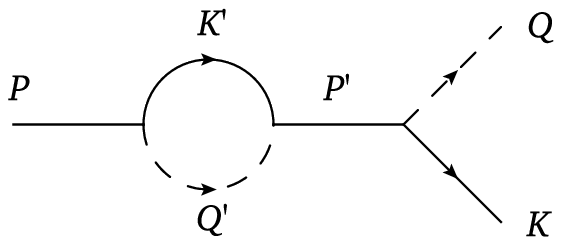}
\includegraphics{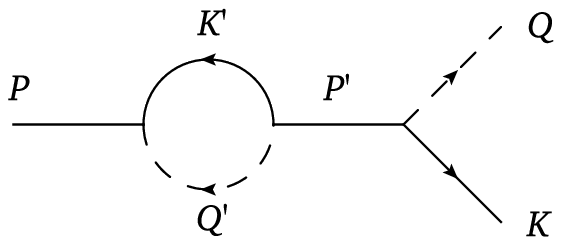}
\caption[The momentum assignments for the self-energy
contribution]{The momentum assignments for the self-energy
  contribution. The solid lines without arrows are neutrinos, the ones
  with arrows the leptons and the dashed lines the Higgs bosons.}
\label{selfenergy}
\end{center}
\end{figure}

For the self-energy contribution, the integral $I_S=I_V$ is the same as
for the vertex contribution, only the momentum relations are different
(cf.~figure~\ref{selfenergy}). The left diagram does not give a
contribution since the combination of couplings, $|(\lambda^\dagger
\lambda)_{jk}|^2$, does not have an imaginary part.
We can use the Saclay representation for the Higgs and the lepton
propagator $\D_{\phi'}$ and $\Delta'_\pm$ as in
Eqns.~\eqref{saclayhiggs} and \eqref{saclaylepton}, remembering the
different momentum relations. The neutrino propagator simply reads
\begin{align}
\D_{N'}=\frac{1}{M_j^2-M_k^2},
\end{align}
since the internal neutrino momentum $P'$ is the same as the external
neutrino momentum $P$.

We can calculate the frequency sum directly using $\rme^{p_0
  \beta}=-1$,
\begin{align}
T \sum_{k_0'} \rme^{q_0' \tau} \rme^{k_0' \tau'} 
= \rme^{p_0 \tau} \delta(\tau'-\tau) \, .
\end{align}
and get
\begin{align}
T \sum_{k_0'} \D_{\phi'} \Delta'(h')=-
  \int_{-\infty}^\infty \rmd \omega' \rho'(h') \frac{1}{2 \omega_{q'}}
  \left( B_N^N - B_N^{\ell'} \right)\, .
\end{align}
Alternatively, we can use equation \eqref{eq:160} and write
\begin{align}
  \label{eq:159}
  T \sum_{k_0'} \Dt_{h',s}^{\rm pole}(k_0',\omega')
\Delta_{s_{\phi'}}(p_0-k_0',\omega_{q'}) = Z_{sh'}
\frac{s_{\phi'}}{2 \omega_{q'}} \frac{1-f_F(s
  \omega_{sh'})+f(s_{\phi'} \omega_{q'})}{p_0-s\omega_{sh'}-s_{\phi'}
  \omega_{q'}} \, .
\end{align}
Both calculations lead to
\begin{align}
T \sum_{k_0'} \sum_{h'}
  \D_{\phi'} \Delta' H_- = \sum_{h'}
  \frac{1}{2 \omega_{q'}} Z_{h'} [(B_N^N - B_N^{\ell'})
    H_- + (B_N^{\phi'} - B_N^0) H_+]\, .
\end{align}

\section{Imaginary Parts}
\label{sec:imaginary-parts}

The terms $B_{N/\phi}^\psi$ and $A_\ell^\psi$ in the vertex
contribution correspond to the three vertices where the denominator
fulfills certain momentum relations when set to zero: the $B_N$-terms
correspond to the vertex with an incoming $N_1$ and $\{\ell',\phi'\}$
in the loop, the $B_\phi$-terms to the vertex with an outgoing $\phi$
and $\{N_2,\ell'\}$ in the loop, and the $A_\ell$-terms to the vertex
with an outgoing $\ell$ and $\{N_2,\phi'\}$ in the loop. As an
example, the term
\begin{align}
B_N^N=\frac{1-f_{\ell'}+f_{\phi'}}{p_0-\omega'-\omega_{q'}}
\end{align}
corresponds to the incoming neutrino decaying into the lepton and
Higgs boson in the loop. Thus, the terms correspond to cuttings
through the two loop lines adjacent to the vertex, however, a
correspondence with the circlings of the
RTF~\cite{LeBellac:1996,Garny:2010nj} is not obvious. Among these
cuts, only the ones which correspond to a $N_1$ or $N_2$ decaying into
a Higgs boson and a lepton are kinematically possible at the
temperatures where neutrino decay is allowed, that is where $M_1 <
m_\phi$. These terms are $B_N^N$, $A_\ell^{N'}$ and $B_\phi^{N'}$.

\subsection{Regarding the $N_2$ cuts}
\label{sec:other-cuts}

\begin{figure}
  \centering
  \includegraphics[scale=0.8]{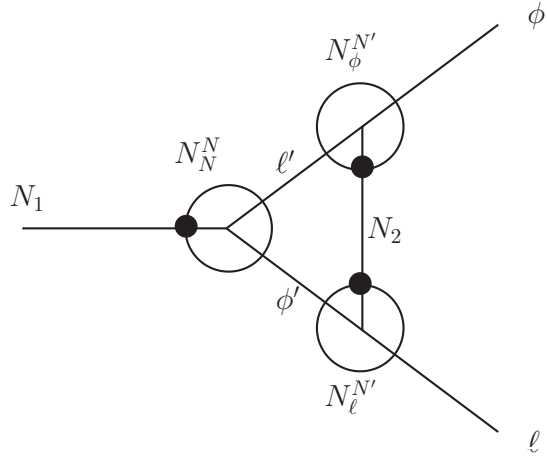}
  \caption[The cuts through the vertex contribution]{The cuts through the vertex contribution at finite
    temperature. The cuts are closed to form circles and the line
    that denotes the decaying particle in the corresponding $1 \to 2$
    process is indicated by a blob.}
  \label{fig:vertexcuts}
\end{figure}

The diagrams develop an imaginary part when one of the denominators of
the relevant terms $B_N^N$, $A_\ell^{N'}$ and $B_\phi^{N'}$
vanishes. The contributions from these denominators, $N_N^N$,
$N_\ell^{N'}$ and $N_\phi^{N'}$, correspond to the three possible cuts
shown in figure~\ref{fig:vertexcuts}. As explained in
section~\ref{sec:cp-asymmetry}, the contribution from $N_N^N$ is the
only possible cut at zero temperature. At finite temperature, the
other two cuts correspond to exchanging energy with the heat
bath. When choosing the imaginary parts corresponding to these two
cuts, the loop momentum $K'$ is of the order of $M_2$. Since we assume
a strong hierarchy $M_2 \gg M_1$, the thermal factors $f_{\phi'},
f_{\ell'}$ and $f_{N'}$ are suppressed by the large loop momentum and
the contributions become very small. In fact, they turn out to be
numerically irrelevant in the hierarchical limit.
The physical interpretation of this is as follows: Consider for
example the cut through $\{\ell',N_2\}$, which is given by a vanishing
denominator $N_\phi^{N'}$. The corresponding thermal weighting factor
is the numerator $Z_\phi^{N'}=f_{\ell'}-f_{N_2} = f_{\ell'}
(1-f_{N_2}) - (1-f_{\ell'}) f_{N_2}$. It corresponds to two processes:
absorption of a neutrino from the thermal bath and induced emission of
a lepton, or absorption of a lepton and induced emission of a
neutrino. The phase space distribution of the $N_2$s in the bath is
suppressed due to their large mass and also the distribution of
$\ell$s that have momenta large enough to fulfill momentum
conservation in the process is suppressed, so the process is
suppressed. Therefore, the thermal factors suppress the contribution
from the $N_2$-cuts. Only when we have degenerate masses $M_2 \gtrsim
M_1$, these cuts will give a contribution similar to the one from
$N_N^N$. In this case\footnote{Note that there is a mass range for
  $M_2$ where we have a contribution from the $N_2$ cuts but no
  resonant enhancement by the self-energy contribution, which becomes
  relevant when $\Delta M \equiv M_2-M_1\sim \Gamma$. This mass range is at $M_1
  \sim \Delta M \gg \G$}, the energy and temperature scales that
correspond to $N_1$ and $N_2$ processes are not clearly separated and
one has to account for the possibility of an asymmetry creation by
$N_2$ as well. Implications of these cuts were discussed in
reference~\cite{Garbrecht:2010sz}. We do not consider the influence of
this cuts, since we are in the hierarchical limit, but we present the
analytical expression in appendix~\ref{cha:other-cuts}.

\subsection[Vertex cut through ${\{\ell',\phi'\}}$]{Vertex cut through $\boldsymbol{\{\ell',\phi'\}}$}
\label{sec:vertex-cut:-external-1}

The imaginary part from $N_N^N$, which implies cutting through the
lepton and Higgs boson in the loop, is the only cut that is also
possible at zero temperature and the only vertex cut that contributes
in the hierarchical limit\footnote{The corresponding $\CP$-asymmetry
  has been calculated in reference~\cite{Giudice:2003jh}, but with a
  thermal factor $1-f_{\ell'}+f_{\phi'}-2f_{\ell'} f_{\phi'}$ instead
  of the correct $1-f_{\ell'}+f_{\phi'}$. For details, see
  reference~\cite{Garny:2010nj}.}. We denote the angle between $\bf p$
and $\bf k'$ with $\eta'$,
\begin{align}
\eta'=\frac{\bf p \cdot k'}{p k'}.
\end{align}
Then
\begin{align}
{\rm Im}  \left(\int_{-1}^1 {\rmd} \eta' \frac{1}{N_N^N} \right) &= -  \pi
\int_{-1}^1 \rmd \eta' \delta(N_N^N) = -  \pi \int_{-1}^1
\frac{\omega_{q'}}{p k'} \delta(\eta' - \eta_0') \\ \nonumber
 &= -  \pi
\frac{\omega_{q'}}{p k'},
\end{align}
where the angle is
\begin{align}
\eta_0'= \frac{1}{2 p k'} \left( 2 p_0 \omega' - \Sigma_{m^2} \right)
\end{align}
and 
\begin{align}
\Sigma_{m^2}=M_j^2 + (\omega'^2 - k'^2) - m_\phi^2.
\end{align}
We get
\begin{align}
\label{imnn}
\rmIm \left( T \sum_{k_0', h'} \int \frac{\rmd^3 k'}{(2 \pi)^3}
  \D_{N'} \D_{\phi'} \Delta' H_- \right)_{N_N^N} = & \frac{1}{4 \pi^3}\rmIm \left(
  T \sum_{k_0',h'} \int_0^\infty \rmd k' k'^2 \rmd \eta' \int_0^\pi \rm
  d\phi' \D_{N'} \D_{\phi'} \Delta' H_-
\right) \nonumber \\ = & - \frac{1}{16 \pi^2} \sum_{h'} \int \rmd k'
\rmd \phi' \frac{k'}{p \omega_{p'}} Z_{h'} Z_N^N (A_\ell^\ell
-A_\ell^{\phi'}) H_-.
\end{align}
It is sufficient to perform the integration over $\phi'$ from $0$ to
$\pi$ since $\cos \phi'$ is the only quantity that depends
on $\phi'$. 

We note that we can write
\begin{align}
  \label{eq:162}
  A_\ell^\ell-A_\ell^{\phi'} = \frac{2
    \omega_{p'}}{(k_0-\omega_{q'})^2-\omega_{p'}^2} \equiv 2
  \omega_{p'} \Delta_{N'}^{VN} \, ,
\end{align}
where $\Delta_{N'}^{VN}$ can be viewed as the propagator of the internal neutrino,
since we can interpret the contribution we are looking at as putting
the internal Higgs boson on-shell and thus we have $k_0-\omega_{q'} = k_0 -
q_0' = p_0'$.

\subsection{Self-energy cut}
\label{sec:self-energy-cut}

For the self-energy diagram, only $N_N^N$ contributes. Taking $\eta'$ as
the angle between $\bf p$ and $\bf k'$, we get
\begin{align}
\rmIm \left( T \sum_{k_0', h'} \int \frac{\rmd^4 k'}{(2 \pi)^4} \sum_{h'}
  \D_{N'} \D_{\phi'} \Delta' H_- \right)_S  = & \frac{1}{4 \pi^3}\rmIm \left(
  T \sum_{k_0',h'} \int_0^\infty \rmd k' k'^2 \rmd \eta' \int_0^\pi \rm
  d\phi' \D_{N'} \D_{\phi'} \Delta' H_-
\right) \nonumber \\ = & - \frac{1}{16 \pi^2} \frac{1}{M_j^2-M_k^2} \sum_{h'} \int \rmd k'
\rmd \phi' \frac{k'}{p} Z_{h'} Z_N^N H_-.
\end{align}
Comparing this expression with the contribution from $N_N$ in
equation~\eqref{imnn}, we see that calculating the self-energy contribution
amounts to replacing $\Delta_{N'}^{VN}$ by $
\Delta_{N'}^{SN}=(M_j^2-M_k^2)^{-1}$ in the $N_N$-vertex contribution. If $M_k \gg M_j$, we get
\begin{align}
  \label{eq:163}
  \Delta_{N'}^{VN} \approx \Delta_{N'}^{SN} \approx -\frac{1}{M_k^2}
  \, ,
\end{align}
so the self-energy contribution is twice as large as the vertex
contribution, $\epsilon_S \approx 2 \, \epsilon_V$,
where the factor two comes from the fact that we have two
possibilities for the components of the $SU(2)$ doublets in the loop of the self-energy
diagram. This resembles the situation in vacuum.

\section{Analytic Expressions for the $\CP$-Asymmetries}
\label{sec:putting-all-together}

\subsection[Vertex cut through ${\{\ell',\phi'\}}$]{Vertex cut through $\boldsymbol{\{\ell',\phi'\}}$}
\label{sec:all-cuts}

We simplify the analytic expression for $\epsilon_\gamma(T)$ in
equation~\eqref{epsilongamma}. For $I_V$ in equation~\eqref{eq:131}
we get
\begin{align}
\rmIm(I_V)_{N_N^N}= \frac{M_j M_k}{16 \pi^2}
\frac{Z_h \omega}{p} \sum_{h'} \int_0^\infty \rmd k' \int_0^{\pi}
\rmd \phi' \frac{k'}{\omega_{p'}} Z_{h'} Z_N^N (A_\ell^\ell -A_\ell^{\phi'}) H_-
,
\end{align}
where it is sufficient to integrate $\phi'$ from $0$ to $\pi$. The
difference of the matrix elements reads for the vertex contribution
\begin{align}
  \label{eq:133}
  \left| \mathcal{M} (N \to \ell_h \phi) \right|^2 
 - \left|
    \mathcal{M}(N \to \barell_h \barphi) \right|^2  =& - g_{SU(2)} \rmIm
  \left\{ \left[ \left( \lambda^\dagger \lambda \right)_{jk} \right]^2
  \right\} \frac{M_j M_k}{4 \pi^2} \frac{Z_h \omega_h}{p}
  \nonumber \\
  & \times \sum_{h'}
  \int_0^\infty \rmd k' \int_0^{\pi} \rmd \phi' \frac{k'}{\omega_{p'}}
  Z_{h'} Z_N^N (A_\ell^\ell -A_\ell^{\phi'}) H_- \, .
\end{align}
Correspondingly, the difference in decay rates reads
\begin{align}
  \label{eq:134}
   \gamma(N \to \ell_h \phi) - \gamma(N \to \barell_h \barphi) = 
& - g_{SU(2)} \rmIm
  \left\{ \left[ \left( \lambda^\dagger \lambda \right)_{jk} \right]^2
  \right\} \frac{M_j M_k}{4 (2 \pi)^5} \nonumber \\
 & \times \sum_{h'} \int \rmd E \rmd k \rmd k'
  \int_0^\pi \rmd \phi' k F_{N_ h}^\rmeq Z_h \frac{k'}{p \o_{p'}}Z_N^N
  Z_{h'}(A_\ell^\ell -A_\ell^{\phi'}) H_- \, ,
\end{align}
where $F_{hN} = f_N^\rmeq (1+f_\phi^\rmeq) (1-f_{\ell h}^\rmeq)$ is
the statistical factor for the decay.

We know from chapter~\ref{decayrate} that
\begin{align}
\sum_{s} \left| \mathcal{M}^s_h(N \to LH) \right|^2 = g_{SU(2)} g_c (\lambda^\dagger \lambda)_{jj}
Z_h \omega (p_0 - h p \eta) \, ,
\end{align}
where $g_c=2$ indicates that we sum over $N \to \phi \ell$ and $N \to
\barphi \barell$. Thus
\begin{align}
\Gamma(N \to L_h H) = g_{SU(2)} g_c \frac{(\lambda^\dagger \lambda)_{jj}}{16 \pi p p_0} \int
\rmd k k Z_D Z_h (p_0 - h p \eta)
\end{align}
and 
\begin{align}
\gamma(N \to L_h H) =  g_{SU(2)} g_c \frac{(\lambda^\dagger
  \lambda)_{jj}}{4 (2 \pi)^3} \int
\rmd E \rmd k k f_N Z_D Z_h (p_0-h p \eta) \, ,
\end{align}
where we have summed over the neutrino degrees of freedom.

We arrive at
\begin{align}
\label{eq:174}
\epsilon_h(T) = &  - g_{SU(2)} \frac{\rmIm
  \{[( \lambda^\dagger \lambda)_{jk}]^2\}}{\gamma(N \to L_h N)} 
\frac{M_j M_k}{4 (2 \pi)^5} \sum_{h'} \int \rmd E \rmd k \rmd k'
  \int_0^\pi \rmd \phi' k F_{N_ h}^\rmeq Z_h \frac{k'}{p \o_{p'}}Z_N^N
  Z_{h'}(A_\ell^\ell -A_\ell^{\phi'}) H_- \nonumber \\
=&- \frac{\rmIm\{[(\lambda^\dagger
  \lambda)_{jk}]^2\}}{g_c (\lambda^\dagger \lambda)_{jj}} \frac{M_j
  M_k}{4 \pi^2}  
\frac{\sum_{h'} \int \rmd E \rmd k \rmd k'
  \int_0^\pi \rmd \phi' k F_{N_ h}^\rmeq Z_h \frac{k'}{p \o_{p'}}Z_N^N
  Z_{h'}(A_\ell^\ell -A_\ell^{\phi'}) H_-}{\int
\rmd E \rmd k k f_N Z_D Z_h (p_0-h p \eta)} \, .
\end{align}

\subsection{Self-energy cut}
\label{sec:self-energy-cut-1}

For the self-energy contribution, we get
\begin{align} 
\rmIm(I_S)_{N_N}=  \frac{M_j M_k}{M_j^2-M_k^2} \frac{1}{4 \pi^2}
\frac{Z_h \omega}{p} \sum_{h h'} \int_0^\infty \rmd k' \int_0^{\pi}
\rmd \phi' k' Z_{h'} Z_N^N H_- \, .
\end{align}
The difference in decay rates reads
\begin{align}
  \label{eq:140}
     \gamma(N \to \ell_h \phi) - \gamma(N \to \barell_h \barphi) = 
& - g_{SU(2)} \rmIm
  \left\{ \left[ \left( \lambda^\dagger \lambda \right)_{jk} \right]^2
  \right\} \frac{M_j M_k}{(M_j^2-M_k^2)}\frac{1}{(2 \pi)^5} \nonumber \\
 & \times \sum_{h'} \int \rmd E \rmd k \rmd k'
  \int_0^\pi \rmd \phi' k F_{N_ h}^\rmeq Z_h \frac{k'}{p}Z_N^N
  Z_{h'} H_- \, .
\end{align}
The $\CP$-asymmetry reads
\begin{align}
  \label{eq:141}
  \epsilon_h(T) = &  - g_{SU(2)} \frac{\rmIm
  \{[( \lambda^\dagger \lambda)_{jk}]^2\}}{\gamma(N \to L_h N)} 
\frac{M_j M_k}{M_j^2-M_k^2}\frac{1}{(2 \pi)^5} \sum_{h'} \int \rmd E \rmd k \rmd k'
  \int_0^\pi \rmd \phi' k F_{N_ h}^\rmeq Z_h \frac{k'}{p}Z_N^N
  Z_{h'} H_- \nonumber \\
=&- \frac{\rmIm\{[(\lambda^\dagger
  \lambda)_{jk}]^2\}}{g_c (\lambda^\dagger \lambda)_{jj}} \frac{M_j
  M_k}{M_j^2-M_k^2}\frac{1}{2 \pi^2}  
\frac{\sum_{h'} \int \rmd E \rmd k \rmd k'
  \int_0^\pi \rmd \phi' k F_{N_ h}^\rmeq Z_h \frac{k'}{p}Z_N^N
  Z_{h'} H_-}{\int
\rmd E \rmd k k f_N Z_D Z_h (p_0-h p \eta)} \, ,
\end{align}

\subsection{Symmetry under lepton-mode exchange}
\label{sec:quant-boltzm-equat}



\begin{figure}
  \centering
  \includegraphics[width=0.8 \textwidth]{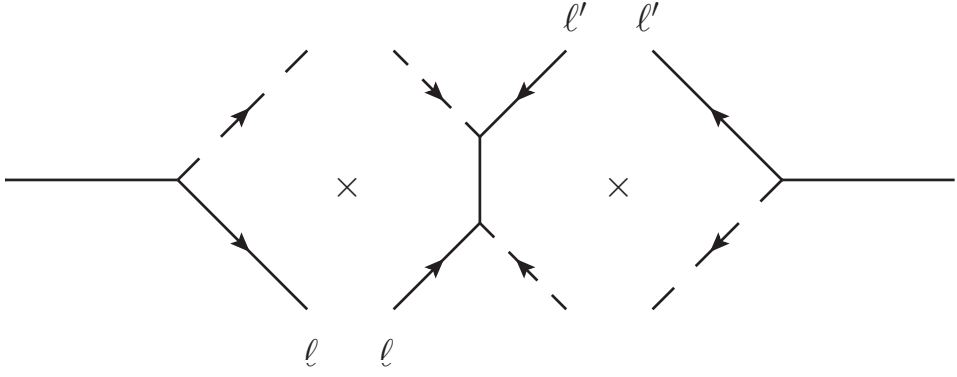}
  \caption[Factorisation of the vertex contribution]{The product of
    diagrams that corresponds to the vertex contribution of the $\CP$
    asymmetry at low temperature. It is symmetric under the exchange
    of the leptons $\ell$ and $\ell'$.}
  \label{fig:exchangeleptons}
\end{figure}

We can use equation~\eqref{eq:167} and collect all factors that depend
on $\bf k$ and $\bf k'$,
\begin{align}
  \label{eq:166}
  (1+f_\phi-f_\ell)(1+f_{\phi'}-f_{\ell'}) Z_h Z_{h'} k k' \Delta_{N'}^{VN}
  H_- \, ,
\end{align}
where we have suppressed the indices for helicity-over-chirality ratios $h$ and
$h'$. The internal neutrino momentum $\bf p = k + k' - p'$ is
symmetric under a replacement of $\bf k$ and $\bf k'$ and likewise the difference
$\omega-\omega_{q'} = \omega + \omega' - p_0$. The Higgs boson momenta
$\bf q = p - k$ and $\bf q' = p - k'$ are also exchanged when we
exchange $\bf k$ and $\bf k'$. Thus, the $\CP$-asymmetry for the
vertex contribution is symmetric under an exchange of the internal and
the external lepton. This can be understood as follows: Taking the imaginary part of $\mathcal{M}_0
\mathcal{M}_1^*$ by putting the internal lepton and Higgs boson
on-shell corresponds to calculating the product of the amplitudes of two
decays and one $\Delta L =2 $ scattering with a neutrino in the
$u$-channel, as shown in figure~\ref{fig:exchangeleptons}.
It can easily be checked that this symmetry also holds for the
self-energy diagram, where the corresponding $\Delta L=2$ scattering
has a neutrino in the $s$-channel.

\section[The $\CP$-Asymmetry at High Temperature]{The $\boldsymbol{\CP}$-Asymmetry at High Temperature}
\label{sec:cp-asymmetry-at}

\begin{figure}
  \centering
    \includegraphics[scale=0.8]{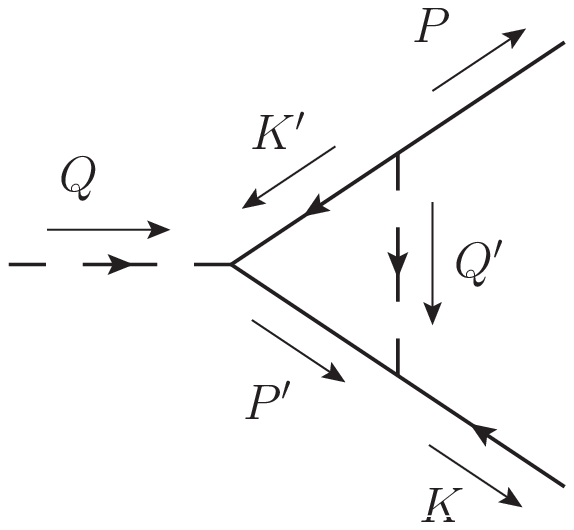} \hspace{0.5cm}
    \includegraphics[scale=0.8]{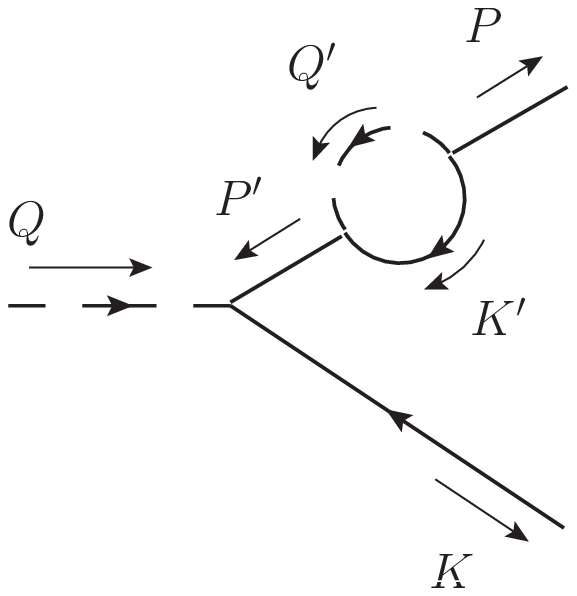}
  \caption{The vertex and the self-energy contribution for the $\phi$ decay.}
  \label{fig:phicuts}
\end{figure}

At high temperature, where we have the decays of Higgs bosons, the
$\CP$-asymmetry on amplitude level is defined as
\begin{align}
  \label{eq:101}
    \e_{h}^\phi \equiv \frac{\left| \mathcal{M}(\barphi \to N \ell_h) \right|^2 - 
\left| \mathcal{M}(\phi \to N \barell_h) \right|^2}
{\left| \mathcal{M}(\barphi \to N \ell_h) \right|^2 + 
\left| \mathcal{M}(\phi \to N \barell_h) \right|^2} \, ,
\end{align}
as in equation~\eqref{eq:b65}. The external momenta are now related as
$q_0 = p_0+k_0$. The momentum assignments are shown in
figure~\ref{fig:phicuts}.  We take $\bf q$ and $\bf p$ as the
three-momenta of the initial-state Higgs boson and the final-state
neutrino as in section~\ref{sec:decays-at-high}, this way we can
directly use the results from the $\CP$-asymmetry in neutrino
decays. The matrix elements are the same as for the low temperature
case, so $\mathcal{M}(\phi \to N \barell_h)$ corresponds to
$\mathcal{M}(N \to \barphi \barell_h)$, just the energy relations are
different and the $\CP$-asymmetry is defined with a minus sign relative
to low temperature. The self-energy contribution from the external
neutrino line is the only $\CP$-asymmetric self-energy, the other self
energies do not exhibit an imaginary part in the combination of the
couplings.
The couplings read
\begin{align}
  \label{eq:142}
  \rmIm \left\{ \lambda_0^\phi \lambda_1^{\phi*} \right\} = g_{SU(2)}
  \rmIm \left\{ \left[ \left( \lambda^\dagger \lambda\right)_{kj}
    \right]^2 \right\} = - g_{SU(2)} \rmIm \left\{ \left[ \left(
        \lambda^\dagger \lambda\right)_{jk} \right]^2 \right\} \, .
\end{align}
The integrals for the vertex and the self-energy-contribution are
\begin{align}
  \label{eq:137}
    I_0^\phi I_1^{\phi*}= - T \sum_{k_0', h'} \int \frac{{\rm d}^3 k'}{(2 \pi)^3} M_k M_j Z_h \o_h
  \D_{N'} \D_{\phi'} \Delta'_{h'} H_-^{h h'} \, ,
\end{align}
where we remember that $\Delta_{N'}=1/(M_j^2-M_k^2)$ for the self-energy graph.

The frequency sum for the vertex diagram reads
\begin{align}
  \label{eq:143}
  \begin{split}
    T \sum_{k_0'}\sum_{h'} &
    \D_{N'}\D_{\phi'}\Delta'H_-=\\
    = \sum_{h'} \frac{Z_{h'}}{4 \omega_{q'} \omega_{p'}} & \left\{
      \left[ \left(B_\phi^\phi-B_N^N\right)A_\ell^{N'}-\left(
          B_\phi^{\ell'}-B_N^N \right) A_\ell^0 - \left(B_\phi^\phi-
          B_N^{\ell'}\right)A_\ell^\ell+ \left(B_\phi^{\ell'}-
          B_N^{\ell'} \right) A_\ell^{\phi'}
      \right] H_+ \right.\\
    & \hspace{-6pt} + \left. \left[ \left(B_\phi^{N'}-B_N^{\phi'}
        \right) A_\ell^{N'}- \left( B_\phi^0-B_N^{\phi'} \right)
        A_\ell^0 - \left( B_\phi^{N'}- B_N^0 \right) A_\ell^\ell+
        \left( B_\phi^0- B_N^0 \right) A_\ell^{\phi'} \right] H_-
    \right\} \, .
\end{split}
\end{align}

Since we have $M_2 \gg M_1$, we also have $M_2 \gg m_\phi$ in the
relevant temperature range, so the possible contributions are from
$N_N^{\phi'}$, $N_\ell^{N'}$ and $N_\phi^{N'}$\footnote{If $m_\phi \gg
  M_2$, we would have contributions from $N_\ell^{\phi'}$ and
  $N_\phi^{\phi}$ instead}. Again, the $N_2$ cuts can be neglected
because they are kinematically suppressed. When taking the
discontinuity of the diagrams, we get for the angle between $\bf p$
and $\bf k'$,
\begin{align}
  \label{eq:103}
  \eta_{\phi,0}' = \frac{1}{2 p k'} (2 p_0 \omega' -  \Sigma_\phi) \, ,
\end{align}
where
\begin{align}
  \label{eq:118}
  \Sigma_\phi = m_\phi^2 - (\omega'^2-k'^2) - M_1^2 \, ,
\end{align}
so we arrive at
\begin{align}
  \label{eq:107}
  \left( \epsilon_{\gamma h}^N \gamma_{\epsilon h}^N \right)_V &=
-  \frac{\rmIm \lambda_{CP}}{4 (2 \pi)^5} M_j M_k \sum_{h'} \int \rmd E
  \rmd k \rmd k' \int_0^{\pi} \rmd \phi' k F_{\phi h} Z_h \frac{1}{p}
  \frac{k'}{\omega_{p'}} Z_{h'} Z_N^{\phi'} (A_\ell^{N'} - A_\ell^0) H_- \, ,
\end{align}
where we can write
\begin{align}
  \label{eq:164}
  A_\ell^{N'} - A_\ell^{0} = \frac{2
    \omega_{p'}}{(k_0+\omega_{q'})^2-\omega_{p'}^2} = 2 \omega_{p'}
  \Delta_{N'}^{V \phi} \, .
\end{align}
Contrary to the $\CP$-asymmetry in neutrino decays, this expression
can not strictly be seen as the propagator of the neutrino since the
contribution does not correspond to a zero temperature cut but is a
pure thermal effect induced by the presence of leptons and Higgs
bosons in the thermal bath. This is illustrated by the factor
$Z_N^{\phi'}=f_{\phi'}+f_{\ell'} = f_{\phi'} (1- f_{\ell '}) + (1+
f_{\phi'}) f_{\ell'}$\footnote{Reference \cite{Giudice:2003jh} obtains
  a different factor $f_{\phi'}-f_{\ell'}-2 f_{\phi'} f_{\ell'}$ due
  to an incorrect choice of cutting rules as explained in
  reference~\cite{Garny:2010nj}.}, which describes the absorption of a
Higgs boson and the stimulated emission of a lepton and the opposite
process, the absorption of a lepton and the stimulated emission of a
Higgs boson. Compared to low temperature, we have replaced
$\Delta_{N'}^{VN} Z_N^N$ by $\Delta_{N'}^{V \phi} Z_N^{\phi'}$.

For the self-energy diagram, the frequency sum is given by
\begin{align}
  \label{eq:144}
  T \sum_{k_0'}\sum_{h'} \D_{N'}\D_{\phi'}\Delta'H_- = \Delta_{N'}
  \frac{1}{2 \omega_{q'}} \sum_{h'} Z_{h'} \left[ H_- \left(
      B_N^0-B_N^{\phi'}\right) + H_+ \left( B_N^{\ell'}- B_N^N
    \right)\right] \, ,
\end{align}
after taking the discontinuity, the $\CP$-asymmetry reads
\begin{align}
  \left( \epsilon_{\gamma h}^N \gamma_{\epsilon h}^N \right)_S &=
  - \frac{\rmIm \lambda_{CP}}{(2 \pi)^5} \frac{M_j M_k}{M_j^2-M_k^2}
  \sum_{h'} \int \rmd E \rmd k \rmd k' \int_0^{\pi} \rmd \phi' k F_{\phi
    h} Z_h \frac{1}{p} k' Z_{h'} (-Z_N^{\phi'})
  H_- \, ,  
\end{align}
where
\begin{align}
  \label{eq:108}
    F_{\phi h}& = f_\phi^\rmeq
(1- f_{\ell h}^\rmeq) (1 -2 f_N^\rmeq).
\end{align}
Compared to low temperature, we have replaced $Z_N^N$ by $Z_N^{\phi'}$.
The self-energy contribution is given by replacing
$\Delta_{N'}^{V \phi}$ by $\Delta_{N'}^{S \phi}=(M_j^2-M_k^2)$ in the
vertex case. For $M_k \gg M_j$, we have 
\begin{align}
  \label{eq:165}
  \Delta_{N'}^{V \phi} \approx \Delta_{N'}^{S \phi} \approx
  -\frac{1}{M_k^2} \, ,
\end{align}
so the relation $\epsilon_S \approx 2 \epsilon_V$ also holds for the
Higgs boson decays.

Using $f_\phi (1-f_\ell) = (f_\phi + f_\ell) f_N$, the terms that
depend on the lepton momenta $\bf k$ and $\bf k'$ are
\begin{align}
  \label{eq:169}
  (f_\phi + f_\ell) (f_{\phi'}+f_{\ell'}) Z_h Z_{h'} k k' \Delta_{N'}^{S/V
    \phi} H_- \, .
\end{align}
where now $\bf p' = k+k'+p$ and $\omega+\omega_{q'} = \omega+\omega'+p_0$,
so the $\CP$-asymmetry in Higgs boson decays is symmetric under exchanging the internal and
external lepton as well.

\section{One-Mode Approach}
\label{sec:one-mode-approach}

We also calculate the $\CP$-asymmetry within the one-mode approach
where we treat the thermal mass like a kinematical mass and use lepton
propagators $(\slashed{k}-m_\ell)^{-1}$ or $(\slashed{k}-\sqrt{2}
\,m_\ell)^{-1}$ as in section~\ref{sec:disc-yukawa-ferm}. The spin sum
corresponding to equation~\eqref{eq:155} then reads
\begin{align}
  \label{eq:156}
  \sum_{s,r} (\overline{u}_\ell^r P_R u_N^s) (\overline{u}_N^s P_R S_{\ell'} P_L
u_\ell^r) = 2 M_j \Delta_{\ell'} K^\mu K_\mu' \, ,
\end{align}
where $\Delta_{\ell'}=(k_0'^2-\omega_{k'}^2)^{-1}$. In the frequency
sums in equations \eqref{eq:157} and \eqref{eq:159}, we replace
$\tilde{\Delta}_{s h'}^{\rm pole}$ by the usual decomposition
$\Delta_{s,\ell'}$ in equation~\eqref{eq:201}, which means replacing $Z_{s h'}$ by $-s/(2
\omega')$ on the right-hand sides. One can check that in the final
expression for the $\CP$-asymmetry, this amounts to
replacing the sum of the helicity contributions 
\begin{align}
  \label{eq:161}
  \sum_{h h'} Z_h
Z_{h'} (1-h h' \xi) \quad \textrm{by} \quad \frac{K^\mu
K_\mu'}{\omega_k \omega_{k'}} = 1- \frac{ k
  k'}{\omega_k \omega_{k'}} \xi \, .
\end{align}
This means that in the two mode treatment, it is forbidden for the
external and internal lepton to be scattered strictly in the same direction if
they have the same helicity or in the opposite direction if they have
opposite helicity. For the one-mode approximation this is not the case
since $\omega_k \omega_{k'}$ is always larger than $ k k'$. This result illustrates
that the leptonic quasiparticles still behave as if they are massless
in terms of the helicity structure of their interactions, while the
one-mode approach is not able to describe this behaviour.

For the $\CP$-asymmetries in the decay densities we get
\begin{align}
  \label{eq:170}
    (\Delta \gamma_m^N)_V &\equiv  \left[\gamma_m(N \to \phi \ell) -
    \gamma_m(N \to \barphi \barell) \right]_V \nonumber \\
&=
 - \frac{\rmIm \lambda_{CP}}{2 (2 \pi)^5} M_j M_k \sum_{h'} \int \rmd E
  \rmd k \rmd k' \int_0^{\pi} \rmd \phi' \frac{F_{N h} Z_N^N}{p} k k'
  \Delta_{N'}^{VN} \frac{K \cdot K'}{\omega_k \omega_{k'}} \, ,
  \nonumber \\
    (\Delta \gamma_m^N)_S
&=
 - \frac{\rmIm \lambda_{CP}}{(2 \pi)^5} M_j M_k \sum_{h'} \int \rmd E
  \rmd k \rmd k' \int_0^{\pi} \rmd \phi' \frac{F_{N h} Z_N^N}{p} k k'
  \Delta_{N'}^{SN} \frac{K \cdot K'}{\omega_k \omega_{k'}} \, ,
  \nonumber \\
    (\Delta \gamma_m^\phi)_V 
&=
 - \frac{\rmIm \lambda_{CP}}{2 (2 \pi)^5} M_j M_k \sum_{h'} \int \rmd E
  \rmd k \rmd k' \int_0^{\pi} \rmd \phi' \frac{F_{\phi h} Z_N^{\phi'}}{p} k k'
  \Delta_{N'}^{V \phi} \frac{K \cdot K'}{\omega_k \omega_{k'}} \, ,
  \nonumber \\
    (\Delta \gamma_m^\phi)_S
&=
 - \frac{\rmIm \lambda_{CP}}{(2 \pi)^5} M_j M_k \sum_{h'} \int \rmd E
  \rmd k \rmd k' \int_0^{\pi} \rmd \phi' \frac{F_{\phi h} Z_N^{\phi'}}{p} k k'
  \Delta_{N'}^{S \phi} \frac{K \cdot K'}{\omega_k \omega_{k'}} \, ,
\end{align}
where $F_{Nh} = f_N (1- f_N) (1+f_\phi -f_\ell)$, $F_{\phi h} = f_N (1
- f_N) (f_\phi +f_\ell)$ .

Let us examine the high temperature behaviour of the one-mode approach
by calculating the $\CP$-asymmetry in the matrix elements of a Higgs
boson at rest, where we assume that $M_j, m_\ell \ll m_\phi \ll
M_k$. For simplicity, we calculate the self-enegy contribution. The
integral that corresponds to equation~\eqref{eq:137} reads
\begin{align}
  \label{eq:172}
    I_0 I_1^* = 2 T \sum_{k_0'} \int \frac{{\rm d}^3 k'}{(2 \pi)^3} M_k M_j \left[
  \Delta_{N'} \Delta_{\phi'} \Delta'_{\ell'} \right]^* K \cdot K' \, .
\end{align}
The part that contributes to the imaginary part of the diagram is
\begin{align}
  \label{eq:173}
  \left. I_0 I_1^* \right|_N^{\phi'} &= 2 \int \frac{{\rm d}^3 k'}{(2
    \pi)^3} M_k M_j \Delta_{N'} \frac{1}{4 \omega_{q'} \omega_{k'}}
  B_N^{\phi'} (K \cdot
  K') \, , \nonumber \\
  \rmIm \, B_N^{\phi'}& = - \pi Z_N^{\phi'} \delta(N_N^{\phi'}) = -\pi
  \frac{\omega_{q'}}{k k'} Z_N^{\phi'} \delta(\xi - \xi_0) \, ,
\end{align}
where $\xi \equiv ({\bf k k'})/(k k')$,
\begin{align}
  \label{eq:175}
  \xi_0 = \frac{m_\phi-k'}{k'} \,
\end{align}
and we have neglected $M_j$ and $m_\ell$.
We get
\begin{align}
  \label{eq:176}
  \rmIm(I_0 I_1^*)_N^{\phi'} = - \frac{1}{8 \pi} \frac{\sqrt{x}}{1-x}
  \int_{k}^\infty \rmd k' Z_N^{\phi '} (2 k' - m_\phi) \, ,
\end{align}
where $x \equiv M_k^2/M_j^2$ and $k=m_\phi/2$. For simplicity, we
make the approximation
\begin{align}
  \label{eq:177}
  Z_N^{\phi'}=f_{\phi'}+f_{\ell'} \approx \rme^{- \omega_{q'} \beta} +
  \rme^{-\omega_{k'} \beta} = (1 + \rme^{- k \beta}) \rme^{-k' \beta}
  \, ,
\end{align}
so 
\begin{align}
  \label{eq:178}
    \rmIm(I_0 I_1^*)_N^{\phi'} = - \frac{1}{8 \pi}
    \frac{\sqrt{x}}{1-x} (1+\rme^{-k \beta})
  \int_{k_1'}^\infty \rmd k' \rme^{-k' \beta} (2 k' - m_\phi) \, .
\end{align}
The integral gives
\begin{align}
  \label{eq:179}
      \rmIm(I_0 I_1^*)_N^{\phi'} = - \frac{1}{4 \pi}
    \frac{\sqrt{x}}{1-x} (1+\rme^{-k \beta}) T^2 \rme^{-k \beta} \, .
\end{align}
We parameterise $m_\phi$ as $m_\phi = g_\phi T$ and obtain
\begin{align}
  \label{eq:180}
\Delta \left| \mathcal{M} \right|^2 \equiv
| \mathcal{M} |^2- | \widetilde{\mathcal{M}} |^2
= - 8 \rmIm \lambda_{CP} \rmIm(I_0 I_1^*)= 
2 \frac{\rmIm \lambda_{CP}}{\pi}
    \frac{\sqrt{x}}{1-x} T^2 \rme^{-g_\phi/2} (1 +\rme^{-g_\phi/2}) \, .
\end{align}
Using the expression
\begin{align}
  \label{eq:182}
  | \mathcal{M}_{\rm tot} |^2 = 4 (\lambda^\dagger \lambda)_{11} K \cdot
  P = 2 (\lambda^\dagger \lambda)_{11} g_\phi^2 T^2
\end{align}
we arrive at
\begin{align}
  \label{eq:181}
  \epsilon = \frac{\Delta | \mathcal{M} |^2}{|\mathcal{M}_{\rm
      tot}|^2} =  \frac{\rmIm \lambda_{CP}}{(\lambda^\dagger
    \lambda)_{11}} \frac{1}{\pi} \frac{\sqrt{x}}{1-x} \frac{1}{g_\phi^2}
  \rme^{-g_\phi/2} (1+ \rme^{-g_\phi/2}) = \frac{8}{g_\phi^2}
  \rme^{-g_\phi/2} (1 + \rme^{-g_\phi/2})
  \epsilon_0 .
\end{align}
Assuming that $g_\phi \ll 1$, we get
\begin{align}
  \label{eq:184}
  \frac{\epsilon_{\rm rf}^{T \gg M_1}}{\epsilon_0} \approx \frac{32}{g_\phi^2}
\end{align}
Taking $g_\phi = m_\phi/T \approx 0.42$ for $T=10^{12} \, {\rm GeV}$
and using the more accurate term in equation \eqref{eq:181}, we get
$\epsilon/\epsilon_0 \approx 70$, while we get $\epsilon/\epsilon_0
\approx 90$ for equation~\eqref{eq:184}. We view this result as a
rough approximation of the value of the $\CP$-asymmetry in Higgs boson
decays at high temperature. Both our approximation and the exact
numerical solution in the next section give a factor of 100 difference
to the $\CP$-asymmetry in vacuum.

\section[Temperature-Dependence of the
${\CP}$-Asymmetries]{Temperature Dependence of the
  $\boldsymbol{\CP}$-Asymmetries}
\label{sec:results}

\begin{figure}[h]
  \centering
  \includegraphics[width=0.8 \textwidth]{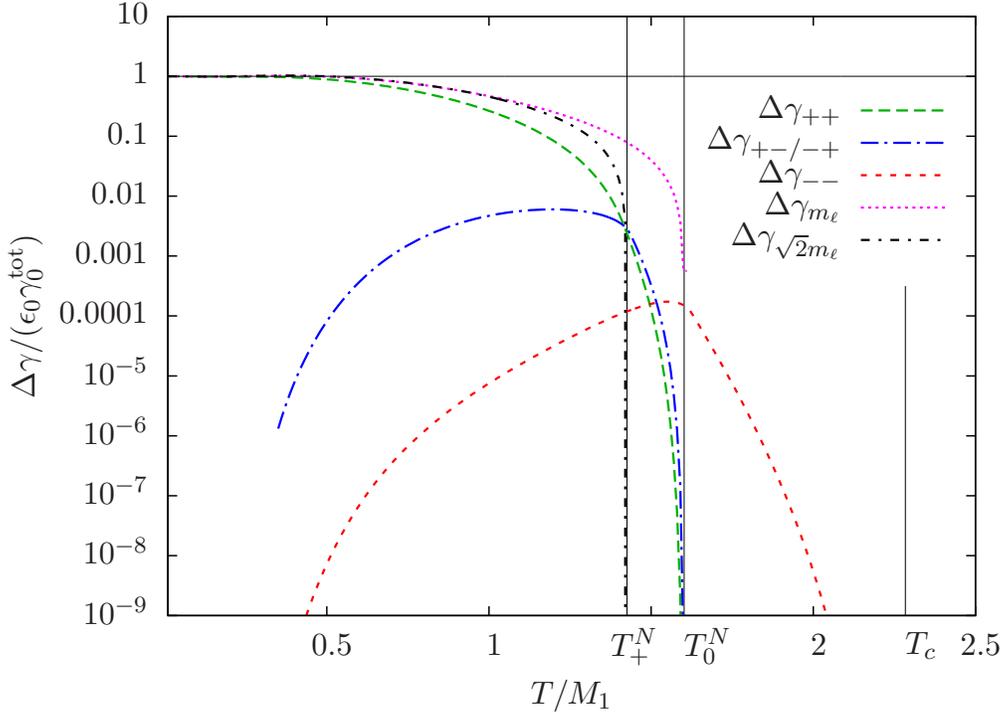}
  \caption[The $\CP$-asymmetries in $N$ decays in units of $\e_0
  \g_0$]{The $\CP$-asymmetries in neutrino decays normalised by the
    $\CP$-asymmetry in vacuum and the total decay density in vacuum,
    $\D \g/(\gamma^{\rm tot}_0 \e_0)$. We choose $M_1=10^{10} \, {\rm
      GeV}$ and $M_2 \gg M_1$. The term $\Delta \gamma_{h_1 h_2}$
    denotes the difference between the decay rate and its $\CP$
    conjugated rate, which is proportional to the
    $\CP$-asymmetry. Here, $h_1$ denotes the mode of the external
    lepton, while $h_2$ denotes the mode of the lepton in the
    loop. For example, $\Delta \gamma_{+-} = \gamma(N \to \phi
    \ell_+)- \gamma(N \to \barphi \barell_+)$, where a minus-mode
    lepton is present in the loop. $\Delta \gamma_{m_\ell}$ and
    $\Delta \gamma_{\sqrt{2} \, m_\ell}$ denote the rate differences
    for the one-mode approach with a thermal mass $m_\ell$ and an
    asymptotic thermal mass $\sqrt{2} \, m_\ell$.}
  \label{fig:cpn0}
\end{figure}

We show the temperature dependence of the $\CP$-asymmetries in neutrino
decays in the full HTL calculation and in the one-mode approach for
$m_\ell$ and $\sqrt{2} \, m_\ell$ in figure~\ref{fig:cpn0}.
We choose $M_1 = 10^{10} {\rm \, GeV}$ and normalise the asymmetries
by the product of the $\CP$-asymmetry at zero temperature and the
total decay density in vacuum, $\epsilon_0 \gamma_0^{\rm tot}$. As
discussed in sections~\ref{sec:self-energy-cut} and
\ref{sec:cp-asymmetry-at}, the vertex contribution and the self-energy
contribution have the same temperature dependence for $M_2 \gg
M_1$. Moreover, as discussed in sections~\ref{sec:quant-boltzm-equat}
and \ref{sec:cp-asymmetry-at}, the asymmetries are the same when we
exchange the internal and the external lepton, therefore the asymmetry
for a plus-mode external lepton combined with a minus-mode internal
lepton is the same as the asymmetry for a minus-mode external lepton
with a plus-mode internal lepton, in short, $\Delta \gamma_{+-} =
\Delta \gamma_{-+}$. We see that generally, the thresholds are the
ones we expect from our analysis of the decay rates in
section~\ref{sec:appl-lept}. For the one-mode calculations we have the
expected thresholds at $T^N_0$ for $m_\ell$ and at $T^N_+$ for
$\sqrt{2} \, m_\ell$. For all asymmetries where a plus-mode lepton is
involved, that is $\Delta \gamma_{++}$, $\Delta \gamma_{+-}$ and
$\Delta \gamma_{-+}$, the phase space is reduced similar to the
$\sqrt{2} \, m_\ell$ case below $T^N_+$ and an additional reduction of
the phase space sets in between $T^N_+$ and $T^N_0$ since large
momenta $k$ or $k'$ that correspond to a large mass $m(k)$ become
kinematically forbidden. Between these two thresholds, $T_+^N$ and
$T_0^N$, the asymmetry for $\Delta \gamma_{+-/-+}$ becomes larger than
the asymmetry for $\Delta \gamma_{++}$. This effect occurs because in
the ($++$)-asymmetry the phase spaces of both the internal and the
external lepton are suppressed, while for the mixed modes, ($+-$) or
($-+$), only the phase space of one momentum is suppressed, while the
phase space of the other momentum is still large. The effect is
similar to the observation that $\gamma_-$ becomes larger than
$\gamma_+$ above $T_+^N$. Relying solely on phase-space arguments, one
would expect that $\Delta \gamma_{\sqrt{2} \, m_\ell}$ is a good
approximation for $\Delta \gamma_{++}$. The fact that $\D \g_{++}$ is
clearly smaller than $\D \g_{\sqrt{2} \, m_\ell}$ is due to two
suppressing factors: One factor is the effect of the two residues
$Z_h(k)$ and $Z_{h'}(k')$, which suppress the rate somewhat for small
momenta $k$ and $k'$. The other, more important factor is the fact
that the helicity structure and angular dependence of the integrals
are different for the ($++$)- and the $\sqrt{2} \, m_\ell$-case as
explained in section~\ref{sec:one-mode-approach}. Since neutrino
momenta are of the order $\sim M_1 \sim T$ for our temperature range,
the lepton momenta will be of the same order, that is $k > m_\ell$,
and the leptons and Higgs bosons will preferentially be scattered
forward. Thus also the angle $\xi$ between the two leptons will be
small and the factor $H_-$ defined in equation~\eqref{eq:117} is
suppressed, while the corresponding one-mode factor $1-\xi (k
k')/(\omega_k \omega_{k'})$ is larger than $H_-$ for small angles and
still finite if both leptons are scattered strictly in the same
direction, that is $\xi=1$. We have checked numerically that this is
the main reason why $\Delta \gamma_{\sqrt{2} \, m_\ell} > \Delta
\gamma_{++}$ in the range $1/2 \, M_1 \lesssim T \lesssim T_+^N$.

\begin{figure}[t]
  \centering
\includegraphics[width=0.77 \textwidth]{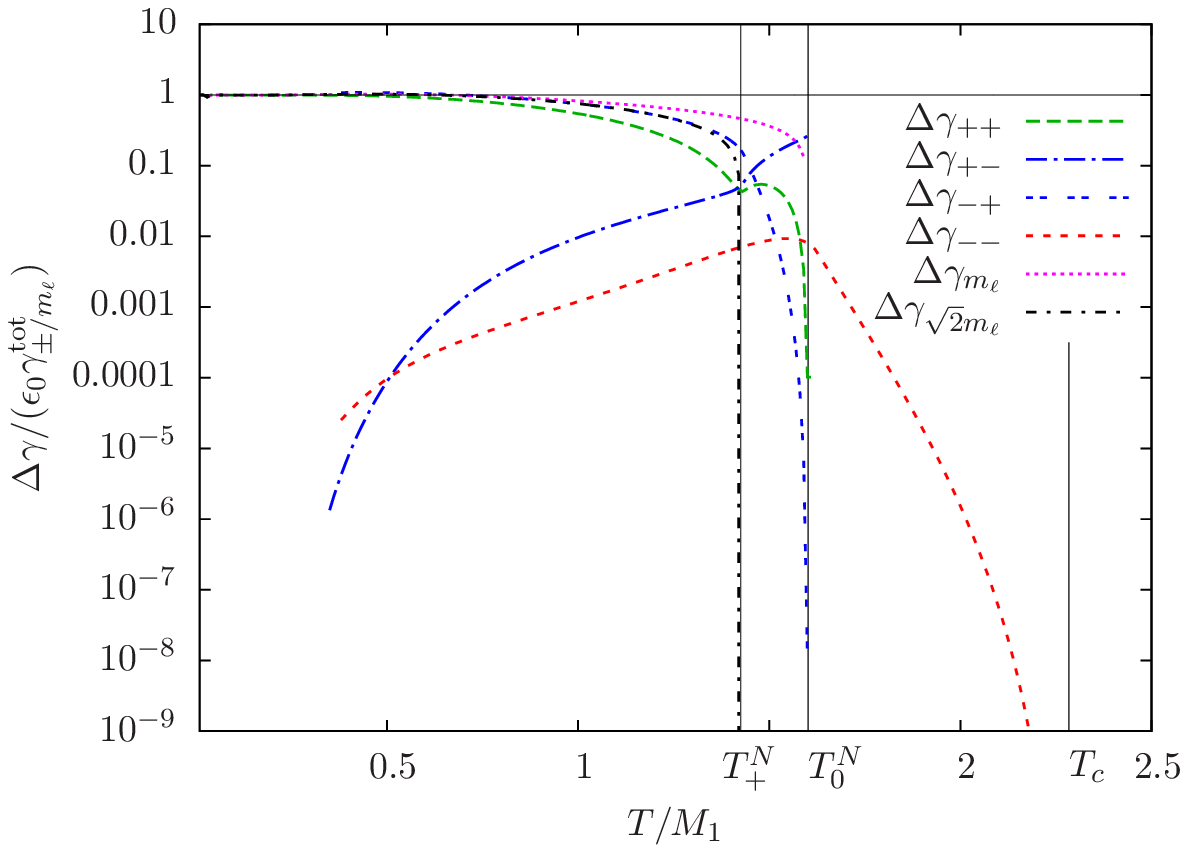}
\caption[The $\CP$-asymmetries in $N$ decays in units of $\e_0
\g^{T>0}$]{The $\CP$-asymmetries in neutrino decays normalised by the
  $\CP$-asymmetry in vacuum and the corresponding total decay density
  at finite temperature, that is $\D \g_{++}/(\gamma^{\rm tot}_+
  \e_0)$, $\D \g_{-+}/(\gamma^{\rm tot}_+ \e_0)$, $\D
  \g_{+-}/(\gamma^{\rm tot}_- \e_0)$, $\D \g_{--}/(\gamma^{\rm tot}_-
  \e_0)$, $\D \g_{m_\ell}/(\gamma^{\rm tot}_{m_\ell} \e_0)$ and $\D
  \g_{\sqrt{2} \, m_\ell}/(\gamma^{\rm tot}_{\sqrt{2} \, m_\ell}
  \e_0)$, where the $\CP$ asymmetries $\D \g$ are explained in
  figure~\ref{fig:cpn0}. We choose $M_1=10^{10} \, {\rm GeV}$ and $M_2
  \gg M_1$.}
  \label{fig:cpn1}
\end{figure}

Since the $\CP$-asymmetries follow the corresponding
finite-temperature decay rates that are in figure \ref{comp}, it is
very instructive to normalise them via these decay rates, that is
$\gamma_+$, $\gamma_-$, $\gamma_{m_\ell}$ and $\gamma_{\sqrt{2}\,
  m_\ell}$. This also gives a more intuitive definition of the
$\CP$-asymmetries at finite temperature. These asymmetries are shown
in figure~\ref{fig:cpn1}, normalised by the zero temperature
$\CP$-asymmetry.
Compared to the normalisation via $\gamma_0$ in figure~\ref{fig:cpn0},
we see that the ($++$)-asymmetry does not fall as steeply as the
corresponding decay rate $\gamma_+$ between $T_+^N$ and $T_0^N$, so
the ratio $\D \g_{++}/\g_+$ is dented at the threshold $T_+^N$. This
illustrates that the ($++$)-$\CP$-asymmetry shows a stronger
suppression below the threshold $T_+^N$, since it suffers from two
phase space reductions and two residues that are smaller than
one. Therefore, the $(++)$-asymmetry is not affected as strongly as
$\g_+$ by the additional suppression above $T_+^N$ when large momenta
$k$ and $k'$ are forbidden and the transition over this threshold is
smoother than for the decay rate $\g_+$. So $\gamma_+$ falls more
steeply than $\D \gamma_{++}$ above the threshold and the ratio of the
two rates has a dent at $T_+^N$. For the ($+-$)-asymmetry, this effect
is even stronger, since it is less suppressed than the
($++$)-asymmetry above $T_+^N$, so the ratio $\D \g_{+-}/(\e_0 \g_-)$
rises up to a value of $\mathcal{O}(0.1)$.
\begin{figure}
  \centering
\includegraphics[width=0.75 \textwidth]{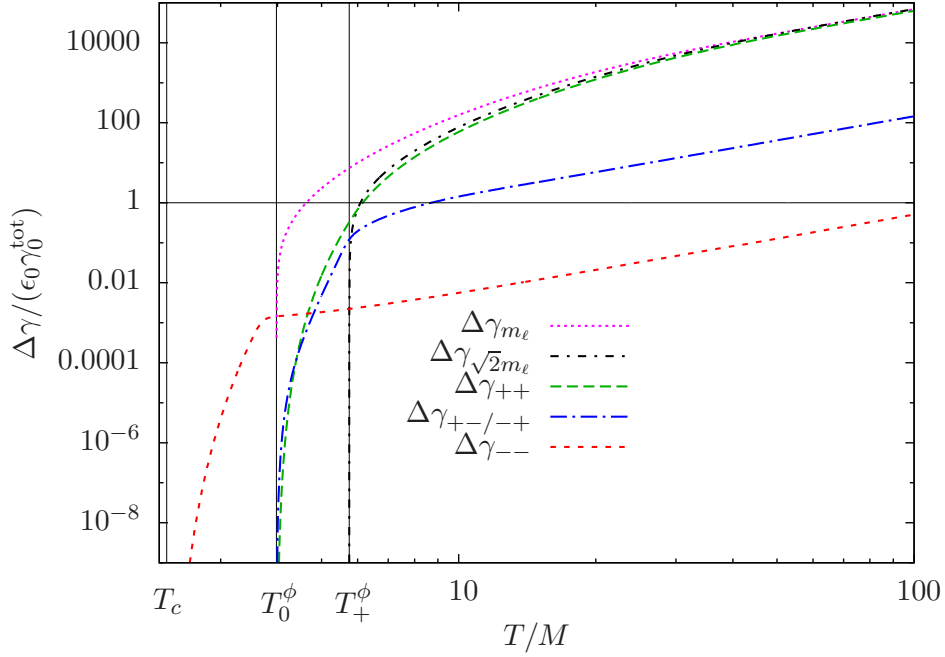}
\caption[The $\CP$-asymmetries in $\phi$ decays in units of $\e_0
\g_0$]{The $\CP$-asymmetries in Higgs boson decays normalised by the
  $\CP$-asymmetry in vacuum and the total decay density in vacuum, $\D
  \g/(\gamma^{\rm tot}_0 \e_0)$, where the asymmetries $\D \g$ are
  explained in figure~\ref{fig:cpn0}. We choose $M_1=10^{10} \, {\rm
    GeV}$ and $M_2 \gg M_1$.}
  \label{fig:cpphi0}
\end{figure}
\begin{figure}
  \centering
\includegraphics[width=0.75 \textwidth]{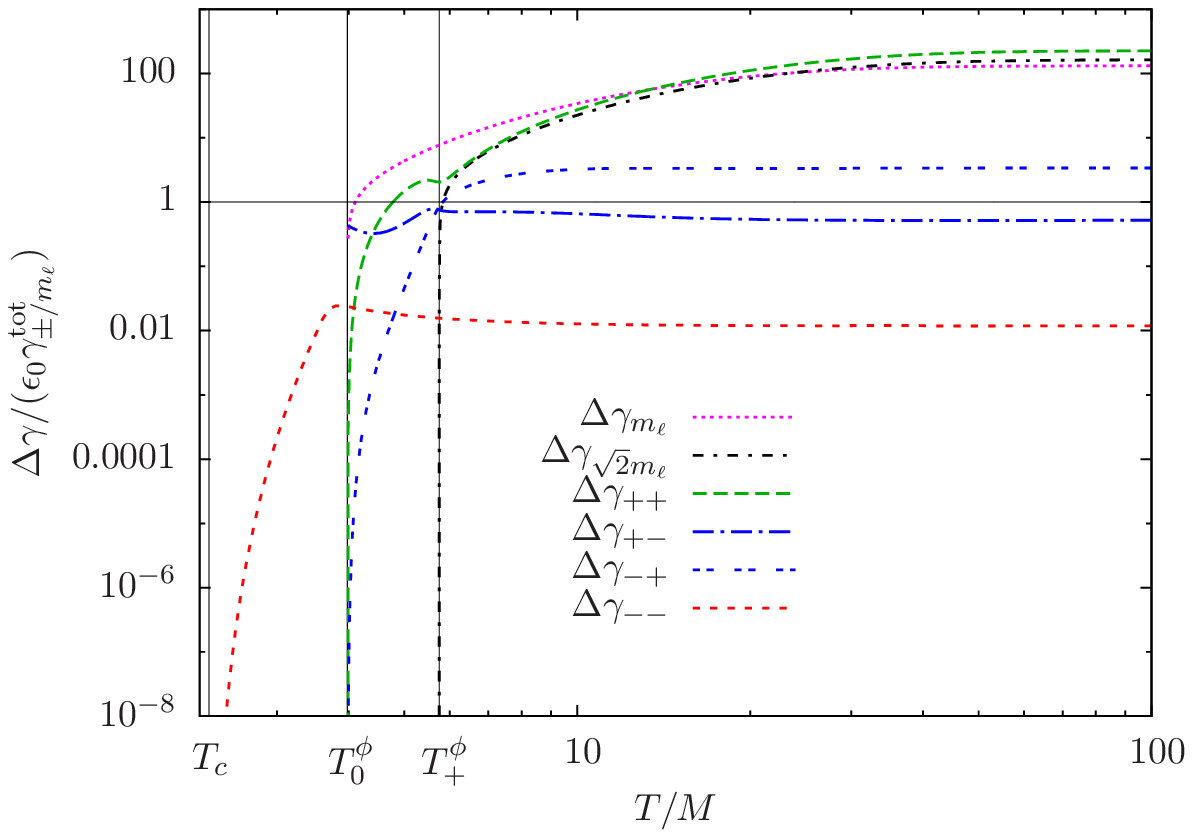}
\caption[The $\CP$-asymmetries in $\phi$ decays in units of $\e_0
\g^{T>0}$]{The $\CP$-asymmetries in Higgs boson decays normalised by
  the $\CP$-asymmetry in vacuum and the corresponding total decay
  density at finite temperature, that is $\D \g_{++}/(\gamma^{\rm
    tot}_+ \e_0)$, $\D \g_{-+}/(\gamma^{\rm tot}_+ \e_0)$, $\D
  \g_{+-}/(\gamma^{\rm tot}_- \e_0)$, $\D \g_{--}/(\gamma^{\rm tot}_-
  \e_0)$, $\D \g_{m_\ell}/(\gamma^{\rm tot}_{m_\ell} \e_0)$ and $\D
  \g_{\sqrt{2} \, m_\ell}/(\gamma^{\rm tot}_{\sqrt{2} \, m_\ell}
  \e_0)$, where the $\CP$ asymmetries $\D \g$ are explained in
  figure~\ref{fig:cpn0}. We choose $M_1=10^{10} \, {\rm GeV}$ and $M_2
  \gg M_1$.}
  \label{fig:cpphi1}
\end{figure}

The $\CP$-asymmetries in Higgs boson decays at high temperature are
shown in figure~\ref{fig:cpphi0}, normalised to $\epsilon_0
\gamma^{\rm tot}$ and we have assumed that $M_2 \gg M_1, T$.  The
behaviour is similar to the neutrino decays, where the
($--$)-asymmetry is strongly suppressed, while the ($+-$)-, the
($-+$)- and the ($++$)-asymmetries have a strict threshold at
$T_0^\phi$ and are suppressed due to the reduced phase space between
$T_0^\phi$ and $T_+^\phi$, as expected. The ($+-$)- and ($-+$)-
asymmetries are the same and are somewhat less suppressed than the
$(++)$-asymmetry between the thresholds $T_0^\phi$ and $T_+^\phi$. In
our approximation in the end of section~\ref{sec:one-mode-approach},
we have seen that the difference of the matrix elements $\Delta
|\mathcal{M} |^2$ rises as $T^2$, so $\Delta \gamma$ rises as $T^4$,
which can be seen in the plot for all finite-temperature
asymmetries. The one-mode asymmetries $\Delta \gamma_{m_\ell}$,
$\Delta \gamma_{\sqrt{2} \, m_\ell}$ and the ($++$)-asymmetry are very
close to each other at high temperature.

We also normalise the asymmetries to the corresponding decay rates in
figure~\ref{fig:cpphi1} and find that they all approach a constant
value at high temperature, as is expected since the decay rates also
rise as $T^4$.  The dents in the ratios $\D \g_{++}/(\e_0 \g_+)$ and
$\D \g_{+-}/(\e_0 \g_+)$ with an external plus-mode lepton are similar
to the ones for the neutrino decays and due to the very strong
suppression of $\gamma_+$ below the threshold $T_+^\phi$. The
numerically dominant asymmetries $\Delta \gamma_{m_\ell}$, $\Delta
\gamma_{\sqrt{2} \, m_\ell}$ and $\Delta \gamma_{++}$ all settle at a
rather high $\CP$-asymmetry, two orders of magnitude higher than at
zero temperature, as we expect from our approximate calculation for a
Higgs boson at rest at the end of
section~\ref{sec:one-mode-approach}. This is partly due to a
suppression of $| \mathcal{M} |^2$ which rises as $m_\phi^2=g_\phi^2
T^2$, but mainly due to the larger difference in matrix elements
$\Delta |\mathcal{M} |^2$ for Higgs boson decays.

%


    \clearpage
\chapter[Boltzmann Equations]{Boltzmann Equations \label{boltzmanneq}}

\section*{Preliminary}
\label{sec:preliminary}

In this chapter, we calculate the Boltzmann equations for
leptogenesis. We include decays and inverse decays involving
neutrinos, leptons and Higgs bosons. Concerning the scatterings, we
argue that we can neglect all of them expect the on-shell contribution
of the $\Delta L=2$ scatterings. We take into account thermal
dispersion relations, but assume distributions close enough to
equilibrium that we can use Boltzmann equations. For the distribution
functions, we use the full quantum statistics, that is, Fermi-Dirac
and Bose-Einstein statistics, but assume the kinetic equilibrium
approximation $f_i = n_i/n_i^\rmeq f_i^\rmeq$. It has been shown by
reference~\cite{HahnWoernle:2009qn} that this is a good approximation.

\section{Particle Kinematics}
\label{sec:particle-kinematics}

The Boltzmann equations describe the time evolution of the
distribution function of a particle species $\psi$. We assume an
isotropic and spatially homogeneous universe described by the
Friedmann-Lemaitre-Robertson-Walker (FLRW) metric~\cite{Kolb:EarlyU},
\begin{equation}
  \label{eq:b1}
 \rmd s^2 = \rmd t^2 - a(t)^2 \left \{ \frac{\rmd r^2}{1-k r^2} + r^2
  \rmd \theta^2 + r^2 \sin^2 \theta \rmd \phi^2 \right \} ,
\end{equation}
where $a(t)$ is the cosmic scale factor, which describes the expansion
of the universe, $k=\pm1,0$ specifies the curvature, and
$(t,r,\theta,\phi)$ are the comoving coordinates.

The trajectory of a particle $\psi$ with mass $m_{\psi}\geq0$ moving
in a gravitational field is given by the geodesic equations of
motion~\cite{Weinberg:Gravitation}:
\begin{align}
  \label{eq:eqm1}
  \frac{\rmd p_{\psi}^{\mu}}{\rmd \tau}+\Gamma^{\mu}_{\nu\alpha}
\,p_{\psi}^{\nu}\,p_{\psi}^{\alpha}&=0, \\
  \label{eq:eqm2}
  \frac{\rmd x_{\psi}^{\mu}}{\rmd \tau}&=p_{\psi}^{\mu}.
\end{align}
Since $s=m_{\psi}\tau$ is the eigen-time of the particle, $\tau$ is
fixed and $p^{\mu}$ is the momentum of a particle $\psi$.

In the FLRW metric the $\mu=0$ component of
Eq~(\ref{eq:eqm1}) is given as
\begin{align}
  \frac{\rmd p_{\psi}^{0}}{\rmd \tau}+ \frac{\dot{a}}{a}
  \mathbf{p}_{\psi}^{2}=0,\ \qquad \textrm{with} \quad \dot{a}=
  \frac{ \partial a}{ \partial t}.
\end{align}
Writing $p_{\psi}^{0}\,\rmd p_{\psi}^{0}=\vert\mathbf{p}_{\psi}\vert
\,\rmd \vert\mathbf{p}_{\psi}\vert$, this leads to:

\begin{align}
  \label{eq:scaling}
  \hphantom{\Leftrightarrow}&\vert\dot{\mathbf{p}}_{\psi}\vert
  a+\dot{a}\vert\mathbf{p}_{\psi}\vert=0
  \nonumber\\
  \Leftrightarrow&\frac{\rmd}{\rmd t}(\vert\mathbf{p}_{\psi}\vert a)=0
  \nonumber \\
  \Leftrightarrow &\vert\mathbf{p}_{\psi}\vert=
  {\rm const.}\times\frac{1}{a}.
\end{align}
Therefore the 3-momentum scales as $1/a$.

In general, the Liouville operator describing the evolution of a point
particle's phase space in a gravitational field is given by
\begin{align}
  \label{eq:liouville}
  L = p^{\alpha} \frac{\partial}{\partial x^{\alpha}} -
  \Gamma{^\alpha_{\beta \gamma}}p^{\beta}p^{\gamma}
  \frac{\partial}{\partial p^{\alpha}}.
\end{align}
With this operator the equations of motion~(\ref{eq:eqm1})
and~(\ref{eq:eqm2}) can be written for the momentum as
\begin{align}
  \label{eq:eqm-liouville1}
  \frac{\rmd p^{\mu}}{\rmd \tau} = L \left[p^{\mu} \right],
\end{align}
and for the space-time as
\begin{align}
  \label{eq:eqm-liouville2}
  \frac{\rmd x^{\mu}}{\rmd \tau} = L \left[x^{\mu} \right].
\end{align}
Furthermore, the time derivative of the phase space
distribution of a non-interacting gas vanishes,
\begin{align}
  \label{eq:derivative-phase}
  \frac{\rmd f(x,p)}{\rmd \tau} = 0.
\end{align}
Using the equations of motion for the particle we obtain the
Boltzmann equations for the non-interacting particle species $\psi$,
\begin{align}
  \label{eq:derivative-liouville}
  L \left[f_{\psi}(x,p) \right] =0.
\end{align}
Since we are assuming a Robertson--Walker universe which is isotropic
and homogeneous, the distribution function $f_{\psi}$ depends only on
$t$ and $\vert \mathbf{p}_{\psi} \vert$. Therefore, the Boltzmann
equation can be written as~\cite{Kolb:EarlyU}
\begin{align}
  \label{eq:boltzmann-without}
  L\left[f_{\psi} \right] = E_{\psi} \frac{\partial f_{\psi}}
  {\partial t} - H \vert \mathbf{p}_{\psi} \vert^{2} \frac{
    \partial f_{\psi}}{\partial E_{\psi}} = 0,
\end{align}
where we have omitted arguments for the sake of notational clarity.

Since $p_{\psi}^{2}=m_{\psi}^{2}$ and because of the spatial isotropy
of the Robertson--Walker--Metric, we have
\begin{align}
  \label{eq:E-p-relation}
 \vert \mathbf{p}_{\psi} \vert^{2}\,\frac{
   \partial f_{\psi}}{\partial E_{\psi}} =E_{\psi}\, \vert
 \mathbf{p}_{\psi} \vert\, \frac{\partial f_{\psi}}{\partial \vert
   \mathbf{p}_{\psi} \vert}.
\end{align}
After dividing by $E_{\psi}$, equation~(\ref{eq:boltzmann-without}) has the form
\begin{align}
  \label{eq:boltzmann-without-2}
  L'\left[f_{\psi} \right] = \frac{\partial f_{\psi}}{\partial t}-H\, \vert
  \mathbf{p}_{\psi} \vert \,\frac{\partial f_{\psi}}{\partial \vert
    \mathbf{p}_{\psi}\vert}.
 \end{align}
Interactions are introduced on the right-hand side 
by a collision term $C \left[f_{\psi} \right]$, which drives
the distribution function towards its equilibrium value.
The complete Boltzmann equation reads
\begin{align}
  \label{eq:boltzmann-with}
   L' \left[f_{\psi} \right] =  \frac{\partial f_{\psi}}
  {\partial t} - H\, \vert \mathbf{p}_{\psi} \vert\, \frac{
    \partial f_{\psi}}{\partial \vert \mathbf{p}_{\psi}\vert} = C
  \left[f_{\psi} \right].
\end{align}
Thus, the Boltzmann equation in a Robertson--Walker universe has the form
of a partial differential equation. However, in the radiation
dominated phase of the universe, in which leptogenesis takes place,
equation~(\ref{eq:boltzmann-with}) can be written as an ordinary
differential equation by transforming to the dimensionless coordinates
$z= m_{\psi}/T$ and $y_{\psi}=\vert \mathbf{p}_{\psi}\vert/T$. Using
the relation $\rmd T/\rmd t = - H T$, the differential operator
$\partial_t-\vert \mathbf{p}_{\psi} \vert H \partial_{\vert
  \mathbf{p}_i\vert}$ is written as $z H \partial_z$, and
consequently~\cite{Kawasaki:1992kg}
\begin{align}
  \label{eq:boltzmann-with-2}
  \frac{\partial f_{\psi}(z,y)}{\partial z} = \frac{z}{H(m_{\psi})}\,
  C_{D}\left[ f_{\psi}(z,y)\right]
\end{align}
with $H \left(m_{\psi}\right)= H \left|_{T=m_\psi} \right.$. 
In this form, the Boltzmann equation can be easily solved numerically
on a grid for specific rescaled momenta $y$. For the right hand side,
we have to sum over the collision terms of all processes which involve
the particle $\psi$ and change the phase space distribution. The
collision term for a process $\psi + a + \cdots \leftrightarrow i + j
+ \cdots$ is given by\cite{Kolb:EarlyU}\footnote{We have chosen a
  normalisation different from Kolb and Turner, so
  $C_\textrm{here}=\frac{1}{2 E_\psi} C_{\rm KT}$}
\begin{align}
  \label{eq:b3}
g_\j  \, C[\psi + a + \cdots \leftrightarrow i + j + \cdots]& = - \frac{1}{2
    E_\psi} \int \prod_\alpha \rmd \tilde{p}_\alpha (2 \pi)^4
  \delta^4(p_\psi+p_a + \cdots - p_i - p_j -\cdots) \nonumber \\
& \times \left[ \left| \mathcal{M}
    (\psi + a + \cdots \rightarrow i + j + \cdots) \right|^2 f_\psi
  f_a \cdots (1\pm f_i)(1\pm f_j) \cdots \right. \nonumber \\
& - \left. \left| \mathcal{M}
 (i + j + \cdots \rightarrow \psi + a + \cdots) \right|^2 f_i
  f_j \cdots (1\pm f_\psi)(1\pm f_a) \cdots \right],
\end{align}
where $\alpha=(a, \cdots,i,j,\cdots)$,
\begin{equation}
  \label{eq:b4}
\rmd \tilde{p}_\alpha= \frac{\rmd^3 p_\alpha}{(2 \pi)^3 2 E_\alpha}.
\end{equation}
The terms $(1\pm f_i)$ hold for fermions ($-$) and bosons ($+$) and are
interpreted as Fermi-blocking ($-$) and Bose-enhancement ($+$). In
practice, we will only look at processes which involve three or four
particles, that is, decays, inverse decays and scatterings. We have
included the internal degrees of freedom, $g_\psi, g_a, \cdots, g_i, g_j,
\cdots$, in the matrix elements, therefore we need to put $g_\psi$ in
front of the collision term since it is not included in the
phase-space density $f_\psi$.

We integrate
equation~\eqref{eq:boltzmann-with-2} over the phase space of the
incoming particle with $g_\psi \int \rmd^3 p_\psi/(2 \pi)^3$ and arrive at
\begin{equation}
  \label{eq:b2}
  \frac{\rmd n_\psi}{\rmd z}= - \frac{z}{H(m_\psi)} \sum_{\rm processes} \left[ 
\gamma(\psi + a + \cdots \rightarrow i + j + \cdots)
- \gamma(i + j + \cdots \rightarrow \psi + a + \cdots)
\right],
\end{equation}
where
\begin{align}
  \label{eq:b5}
  \gamma(\psi + a + \cdots \rightarrow i + j + \cdots) &= - g_\psi \int
  \frac{\rmd^3 p_\psi}{(2 \pi)^3} C[\psi + a + \cdots \rightarrow i +
  j + \cdots] \nonumber \\
& =   \int \prod_\beta \rmd \tilde{p}_\beta (2 \pi)^4
  \delta^4(p_\psi+p_a + \cdots - p_i - p_j -\cdots) \nonumber \\
& \times \left| \mathcal{M}
    (\psi + a + \cdots \rightarrow i + j + \cdots) \right|^2 f_\psi
  f_a \cdots (1\pm f_i)(1\pm f_j) \cdots,
\end{align}
where we now integrate over $p_\psi$ as well, that is,
$\beta=(\psi,a,\cdots,i,j,\cdots)$. The analogous equation holds for
$\gamma (i + j + \cdots \rightarrow \psi + a + \cdots)$.

\section{Low Temperature}
\label{sec:low-temperature}

\subsection{Neutrino evolution}
\label{sec:neutrino-evolution}

The Boltzmann equation for the evolution of the lightest right-handed
neutrino is
\begin{align}
  \label{eq:b6}
  \frac{\rmd n_{N_1}}{\rmd z}= - \frac{z}{H(M_{N_1})}
    & \left[ \gamma(N_1 \rightarrow \phi \ell_+) + \gamma(N_1 \to \barphi
    \barell_+) 
    - \gamma(\phi \ell_+ \to N_1) - \gamma(\barphi \barell_+ \to N_1) 
  \right. \nonumber \\
+   & \left. \gamma(N_1 \rightarrow \phi \ell_-) + \gamma(N_1 \to \barphi
    \barell_-) 
    - \gamma(\phi \ell_- \to N_1) - \gamma(\barphi \barell_- \to N_1) 
  \right]
\end{align}
where $\ell_\pm$ denote the two lepton modes.  We neglect scatterings
since they are of higher order in the coupling constant.  We will from
now on omit the subscript 1 for the neutrino and write $N$. The
$CP$-asymmetry in the matrix element is not relevant for neutrino
decay, so we calculate the matrix element, which is the same for the
above processes and define
\begin{align}
  \label{eq:b68}
  \left|\mathcal{M}^0_\pm \right|^2 \equiv \left| \mathcal{M} (N \to
    HL_\pm) \right|^2 =  \left| \mathcal{M}
    (HL_\pm \to N) \right|^2, 
\end{align}
where now $HL_\pm$ denotes the sum of leptons and Higgs doublets,
$\ell_\pm$ and $\phi$, and their charge conjugated states
$\bar{\ell}_\pm$ and $\bar{\phi}$. The matrix elements are, however,
different for the different lepton modes and also the
momentum-conserving delta functions differ from each other. The
subscript $\pm$ means ($+$) or ($-$), not the sum. When summing an
expression $A_\pm$ that is dependent on the kind of lepton dispersion
relation over the lepton modes, we write $\sum_\pm A_\pm$. We have
\begin{align}
  \label{eq:b69}
    \frac{\rmd n_{N}}{\rmd z}= - \frac{z}{H(M_{N_1})} \sum_\pm
    \left[ \gamma(N \rightarrow HL_\pm) - \gamma(HL_\pm \to N)  \right]
\end{align}
For each of the two lepton modes, we have now
\begin{align}
  \label{eq:b7}
  \gamma(N \rightarrow H L_\pm) - \gamma(HL_\pm \rightarrow N) 
& =  \int
  \rmd \tilde{p}_N \rmd \tilde{p}_{L \pm} \rmd \tilde{p}_H (2 \pi)^4 \delta^4(p_{N} -p_H-p_{L \pm}) \nonumber \\
& \times
  \left[ \left|\mathcal{M}(N \rightarrow H L_\pm)\right|^2 f_{N}
    (1+f_H)(1-f_{L \pm}) \right. \nonumber \\ 
&- \left. \left|\mathcal{M}(HL_\pm \rightarrow N)\right|^2
    (1-f_{N}) f_H f_{L \pm} \right].
\end{align}
The term in square brackets in equation~\eqref{eq:b7} reduces to
\begin{equation}
  \label{eq:b10}
  \left|\mathcal{M}^0_\pm \right|^2  \left[ f_{N}
    (1+f_H)(1-f_{L \pm})- (1-f_{N}) f_H f_{L \pm} \right] =
  \left|\mathcal{M}^0_\pm \right|^2 \left[ c_{N \rightarrow HL\pm}- c_{HL\pm
      \rightarrow N} \right],
\end{equation}
where
\begin{align}
  \label{eq:b11}
  c_{N \rightarrow HL\pm} &= f_N (1+f_H)(1-f_{L \pm}) \, , \nonumber \\
  c_{HL\pm \rightarrow N} &= (1-f_N) f_H f_{L \pm} \, .
\end{align}
Throughout this chapter, we make the kinetic equilibrium assumption,
that is, the phase space densities can be written as
\begin{equation}
  \label{eq:b9}
  f_i=\frac{n_i}{n_i^{\rm eq}} f_i^{\rm eq}= x_i f_i^\rmeq,
\end{equation}
where
\begin{align}
  \label{eq:b12}
  x_i \equiv \frac{n_i}{n_i^\rmeq} \, ,
\end{align}
where $n^\rmeq$ and $f^\rmeq$ are the equilibrium number densities and
distributions.  For the neutrino evolution, we can assume that the
Higgs bosons are in equilibrium since they couple very strongly to the
thermal bath, $f_H=f_H^\rmeq$. The lepton distributions of the two
modes are, strictly speaking, out of equilibrium since leptons and
antileptons are created asymmetrically. However, the leptons are much
closer to equilibrium than the neutrinos, so the neutrino evolution is
not influenced by the lepton asymmetry. Therefore we approximate the
lepton densities with their equilibrium density, $f_{L \pm}=f_{L
  \pm}^\rmeq$. We will relax this assumption in the section on the
lepton asymmetry evolution.

Using the relation
\begin{equation}
  \label{eq:b13}
  f_N^\rmeq (1+f_H^\rmeq)(1-f_{L \pm}^\rmeq)=(1-f_N^\rmeq) f_{L \pm}^\rmeq f_H^\rmeq,
\end{equation}
we can write
\begin{align}
  \label{eq:b14}
  c_{N \rightarrow HL\pm} &=x_N f_N^\rmeq (1+f_H^\rmeq)(1-f_{L
    \pm}^\rmeq)=
  (x_N- x_N f_N^\rmeq) f_H^\rmeq f_{L \pm}^\rmeq \, , \nonumber \\
  c_{HL\pm \rightarrow N} &= (1- x_N f_N^\rmeq) f_H^\rmeq f_{L
    \pm}^\rmeq \, ,
\end{align}
and
\begin{align}
\label{eq:158}
c_{N \rightarrow HL\pm}- c_{N \rightarrow HL\pm} &=
(x_N-1) f_H^\rmeq f_{L \pm}^\rmeq.
\end{align}
The decay densities are
\begin{align}
  \label{eq:b16}
  \gamma(N \rightarrow H L_\pm) - \gamma(HL_\pm\rightarrow N) & = \int \rmd
  \tilde{p}_N \rmd \tilde{p}_{L \pm} \rmd \tilde{p}_H (2 \pi)^4
  \delta^4(p_{N} -p_H-p_{L \pm}) \nonumber \\ 
& \times \left|\mathcal{M}_\pm^0\right|^2 (x_N-1)
  f_H^\rmeq f_{L \pm}^\rmeq \nonumber \\
& = (x_N-1) \gamma_{D \pm}^N,
\end{align}
where
\begin{align}
  \label{eq:b17}
  \gamma_{D \pm}^N=  \int \rmd
  \tilde{p}_N \rmd \tilde{p}_{L \pm} \rmd \tilde{p}_H (2 \pi)^4
  \delta^4(p_{N} -p_H-p_{L \pm})  \left|\mathcal{M}_\pm^0\right|^2
  f_H^\rmeq f_{L \pm}^\rmeq.
\end{align}
Note that $\gamma_{D \pm}^N$ is not the same as the equilibrium decay
density in equation \eqref{eq:128}, but differs from the latter
through the thermal factor $f_H f_L$. It is an effective decay
density, which enters the Boltzmann equations. The Boltzmann equation
for the neutrinos reads
\begin{align}
  \label{eq:b18}
    \frac{\rmd n_{N}}{\rmd z}= - \frac{z}{H(M_{N})} (x_N-1)
    \gamma_D^N,
\end{align}
where $\gamma_D^N \equiv \gamma_+ + \gamma_-$, or, in analogy to
equation~\eqref{eq:49},
\begin{align}
  \label{eq:b19}
   \frac{\rmd n_{N}}{\rmd z}= - D^N (n_N-n_N^\rmeq),
\end{align}
where
\begin{align}
  \label{eq:b20}
  D^N=\frac{\gamma_D^N}{n_N^\rmeq} \frac{1}{H z}
\end{align}
and we have used $H(M_N)=H z^2$.  Most conveniently, the number
densities are normalised by the entropy density $s$ in order to
factorise their dependence on the expansion of the universe. The
entropy density scales as
\begin{align}
  \label{eq:66}
  s=g_* \frac{2 \pi^2}{45} T^3, 
\end{align}
where $g_*$ counts the total number of effectively massless degrees of
freedom and is defined in equation~\eqref{eq:212}. We define all
number densities in terms of the entropy density as
\begin{align}
  \label{eq:70}
  Y_i \equiv \frac{n_i}{s},
\end{align}
then
\begin{align}
  \label{eq:76}
  x_i = \frac{Y_i}{Y_i^\rmeq}.
\end{align}
The Boltzmann equation reads
\begin{align}
  \label{eq:71}
      \frac{\rmd Y_{N}}{\rmd z}= - \frac{z}{s H_1} (x_N-1)
    \gamma_D^N,
\end{align}
where
\begin{align}
  \label{eq:72}
  H_1 \equiv H(T=M_N) = \sqrt{\frac{4 \pi^3 g_*}{45}} \frac{M_N}{M_{\rm Pl}},
 \end{align}
and the Planck mass is
\begin{align}
  \label{eq:73}
  M_{\rm Pl}= 1.221 \cdot 10^{19} \, {\rm GeV}.
\end{align}

\subsection{Lepton asymmetry evolution}
\label{sec:lept-antil-evol}

We set up evolution equations for the two different lepton modes
separately and define the phase space density of the lepton asymmetry
in the respective mode as
\begin{align}
  \label{eq:b21}
  f_{\mathcal{L}h}
  =f_{\ell h}-f_{\bar{\ell} h}.
\end{align}
where $h=\pm1$ denotes the helicity-over-chirality ratio of the
leptons. The final lepton asymmetry is then
$n_{\mathcal{L}}^{\rm fin}=n_{\mathcal{L}+}^{\rm fin}+n_{\mathcal{L}-}^{\rm fin}$ after evaluating
the Boltzmann equations for each mode separately. 
\pagebreak
The Boltzmann
equations for leptons and antileptons read
 \begin{align}
   \label{eq:b22}
   \frac{\rmd n_{\ell h_1}}{\rmd z} = - \frac{z}{H(M_{N})} \Big\{
     & \gamma(\ell_{h_1} \phi \rightarrow N)
     - \gamma(N \rightarrow  \ell_{h_1} \phi) \nonumber \\
     & + \sum_{h_2} \left[ \gamma(\ell_{h_1} \phi \rightarrow
       \bar{\ell}_{h_2} \bar{\phi})- \gamma(\bar{\ell}_{h_2}
       \bar{\phi} \rightarrow \ell_{h_1} \phi) \right]
   \Big\} \, , \nonumber \\
   \frac{\rmd n_{\bar{\ell} h_1}}{\rmd z} = - \frac{z}{H(M_{N})}
   \Big\{ & \gamma(\bar{\ell}_{h_1} \bar{\phi} \rightarrow N)
     - \gamma(N \rightarrow  \bar{\ell}_{h_1} \bar{\phi}) \nonumber \\
     & + \sum_{h_2} \left[ \gamma(\bar{\ell}_{h_1} \bar{\phi}
       \rightarrow \ell_{h_2} \phi) - \gamma(\ell_{h_2} \phi
       \rightarrow \bar{\ell}_{h_1} \bar{\phi}) \right] \Big\},
\end{align}
where we have $(h_1,h_2)= \pm 1$ to account for the second lepton
involved in the scatterings and we have only included $\Delta L=2$
scatterings since the other scatterings involving neutrinos are
negligible.
For the evolution of the lepton asymmetry, we have
\begin{align}
  \label{eq:b23}
  \frac{\rmd n_{\mathcal{L} h_1}}{\rmd z} =& - \frac{1}{H z}  \Big\{
    \gamma(\ell_{h_1} \phi \to N) - \gamma(\barell_{h_1} \barphi \to N) - \gamma(N
    \to \ell_{h_1} \phi) + \gamma(N \to \barell_{h_1} \barphi)  \nonumber \\
  &+ \sum_{h_2}  \left[ 
\gamma(\ell_{h_1} \phi\to \barell_{h_2} \barphi)
 - \gamma(\barell_{h_1} \barphi \to \ell_{h_2} \phi)
+ \gamma(\ell_{h_2}\phi\to \barell_{h_1} \barphi) 
 - \gamma(\barell_{h_2} \barphi \to \ell_{h_1} \phi) \right]
\Big\} \, .
\end{align}

\begin{figure}
  \centering
  \includegraphics{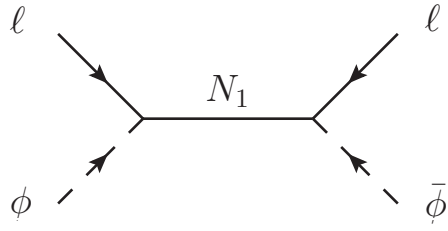}
  \caption{The $s$-channel contribution to the $\D L=2$ scattering $\ell
    \phi\to \barell \barphi$.}
  \label{fig:schannel}
\end{figure}

At leading order in the couplings, the $\D L=2$ scatterings are
computed at tree level and are consequently $CP$-conserving. However,
from these scatterings we must subtract the $CP$-violating
contribution where an on-shell $N_1$ is exchanged in the $s$ channel,
shown in figure~\ref{fig:schannel}. This is because in the Boltzmann equations
the process is already taken into account by inverse decays with
successive decays, $\ell \phi \to N \to \barell
\barphi$~\cite{Kolb:1979qa,Giudice:2003jh,Buchmuller:2004nz}.
We must therefore
replace the scattering rate by the subtracted rate,
\begin{equation}
  \label{eq:b24}
  \gamma(\ell_{h_1} \phi \to \barell_{h_2} \barphi) \to \gamma^{\rm
    sub}(\ell_{h_1} \phi\to \barell_{h_2}
  \barphi) \equiv \gamma(\ell_{h_1} \phi \to \barell_{h_2} \barphi) - \gamma^{\rm
    on-shell}(\ell_{h_1} \phi 
  \to \barell_{h_2} \barphi)
\end{equation}
and $\gamma(\barell_{h_1} \barphi \to \ell_{h_2} \phi)$ accordingly,
where $\gamma^{\rm on-shell}$ is the on-shell contribution.  The
Boltzmann equations then read
\begin{align}
  \label{eq:b30}
  \frac{\rmd n_{\mathcal{L} h_1}}{\rmd z} =& - \frac{1}{H z}  \Big\{
    \gamma(\ell_{h_1} \phi \to N) - \gamma(\barell_{h_1} \barphi \to N) - \gamma(N
    \to \ell \phi_{h_1}) + \gamma(N \to \barell_{h_1} \barphi)  \nonumber \\
  &+ \sum_{h_2}  \big[ 
\gamma^{\rm sub}(\ell_{h_1} \phi\to \barell_{h_2} \barphi)
 - \gamma^{\rm sub}(\barell_{h_1} \barphi \to \ell_{h_2} \phi)
 \nonumber \\
& + \gamma^{\rm sub}(\ell_{h_2}\phi\to \barell_{h_1} \barphi) 
 - \gamma^{\rm sub}(\barell_{h_2} \barphi \to \ell_{h_1} \phi) \big]
\Big\} \, .
\end{align}

Since the $CP$-asymmetry in neutrino decays is defined on amplitude
level as
\begin{equation}
  \label{eq:b26}
  \epsilon^N_{h}=\frac{\left| \mathcal{M}(N \to \phi\ell_{h}) \right|^2 - 
\left| \mathcal{M}(N \to \barphi \barell_{h}) \right|^2}
{\left| \mathcal{M}(N \to \phi\ell_{h}) \right|^2 + 
\left| \mathcal{M}(N \to \barphi \barell_{h}) \right|^2}
\end{equation}
and $\left| \mathcal{M}(N \to \phi\ell_{h}) \right|^2 + \left|
    \mathcal{M}(N \to \barphi \barell_{h}) \right|^2= \left| \mathcal{M}_{0 h}
  \right|^2$, we write
  \begin{align}
    \label{eq:b27}
    \left| \mathcal{M}(N \to \phi\ell_h) \right|^2=
\left| \mathcal{M}(\barphi \barell_h \to N) \right|^2 &= \frac{1+\epsilon_h^N}{2} 
\left| \mathcal{M}_{0 h} \right|^2, \nonumber \\
    \left| \mathcal{M}(N \to \barphi \barell_h) \right|^2=
\left| \mathcal{M}(\phi\ell_h\to N) \right|^2 &= \frac{1-\epsilon_h^N}{2} 
\left| \mathcal{M}_{0 h} \right|^2.
  \end{align}

  It is useful to write the decay rates for the above $1
  \leftrightarrow 2$ processes as in
  section~\ref{sec:neutrino-evolution},
\begin{equation}
  \label{eq:b25}
  \gamma({\rm process})=\int \prod_j \rmd \tilde{p}_j (2 \pi)^4
  \delta^4\left(\sum p_j\right) c({\rm process}),
\end{equation}
where $p_j$ denotes the relevant momenta $p_N$, $p_{\ell h}=p_{\barell h}$ and
$p_\phi=p_{\barphi}$ and $\delta^4(\sum p_j)$ the momentum conservation
$\delta^4(p_N-p_{\ell h}-p_\phi)$. The information about the specific
process is encoded in $c({\rm process})$ and we have
\begin{align}
  \label{eq:b28}
  c_{N \to \ell_h \phi} &= \left| \mathcal{M}(N \to \phi\ell_h)
\right|^2 f_N (1-f_{\ell h}) (1+f_\phi)
  \nonumber \\
  c_{\ell_h \phi \to N} &= \left| \mathcal{M}(\phi\ell_h\to N) \right|^2
 (1-f_N) f_{\ell h} f_\phi
  \nonumber \\
  c_{N \to \barell_h \barphi} &= \left| \mathcal{M}(N \to \barphi
    \barell_h) \right|^2 f_N (1-f_{\barell h})
  (1+f_{\barphi})
  \nonumber \\
  c_{\barell_h \barphi \to N} &= \left| \mathcal{M}(\barphi \barell_h \to
    N) \right|^2 (1-f_N) f_{\barell h} f_{\barphi}.
\end{align}
Since we are looking at the lepton asymmetry, the lepton distributions
have to be out of equilibrium, 
\begin{align}
  \label{eq:b29}
  f_{\ell/\barell h}&=x_{\ell/\barell h} f_{\ell h}^\rmeq, \nonumber \\
f_{\mathcal{L} h}&= x_{\mathcal{L}.h} f_{\ell h}^\rmeq, \nonumber \\
f_{\ell h}+f_{\barell h}& \approx
 2 f_{\ell h}^\rmeq,
\end{align}
while the Higgs bosons can be assumed to be in equilibrium.

As explained in appendix~\ref{sec:subtr-shell-prop}, the scattering
rates can be written as
\begin{align}
  \label{eq:b31}
  \sum_{h_f} \left[ \gamma^{\rm sub} (\ell_{h_i}\phi \right.
&
\left. \to \barell_{h_f}
    \barphi) - \gamma^{\rm sub}(\barell_{h_i} \barphi \to \ell_{h_f} \phi)
  \right] \nonumber \\
& =   \sum_{h_f} \left[ \gamma^{\rm sub} (\ell_{h_f}\phi\to \barell_{h_i}
    \barphi) - \gamma^{\rm sub}(\barell_{h_f} \barphi \to \ell_{h_i} \phi)
  \right]= \nonumber \\
& = \int \rmd \tilde{p}_N \rmd \tilde{p}_{\ell h_i} \rmd
  \tilde{p}_\phi (2 \pi)^4 \delta^4(p_N-p_{\ell h_i}-p_\phi)
  \e_{h_i}^N \left| \mathcal{M}_{h_i}^0 \right|^2 f_{\ell h_i}^\rmeq f_\phi^\rmeq
  (1-f_N^\rmeq).
\end{align}
so we define\footnote{Note that our factor $c^{\rm sub}$ differs from
  reference~\cite{HahnWoernle:2009qn}, where they have the
  out-of-equilibrium distribution $(1-f_N)$ instead of
  $(1-f_N^\rmeq)$. However, as derived in
  appendix~\ref{sec:subtr-shell-prop}, we must employ $f_N^\rmeq$,
  even if we had only one lepton mode, which also results in a
  Boltzmann equation for $(\ell - \barell)$ which is slightly
  different from theirs.}
\begin{align}
  \label{eq:b32}
  c^{\rm sub}_h=2 \e_h^N \left| \mathcal{M}^0_h \right|^2
    f_{\ell h}^\rmeq f_\phi^\rmeq (1-f_N^\rmeq)
\end{align}
\noindent and calculate the integrand for the right-hand side of the Boltzmann
equation~\eqref{eq:b30},
\begin{align}
  \label{eq:b33}
  c(N \to \ell_h \phi)  - c(\ell_h \phi \to N) - 
 & 
c(N  \to \barell_h
    \barphi) + c(\barell_h \barphi \to N) + c^{\rm sub}_h 
= 
\nonumber \\[1ex]
& =
  x_{\mathcal{L} h} f_{\ell h}^\rmeq (f_\phi^\rmeq + x_N f_N^\rmeq )
               - 2 \e_h^N
  f_{\ell h}^\rmeq f_\phi^\rmeq \left( x_N -1 \right) \left( 1 - 2
    f_N^\rmeq \right).
\end{align}
We can easily check that this term vanishes when the neutrinos are in
equilibrium, $x_N=1$, and there is no previous lepton asymmetry,
$x_{\mathcal{L}h}=0$.

The Boltzmann equation reads now
\begin{align}
  \label{eq:b34} 
\frac{\rmd n_{\mathcal{L}h}}{\rmd z} = - \frac{1}{H
z} \left[ - \e^N_{\g h} \g_{\e h}^N \left( x_N-1 \right) +
\frac{x_{\mathcal{L}h}}{2} \left( \gamma_{W h}^N + x_N \g_{N
h}^N \right) \right] ,
\end{align}
where $\g_{W h}^N=\g_{D h}^N$ is defined in equation~\eqref{eq:b17} and
\begin{align}
  \label{eq:b35}
  \g_{\e h}^N  &= \int \rmd
  \tilde{p}_N \rmd \tilde{p}_{\ell h} \rmd \tilde{p}_\phi (2 \pi)^4
  \delta^4(p_{N} -p_\phi-p_{\ell h})  \left|\mathcal{M}_0\right|^2
  f_\phi^\rmeq f_{\ell h}^\rmeq (1 - 2 f_N^\rmeq) \nonumber \\ 
\g_{N h}^N &=  \int \rmd
  \tilde{p}_N \rmd \tilde{p}_{\ell h} \rmd \tilde{p}_\phi (2 \pi)^4
  \delta^4(p_{N} -p_\phi-p_{\ell h})  \left|\mathcal{M}_0\right|^2
  f_{\ell h}^\rmeq f_N^\rmeq, \nonumber \\
\e_{\g h}^N &=  \frac{1}{\g_{\e h}} \int \rmd
  \tilde{p}_N \rmd \tilde{p}_{\ell h} \rmd \tilde{p}_\phi (2 \pi)^4
  \delta^4(p_{N} -p_\phi-p_{\ell h}) \e_h^N \left|\mathcal{M}_0\right|^2
  f_\phi^\rmeq f_{\ell h}^\rmeq (1 - 2 f_N^\rmeq).
\end{align}
We see that the rates and the $\CP$-asymmetries that enter the
Boltzmann equations have slightly different thermal factors than the
equilibrium rates and $\CP$-asymmetries in equations~\eqref{eq:128}
and~\eqref{eq:174}, which employ the factor $f_N (1-f_\ell) (1+f_\phi)$
for $N$ decays.

We may also write
\begin{align}
  \label{eq:74}
  \frac{\rmd Y_{\mathcal{L}h}}{\rmd z} = - \frac{z}{s H_1} \left[ -
    \e^N_{\g h} \g_{\e h}^N \left( x_N-1 \right) +
    \frac{x_{\mathcal{L}h}}{2} \left( \gamma_{W h}^N + x_N \g_{N h}^N
    \right) \right]
\end{align}
or, corresponding to equation~\eqref{eq:50},
\begin{align}
  \label{eq:b36}
    \frac{\rmd n_{\mathcal{L}h}}{\rmd z} = \epsilon_{\g h}^N D_{\e h}^N
    (n_N-n_N^\rmeq) - (W_{0 h}^N+ W_{N h}^N x_N)  n_{\mathcal{L}
      h},
\end{align}
where
\begin{align}
  \label{eq:b37}
  D_{\e h}^N&=\frac{1}{Hz} \frac{\g_{\e h}^N}{n_N^\rmeq} \nonumber \\
  W_{0 h}&=\frac{1}{Hz} \frac{\g_{W h}^N}{2 n_{\ell h}^\rmeq} \nonumber \\
  W_{N h} & = \frac{1}{Hz} \frac{\g_{N h}^N}{2 n_{\ell h}^\rmeq}.
\end{align}

\section{High temperature}
\label{sec:high-temperature}

As discussed in chapter~\ref{decayrate}, the neutrino processes $N \leftrightarrow
\ell \phi$ are forbidden when the thermal masses of the Higgs bosons
and leptons become too large, that is, when $m_\phi > M_N$. However,
new processes with the Higgs as single initial or final state are then
allowed, $\phi \leftrightarrow N \ell$. These are the dominant
contributions to the neutrino and lepton evolution and they can be
$CP$-violating as well, so they contribute to generating a lepton
asymmetry. We derive the Boltzmann equations for this high temperature
regime in the following.

\subsection{Neutrino evolution}
\label{sec:neutrino-evolution-1}

We derive the Boltzmann equation analogously to
section~\ref{sec:neutrino-evolution},
\begin{align}
  \label{eq:b44}
    \frac{\rmd n_{N}}{\rmd z}= - \frac{1}{Hz} \sum_h \left[ 
\gamma(NL_h \rightarrow H)
- \gamma(H\rightarrow NL_h)
\right].
\end{align}
We have
\begin{align}
  \label{eq:b52}
  \gamma(NL_h \rightarrow H ) - \gamma(H\rightarrow NL_h) & = \int
  \rmd \tilde{p}_N \rmd \tilde{p}_{Lh} \rmd \tilde{p}_H (2 \pi)^4
  \delta^4 (p_H -p_N-p_{Lh}) \nonumber \\
  & \times \left[ \left|\mathcal{M}(NL_h \rightarrow H)\right|^2 f_N
    f_{Lh} (1+f_H) \right. \nonumber \\
  &- \left. \left|\mathcal{M}(H \rightarrow NL_h)\right|^2 (1-f_N)
    (1-f_{Lh}) f_H \right].
\end{align}
The tree-level matrix elements $\left| \mathcal{M}^0_h \right|^2$ are the
same at high temperature for the Higgs-processes, just the kinematics
differ. So we have
\begin{align}
  \label{eq:b53}
    \left|\mathcal{M}^0_h\right|^2 \equiv
  \left|\mathcal{M}(NL_h \rightarrow H)\right|^2 = \left|\mathcal{M}(H
    \rightarrow NL_h)\right|^2.
\end{align}
Again, we assume the Higgs bosons and leptons to be in equilibrium. We
write
\begin{align}
  \label{eq:b54}
    \left|\mathcal{M}_0\right|^2  \left[ f_N
    f_{Lh} (1+f_H)- (1-f_N) (1-f_{Lh}) f_H \right] =
  \left|\mathcal{M}_0\right|^2 \left[ c(NL_h \rightarrow H)- c(H
      \rightarrow NL_h) \right] \, .
\end{align}
Using the relation
\begin{align}
  \label{eq:b55}
  f_N^\rmeq f_{Lh}^\rmeq (1+f_H^\rmeq) = (1-f_N^\rmeq) (1-f_{Lh}^\rmeq) f_H^\rmeq,
\end{align}
we get
\begin{align}
  \label{eq:b56}
   c(NL_h \rightarrow H)- c(H
      \rightarrow NL_h) = (x_N-1) (1-f_{Lh}^\rmeq) f_H^\rmeq.
\end{align}
The Boltzmann equation then reads
\begin{align}
  \label{eq:b57}
      \frac{\rmd n_{N}}{\rmd z} = - \frac{1}{Hz} (x_N-1) 
    \gamma_{D}^\phi,
\end{align}
\begin{align}
  \label{eq:77}
    \frac{\rmd Y_{N}}{\rmd z} = - \frac{z}{s H_1} (x_N-1) 
    \gamma_{D}^\phi,
\end{align}
or
\begin{align}
  \label{eq:b60}
  \frac{\rmd n_{N}}{\rmd z} = - D^\phi (n_N-n_N^\rmeq),
\end{align}
where $\gamma_{D}^\phi= \gamma_{D+}^\phi+ \gamma_{D-}^\phi$,
\begin{align}
  \label{eq:b58}
    \gamma_{Dh}^\phi=  \int \rmd
  \tilde{p}_N \rmd \tilde{p}_{Lh} \rmd \tilde{p}_H (2 \pi)^4
  \delta^4(p_H -p_N-p_{Lh})  \left|\mathcal{M}^0_h\right|^2
  f_H^\rmeq (1-f_{Lh}^\rmeq)
\end{align}
and
\begin{align}
  \label{eq:b59}
  D^\phi=\frac{\gamma_{D}^\phi}{n_N^\rmeq} \frac{1}{H z} \, .
\end{align}

\subsection{Lepton asymmetry evolution}
\label{sec:lept-antil-evol-1}

\begin{figure}
  \centering
  \includegraphics{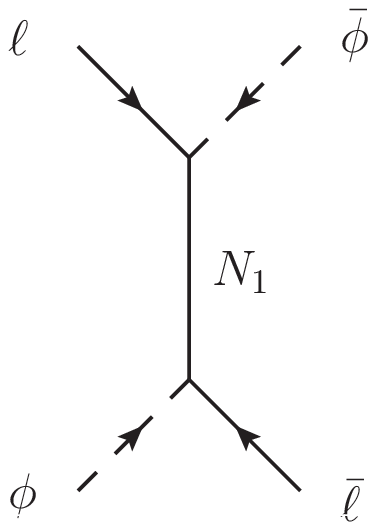}
  \caption{The $u$-channel contribution to the $\D L=2$ scattering $\ell
    \phi\to \barell \barphi$.}
  \label{fig:uchannel}
\end{figure}

The Boltzmann equations for leptons and antileptons read
\begin{align}
  \label{eq:b61}
    \frac{\rmd n_{\ell h_1}}{\rmd z}  = 
- \frac{1}{Hz} \Big\{
   & \gamma(\ell_{h_1} N \rightarrow \barphi) 
- \gamma(\barphi \rightarrow \ell_{h_1} N) \nonumber \\
  &  + \sum_{h_2} \left[\gamma(\ell_{h_1} \phi \rightarrow
      \bar{\ell}_{h_2} \bar{\phi})-
    \gamma(\bar{\ell}_{h_2} \bar{\phi} \rightarrow \ell_{h_1} \phi)
  \right] \Big\}, \\
  \frac{\rmd n_{\bar{\ell h_1}}}{\rmd z}   =
- \frac{z}{H(M_{N})}
 \Big\{
 &  \gamma(\bar{\ell}_{h_1} N \rightarrow \phi) 
- \gamma(\phi \rightarrow \barell_{h_1} N) \nonumber \\
& + \sum_{h_2} \left[ \gamma(\bar{\ell}_{h_1} \bar{\phi} \rightarrow \ell_{h_2} \phi)
 - \gamma(\ell_{h_2} \phi \rightarrow \bar{\ell}_{h_1} \bar{\phi})
\right] \Big\}, 
\end{align}
combined we get
\begin{align}
  \label{eq:b62}
  \frac{\rmd n_{\mathcal{L} h_1}}{\rmd z} =& - \frac{1}{H z} \Big\{
  \gamma(\ell_{h_1} N \to \barphi) - \gamma(\barell_{h_1} N \to
  \phi) - \gamma(\barphi
  \to \ell_{h_1} N) + \gamma(\phi \to \barell_{h_1} N)  \nonumber \\
  &+ \sum_{h_2} \left[ \gamma(\ell_{h_1} \phi\to \barell_{h_2}
    \barphi) - \gamma(\barell_{h_1} \barphi \to \ell_{h_2} \phi) +
    \gamma(\ell_{h_2}\phi\to \barell_{h_1} \barphi) -
    \gamma(\barell_{h_2} \barphi \to \ell_{h_1} \phi) \right] \Big\} \, .
\end{align}
At high temperature, there can be no on-shell neutrino in the
$s$-channel of the $\D L=2$ scatterings, but there can be an on-shell
neutrino exchange in the $u$-channel as shown in figure~\ref{fig:uchannel}.
Again, we need to subtract the $\D L=2$ rates since the $u$-channel
on-shell neutrino exchange corresponds to a Higgs decay followed by an
inverse decay, $\ell \phi \to \ell N \barell \to \barphi
\barell$. We replace
\begin{align}
  \label{eq:b63}
  \gamma(\ell_{h_1} \phi \to \barell_{h_2} \barphi) \to \gamma^{\rm
    sub}_u(\ell_{h_1}\phi\to \barell_{h_2} \barphi) \equiv \gamma(\ell_{h_1} \phi \to
  \barell_{h_2} \barphi) - \gamma^{\rm on-shell}_u(\ell_{h_1} \phi \to \barell_{h_2}
  \barphi)
\end{align}
and get
\begin{align}
  \label{eq:b64}
  \frac{\rmd n_{\mathcal{L} h_1}}{\rmd z} =& - \frac{1}{H z} \Big\{
  \gamma(\ell_{h_1} N \to \barphi) - \gamma(\barell_{h_1} N \to
  \phi) - \gamma(\barphi
  \to \ell_{h_1} N) + \gamma(\phi \to \barell_{h_1} N)  \nonumber \\
  &+ \sum_{h_2} \left[ \gamma^{\rm sub}(\ell_{h_1} \phi\to \barell_{h_2}
    \barphi) - \gamma^{\rm sub}(\barell_{h_1} \barphi \to \ell_{h_2} \phi) +
    \gamma^{\rm sub}(\ell_{h_2}\phi\to \barell_{h_1} \barphi) -
    \gamma^{\rm sub}(\barell_{h_2} \barphi \to \ell_{h_1} \phi) \right] \Big\} \, .
\end{align}
We define the $CP$-asymmetry in Higgs decays as
\begin{align}
  \label{eq:b65}
  \e_{h}^\phi \equiv \frac{\left| \mathcal{M}(\barphi \to N \ell_h) \right|^2 - 
\left| \mathcal{M}(\phi \to N \barell_h) \right|^2}
{\left| \mathcal{M}(\barphi \to N \ell_h) \right|^2 + 
\left| \mathcal{M}(\phi \to N \barell_h) \right|^2},
\end{align}
thus
\begin{align}
  \label{eq:b66}
  \left| \mathcal{M}(\barphi \to N \ell_h) \right|^2 = 
  \left| \mathcal{M}(\barell_h N \to \phi) \right|^2 =
  \frac{1+\e_{h}^\phi}{2} \left| \mathcal{M}_h \right|^2, \nonumber \\
  \left| \mathcal{M}(\phi \to N \barell_h) \right|^2 = 
  \left| \mathcal{M}(\ell_h N \to \barphi) \right|^2 =
  \frac{1-\e_{h}^\phi}{2} \left| \mathcal{M}_h \right|^2 \, .
\end{align}
As explained in appendix~\ref{sec:subtr-shell-prop}, the scattering
rates are written as
\begin{align}
  \label{eq:b67}
  \sum_{h_f} \left[ \gamma^{\rm sub} (\ell_{h_i}\phi \right.  &
  \left. \to \barell_{h_f} \barphi) - \gamma^{\rm sub}(\barell_{h_i}
    \barphi \to \ell_{h_f} \phi)
  \right] \nonumber \\
  & = \sum_{h_f} \left[ \gamma^{\rm sub} (\ell_{h_f}\phi\to
    \barell_{h_i} \barphi) - \gamma^{\rm sub}(\barell_{h_f} \barphi
    \to \ell_{h_i} \phi)
  \right] \nonumber \\
  & = \int \rmd \tilde{p}_N \rmd \tilde{p}_{\ell h_i} \rmd
  \tilde{p}_\phi (2 \pi)^4 \delta^4(p_N-p_{\ell h_i}-p_\phi) \e_h^\phi \left|
    \mathcal{M}_{h_i}^0 \right|^2 f_{\ell h_i}^\rmeq (1+f_\phi^\rmeq)
  f_N^\rmeq \, ,
\end{align}
so 
\begin{align}
  \label{eq:62}
  c_h^{\rm sub}= 2 \e^\phi_h \left| \mathcal{M}_h^0 \right|^2 f_{\ell
    h}^\rmeq (1+ f_\phi^\rmeq) f_N^\rmeq
\end{align}
and we get for the right-hand side of equation~\eqref{eq:b64}, 
\begin{align}
  \label{eq:63}
  c(\ell_{h} N \to \barphi) - & c(\barell_{h} N \to
  \phi) - 
c(\barphi
  \to \ell_{h} N) + c(\phi \to \barell_{h} N) +  c_h^{\rm sub}
\nonumber \\[1ex]
& =
  x_{\mathcal{L} h} f_{\ell h}^\rmeq (f_\phi^\rmeq + x_N f_N^\rmeq )
               + 2 \e^\phi_h
  (1-f_{\ell h}^\rmeq) f_\phi^\rmeq \left( x_N -1 \right) \left( 1 - 2
    f_N^\rmeq \right).
\end{align}

The Boltzmann equation reads now
\begin{align}
\label{eq:64}
   \frac{\rmd n_{\mathcal{L}h}}{\rmd z} = - \frac{1}{H z} \left[
    - \e_{\g h}^\phi
     \g_{\e h}^\phi \left( x_N-1 \right) + \frac{x_{\mathcal{L}h}}{2} \left(
       \gamma_{W h}^\phi + x_N \g^\phi_{N h} \right) \right] ,
\end{align}
where
\begin{align}
  \label{eq:65}
    \g^\phi_{\e h}  &= \int \rmd
  \tilde{p}_N \rmd \tilde{p}_{\ell h} \rmd \tilde{p}_\phi (2 \pi)^4
  \delta^4(p_{N} -p_\phi+p_{\ell h})  \left|\mathcal{M}_h^0\right|^2
  f_\phi^\rmeq (1-f_{\ell h}^\rmeq) (1 - 2 f_N^\rmeq) \nonumber \\ 
\g_{W h}^\phi & = \int \rmd
  \tilde{p}_N \rmd \tilde{p}_{\ell h} \rmd \tilde{p}_\phi (2 \pi)^4
  \delta^4(p_{N} -p_\phi+p_{\ell h})  \left|\mathcal{M}_h^0\right|^2
  f_\phi^\rmeq f_{\ell h}^\rmeq
\nonumber \\
\g^\phi_{N h} &=  \int \rmd
  \tilde{p}_N \rmd \tilde{p}_{\ell h} \rmd \tilde{p}_\phi (2 \pi)^4
  \delta^4(p_{N} -p_\phi+p_{\ell h})  \left|\mathcal{M}_h^0\right|^2
  f_{\ell h}^\rmeq f_N^\rmeq, \nonumber \\
\e^\phi_{\g h} &=  \frac{1}{\g^\phi_{\e h}} \int \rmd
  \tilde{p}_N \rmd \tilde{p}_{\ell h} \rmd \tilde{p}_\phi (2 \pi)^4
  \delta^4(p_{N} -p_\phi+p_{\ell h}) \e^\phi_h \left|\mathcal{M}_h^0\right|^2
  f_\phi^\rmeq (1-f_{\ell h}^\rmeq) (1 - 2 f_N^\rmeq).
\end{align}

Analogous to equation~\eqref{eq:50}, we may also write
\begin{align}
\label{eq:67}
\frac{\rmd n_{\mathcal{L}h}}{\rmd z} = - \epsilon^\phi_{\g h} D^\phi_{\e
  h} (n_N-n_N^\rmeq) - (W^\phi_{0 h}+ W^\phi_{N h} x_N) n_{\mathcal{L}
  h},
\end{align}
where
\begin{align}
\label{eq:68}
D^\phi_{\e h}&=\frac{1}{Hz} \frac{\g^\phi_{\e h}}{n_N^\rmeq} \nonumber \\
W^\phi_{0 h}&=\frac{1}{Hz} \frac{\g^\phi_{D h}}{2 n_{\ell h}^\rmeq}
\nonumber \\
W^\phi_{N h} & = \frac{1}{Hz} \frac{\g^\phi_{\mathcal{L}N h}}{2
  n_{\ell h}^\rmeq}.
\end{align}
Normalised by the entropy density, the equation reads
\begin{align}
  \label{eq:78}
     \frac{\rmd Y_{\mathcal{L}h}}{\rmd z} = - \frac{z}{s H_1} \left[ - 
     \e_{\g h}^\phi
     \g_{\e h}^\phi \left( x_N-1 \right) + \frac{x_{\mathcal{L}h}}{2} \left(
       \gamma_{W h}^\phi + x_N \g^\phi_{N h} \right) \right].
\end{align}

\section{Interacting Modes}
\label{sec:interacting-modes}

The Boltzmann equations in the previous sections were derived under
the assumption that the only interactions in which the leptons take
part are the Yukawa interactions with Higgs bosons and heavy
neutrinos, which have very small coupling constants $\lambda$. This
scenario would imply that the two modes only interact with each other
via intermediate neutrinos or Higgs bosons, where the distributions
and also the asymmetries in each mode are to first approximation
decoupled.  In a more realistic model, the lepton modes will couple to
each other via the $SU(2)$ and $U(1)$ gauge bosons $W^a_\mu$ and
$B_\mu$ in processes like $\ell_\pm \to \ell_\mp B$. While it is
conceptually interesting to consider the case that the two modes are
completely decoupled, it might be more realistic to study the scenario
where the interactions between the lepton modes are fast enough to
keep them in chemical equilibrium.

Chemical equilibrium implies that for species that interact via
processes $a+b \to i+j$, the corresponding chemical potentials are
related as
\begin{align}
  \label{eq:183}
  \mu_a+\mu_b=\mu_i+\mu_j \, .
\end{align}
When the processes which create or annihilate the particles and
antiparticles of some species are fast, for example via $a + \bar{a}
\to i+j$, where $i$ and $j$ are in equilibrium and their chemical
potentials vanish, then the chemical potentials of $a$ and $\bar{a}$
behave as
\begin{align}
  \label{eq:185}
  \mu_a+\mu_{\bar{a}} &= \mu_i + \mu_j = 0 \, , \nonumber \\
  \Rightarrow \mu_a &= -\mu_{\bar{a}} \, .
\end{align}

In order to derive the corresponding Boltzmann equation, we introduce
a chemical potential $\mu_h$ for the lepton mode $\ell_h$. For
simplicity, we approximate the distribution with Maxwell-Boltzmann
statistics, an approximation which is sufficient to derive the final
Boltzmann equations. The distribution functions in kinetic equilibrium
are
\begin{align}
  \label{eq:171}
  f_{\ell h}(k)&=\rme^{-\beta (\omega_h - \mu_h)} \, , \nonumber \\
  f_{\barell h}(k)&=\rme^{-\beta (\omega_h + \mu_h)} \, , \nonumber \\
  f_{\ell h}(k)-f_{\barell h}(k) &= \rme^{-\beta \omega_h}
  (\rme^{\beta \mu_h} - \rme^{-\beta \mu_h}) \approx 2 \beta \mu_h
  f_{\ell h}^{\rmeq} \, ,
\end{align}
for $\mu_h \ll T$. We assume chemical equilibrium between the plus-
and the minus-mode,
\begin{align}
  \label{eq:188}
  \mu_+ = \mu_- \equiv \mu_\ell \, .
\end{align}
Moreover, we can make the approximation that the equilibrium densities
are about the same since the thermal mass $m_\ell \approx 0.2 \, T$ is
too small to affect the momentum integration considerably in
\begin{align}
  \label{eq:189}
  n_{\ell_h}^\rmeq &= \int \frac{\rmd^3 k}{(2 \pi)^3}
  f_{\ell_h}^\rmeq(k) \, \nonumber \\, 
\Rightarrow n_{\ell_+}^\rmeq
  &\approx n_{\ell_-}^\rmeq \approx n_{\ell, 0}^\rmeq \, ,
\end{align}
where $n_{\ell,0}^\rmeq$ is the distribution for massless leptons. With these approximations, we have
\begin{align}
  \label{eq:192}
  n_{\mathcal{L}_+} &= 2 \beta \mu_\ell n_{\ell_0}^\rmeq = n_{\mathcal{L}_-} \, , \nonumber \\
  n_{\mathcal{L}  \pm} &\equiv n_{\mathcal{L}_+} +
  n_{\mathcal{L}_-} \equiv 2 n_{\mathcal{L}_h}\, 
, \nonumber \\
  x_{\mathcal{L}  \pm} &\equiv \frac{n_{\mathcal{L} 
      \pm}}{n_{\ell 0}^\rmeq} \, ,
\end{align}
where the subscript $\pm$ indicates that we sum over the two modes,
contrary to its use in the previous sections.  We can now add the
Boltzmann equations for the two modes in equations~\eqref{eq:74} and
\eqref{eq:78} and arrive at
\begin{align}
  \label{eq:191}
       \frac{\rmd Y_{\mathcal{L} \pm}}{\rmd z} = - \frac{z}{s H_1} \left[
     \Delta \gamma_{ \pm} \left( x_N-1 \right) +
     \frac{x_{\mathcal{L} \pm}}{4} \left(
       \gamma_{W  \pm} + x_N \g_{N  \pm} \right) \right],
\end{align}
where 
\begin{align}
  \label{eq:193}
  Y_{\mathcal{L} \pm} &= Y_{\mathcal{L}+} +  Y_{\mathcal{L}-} \, , \nonumber \\
\Delta \gamma_{ \pm} &= \epsilon_{\g +} \g_{\e +} + \epsilon_{\g -} \g_{\e -} \, , \nonumber \\
\g_{W  \pm} &= \g_{W+}+\g_{W-} \, , \nonumber \\
\g_{N  \pm} &= \g_{N+}+\g_{N-} \, .
\end{align}
The factor $1/4$ comes from the fact that $x_{\mathcal{L}h} =
x_{\mathcal{L} \pm}/2$. Depending on the temperature regime, we either
have to employ the Higgs boson or the neutrino rates in the Boltzmann
equations.

\section{One-Mode Approximation}
\label{sec:one-mode-appr}

As we did in sections~\ref{sec:disc-yukawa-ferm} and
\ref{sec:one-mode-approach} for the decay rates and the $\CP$
asymmetries, we also employ the one-mode approach for the Boltzmann
equations. The equations are derived in analogy to
sections~\ref{sec:low-temperature} and~\ref{sec:high-temperature} and
read
\begin{align}
  \label{eq:194}
  \frac{\rmd Y_{N}}{\rmd z}&= - \frac{z}{s H_1} (x_N-1)
  \gamma_{D m_\ell}, \nonumber \\
  \frac{\rmd Y_{\mathcal{L}}}{\rmd z} &= - \frac{z}{s H_1} \left[ - \D
    \g_{m_\ell} \left( x_{N}-1 \right) + \frac{x_{\mathcal{L}}}{2}
    \left( \gamma_{W m_\ell} + x_N \g_{N m_\ell} \right) \right],
\end{align}
where $\gamma_{D m_\ell}$, $\gamma_{W m_\ell}$, $\gamma_{N m_\ell}$
and $\Delta \gamma_{m_\ell}$ are defined in equations~\eqref{eq:b17},
\eqref{eq:b35}, \eqref{eq:b58} and \eqref{eq:65} and one has to make
the appropriate replacements for the matrix elements and the lepton
dispersion relations of the one-mode approach for $m_\ell$ and
$\sqrt{2} \, m_\ell$.

\section{Evaluation of the Boltzmann Equations}
\label{sec:eval-boltzm-equat}

We solve the Boltzmann equations for five different scenarios:
\begin{enumerate}
\item
 the
zero temperature case with Maxwell-Boltzmann statistics,
\item the
two-lepton-mode approach where the two modes do not interact with each
other, 
\item the two-mode approach where the modes couple strongly to each
  other,
\item the one-mode approach for a thermal mass $m_\ell$,
\item and the one-mode approach for an asymptotic thermal mass
  $\sqrt{2} \, m_\ell$. 
\end{enumerate}
In the decoupled case, the lepton asymmetries for the plus- and the
minus-mode evolve separately from each other. When solving the
equations, one has to specify the initial conditions for the neutrino
abundance and the lepton asymmetry. We assume a vanishing initial
lepton asymmetry and distinguish between three cases for the neutrino
abundances:
\begin{enumerate}
\item Zero initial abundance: this is the case, for example, when an
  inflaton field decays mostly into SM particles and not into the
  heavy neutrinos.
\item Thermal initial abundance: this can be realised when some
  additional interactions keep the neutrinos in equilibrium at $T \gg
  M_1$, for example via a heavy $Z'$ boson related to $SO(10)$
  unification~\cite{Plumacher:1996kc}.
\item Dominant initial abundance: this is the case, for example, when
  an inflaton decays predominantly into $N_1$.
\end{enumerate}
The coupling $(\lambda^\dagger \lambda)_{11}$, which enters the
neutrino decay rate, is parameterised by the so-called decay parameter
$K$, defined as
\begin{align}
  \label{eq:195}
  K \equiv \frac{\widetilde{m}_1}{m^*} \, ,
\end{align}
where $\widetilde{m}_1$ and $m^*$ are the effective neutrino mass and
the so-called equilibrium neutrino mass, defined in
equations~\eqref{eq:196} and \eqref{eq:197}. The case $K > 1$ is
called strong washout regime and the case $K < 1$ weak washout regime.

We want to analyse the evolution of the neutrino abundance and lepton
asymmetries for the weak and strong washout regimes and different
initial abundances. To this end, we write the Boltzmann equations for
the different scenarios in the form of equations~\eqref{eq:b19},
\eqref{eq:b36}, \eqref{eq:b60} and \eqref{eq:67},
\begin{align}
  \label{eq:168}
  \frac{\rmd Y_N}{\rmd z} &= - D (Y_N - Y_N^\rmeq) \, . \nonumber \\
\frac{\rmd Y_{\mathcal{L}}}{\rmd z} &= \epsilon_0 D_\epsilon (Y_N - Y_N^\rmeq) -
 (W + W_N x_N) Y_{\mathcal{L}} \, ,
\end{align}
where
\begin{align}
  \label{eq:198}
  D &= \frac{z}{H_1} \frac{\g_D}{s Y_N^\rmeq} \, ,
  & D_\epsilon &= \frac{z}{H_1} \frac{\D \g}{\epsilon_0 s Y_N^\rmeq}\, , \nonumber \\
  W & = \frac{z}{H_1} \frac{\g_W}{2 s Y_{\mathcal{L}}^\rmeq} \, ,
  \quad & W_N & = \frac{z}{H_1} \frac{\g_N}{2 s Y_{\mathcal{L}}^\rmeq}
  \, .
\end{align}
One usually refers to $D_\e (Y_N-Y_N^\rmeq)$ as source term since this
term is responsible for the production of a lepton asymmetry. The term
$(W+W_N x_N) Y_\ml^\rmeq$ is called washout term since it usually has
the opposite sign as the source term and reduces the production of the
lepton asymmetry. The terms $D$, $D_\e$, $W$ and $W_N$ are different
for the different scenarios. Note that for the finite temperature
cases, $D_\epsilon$ is not the same as $D$ and there is an additional
washout term $W_N$ due to the quantum statistics. Our analysis closely
follows the arguments and explanations in
reference~\cite{Buchmuller:2004nz} and the interested reader will find
a comprehensive explanation of leptogenesis dynamics in the vacuum
case therein.

\subsection{Weak washout for zero initial abundance}
\label{sec:weak-washout}

\begin{figure}
  \centering
  \includegraphics[width=\textwidth]{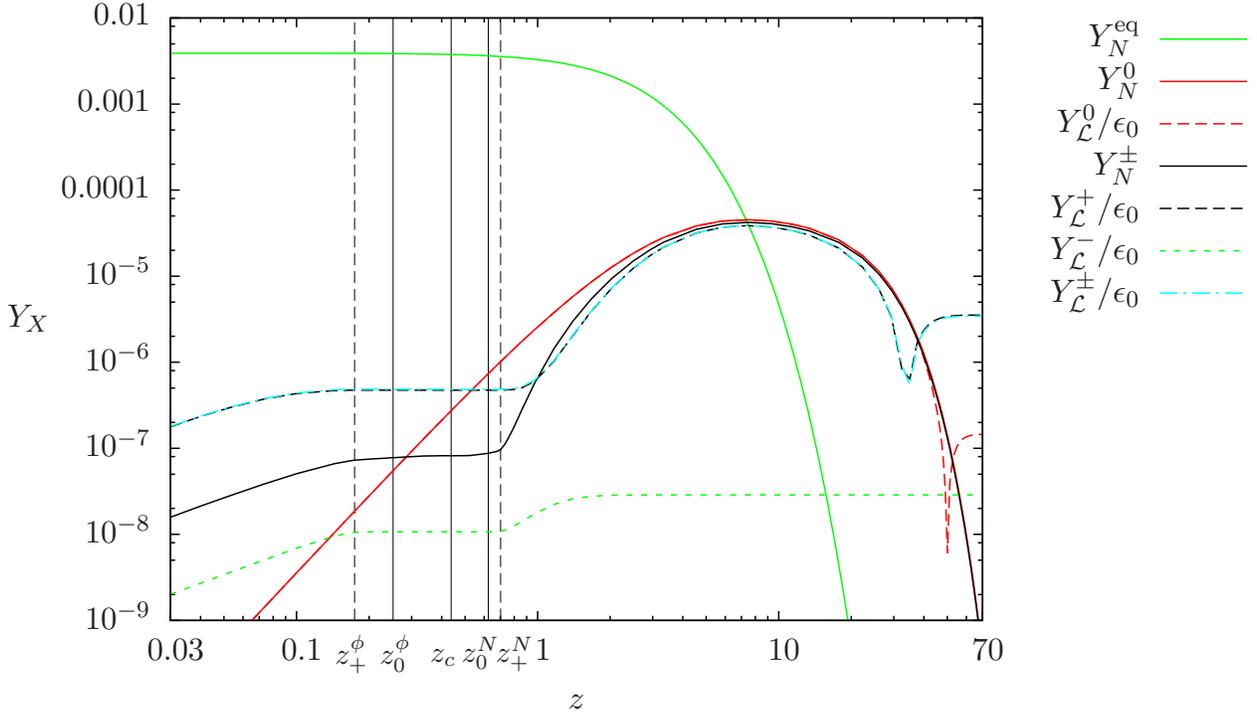}
  \caption[$Y_N(z)$ and $Y_\ml(z)$ for weak washout, zero initial
  $N$ abundance and two-mode cases.]{Evolution of neutrino abundance $Y_N(z)$ and
    lepton asymmetry $Y_\ml(z)$ for $K=0.005$ and zero initial
    neutrino abundance. We show the the two-mode cases and the vacuum
    case.}
  \label{fig:nell2k0.005ztd0}
\end{figure}

\begin{figure}
  \centering
  \includegraphics[width=\textwidth]{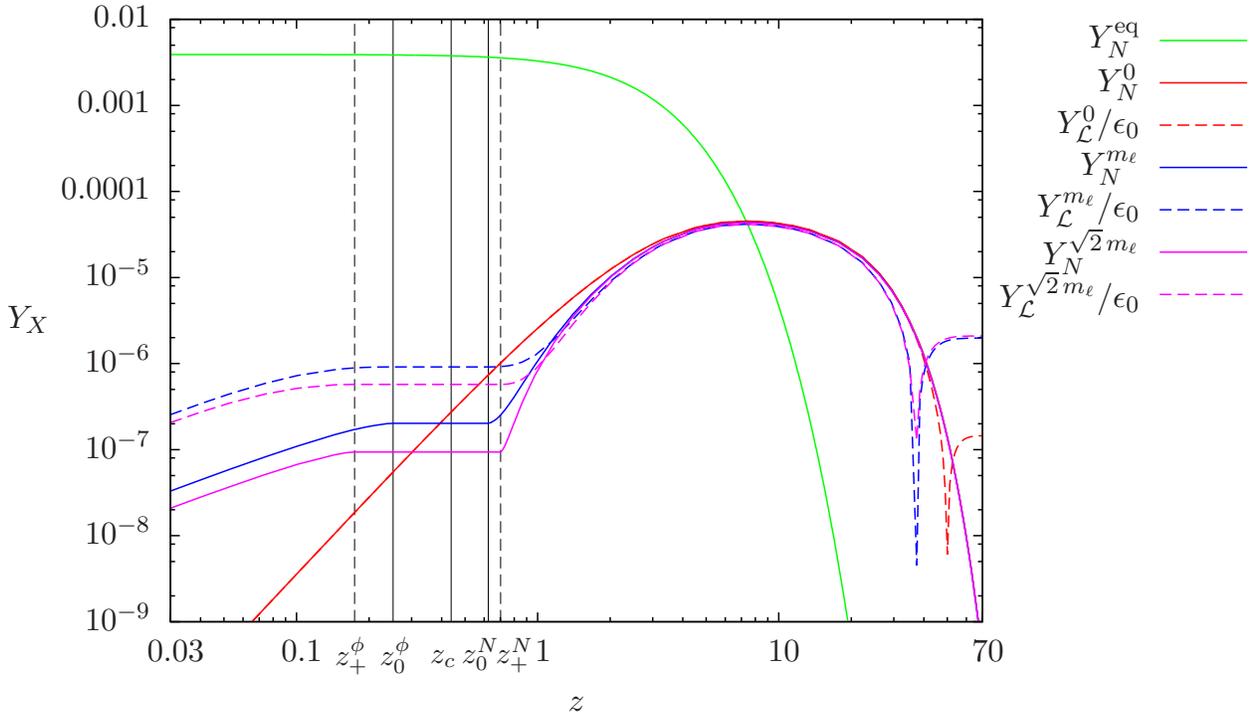}
  \caption[Same as above; weak washout, zero $N$ abundance,
  one-mode cases]{Evolution of neutrino density and lepton asymmetry
    for $K=0.005$ and zero initial neutrino abundance. We show the
    one-mode cases and the vacuum case.}
  \label{fig:nell1k0.005ztd0}
\end{figure}

Let us start with the weak washout regime and zero initial
abundance.  We define a value $z_\rmeq$ by the condition
\begin{align}
  \label{eq:200}
  Y_N(z_\rmeq) = Y_N^\rmeq(z_\rmeq) \, .
\end{align}
For $z \ll 1$, the neutrino abundance is negligible compared to
$Y_N^\rmeq$,
\begin{align}
  \label{eq:199}
  \frac{\rmd Y_N}{\rmd z} \simeq D Y_N^\rmeq \, ,
\end{align}
where $Y_N^\rmeq$ is approximately constant for $z \ll 1$. The entropy
density $s$ is proportional to $z^{-3}$ and $\gamma_D$ is proportional
to $z^{-2}$ in vacuum and $z^{-4}$ for the Higgs boson decays at high
temperature in the finite temperature cases. Thus $D \sim z^2$ in the
vacuum case and $D \sim {\rm const.}$ at finite
temperature. Neglecting $Y_{N}^{\rm initial}$ and $z^{\rm initial}$,
the integration yields $Y_N \simeq z D(z)/3 \sim z^3 $ for the vacuum
case and $Y_N \simeq z D \sim z$ for the finite temperature cases. We
show the numerical results for $K=0.005$ and zero initial abundance in
figures~\ref{fig:nell2k0.005ztd0} and~\ref{fig:nell1k0.005ztd0}, where
these power laws for $Y_N(z)$ can be observed for $z \lesssim 0.1$.
We also see that $Y_N^{m_\ell} > Y_N^{\sqrt{2} \, m_\ell} > Y_N^\pm
\gg Y_N^0$, which reflects $\gamma_D^{m_\ell} > \gamma_D^{\sqrt{2} \,
  m_\ell} > \g_D^\pm \gg \g_0$. Between the thresholds $z_+^\phi$ and
$z_+^N$, the finite-temperature abundances do not evolve much, which
reflects that the decay rates are very low or vanishing in this
regime. The neutrino abundance for the asymptotic mass $\sqrt{2} \,
m_\ell$ does not rise at all at high temperature, while $Y_N^{m_\ell}$
rises slightly between $z_0^{\phi/N}$ and $z_+^{\phi/ N}$, where the
rate is non-zero. The two-mode rate $\g_\pm$ is, though very
suppressed, present over the whole threshold range between $z_+^\phi$
and $z_+^N$ due to the minus-modes, so $Y_N^\pm$ rises slightly. At
low temperature, $z > z_\rmeq$, the neutrino abundances of the
different scenarios are very close to each other, since for $z \gtrsim
2$, the rates are very close to the vacuum rate, $\gamma_{D,W,N}^{T>0}
\simeq \D \g^{T>0} / \e_0 \simeq \g_0$. In this regime, $Y_N$ is much
larger than $Y_N^\rmeq$ since the coupling is too small to keep the
abundance close to equilibrium.

Having outlined the evolution of the neutrino abundance at high
temperature, we can understand the evolution of the lepton asymmetry
at high temperature. The washout term in the
Boltzmann equations proportional to $Y_\ml$ is much smaller than the
source term $D_\e (Y_N-Y_N^\rmeq)$, since $Y_\ml/\e_0 \ll Y_N^\rmeq$,
and we have for $z \ll z_\rmeq$,
\begin{align}
  \label{eq:203}
  Y_\ml^{T>0} &\simeq - z \e_0 D_\e \, \nonumber \\
  Y_\ml^{T=0} &\simeq - z \e_0 \frac{D_\e(z)}{3} \, .
\end{align}
In the high temperature regime, the lepton asymmetry is negative and
follows the neutrino abundance in its absolute value,
\begin{align}
  \label{eq:204}
  Y_\ml(z) \simeq - \e_0 \frac{D_\e}{D} Y_N(z) = -\frac{\D \g}{\g_D}
  Y_N(z) \, .
\end{align}
For the vacuum case, $D_\e \equiv D$ and $Y_\ml/\e_0\simeq -Y_N$,
while for the finite temperature cases, $D_\e/D = \D \g/(\e_0 \g_D)
\sim \mathcal{O}(10^1)$.\footnote{We saw in chapter~\ref{cpas} that
  $\D \g / (\e_0 \g_D) \sim \mathcal{O}(10^2)$ at high temperature
  instead of $\mathcal{O}(10^1)$. The discrepancy is due to rates and
  asymmetries that occur in the Boltzmann equations and that are
  slightly different from the usual rates and $\CP$-asymmetries in
  chapters~\ref{decayrate} and~\ref{cpas} due to the different
  statistical factors they employ, for example $1$ or $(1-2
  f_N^\rmeq)$ instead of the usual factor $(1-f_N^\rmeq)$. See the
  discussion in section~\ref{sec:quant-boltzm-equat} and compare the
  thermal factors in equations~\eqref{eq:129} and~\eqref{eq:b58}.}
Both behaviours can be observed in figures~\ref{fig:nell2k0.005ztd0}
and~\ref{fig:nell1k0.005ztd0}. The lepton asymmetry in the minus-mode
obeys the same power law as the other finite-temperature modes, but is
about a factor $100$ lower due to the lower rates. The combined
($\pm$)-abundance closely follows the plus-abundance since the
influence of the minus-mode rates is very suppressed and also the
different washout term with factor $1/4$ instead of $1/2$ can be
neglected.

Before turning to the intermediate temperatures $z \sim z_\rmeq$, let
us discuss the low temperature regime. For $z > z_\rmeq$, $Y_N \gg
Y_N^\rmeq$, so the source term dominates and washout can be neglected.
Since in this regime, $D_\e^{T>0} \simeq D^{T>0} \simeq D^{T=0}$, we
can write
\begin{align}
  \label{eq:205}
  \frac{\rmd Y_\ml}{\rmd z} \simeq \e_0D (Y_N- Y_N^\rmeq) = - \e
  \frac{\rmd Y_N}{\rmd z} \, .
\end{align}
To first order, the negative lepton asymmetry created below $z_\rmeq$
and the positive contribution from above $z_\rmeq$ have the same
magnitude and cancel each other. For the remaining asymmetry that did
not cancel, the washout contribution up to $z_\rmeq$ and the exact
behaviour of the abundances around $z_\rmeq$ are crucial.

Assuming that $Y_N(z=\infty) = 0$, we get 
\begin{align}
  \label{eq:206}
  Y_\ml^{\rm fin} \simeq \e_0 Y_N(z_\rmeq) - |Y_\ml(z_\rmeq)| \, ,
\end{align}
so we see that the evolution of the difference $\D Y(z) \equiv Y_N(z)
- |Y_\ml(z)|/\e_0$ below $z_\rmeq$ is crucial for the final lepton
asymmetry. For the regime $1 \lesssim z \lesssim z_\rmeq$, we can
write
\begin{align}
  \label{eq:207}
  \frac{\rmd \D Y}{\rmd z} &\simeq Y_N^\rmeq (D -D_\e) \, , \nonumber
  \\
  D-D_\e &= \frac{z}{H_1} \frac{1}{s Y_N^\rmeq} \left(\g_D-\frac{\D
      \g}{\e}\right) \, .
\end{align}
For the finite-temperature cases, the $\CP$-asymmetry in the decay
rates, $\D \g/\e_0$, is considerably smaller than the decay rate
$\g_D$ in the range $z_0^N \lesssim z \lesssim 2$, which can be seen
in figure~\ref{fig:cpn1}. Moreover, for the one-mode cases, $\D
\g_m/\e_0$ approaches $\g_m$ faster than $\D \g_{++}$ approaches
$\g_+$ for the plus-mode. Above $z_+^N$, the ratio $\D \g/(\e_0 \g_D)$
for the two one-mode cases is about the same. Thus, the difference $\D
D \equiv D-D_\e$ is largest for the two-mode approach, smaller and
about the same for the two one-mode approaches and vanishing for the
vacuum approach, $\D D^+ > \D D^{m_\ell} \simeq \D D^{\sqrt{2} \,
  m_\ell}$ and $\D D^0=0$. As a result, $\D Y^+ \gtrsim \D Y^{m_\ell}
\simeq \D Y^{\sqrt{2} \, m_\ell}$ at $z_\rmeq$ and therefore the final
asymmetries are related as $Y_\ml^+ > Y_\ml^{m_\ell} \simeq
Y_\ml^{\sqrt{2}\, m_\ell} \gg Y_\ml^0$. Note that the final asymmetry
is non-vanishing for the vacuum case, since the washout at higher
temperature is larger than at lower temperature due to the larger
decay rate. For the finite temperature cases, the difference $D-D_\e$
bin the crucial regime $z \simeq z_\rmeq$ governs the final asymmetry.

The evolution of the decoupled minus-mode at low temperature is very
different but not hard to understand. The asymmetry rises at $z
\gtrsim z_+^N$, because this is the regime above $z_c$ where $\Delta
\g_{-+}$ is maximal and therefore $D_\e^-$ is maximal as well. Between
the thresholds $z_+^\phi$ and $z_+^N$, $\D \g_{-+}$ is suppressed by
the internal plus-lepton and $\D \g_{--}$ is suppressed by the residue
of the internal minus-lepton, so $D_\e$ is negligible and $Y_\ml^-$
does not rise. Above $z \sim 1$, $\D \g_{-+}$ falls due to the residue
of the external minus-mode and $Y_\ml^-$ does not change. The final
lepton asymmetry therefore does not change its sign above $z \gtrsim
z_\rmeq$ and keeps the value it achieves at around $1 \lesssim z
\lesssim 2$ where the $\CP$-asymmetry $\D \g_{-+}$ becomes small.

\begin{figure}
  \centering
  \includegraphics[width=\textwidth]{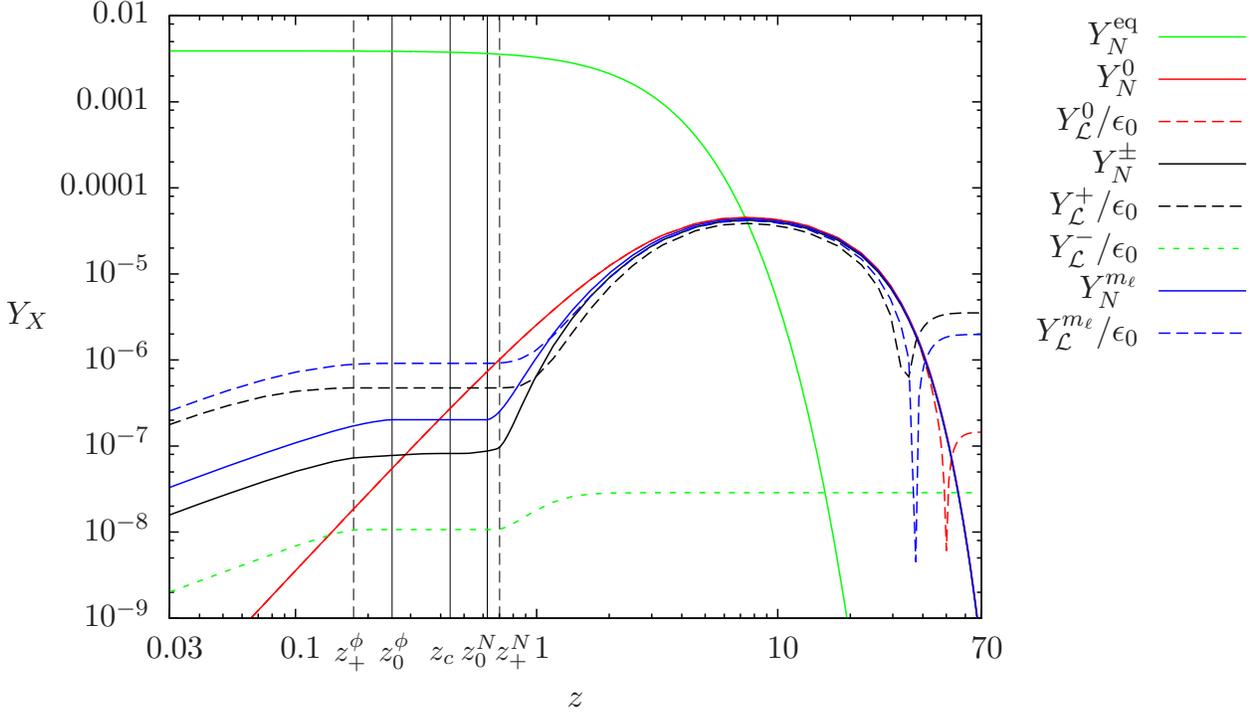}
  \caption[Same as above; weak washout, zero $N$ abundance, ($+$)-,
  ($-$)-, $m_\ell$-cases]{Evolution of neutrino density and lepton
    asymmetry for $K=0.005$ and zero initial neutrino abundance. We
    display the four modes from figures~\ref{fig:nell2k0.005ztd0}
    and~\ref{fig:nell1k0.005ztd0} that give different final lepton
    asymmetries, that is, the plus-mode the minus-mode, the
    $m_\ell$-mode and the vacuum case.}
  \label{fig:nellpmmlk0.005ztd0}
\end{figure}

The combined ($\pm$)-mode does not evolve differently from the
plus-mode since the influence from the $\g^-$ rates can be neglected
and also the washout term with the additional factor $1/2$ is not
noticeable since washout is very small in all temperature
regimes. Summarising, there are four differing lepton asymmetries in
this regime: the vacuum case, the $m_\ell$-case, the ($+$)-case and
the ($-$)-case. The $\sqrt{2} \, m_\ell$-case yields the same
asymmetry as the $m_\ell$-case and the ($\pm$)-case yields the same
asymmetry as the ($+$)-case. We show the four differing lepton
asymmetries together in figure~\ref{fig:nellpmmlk0.005ztd0}.

\subsection{Strong and intermediate washout for zero initial abundance}
\label{sec:strong-washout}

\begin{figure}
  \centering
  \includegraphics[width=\textwidth]{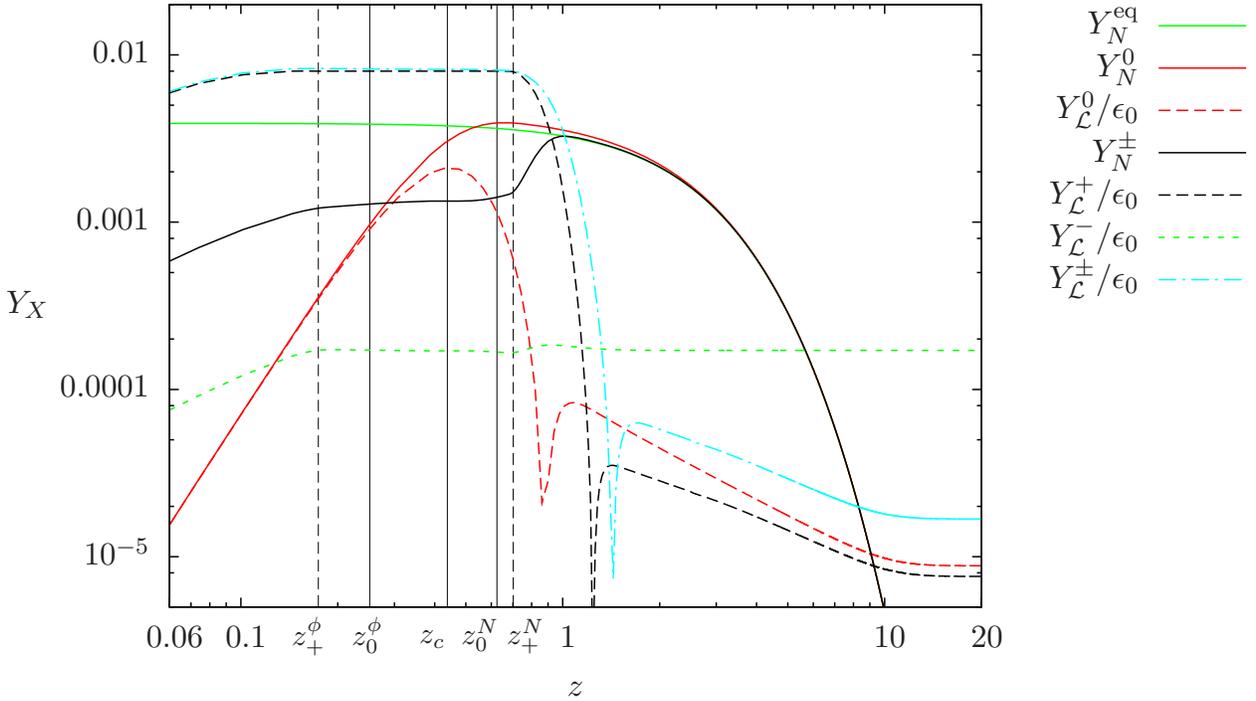}
  \caption [Same as above; strong washout, zero $N$ abundance, two-mode
  cases]{Evolution of neutrino abundance $Y_N(z)$ and lepton asymmetry
    $Y_\ml(z)$ for $K=100$ and zero initial neutrino abundance. We
    show the two-mode cases and the vacuum case.}
  \label{fig:nell2k100.000ztd0}
\end{figure}
\begin{figure}
  \centering
  \includegraphics[width=\textwidth]{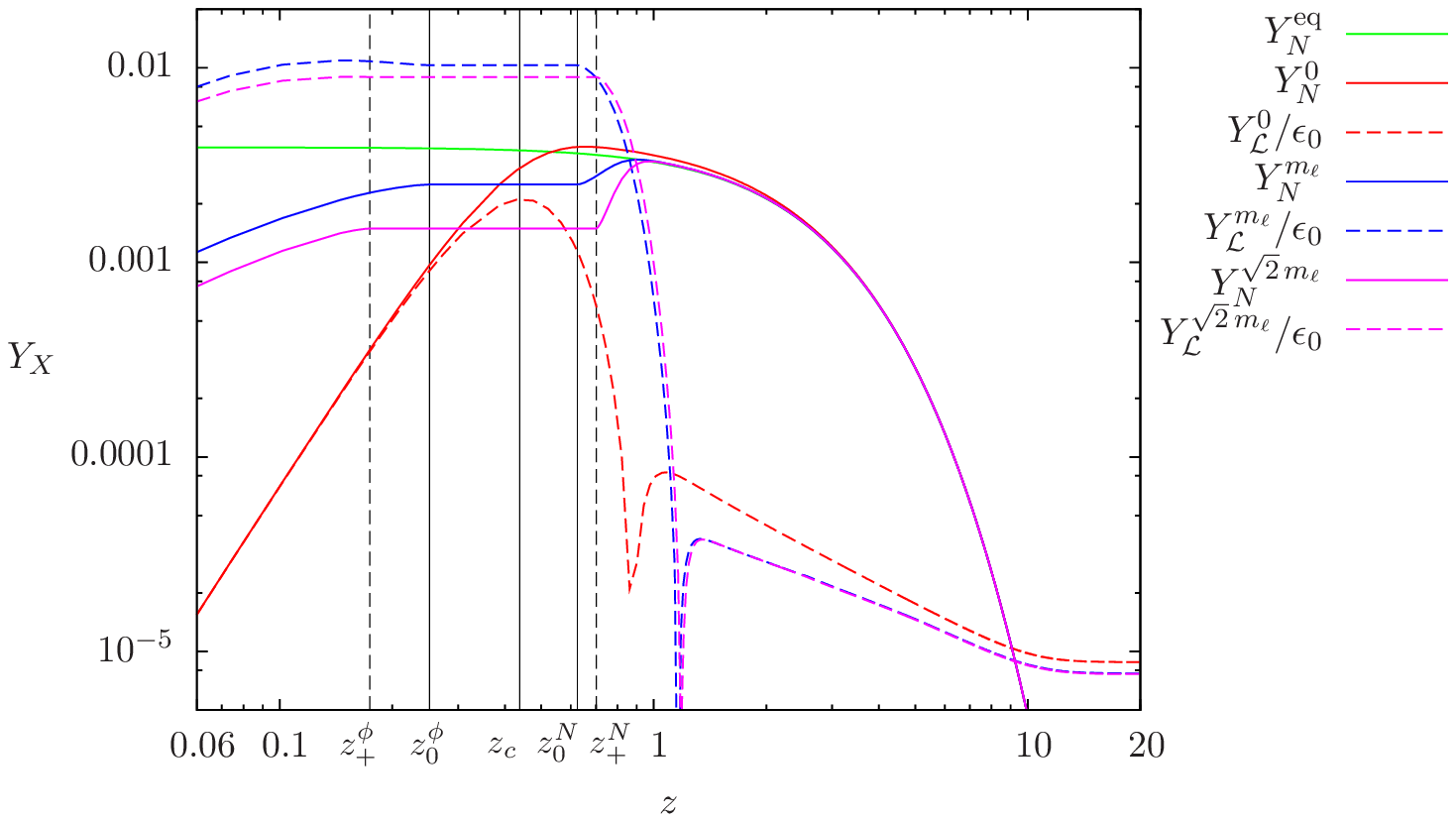}
  \caption[Same as above; strong washout, zero $N$ abundance, one-mode
  cases]{Evolution of neutrino abundance $Y_N(z)$ and
    lepton asymmetry $Y_\ml(z)$ for $K=100$ and zero initial
    neutrino abundance. We show the one-mode cases and the vacuum
    case.}
  \label{fig:nell1k100.000ztd0}
\end{figure}
For strong washout, the evolution of the neutrino abundance is
analogous to the weak washout regime, with $Y_N^T \sim z$ and $Y_N^0 \sim
z^3$, as shown in figures~\ref{fig:nell2k100.000ztd0} and~\ref{fig:nell1k100.000ztd0}. 
The couplings are stronger, therefore the abundances rise faster and
meet $Y_N^\rmeq$ earlier at $z_\rmeq \sim 1$. For larger $z$, the
couplings are strong enough to keep $Y_N$ close to equilibrium. The
evolution of the lepton asymmetry is nicely explained in
reference~\cite{Buchmuller:2004nz} for the vacuum case with some
rather accurate analytical approximations. In this work, we are only
interested in the difference of the vacuum case to the finite
temperature case. In the strong washout regime, the lepton asymmetries
rise rather fast and the washout term, which competes with the source
term, becomes larger than the latter at some temperature $z_{\rm
  min}$, where the lepton asymmetry reaches its most negative
value. The source term becomes small when $Y_N$ approaches its
equilibrium value, so the washout term drives the asymmetry evolution
back to zero. At $z \gtrsim z_\rmeq$, the neutrino abundance slightly
overshoots $Y_N^\rmeq$, so the source term changes sign and adds to
the washout term until the lepton asymmetry becomes positive, where
the washout term changes its sign as well and is competing again. At
low temperature, source term and washout term have the same magnitude
when the lepton asymmetry reaches a maximum at $z_{\rm max}$. Above
$z_{\rm max}$, the lepton asymmetry is again driven to zero by the
larger washout. At very low temperature, washout and source term
become very small and do not influence the asymmetry further, which
settles at a final value $Y_\ml^{\rm fin}$. We see that in the strong
washout regime, the dynamics are governed by the washout term. The
evolution of the finite temperature lepton asymmetries is analogous to
the vacuum case, but they settle to a different final value. For the
finite temperature cases, the equilibrium density of leptons is
smaller than in the vacuum case due to the thermal mass $m_\ell \sim
0.2 \, T$. The washout term is effectively larger than for the vacuum
case and competes with the source term in a stronger way. Therefore,
the asymmetry evolution appears slightly damped compared to the vacuum
case and the final asymmetry is marginally lower. The evolution of the
minus-mode is analogous to the evolution at weak washout, rises fast
below $z_+^\phi$ and does not change above the thresholds since $Y_N
\sim Y_N^\rmeq$ in this regime. The combined ($\pm$)-mode tracks the
plus-mode until washout becomes relevant at $z \gtrsim z_\rmeq$. The
washout term for the ($\pm$)-mode is always about a factor two smaller
than for the plus-mode, since we add the minus- and plus-washout
rates, where the minus-rate is always negligible compared to the
plus-rate. Thus, the ($\pm$)-abundance is less affected by washout,
so the dynamics are affected by the source term in a stronger way and the
final asymmetry is larger. We can view this behaviour as always
distributing half the asymmetry in a mode $\ell_-$ which couples
strongly to $\ell_+$ and is not affected by washout. The final
asymmetry is about a factor two larger than for the other scenarios in
the strong washout regime.

\begin{figure}
  \centering
  \includegraphics[width=\textwidth]{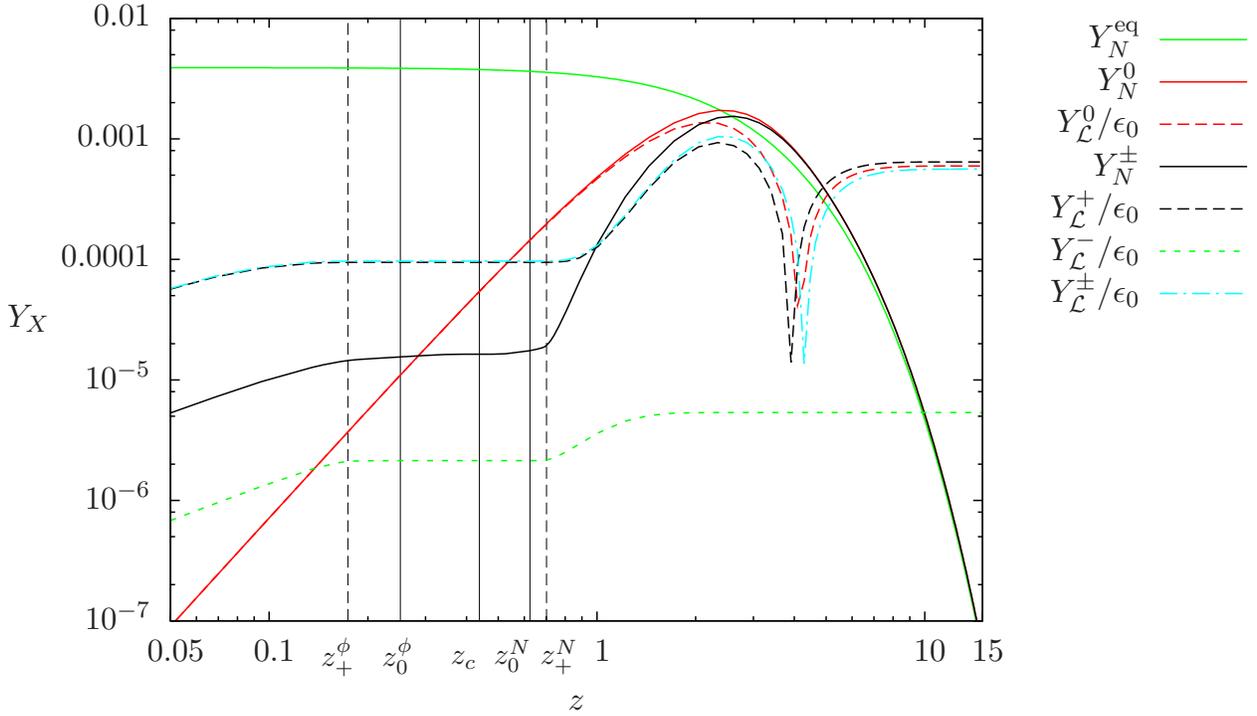}
  \caption[Same as above; intermediate washout, zero $N$ abundance, two-mode
  cases]{Evolution of neutrino abundance $Y_N(z)$ and
    lepton asymmetry $Y_\ml(z)$ for $K=1$ and zero initial
    neutrino abundance. We show the two-mode cases and the vacuum
    case.}
  \label{fig:nell2k1.000ztd0}
\end{figure}

The case of intermediate washout is shown in
figure~\ref{fig:nell2k1.000ztd0}, where we only show the two-mode
cases since in this regime, the final lepton asymmetries of the one-mode cases are the
same as for the plus-mode. The dynamics can be viewed as an
interpolation between the strong and weak washout regimes and the
final asymmetries are very similar to each other.

\subsection{Non-zero initial abundance}
\label{sec:non-zero-initial}

\begin{figure}
  \centering
  \includegraphics[width=\textwidth]{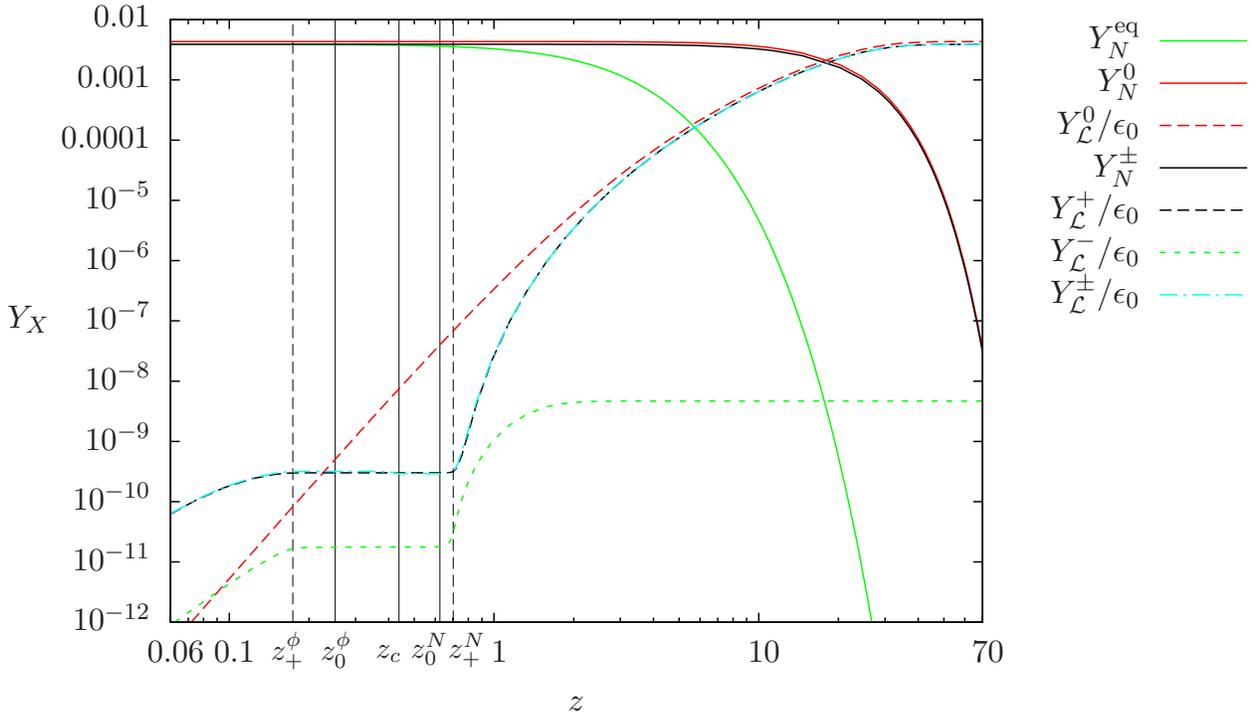}
  \caption[Same as above; weak washout, thermal $N$ abundance, two-mode
  cases]{Evolution of neutrino abundance $Y_N(z)$ and
    lepton asymmetry $Y_\ml(z)$ for $K=0.005$ and thermal initial
    neutrino abundance. We show the two-mode cases and the vacuum
    case.}
  \label{fig:nell2k0.005ztd1}
\end{figure}

\begin{figure}
  \centering
  \includegraphics[width=\textwidth]{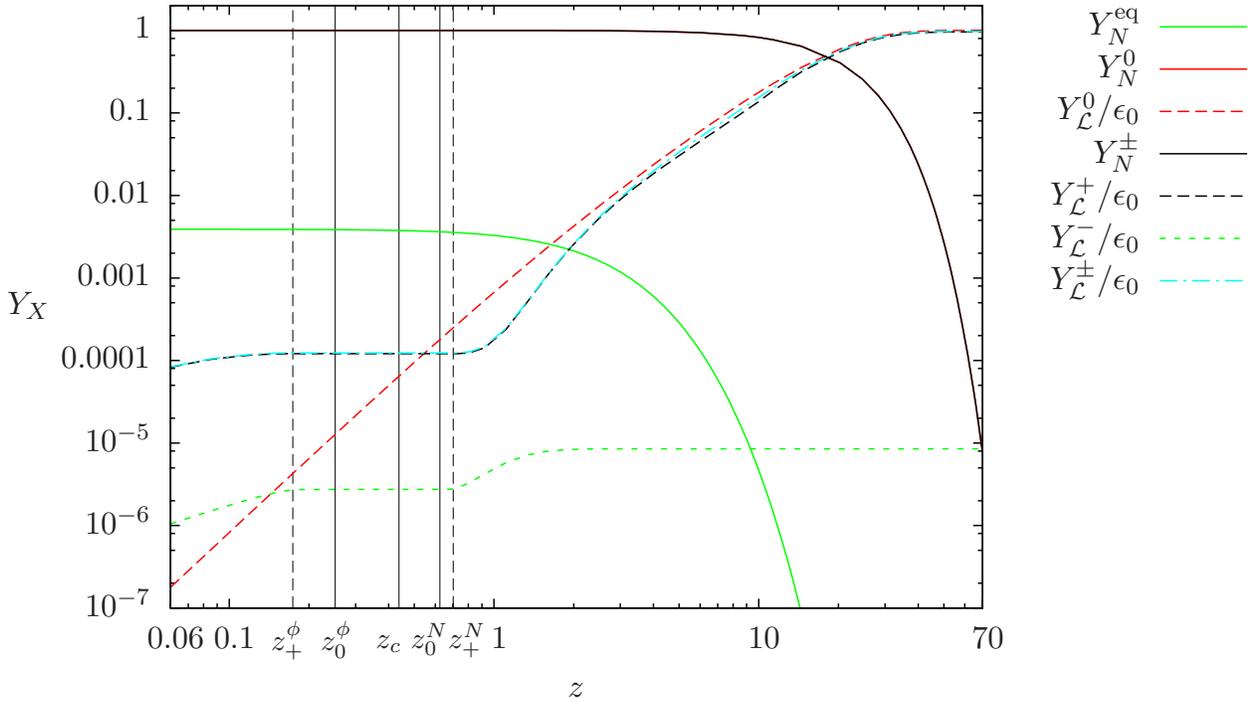}
  \caption[Same as above; weak washout, dominant $N$ abundance, two-mode
  cases]{Evolution of neutrino abundance $Y_N(z)$ and
    lepton asymmetry $Y_\ml(z)$ for $K=0.005$ and dominant initial
    neutrino abundance. We show the two-mode cases and the vacuum
    case.}
  \label{fig:nell2k0.005ztd2}
\end{figure}

\begin{figure}
  \centering
  \includegraphics[width=\textwidth]{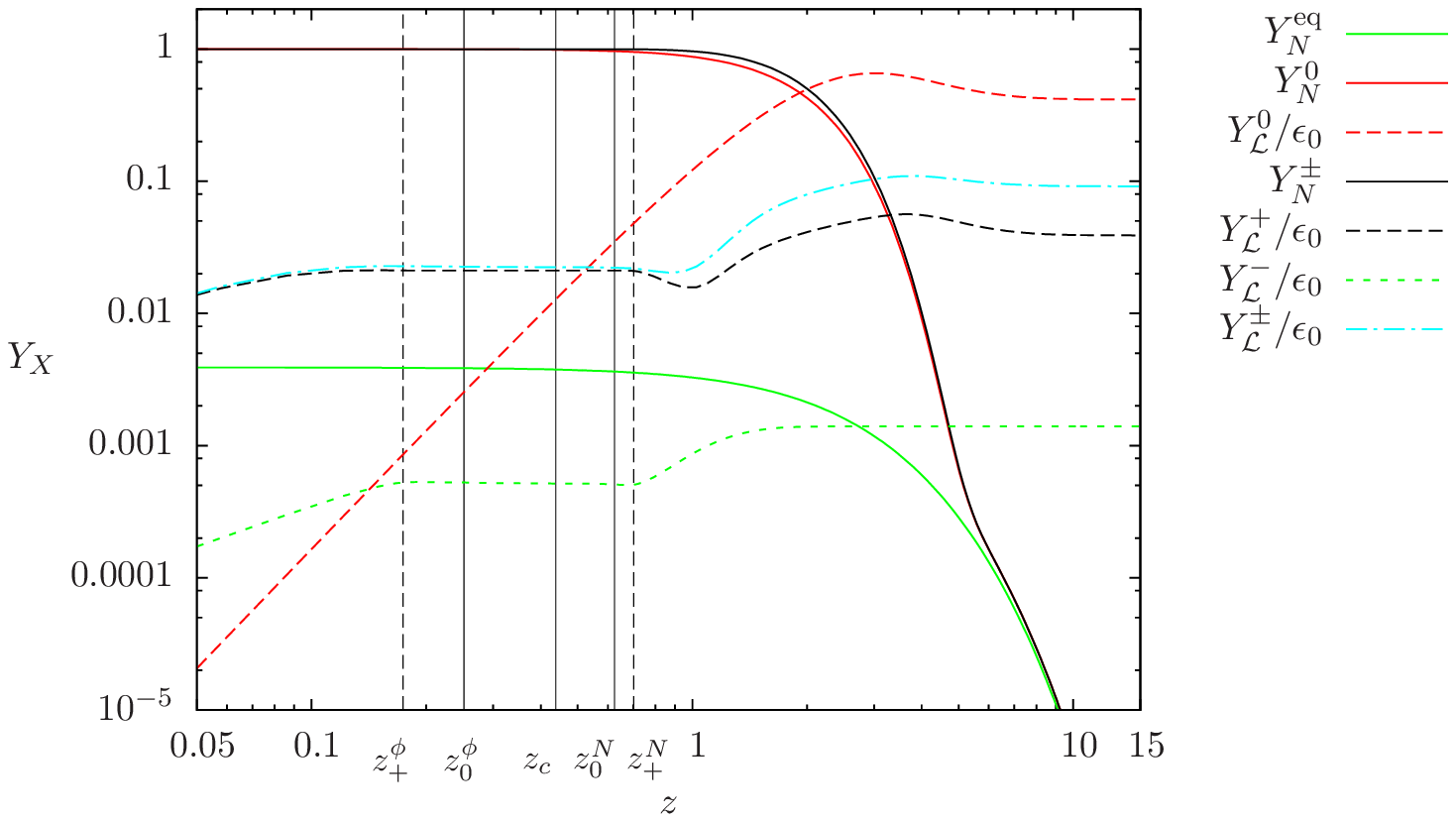}
  \caption[Same as above; intermediate washout, dominant $N$ abundance, two-mode
  cases]{Evolution of neutrino abundance $Y_N(z)$ and
    lepton asymmetry $Y_\ml(z)$ for $K=1$ and dominant initial
    neutrino abundance. We show the two-mode cases and the vacuum
    case.}
  \label{fig:nell2k1.000ztd2}
\end{figure}

\begin{figure}
  \centering
  \includegraphics[width=\textwidth]{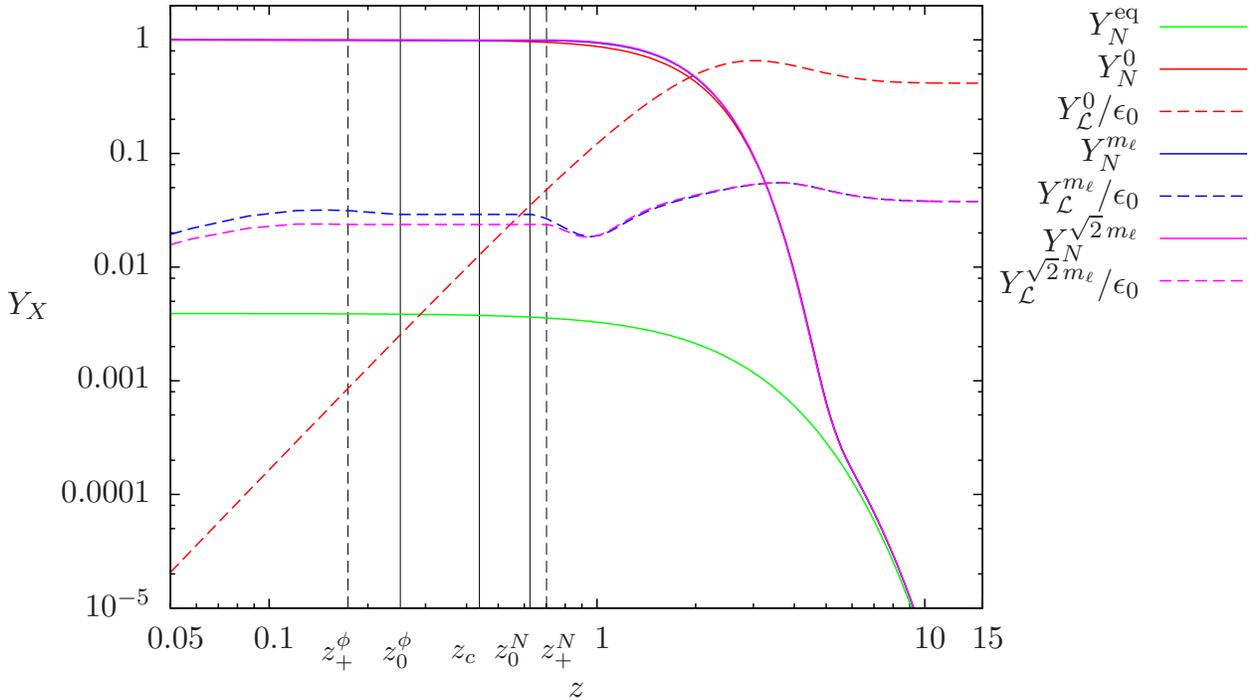}
  \caption[Same as above; intermediate washout, dominant $N$ abundance, one-mode
  cases]{Evolution of neutrino abundance $Y_N(z)$ and
    lepton asymmetry $Y_\ml(z)$ for $K=1$ and dominant initial
    neutrino abundance. We show the one-mode cases and the vacuum
    case.}
  \label{fig:nell1k1.000ztd2}
\end{figure}

\begin{figure}
  \centering
  \includegraphics[width=\textwidth]{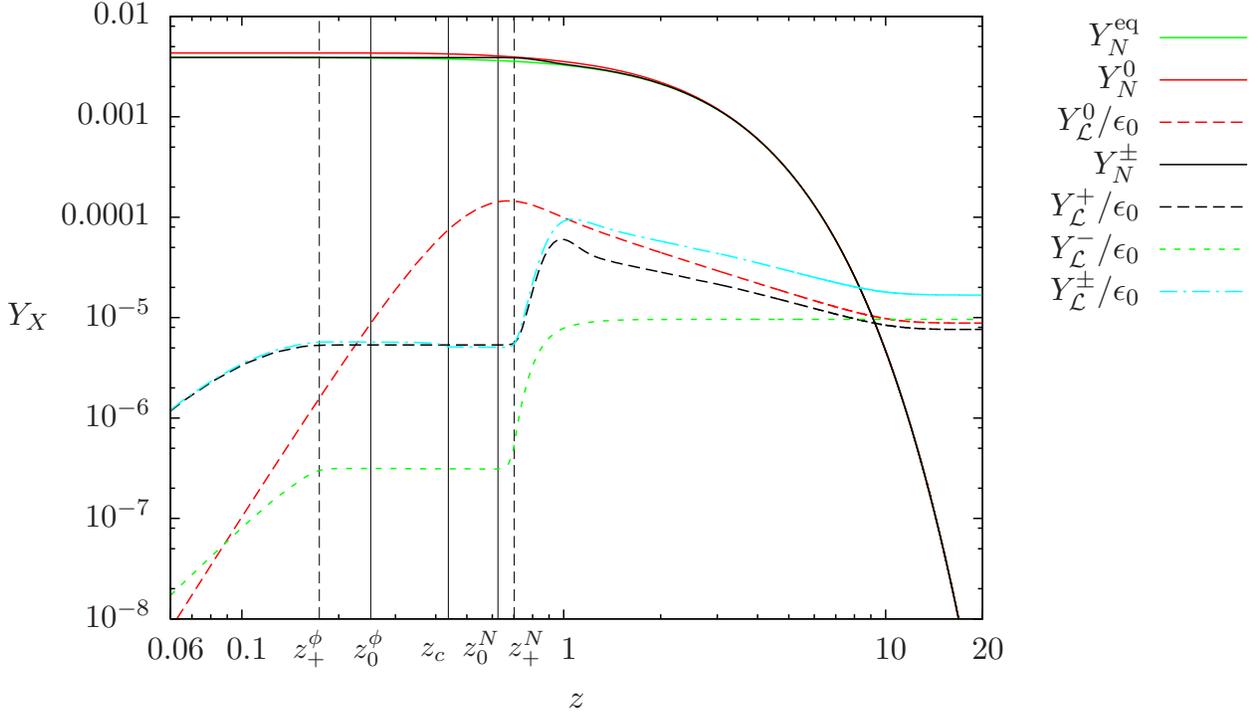}
  \caption[Same as above; strong washout, thermal $N$ abundance, two-mode
  cases]{Evolution of neutrino abundance $Y_N(z)$ and
    lepton asymmetry $Y_\ml(z)$ for $K=100$ and thermal initial
    neutrino abundance. We show the two-mode cases and the vacuum
    case.}
  \label{fig:nell2k100.000ztd1}
\end{figure}

\begin{figure}
  \centering
  \includegraphics[width=\textwidth]{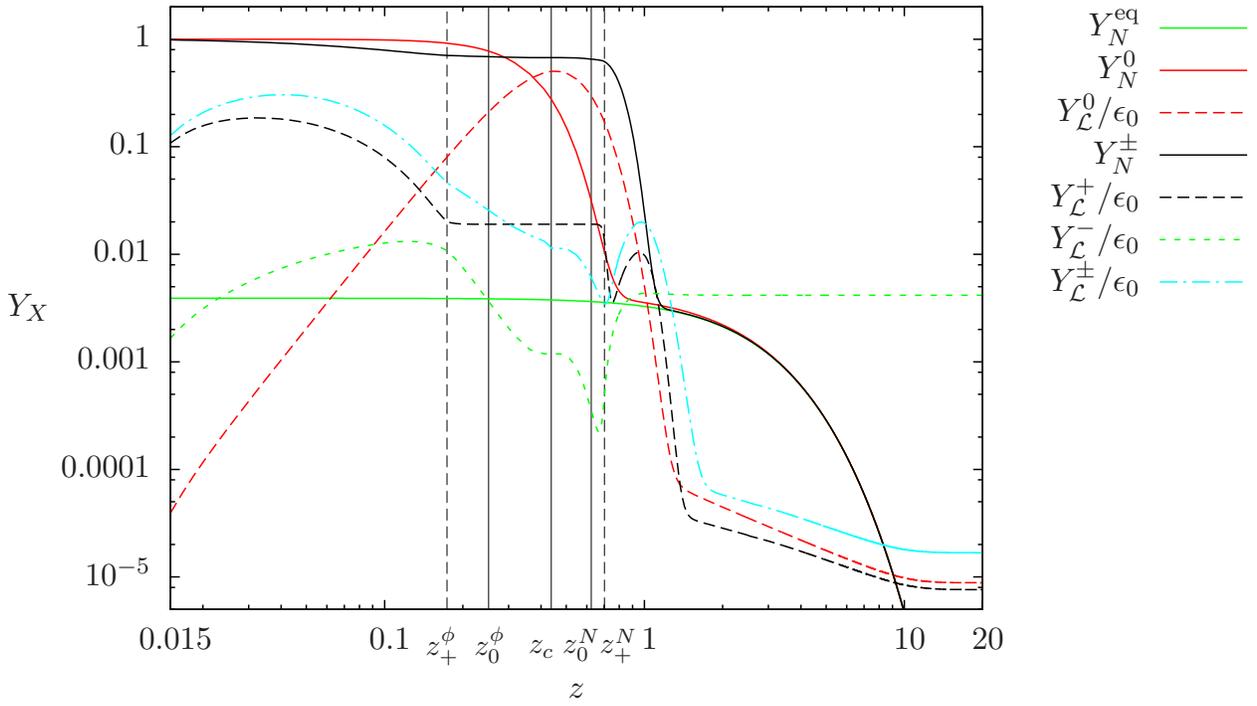}
  \caption[Same as above; strong washout, dominant $N$ abundance, two-mode
  cases]{Evolution of neutrino abundance $Y_N(z)$ and
    lepton asymmetry $Y_\ml(z)$ for $K=100$ and dominant initial
    neutrino abundance. We show the two-mode cases and the vacuum
    case.}
  \label{fig:nell2k100.000ztd2}
\end{figure}

We also present the dynamics for thermal and dominant initial
abundance in
figures~\ref{fig:nell2k0.005ztd1}--\ref{fig:nell2k100.000ztd2}.
For weak washout and thermal initial abundance, $Y_N \gg |Y_\ml|$ for
low temperatures, and according to equation~\eqref{eq:206},
$Y_\ml^{\rm final}/\e_0 \sim Y_N^{\rm initial}$. For weak washout and
dominant initial abundance, this equation holds as well, as can be
seen in figures~\ref{fig:nell2k0.005ztd1}
and~\ref{fig:nell2k0.005ztd2}.  For intermediate washout $K \sim 1$
and dominant abundance, shown in figures~\ref{fig:nell2k1.000ztd2} and
\ref{fig:nell1k1.000ztd2}, the lepton asymmetry production is stopped
between the thresholds for the thermal cases and the production above
$z \sim 1$ does not succed in producing an asymmetry as high as in the
vacuum case.  For the ($\pm$)-case, the asymmetry production is larger
since it is not as much affected by washout.

For the strong washout regime and large initial neutrino abundances,
the dynamics at high temperature are interesting, as shown in
figures~\ref{fig:nell2k100.000ztd1} and~\ref{fig:nell2k100.000ztd2},
but the interplay between source term and washout term at low
temperature governs the final asymmetry as in the zero-abundance
case. We reproduce the well-known fact that the initial conditions do
not influence the final asymmetry in the strong washout regime, while
the arguments concerning the equilibrium distribution of leptons with
thermal mass and the reduced washout of the ($\pm$)-mode still hold
and lead to the same lepton asymmetry as for zero initial
abundance. The decoupled minus-mode is very much affected by the
coupling, that is the decay parameter $K$, and the initial conditions,
since the final lepton asymmetry is produced at high temperatures. The
stronger the coupling, the larger the asymmetry production of the
minus-mode at high temperatures and the larger the final
value. Moreover, the larger the initial deviation of the neutrino
abundance from equilibrium, the larger the asymmetry production and
the final asymmetry. The final lepton asymmetry in this mode is thus
lowest for neutrinos with thermal initial abundance.

\subsection{Final lepton asymmetries}
\label{sec:final-asymmetries}

\begin{figure}
  \centering
  \includegraphics[width=\textwidth]{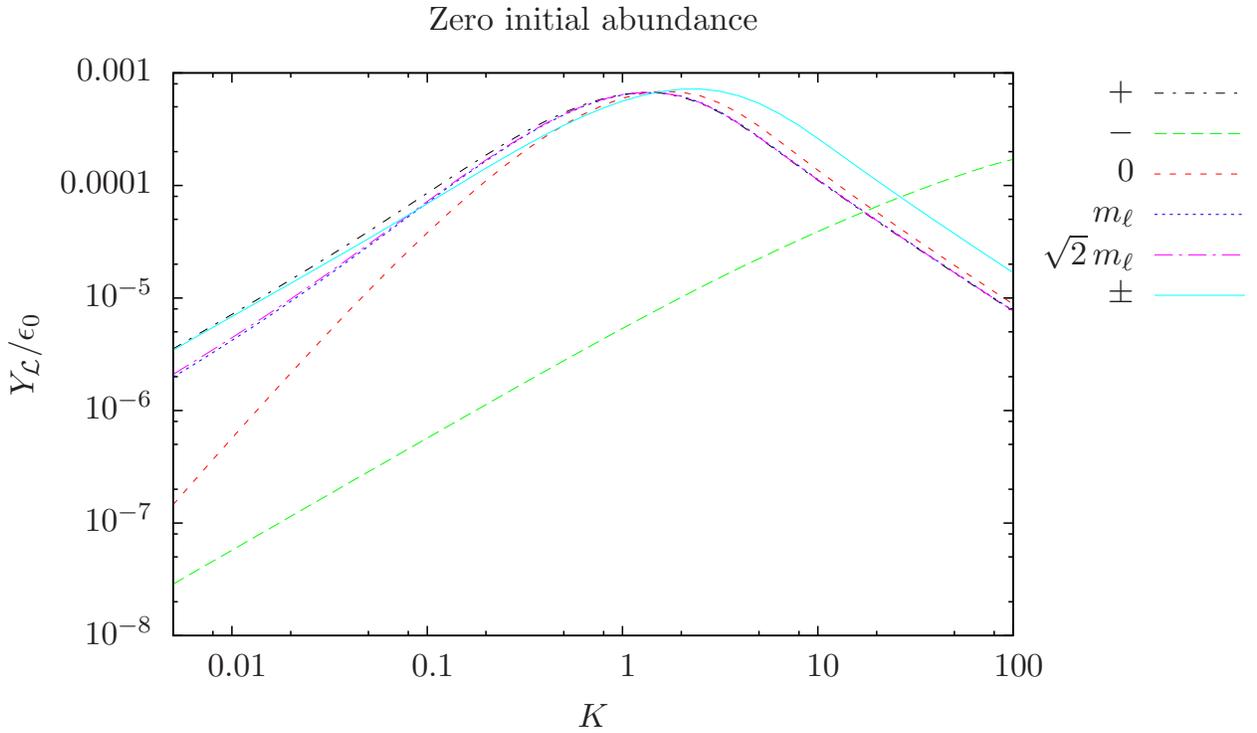}
  \caption[$K$-dependence of $|Y_\ml^{\rm fin}|$ for zero initial
  neutrino abundance]{Final value of the lepton asymmetry for
    different values of $K$ for zero initial neutrino abundance.}
  \label{fig:final0}
\end{figure}

\begin{figure}
  \centering
  \includegraphics[width=0.98\textwidth]{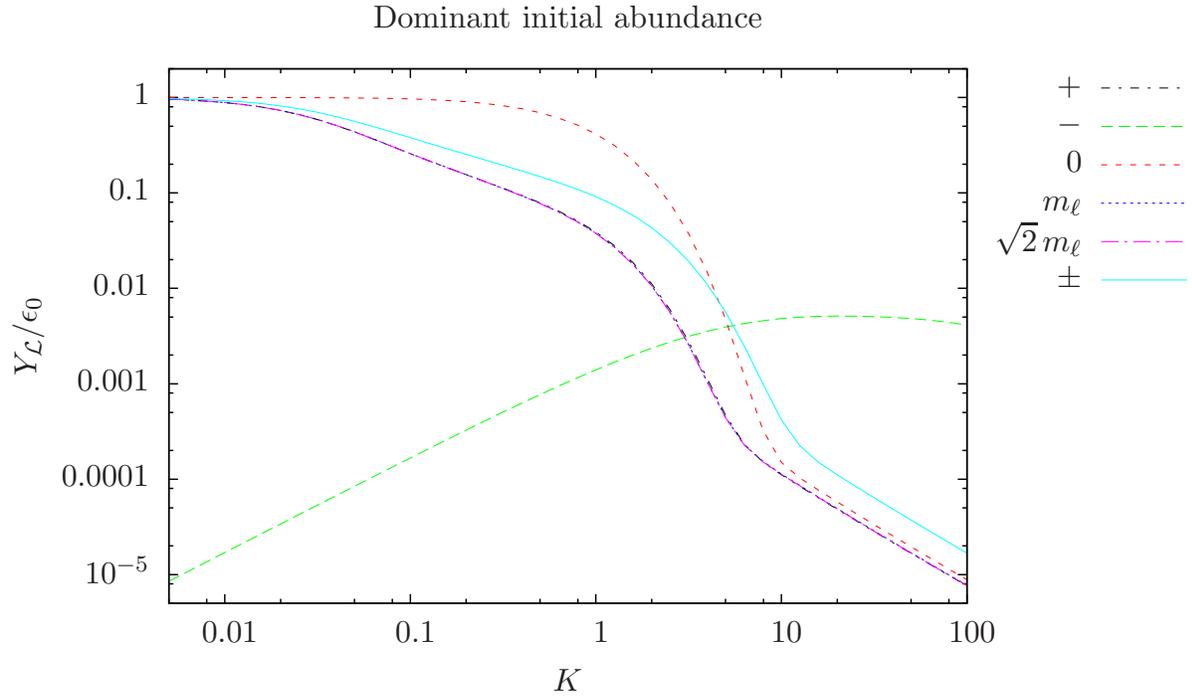}
  \caption[$K$-dependence of $|Y_\ml^{\rm fin}|$ for thermal initial
  neutrino abundance]{Final value of the lepton asymmetry for
    different values of $K$ for thermal initial neutrino abundance.}
  \label{fig:final1}
\end{figure}

The values of the final asymmetries are shown in
figures~\ref{fig:final0}--\ref{fig:final2} for different initial
abundances.
For zero initial abundance and weak washout, shown in
figure~\ref{fig:final0}, the asymmetry for the finite-temperature
cases are larger than for the vacuum case by about one order of
magnitude due to the difference of the thermal rates $\g_D$ and the
$\CP$-asymmetries $\D \g$ at $z \gtrsim z_+^N$. The lepton asymmetry
for the plus-mode is also slightly larger than for the one-mode cases
due to a suppression of the $\CP$ asymmetry compared to the one-mode
approaches. For strong washout, the asymmetry production in the vacuum
case is marginally more efficient than in the thermal cases due to a
smaller lepton equilibrium distribution, while the lepton asymmetry in
the ($\pm$)-approach is by a factor $2$ larger than in the other cases
since half of the asymmetry is in the $\ell_-$-modes and not affected
by washout. The minus-mode is completely decoupled, the lepton
asymmetry bears the opposite sign as the other lepton asymmetries and
rises with stronger couplings, that is with larger decay parameter
$K$. As discussed in section~\ref{sec:interacting-modes}, this
scenario might not be realistic since the modes will couple to each
other via gauge bosons, so an evolution similar to the ($\pm$)-case
seems more likely.
{\begin{figure}
  \centering
  \includegraphics[width=0.95\textwidth]{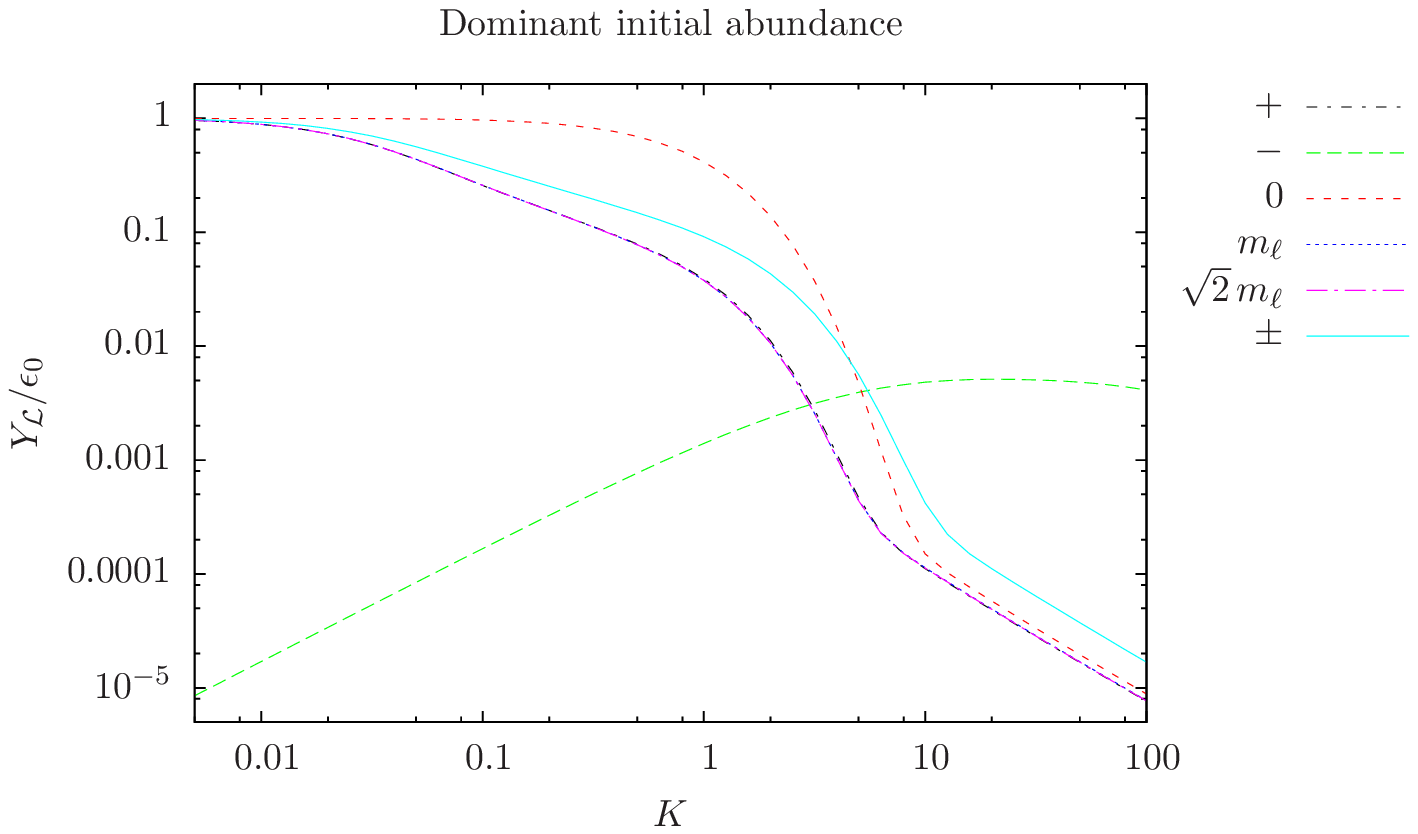}
  \caption[$K$-dependence of $|Y_\ml^{\rm fin}|$ for dominant initial
  neutrino abundance]{Final value of the lepton asymmetry for different values of
    $K$ for thermal initial neutrino abundance.}
  \label{fig:final2}
\end{figure}
\begin{figure}
  \centering
  \includegraphics[width=0.95\textwidth]{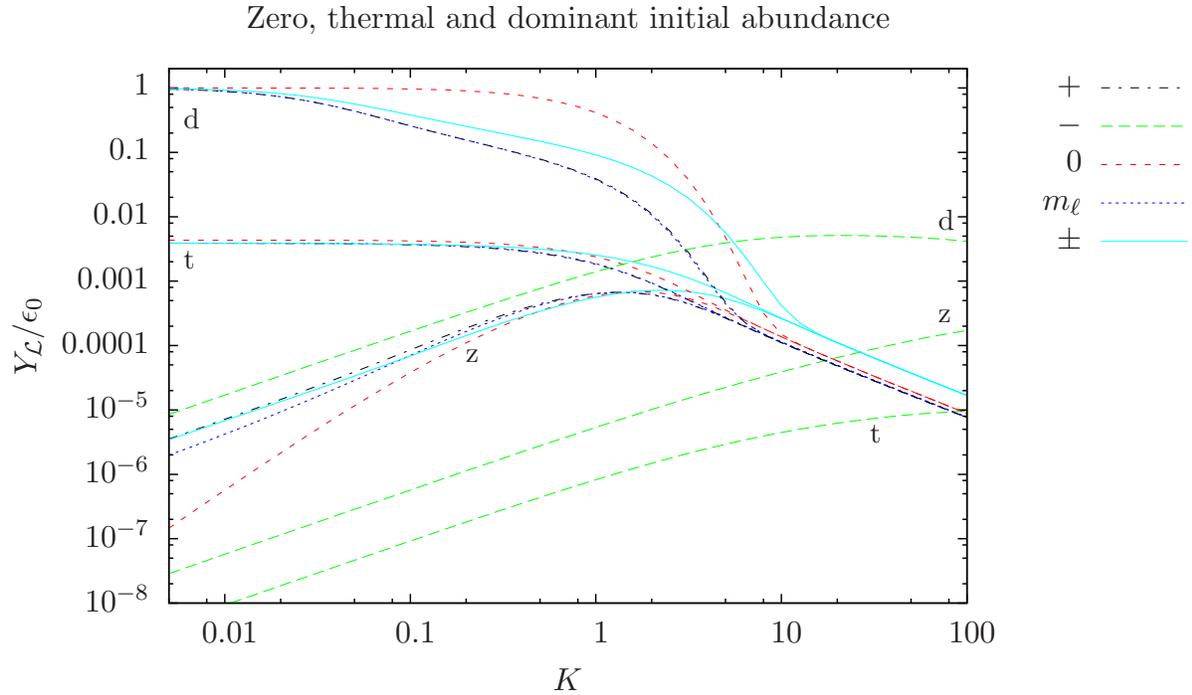}
  \caption[$K$-dependence of $|Y_\ml^{\rm fin}|$ for different initial
  neutrino abundances]{Final value of the lepton asymmetry for
    different values of $K$ for zero, thermal and dominant initial
    neutrino abundance. The letters z, t and d denote the curves for
    zero, thermal and dominant abundance. Note that the final
    asymmetry of the minus-mode has opposite sign for zero initial
    neutrino abundance, compared to the asymmetries of all other
    cases.}
  \label{fig:finaltot}
\end{figure}}

For thermal initial abundance and weak washout, shown in
figure~\ref{fig:final1}, the final asymmetry equals the equilibrium
abundance $Y_\ml/\e_0$, while in the strong washout regime, it shows
the same behaviour as in the case of zero initial neutrino
abundance. The minus-mode asymmetry is very low for thermal initial
neutrino abundance since the neutrinos are close to equilibrium at
high temperatures. Contrary to the zero initial abundance case, it
bears the same sign as the lepton asymmetries of the other scenarios.

For dominant initial abundance, shown in figure~\ref{fig:final2}, the
final asymmetries assume their maximal value in the weak washout
regime, when the coupling is weak enough not to wash them out at low
temperature. For larger couplings $K \sim 1$, the thresholds lead to a
halted asymmetry production for the finite temperature cases and not
as much asymmetry can be produced as for the vacuum case. The
($\pm$)-case shows a larger asymmetry compared to the other thermal
cases due to the weaker washout. At strong coupling $K \gg 1$, the
asymmetries are the same as for thermal and zero initial neutrino
abundance.
The minus-mode asymmetry is large in all washout regimes, since the
neutrinos are far from equilibrium at high temperatures when the
$\ell_-$-asymmetry is produced.

A summary of the several initial conditions can be seen in
figure~\ref{fig:finaltot}, where we have omitted the $\sqrt{2} \,
m_\ell$ case since it is very close to the $m_\ell$ case in all
scenarios.  In the weak washout regime, the case with zero initial
abundance is most strongly affected by thermal corrections which
amount to one order of magnitude, and the plus-mode asymmetry is
additionally enhanced by a factor of about two. In the intermediate
regime, the dominant-initial-abundance case is influenced very much by
thermal corrections. Therefore, a production mechanism for dominant
neutrino abundance has to take into account such thermal effects. In
the strong washout regime, one would naturally expect that thermal
corrections can be neglected. We see that this is not the case since
for strongly interacting leptonic quasiparticles, a part of the lepton
asymmetry can be hidden in the $\ell_-$-mode which is unaffected by
washout, thus producing an asymmetry by up to a factor of two larger
than at zero temperature. The effect of the thermal lepton mass on the
equilibrium distribution of the leptons is an interesting feature, but
very small and can be neglected for all practical purposes.


    \clearpage\chapter*{Conclusions \label{conclusions}}
\addcontentsline{toc}{chapter}{Conclusions}
\markboth{Conclusions}{Conclusions}

We see this work as a contribution to the revived quest for
calculating and capturing effects that the finite density and
temperature of the medium has on early universe dynamics. For a
minimal and self-consistent toy model of leptogenesis, which consists
only of neutrinos, leptons and Higgs bosons, we have performed an
extensive analysis of the effects of HTL corrections. This implies
capturing the effects of thermal masses, modified dispersion relations
and modified helicity structures. We put special emphasis on the
influence of the two fermionic quasiparticles, which show a different
behaviour than particles in vacuum, notably through their dispersion
relations, but also the helicity structure of their interactions.  Our
work is thus similar to the work done in
reference~\cite{Giudice:2003jh}, where the authors of the latter work
did not include the effects of fermionic quasiparticles and get a
different result for the $\CP$-asymmetries, which are crucial for the
evolution of the lepton asymmetry. Our toy model produces two lepton
asymmetries stored in the two different lepton modes without the
possibility of an equilibration of these asymmetries by SM
processes. Since we expect the lepton modes to interact via gauge
bosons in the bath, we examine a second case where the modes are
strongly coupled to each other. As a third and fourth case, we
approximate the lepton propagators by zero temperature propagators
with the zero temperature mass replaced by the thermal lepton mass in
one case and the asymptotic mass in the other case. We refer to these
cases as one-mode approach. All four thermal cases are compared to
the zero-temperature case.

Our analysis proceeded in three steps: Calculating interaction rates,
calculating the necessary $\CP$-asymmetries and deriving and solving
the corresponding Boltzmann equations. We calculated the decay rates
in chapter~\ref{decayrate}. Thermal decay rates in the context of
leptogenesis have been calculated using vacuum states as external
states and replacing the zero-temperature mass by the respective
thermal mass\cite{Giudice:2003jh}. We calculated the decay rates with
the full HTL-dependence via the neutrino self-energy and the optical
theorem. While for scalars, that is the Higgs bosons in our case, this
gives the same result as inserting thermal masses in the kinematics,
the results for fermions are somewhat different, both conceptually and
numerically. Neglecting the zero-temperature fermion mass, the
resummation of HTL fermion self-energies results in an effective
fermion propagator that does not break chiral invariance and is split
up in two helicity modes. The external fermion states therefore behave
conceptually different from the ones with chirality-breaking thermal
masses that have been inserted in the kinematics by hand. Moreover,
one has to take care of one additional mode, which has implications
for the Boltzmann equations.

In order to understand and describe the behaviour of the modes, we
have assigned momentum-dependent quasi-masses to them in
section~\ref{sec:effect-prop-disp}. Numerically, the presence of the
two modes with momentum-dependent masses leads to different thresholds
corresponding to the asymptotic masses of the two modes, $\sqrt{2} \,
m_\ell$ and zero mass, and the zero momentum mass $m_\ell$. It is a
well known fact, that for large temperatures, the thermal mass of the
Higgs boson becomes so large that neutrino decays are no longer
allowed, but rather the Higgs bosons decay into neutrinos and
leptons. While reference~\cite{Giudice:2003jh} finds that decays are
possible above or below their respective thresholds, we find that the
decay into the negative mode is always possible, while the decay into
the plus-mode is possible up to the thresholds corresponding to
the thermal mass $m_\ell$, but becomes suppressed by the reduction of
possible lepton momenta at the thresholds with the asymptotic mass
$\sqrt{2} \, m_\ell$.  From pure kinematical arguments, one would
expect the one-mode rate with the asymptotic mass to be a much better
approximation to the plus-rate than the one-mode rate for the thermal
mass. The fact that this is not the case shows the importance of the
helicity structure of the leptonic quasiparticles, which suppresses
the rate for the plus-mode such that it is much lower than
$\gamma_{\sqrt{2} \, m_\ell}$. For the Higgs boson decays at high
temperature, the thresholds are analogous to the neutrino decays at
low temperature. We observe the same $T^4$-proportionality for Higgs
boson decays as reference~\cite{Giudice:2003jh}.

We calculated the $\CP$-asymmetries in chapter~\ref{cpas}. To our
knowledge, this is the first correct calculation of a $\CP$-asymmetry
in leptogenesis that includes
HTL-corrections.\footnote{Reference\cite{Giudice:2003jh} has an
  incorrect calculation for the $\CP$-asymmetry of the neutrino and
  Higgs decays with HTL corrections in the propagators and thermal
  masses in the external states. The discrepancy is discussed in
  detail in reference\cite{Garny:2010nj}. The authors of the first
  reference get a factor $1-f_{\ell'}+f_{\phi'}-2 f_{\ell'} f_{\phi'}$
  for the neutrino decays and $f_{\phi'}-f_{\ell'}-2 f_{\phi'}
  f_{\ell'}$ for the Higgs boson decays due to an erroneous choice of
  cutting rules in the real time formalism. The correct calculation
  gives $1-f_{\ell'}+f_{\phi'}$ for the neutrino decays and
  $f_{\phi'}+f_{\ell'}$ for the Higgs boson decays. This discrepancy
  is also responsible for our $\CP$-asymmetry in Higgs decays being a
  factor ten larger than their result.}  We present rules for the
product of spinors that are related to the fermionic quasiparticles in
equations~\eqref{eq:202} and~\eqref{eq:208} and derive frequency sums
for the HTL fermion propagator in equation~\eqref{eq:160}. We find
four different $\CP$-asymmetries corresponding to the four different
choices of lepton modes both in the loop and as external states. We
find the $\CP$-asymmetry to be symmetric under an exchange of the
lepton mode in the loop and the external lepton mode, such that $\D
\g_{+-}= \D \g_{-+}$. At finite temperature, there are three possible
cuttings for the vertex contribution, the $\{\ell',\phi' \}$-cut that
corresponds to zero temperature and two additional cuts involving the
internal $N'$, namely through $\{N',\ell'\}$ and $\{N',\phi'\}$, which
have been found by~\cite{Giudice:2003jh, Beneke:2010wd, Garny:2009rv}
and examined more closely in~\cite{Garbrecht:2010sz}, using the
real-time formalism. We obtain the same cuts using the imaginary time
formalism and concentrate on the $\{\ell',\phi'\}$-cut, assuming the
hierarchical limit of $M_2 \gg M_1$. As expected from the
zero-temperature result, we find the vertex contribution proportional
to the self-energy contribution in this limit. Contrary to
reference~\cite{Giudice:2003jh}, we find that the $\CP$-asymmetry in
Higgs boson decays is larger than the asymmetry in neutrino decays by
about a factor of $100$ if appropriately normalised. This is due to a
suppression of the Higgs boson decay rate and a thermal enhancement of
the $\CP$-asymmetry by the distribution functions of the Higgs bosons
and leptons. We compare the $\CP$-asymmetries in the two-mode approach
to the $\CP$-asymmetries in the one-mode approach. We find that for
the two-mode approach, the helicity structure of the modes prohibits
the two leptons to be scattered strictly in the same direction while
for the one-mode approach, this direction is only mildly
suppressed. Notably this fact is responsible for suppressing the
($++$)-$\CP$-asymmetry compared to the asymmetries of the one-mode
approach, as well as the residues of the plus-modes to less extent.

We derive and evaluate the Boltzmann equations for our approaches in
chapter~\ref{boltzmanneq}, performing the crucial subtraction of
on-shell intermediate states in
appendix~\ref{sec:subtr-shell-prop}\footnote{The thermal factor
  $(1-f_N^\rmeq)$ we find is different from the factor $(1-f_N)$ that
  reference~\cite{HahnWoernle:2009qn} uses, where they employ the
  out-of equilibrium distribution function for neutrinos without
  deriving their result explicitly.}. We compare the results of the
Boltzmann equations for our five cases, that is, decoupled lepton
modes, strongly coupled lepton modes, the one-mode approach with
$m_\ell$, the one-mode approach with $\sqrt{2} \, m_\ell$, and the
vacuum case. We assume three different initial values for the
abundance of neutrinos: zero, thermal and dominant abundance,
motivated by different scenarios for the production of heavy neutrinos
after inflation~\cite{Giudice:2003jh}. In the weak washout regime, we
find that using thermal masses enhances the final lepton asymmetry by
about one order of magnitude for zero initial neutrino abundance. This
is due to the fact that the $\CP$-asymmetry and the decay rate evolve
differently at $z \gtrsim 1$ when using thermal masses, since the
$\CP$-asymmetry suffers from an additional suppression by thermal
masses through the leptons and Higgs bosons in the loop. Due to the
helicity structure of the modes, the $\CP$-asymmetry of the plus-mode
is additionally suppressed, which results in an additional enhancement
of the plus-mode lepton asymmetry compared to the final asymmetries of
the one-mode approaches. The enhancement we find is similar to the one
found in reference~\cite{Giudice:2003jh} in this regime, but hard to
compare quantitatively due to their different approach which includes
scatterings and the discrepancy in the $\CP$-asymmetry.  In the strong
washout regime, thermal masses do not show an influence as
expected\footnote{There is a slight suppression of the lepton
  asymmetry for thermal masses, since the thermal mass suppresses the
  equilibrium distribution of the leptons somewhat and thereby
  enhances the washout term.}. However, when we couple the plus- and
minus-mode strongly, we observe an enhancement of the lepton asymmetry
by a factor of about two, since we stored half of the asymmetry in a
mode that essentially does not interact with the neutrinos and is
therefore not affected by washout.  For intermediate washout, that is
$K \sim 1$, we find that the lepton asymmetries with thermal masses
are about one magnitude lower than in the vacuum case when we assume
dominant initial neutrino abundance. This is due to the fact that the
lower $\CP$-asymmetry does not succeed in producing as much lepton
asymmetry at $z \gtrsim 1$ when using thermal masses.  A decoupled
minus-mode would show a behaviour completely different from the other
thermal cases and the vacuum case for all initial values of the
neutrino abundance. The lepton asymmetry in such a decoupled mode is
produced mainly at high temperature and only slightly affected by the
development at $z \gtrsim 1$, where it decouples from the evolution of
the other abundances. Therefore, the washout parameter $K$, which
determines the coupling strength and thereby the asymmetry production
at high temperatures, is crucial for the final value of the lepton
asymmetry in this mode, as is the initial neutrino abundance.

Summarising, we argue that for an accurate description of medium
effects on leptogenesis, the influence of thermal quasiparticles,
notably the effects of the two fermionic modes, cannot be
neglected. Similar to reference~\cite{Giudice:2003jh}, our study shows
that thermal masses have a strong effect in the weak washout regime,
while the effect of fermionic modes has an additional influence on the
final lepton asymmetry in this regime. We also showed that notably in
the strong washout regime, the presence of a quasi-sterile lepton mode
that is not affected by washout can have a non-negligible effect on
the final lepton asymmetry. Future studies should clarify the dynamics
of the interaction between the two fermionic modes and determine
whether the evolution of the asymmetries in the two modes is closer to
the decoupled or the strongly coupled case.

Another important aspect that might be studied in future works is the
influence of the finite width of the fermionic
modes~\cite{Drewes:2010yy}, notably the minus-mode. Such effects could
be studied using Kadanoff-Baym equations~\cite{Anisimov:2008dz,
  Garny:2009rv, Garny:2009qn, Anisimov:2010aq, Garny:2010nj,
  Beneke:2010wd, Garny:2010nz, Anisimov:2010dk, Beneke:2010dz,
  Drewes:2010pf, Drewes:2010zz} or some other formalism that takes
into account non-equilibrium quantum effects. In the quest for a
unified description of finite-temperature effects on leptogenesis, it
is important to include SM interactions in the Kadanoff-Baym studies
that are under way. To this end, quasiparticle excitations of fermions
and gauge-bosons should be taken into account. Last but not least, the
fermionic modes might have an influence on other related dynamics in
the early universe that involve fermions, such as thermal production
of axions, axinos or gravitinos. Notably our introduction of sum rules
for quasiparticle spinors in equations~\eqref{eq:202} and
\eqref{eq:208}, as well as the derivation of frequency sums in
equation~\eqref{eq:160} can be a valuable contribution to future
calculations.

Future experimental results from the neutrino sector and from
high-temperature dynamics, such as RHIC or observations at the LHC,
will provide interesting input for these theoretical efforts. Whatever
the outcome of such future experiments and calculations, the quest for
a sufficiently complete and unified description of the influence of
the hot, dense medium on early-universe dynamics, such as leptogenesis
or the production of dark matter particles, is guaranteed to remain an
attractive and vibrant field of research with much insight to be
gained both from theoretical and experimental side.

    \appendix
    \clearpage
\chapter{Green's Functions at Zero Temperature}


\label{sec:green0}
We consider the case of a scalar field $\phi(x)$ with mass $m$, where
$x \equiv (x_0, {\bf x})$, $x_0 \equiv t$, we are using natural units
$\hbar = c = k_B = 1$ and are in the relativistic regime where we use
the Minkowski metric $x^2 = x_\mu x^\mu = g_{\mu \nu}x^\nu x^\mu =
x_0^2 - {\bf x}^2$. The amplitude of a particle to travel from $x$ to
$y$ if $x_0 < y_0$ or from $y$ to $x$ if $y_0 < x_0$ is called Feynman
propagator and is equal to the two-point correlation function defined
as
\begin{equation}
{\rm i} \, \Delta_F (x-y) \equiv \langle 0 | T\{\phi(x) \phi(y) \} | 0 \rangle \, ,
\end{equation}
where $T$ denotes the time ordering operator which acts on two fields as
\begin{equation}
T\{\phi(x) \phi(y) \} = \left \{
\begin{matrix} \; \phi(x) \phi(y) & & x_0>y_0 & \\ \phi(y) \phi(x) & & x_0<y_0 &
\end{matrix} \right. \, .
\end{equation}
We can express the scalar field by its Fourier transform
\begin{equation}
\label{eq:fourier}
\phi(x) = \left. \int \frac{{\rm d^3} k}{(2 \pi)^{3/2}} \frac{1}{(2 \omega_k)^{1/2}} [
a({\bf k}) {\rm e}^{-{\rm i}\, K \cdot x} + a^\dagger({\bf k}) {\rm
e}^{\, {\rm i}\, K \cdot x} ] \right |_{k_0=\omega_k} \, ,
\end{equation}
where $K \equiv (k_0, {\bf k})$ is the four-momentum of the field,
$\omega_k \equiv \sqrt{{\bf k}^2 +m^2}$ the energy and the Fourier
coefficients represent creation $a({\bf k})$ and destruction operators
$a^\dagger({\bf k})$, which create or destroy a boson with momentum
$\bf k$. They act on a state $|n({\bf k}) \rangle$ with $n$ particles
of momentum $\bf k$ as
\begin{eqnarray}
\label{eq:ladder}
a({\bf k}) | n({\bf k}) \rangle & = & \sqrt{n({\bf k})}\; | n({\bf
k})-1 \rangle \\ a^\dagger({\bf k}) | n({\bf k}) \rangle & =
& \sqrt{n({\bf k})+1} \; | n({\bf k})+1 \rangle \, ,
\end{eqnarray}
in particular $a({\bf k}) | 0 \rangle = 0$ for all $\bf k$.

Using the Fourier representation of $\phi$, the Feynman propagator can be written as
\begin{equation}
\Delta_F (x-y) = \int \frac{{\rm d}^4 K}{(2 \pi)^4} \frac{{\rm e}^{- {\rm i} \, K \cdot 
(x - y)}}{K^2-m^2+ {\rm i} \, \epsilon} \, ,
\end{equation}
where the term $+ {\rm i} \, \epsilon$ accounts for the so-called
Feynman prescription, that is in the complex $k_0$-plane we integrate
above the pole at $k_0=\omega_k$ and below the pole at $k_0 =
-\omega_k$. When we carry out the $k_0$-integration, we find for $x_0
> y_0$
\begin{equation}
\label{eq:prop0}
\Delta_F(x-y) = - {\rm i} \, \left. \int \frac{{\rm d}^3k}{(2 \pi)^3} \frac{1}{2 
\omega_k} {\rm e}^{- {\rm i} K \cdot (x-y)} \right |_{k_0=\omega_k} \, .
\end{equation}




\clearpage





\chapter{Analytic Solution of the Dispersion Relations for HTL Fermions}
\label{cha:analyt-solut-htl}

The dispersion relations of the two lepton modes are given by the
poles of the corresponding propagator. Hence, we seek the zeros of
\begin{equation}
\label{D}
D_\pm(K)=\Delta_\pm(K)^{-1}=\left [ -k_0 \pm k + \frac{m_\ell^2}{k}
  \left ( \pm1 - \frac{\pm k_0 - k}{2k} \ln \frac{k_0+k}{k_0-k} \right
  ) \right ]^{-1}
\end{equation}
The equations $D_\pm=0$ can be transformed by the substitutions
\begin{align}
x_+  &:=  \frac{k_0+k}{k_0-k} \\
x_- &:= \frac{k_0-k}{k_0+k}=\frac{1}{x_+} \\
c &:= \frac{k^2}{m_\ell^2}.
\end{align}
This yields
\begin{equation}
D_\pm=\pm \frac{k}{c} \frac{1}{x_\pm-1} \left (-2 c-1+x_\pm-\ln x_\pm \right).
\end{equation}
Further introducing
\begin{equation}
\label{s}
s := - \exp (-2 c -1 )
\end{equation}
leads to
\begin{equation}
D_\pm= \frac{\mp 2 k}{1+\ln(-s)} \frac{1}{x_\pm-1} \left [ x_\pm +
  \ln(-s) - \ln x_\pm \right ].
\end{equation}
Since the prefactor does not have poles for the values of $K$ we are
looking at, solving $D_\pm=0$ amounts to solving 
\begin{equation} x_\pm
+\ln(-s)-\ln x_\pm=0, 
\end{equation} 
which in turn means 
\begin{equation} s=-x_\pm
\e^{-x_\pm}.  
\end{equation}

This is the defining equation of the Lambert W function
\cite{Corless:1996zz,Chapeau}, thus the solution reads
\begin{equation}
x_\pm=-W(s).
\end{equation}
According to the definition in Eq.~\eqref{s}
\begin{equation}
\label{srange}
-1/\e \leq s \leq 0,
\end{equation}
thus the two real branches of the Lambert function, $W_0$ and
$W_{-1}$, correspond to the two solutions we seek. In the range given
by Eq.~\eqref{srange} $W_0 \geq -1$ and $W_{-1} \leq -1$. For $k_0 \geq
k$ we have $x_+ \geq 1$ and $x_- \leq 1$. Hence, the physical solutions for $x_\pm$ read
\begin{equation}
x_+=-W_{-1}(s) \hspace{1cm} {\rm and} \hspace{1cm} x_-=-W_0(s).
\end{equation}

The corresponding results for $\omega_\pm$ are then given by
\begin{align}
\omega_+&= k \; \frac{W_{-1}(s) -1}{W_{-1}(s)+1} \label{op} \\
\omega_-&=- k \; \frac{W_{0}(s) -1}{W_{0}(s)+1}. \label{om}
\end{align}
Making use of the relations \cite{Jeffrey}
\begin{equation}
W_{0,-1}(z)+\ln (W_{0,-1}(z))=\ln z,
\end{equation}
one can directly prove the result by plugging Eqs.~\eqref{op} and
\eqref{om} into Eq.~\eqref{D}.

\clearpage

\chapter{Quantities at Zero Temperature}
\label{cha:quantities-at-zero}

\section[Decay Rate]{Decay Rate at Zero Temperature}
\label{sec:decay-rate-at}

We know that at $T=0$, the decay rate of a heavy neutrino with
definite spin at rest is given by
\begin{equation}
\Gamma_{\rm rf}(N \to \phi \ell) = \frac{(\lambda^\dagger \lambda)_{jj}}{16 \pi} M_j \, ,
\end{equation}
where we have not summed over neutrino spins. Then
\begin{equation}
\Gamma_D = \frac{M_j}{E} \Gamma_{\rm rf} = \frac{M_j}{E}
\frac{(\lambda^\dagger \lambda)_{jj}}{16 \pi} M_j
\end{equation}
and the decay density for definite neutrino spin is
\begin{eqnarray}
  \gamma_D(N \to \phi \ell)&=&\frac{1}{2 \pi^2} \int_{M_j}^\infty \rmd E E p f_N
  \Gamma_D\\ 
  &=& \frac{(\lambda^\dagger \lambda)_{jj}}{4 (2 \pi)^3}
  M_j^2 \int_{M_j}^\infty \rmd E p f_N \\
  &=& \frac{(\lambda^\dagger \lambda)_{jj}}{4 (2 \pi)^3} M_j^3 T K_1 \left(
    \frac{M_j}{T} \right),
\end{eqnarray}
where $K_1(z)$ is a Bessel function of second kind and $z=M_j/T$.


\section[${C\!P}$-Asymmetry]{$\boldsymbol{C\!P}$-Asymmetry at Zero
  Temperature}
\label{sec:boldsymb-asymm-vacu}

For the $\CP$-asymmetry at zero temperature, the integrals are the
same as for $T>0$, just the thermal masses are zero and the lepton
propagator is
\begin{equation}
S_\ell'=\frac{\slashed{K'}}{K'^2},
\end{equation}
so the spin sum is
\begin{equation}
\frac{1}{2} \sum_{r,s} (\overline{u}_\ell P_R u_N) (\overline{u}_N P_R S_\ell' P_L
u_\ell) = M_j K^\mu K'_\mu \Delta_\ell',
\end{equation}
where
\begin{equation}
\Delta_\ell'=\frac{1}{K'^2} \, ,
\end{equation}
and we have averaged over the neutrino degrees of freedom.
 At zero temperature, we can use the Cutkosky rules to compute
 discontinuities, so we replace propagators by
 $\delta$-functions. There are three possible cuttings, however, the
 cuttings which involve $N_2$ correspond to a process where one
 massless particle decays into $N_2$ and the other massless particle
 which is not possible kinematically. The only possible cutting is
 through the Higgs boson and lepton propagator, so
\begin{eqnarray}
\Delta_\ell' \rightarrow - 2 \pi \rmi \delta(K'^2),\\
\Delta_\phi' \rightarrow - 2 \pi \rmi \delta(Q'^2).
\end{eqnarray}

We get
\begin{eqnarray}
\rmIm (I_0 I_1^*)&=& - \frac{\rmi}{2} {\rm Disc} (I_0 I_1^*) = - \frac{1}{2} M_j M_k {\rm Disc} 
\left(\int \frac{\rmd^4 
k'}{(2 \pi)^4} \Delta_N' \Delta_\phi' \Delta_\ell' K_\mu K'^\mu \right) \\
&=& \frac{M_j M_k}{16 \pi} \left[ 1- \left(1+x \right) \ln \frac{1+x}{x} 
\right],
\end{eqnarray}
where
\begin{equation}
x=\frac{M_k^2}{M_j^2} \, .
\end{equation}

Analogously, we get for the self-energy
\begin{equation}
\rmIm (I_0 I_1^*) = \frac{M_j M_k}{16 \pi} \frac{1}{1-x} \, .
\end{equation}
The CP-Asymmetry at zero temperature reads
\begin{equation}
  \epsilon = 
  \frac{1}{8 \pi} \frac{\rmIm [(\lambda^\dagger \lambda)_{jk}^2]}{(\lambda^\dagger \lambda)_{jj}} 
  g(x) \, ,
\end{equation}
where
\begin{align}
g(x)= \sqrt{x} \left[ \frac{1}{1-x} +1 - \left( 1+x \right) \ln \frac{1+x}{x} \right] \, .
\end{align}

\section[Boltzmann Equations]{Boltzmann Equations at Zero Temperature}
\label{sec:boltzm-equat-at}

We can derive the Boltzmann equations for the neutrino and lepton
evolution at zero temperature, approximating the phase-space densities
with Maxwell-Boltzmann distributions,
\begin{align}
  \label{eq:75}
  f_i(E_i) = \exp(- E_i \beta) \, ,
\end{align}
where energy conservation in scatterings and decays implies
\begin{align}
  \label{eq:79}
  f_N &= f_L f_H \,
\end{align}
and there are no Higgs decays at high temperature. For the neutrino
evolution, we get, analogous to equation~\eqref{eq:71},
\begin{align}
  \label{eq:80}
  \frac{\rmd Y_{N}}{\rmd z}= - \frac{z}{s H_1} (x_N-1) \gamma_0,
\end{align}
where 
\begin{align}
  \label{eq:81}
  \gamma_0 =  \int \rmd
  \tilde{p}_N \rmd \tilde{p}_{L} \rmd \tilde{p}_H (2 \pi)^4
  \delta^4(p_{N} -p_H-p_{L})  \left|\mathcal{M}_0\right|^2
  f_N^\rmeq.
\end{align}
The matrix element evaluated at zero temperature reads
\begin{align}
  \label{eq:82}
  \left| \mathcal{M}_0 \right|^2 = 4 \times 2 \, P_N \cdot P_L,
\end{align}
where the factor $4$ comes from summing over $\ell$ and $\barell$, as
well as over the doublets $(e^-,\phi^+)$ and $(\nu, \phi^0)$.

We can express the decay rate $\gamma_0$ in terms of the total decay width
$\G_{\rm rf}^{\rm tot}$ in the rest-frame of the neutrino,
\begin{align}
  \label{eq:83}
  \gamma_0  =  g_N \int \frac{\rmd p_N^3}{(2 \pi)^3} \frac{M}{E_N} \G_{\rm rf}^{\rm tot} f_N^\rmeq,
\end{align}
where $g_N=2$ accounts for the internal degrees of freedom of the
neutrino, the two spins, and
\begin{align}
  \label{eq:84}
  \G_{\rm rf}^{\rm tot}(N \to H L)= \frac{(\lambda^\dagger \lambda)_{11} M_1}{4 \pi g_N}
\end{align}
describes the decay of a neutrino with a definite spin into $(\phi \ell)$ and $(\barphi \barell)$.

Evaluating equation~\eqref{eq:83}, we get
\begin{align}
  \label{eq:85}
  \gamma_0= g_N \frac{M^2}{2 \pi^2} T \G_{\rm rf}^{\rm tot} K_1(z),
\end{align}
where $z=M/T$ and $K_1(z)$ is a Bessel function of second kind.
For the equilibrium density of the neutrinos, we get
\begin{align}
  \label{eq:86}
  n_N^\rmeq = g_N \int \frac{d^3 p_N}{(2 \pi)^3} f_N^\rmeq = g_N
  \frac{M^2}{2 \pi^2} T K_2(z), 
\end{align}
so that
\begin{align}
  \label{eq:87}
  \frac{\gamma_0}{n_N^\rmeq}= \G_{\rm rf}^{\rm tot} \frac{K_1(z)}{K_2(z)}.
\end{align}
We can write the Boltzmann equation as
\begin{align}
  \label{eq:88}
  Y_N' = -D (Y_N-Y_N^\rmeq),
\end{align}
where
\begin{align}
  \label{eq:91}
  Y_X' \equiv \frac{\rmd Y_X}{\rmd z},
\end{align}
\begin{align}
  \label{eq:89}
  D= \frac{z}{H_1} \frac{\gamma_0}{n_N^\rmeq} = z K \frac{K_1(z)}{K_2(z)} \, ,
\end{align}
\begin{align}
  \label{eq:99}
  Y_N^\rmeq= \frac{45}{4 \pi^4} \frac{g_N}{g_*} z^2 K_2(z)
\end{align}
and
\begin{align}
  \label{eq:90}
  K = \frac{\G_{\rm rf}^{\rm tot}}{H_1}= \frac{\tilde{m}}{m^*}
\end{align}
is called decay parameter.

For the lepton evolution, the subtraction of on-shell propagators can
be performed analogously to the finite temperature case, so that
\begin{align}
  \label{eq:92}
  \gamma^{\rm sub}(\ell \phi \to \barell \barphi) - \gamma^{\rm
    sub}(\barell \barphi \to \ell \phi) = \epsilon_0 \gamma_0,
\end{align}
where 
\begin{align}
  \label{eq:93}
  \epsilon_0 \equiv \frac{\Gamma(N \to \ell \phi)- \Gamma(N \to
    \barell \barphi)}{\Gamma(N \to \ell \phi) + \Gamma(N \to
    \barell \barphi)}
\end{align}
is calculated in appendix~\ref{sec:boldsymb-asymm-vacu}. 
Analogous to equation~\eqref{eq:74}, we get
\begin{align}
  \label{eq:94}
    \frac{\rmd Y_{\mathcal{L}}}{\rmd z} = - \frac{z}{s H_1} \left( -
    \e_0 \left( x_N-1 \right) +
    \frac{x_{\mathcal{L}}}{2} \right) \gamma_0,
\end{align}
which can be rewritten as
\begin{align}
  \label{eq:95}
  Y_{\mathcal{L}}'=\epsilon_0 D (Y_N-Y_N^\rmeq) -W Y_{\mathcal{L}},
\end{align}
where 
\begin{align}
  \label{eq:96}
  W \equiv \frac{z}{H_1} \frac{\gamma_0}{2 n_{\ell}^\rmeq}.
\end{align}
We have
\begin{align}
  \label{eq:97}
  n_\ell^\rmeq= g_\ell \int \frac{d^3 p_\ell}{(2 \pi)^3} f_\ell^\rmeq = g_\ell
  \frac{T^3}{\pi^2},
\end{align}
where $g_\ell=2$ accounts for the lepton doublet
components, so we get
\begin{align}
  \label{eq:98}
  W = \frac{1}{4} \frac{g_N}{g_\ell} z^3 K K_2(z).
\end{align}

\chapter{The Other Cuts}
\label{cha:other-cuts}

\section{Imaginary Parts}
\label{sec:imaginary-parts-1}

\subsection[Vertex cut through ${\{N_2,\phi'\}}$]{Vertex cut through $\boldsymbol{\{N_2,\phi'\}}$}
\label{sec:vertex-cut:-external-2}

We use the conventions for the vertex contribution in $N$-decays in
chapter~\ref{cpas}.  For $N_\ell^{N'}$, we shift integration variables
to $\rmd^3 q'$ after carrying out the Matsubara sum over $k_0'$. We
consider the angle
\begin{align}
\eta_{q'}= \frac{\bf k \cdot q'}{k q'}
\end{align}
between $\bf k$ and $\bf q'$ and write
\begin{align}
\rmIm  \left( \int_{-1}^1 \rmd \eta_{q'} \frac{1}{N_\ell^{N'}} \right) = -\pi
\frac{\omega_{p'}}{k q'},
\end{align}
where the angle is
\begin{align}
\eta_{k q', 0}= \frac{1}{2 k q'} \left(- 2 \omega \omega_{q'} +
\Sigma_k \right),
\end{align}
and
\begin{align}
\Sigma_k=M_k^2-(\omega^2-k^2)-m_\phi^2 \, .
\end{align}
The imaginary part reads
\begin{align}
\label{imnn'}
 \rmIm  \Big( T \sum_{k_0', h'}\int \frac{\rmd^3 k'}{(2 \pi)^3}
 \D_{N'} \D_{\phi'} & \D_{h'} H_-
  \Big)_{N_\ell^{N'}} =  \frac{1}{4 \pi^3}\rmIm \left(
  T \sum_{k_0',h'} \int_0^\infty \rmd q' q'^2 \rmd \eta_{q'}
  \int_0^\pi \rmd \phi_{q'} \D_{N'} \D_{\phi'} \D_{h'} H_-
\right) \nonumber \\
 =& - \frac{1}{16 \pi^2} \sum_{h'} \int \rmd q'
\rmd \phi' \frac{q'}{k \omega_{q'}} Z_{h'} 
\left[ \left( B_\phi^{\ell'} - B_N^{\ell'} \right) H_- + \left(
    B_\phi^0  - B_N^0 \right) H_+ \right] \, .
\end{align}

\subsection[Vertex cut through ${\{N_2,\ell'\}}$]{Vertex cut through $\boldsymbol{\{N_2,\ell'\}}$}
\label{sec:vertex-cut:-external-3}

For the $N_\phi^{N'}$-term, we integrate
over $k'$, and choose the polar angle between $\bf q$ and $\bf k'$,
\begin{align}
\eta_{qk'}=\frac{\bf q \cdot k'}{q k'} \, .
\end{align}
We write
\begin{align}
{\rm Im} \left( \int_{-1}^1 \rmd \eta_{qk'} \frac{1}{N_{N'}}\right)=- \pi
\frac{\omega_{p'}}{q k'},
\end{align}
where the angle is
\begin{align}
\eta_{qk'0}= \frac{1}{2 q k'} \left( -2 \omega_q \omega'+\Sigma_{q k'}
\right),
\end{align}
and
\begin{align}
\label{eq:218}
\Sigma_{q k'}=M_k^2-(\omega'^2-k'^2)-m_\phi^2.
\end{align}
The imaginary part is given by
\begin{align}
\label{imnn''}
\rmIm \left( T \sum_{k_0', h'} \int \frac{\rmd^3 k'}{(2 \pi)^3}
  \D_{N'} \D_{\phi'} \D_{h'} H_- \right)_{N_\phi^{N'}} = & \frac{1}{4 \pi^3}\rmIm \left(
  T \sum_{k_0',h'} \int_0^\infty \rmd k' k'^2 \rmd \eta_{qk'}
  \int_0^\pi \rmd\phi_{qk'} 
\D_{N'} \D_{\phi'} \D_{h'} H_-
\right) \nonumber \\ = & - \frac{1}{16 \pi^2} \sum_{h'} \int \rmd k'
\rmd \phi' \frac{k'}{q \omega_{q'}} Z_{h'} Z_\phi^{N'} (A_\ell^{\phi'}
-A_\ell^0) H_+.
\end{align}

\section[Analytic Expressions for the $\CP$-Asymmetries]{Analytic
  Expressions for the $\boldsymbol{\CP}$-Asymmetries}
\label{sec:integr-expr-cp}

\subsection[Vertex cut through ${\{N_2,\phi'\}}$]{Vertex cut through
  $\boldsymbol{\{N_2,\phi'\}}$}
\label{sec:vertex-cut:-external}

We get for $N_\ell^{N'}$
\begin{align}
\rmIm(I_V)_{N_\ell^{N'}}=  \frac{M_j M_k}{16 \pi^2}
\frac{Z_h\omega}{k} \sum_{h h'} \int_0^\infty \rmd q' \int_0^{\pi}
\rmd \phi_{q'} \frac{q'}{\omega_{q'}} Z_{h'}  
 \left[ \left( B_\phi^{\ell'} - B_N^{\ell'} \right) H_-
  + \left( B_\phi^0 - B_N^{\phi'} \right) H_+ \right] \, .
\end{align}
The difference in decay rates reads
\begin{align}
\label{eq:139}
   \gamma(N \to \ell_h \phi)& - \gamma(N \to \barell_h \barphi) = 
 - g_{SU(2)} \rmIm
  \left\{ \left[ \left( \lambda^\dagger \lambda \right)_{jk} \right]^2
  \right\} \frac{M_j M_k}{4 (2 \pi)^5} \nonumber \\
 & \times  \sum_{h'} \int \rmd E \rmd k \rmd q'
  \int_0^\pi \rmd \phi_{q'} k F_{N_ h}^\rmeq Z_h \frac{q'}{k \o_{q'}}
  Z_{h'} \left[ \left( B_\phi^{\ell'} - B_N^{\ell'} \right) H_-
  + \left( B_\phi^0 - B_N^{\phi'} \right) H_+ \right] 
\end{align}
and the $\CP$-asymmetry reads
\begin{align}
\label{eq:136}
\epsilon_h(T) = & - g_{SU(2)} \frac{\rmIm \{[( \lambda^\dagger
  \lambda)_{jk}]^2\}}{\gamma(N \to L_h N)} \frac{M_j M_k}{4 (2 \pi)^5}
\sum_{h'} \int \rmd E \rmd k \rmd q' \int_0^\pi \rmd \phi_{q'} k F_{N_
  h}^\rmeq Z_h \frac{q'}{k \o_{q'}}
Z_{h'} \nonumber \\
& \times \left[ \left( B_\phi^{\ell'} - B_N^{\ell'} \right) H_-
  + \left( B_\phi^0 - B_N^{\phi'} \right) H_+ \right] \nonumber \\
=&- \frac{\rmIm\{[(\lambda^\dagger \lambda)_{jk}]^2\}}{g_c
  (\lambda^\dagger \lambda)_{jj}} \frac{M_j
  M_k}{4 \pi^2}  \nonumber \\
& \times \frac{\sum_{h'} \int \rmd E \rmd k \rmd k' \int_0^\pi \rmd
  \phi' k F_{N_ h}^\rmeq Z_h \frac{k'}{p \o_{p'}} Z_{h'} [(
  B_\phi^{\ell'} - B_N^{\ell'}) H_- + ( B_\phi^0 - B_N^{\phi'}) H_+
  ]}{\int \rmd E \rmd k k f_N Z_D Z_h(p_0-h p \eta)} \, ,
\end{align}

\subsection[Vertex cut through ${\{N_2,\ell'\}}$]{Vertex cut through
  $\boldsymbol{\{N_2,\ell'\}}$}
\label{sec:vertex-cut:-external-4}

For $N_\phi^{N'}$, we get
\begin{align}
\rmIm(I_V)_{N_\phi^{N'}}= \frac{M_j M_k}{16 \pi^2}
\frac{Z_h\omega}{q} \sum_{h'} \int_0^\infty \rmd k' \int_0^{\pi}
\rmd \phi_{qk'} \frac{k'}{\omega_{q'}} Z_{h'} Z_\phi^{N'}
(A_\ell^{\phi'}-A_\ell^0) H_+ \, .
\end{align}
The difference in decay rates reads
\begin{align}
\label{eq:138}
   \gamma(N \to \ell_h \phi) - \gamma(N \to \barell_h \barphi) = 
& - g_{SU(2)} \rmIm
  \left\{ \left[ \left( \lambda^\dagger \lambda \right)_{jk} \right]^2
  \right\} \frac{M_j M_k}{4 (2 \pi)^5} \nonumber \\
 & \times \sum_{h'} \int \rmd E \rmd k \rmd k'
  \int_0^\pi \rmd \phi_{qk'} k F_{N_ h}^\rmeq Z_h \frac{k'}{q \o_{q'}}Z_\phi^{N'}
  Z_{h'}(A_\ell^{\phi'} -A_\ell^0) H_+ \, .
\end{align}
 The $\CP$-asymmetry reads
\begin{align}
\epsilon_h(T) = &  - g_{SU(2)} \frac{\rmIm
  \{[( \lambda^\dagger \lambda)_{jk}]^2\}}{\gamma(N \to L_h N)} 
\frac{M_j M_k}{4 (2 \pi)^5} \sum_{h'} \int \rmd E \rmd k \rmd k'
  \int_0^\pi \rmd \phi_{qk'} k F_{N_ h}^\rmeq Z_h \frac{k'}{q \o_{q'}}Z_\phi^{N'}
  Z_{h'}(A_\ell^{\phi'} -A_\ell^0) H_- \nonumber \\
=&- \frac{\rmIm\{[(\lambda^\dagger
  \lambda)_{jk}]^2\}}{g_c (\lambda^\dagger \lambda)_{jj}} \frac{M_j
  M_k}{4 \pi^2}  
\frac{\sum_{h'} \int \rmd E \rmd k \rmd k'
  \int_0^\pi \rmd \phi_{qk'} k F_{N_ h}^\rmeq Z_h \frac{k'}{q \o_{q'}}Z_\phi^{N'}
  Z_{h'}(A_\ell^{\phi'} -A_\ell^0) H_-}{\int
\rmd E \rmd k k f_N Z_D Z_h (p_0-h p \eta)} \, ,
\end{align}
where we integrate over $\phi_{qk'}$ and we take the coordinate system
differently than for $N_N^N$.

\chapter{Subtraction of On-Shell Propagators}
\label{sec:subtr-shell-prop}

\section{Low Temperature}
\label{sec:low-temperature-2}

We verify the relation in equation~\eqref{eq:b31}. The scattering rate
$\gamma(\ell \phi \to \barell \barphi)$ can be split up into four
scatterings with different kinematics, corresponding to the four
possibilities of combining the in- and outgoing lepton modes. The
scattering rates read
\begin{align}
  \label{eq:b38}
  \g(\ell_{h_i} \phi \to \barell_{h_f} \barphi)= \int \rmd
  \tilde{p}_{\ell h_i} \rmd \tilde{p}_\phi \rmd
  \tilde{p}_{\barell h_f} \tilde{p}_{\barphi} & (2 \pi)^4 \delta^4
  (p_{\ell h_i} + p_\phi- p_{\barell h_f} -p_{\barphi})
  \nonumber \\
  & \times \left| \mathcal{M}(\ell_{h_i} \phi \to \barell_{h_f}
      \barphi) \right|^2 f_{\ell h_i} f_\phi (1- f_{\barell h_f}) (1+
    f_{\barphi})  ,
\end{align}
where $(h_i, h_f) = \pm 1$ denote the helicity-to-chirality ratio of
the initial- and final-state leptons (or antileptons). We will drop
the subscript for this appendix part, unless it is necessary, and all
equations are valid for one specific mode for each involved lepton,
unless otherwise noted. With this simplified notation, each of the
four matrix elements is evaluated as\\
\begin{align}
  \label{eq:b39}
  \sum_{s_\ell,s_{\barell}} \left| \mathcal{M}(\ell_{h_i} \phi \to \barell_{h_f}
    \barphi) \right|^2 = \left[(\l^\dagger \l)_{11} \right]^2
 \left| D_N
  \right|^2 2 \left[2 (p_N \cdot p_{\ell h_i}) (p_N \cdot p_{\barell h_f}) -
    (p_N \cdot p_N) (p_{\ell h_i} \cdot p_{\barell h_f}) \right],
\end{align}
where we sum over the lepton spins $s_\ell$ and $s_{\barell}$ and the
lepton flavours and $D_N=1/[P_N^2-M_N^2+\rmi p_N^0 \Gamma_N(p_N^0)]$
is the neutrino propagator in the narrow-width approximation and
$\Gamma_N(p_N^0)$ the total width of the neutrino, which equals the
total interaction rate, including both lepton modes. Putting the
propagator on its mass shell, $P_N^2=M_N^2$, we get
\begin{align}
  \label{eq:b40}
  \sum_{s_\ell,s_{\barell}} \left| \mathcal{M}^{\rm os}(\ell_{h_i}
    \phi \to \barell_{h_f} \barphi) \right|^2 = \left[(\l^\dagger
    \l)_{11} \right]^2 \left| D_N^{\rm os} \right|^2 2 \left[2 (p_N
    \cdot p_{\ell h_i}) (p_N \cdot p_{\barell h_f}) - M_N^2 (p_{\ell h_i} \cdot
    p_{\barell h_f}) \right],
\end{align}
where
\begin{align}
  \label{eq:b42}
  \left| D_N^{\rm os} \right|^2= \frac{ \pi
    \delta(P_N^2-M^2)}{p_N^0 \Gamma_N(p_N^0)}
\end{align}
In vacuum without thermal masses, this reads
\begin{align}
  \label{eq:b41}
   \sum_{s_\ell,s_{\barell}} \left| \mathcal{M}^{\rm os}(\ell \phi \to \barell
    \barphi) \right|^2 = \left[(\l^\dagger \l)_{11} \right]^2
  \left| D_N^{\rm os} \right|^2
 2 \left[
\frac{M_N^4}{4} (1+\eta)
\right],   
\end{align}
where the dependence on the angle $\eta$ between the external leptons
cancels out in the integration for symmetry reasons, so we can neglect
it and write
\begin{align}
  \label{eq:b43}
   \sum_{s_\ell,s_{\barell}} \left| \mathcal{M}^{\rm os}(\ell \phi \to \barell
    \barphi) \right|^2 =
   \sum_{s_\ell,s_{\barell}}
 \left| \mathcal{M}(\ell \phi \to N) \right|^2 
  \left| D_N^{\rm os} \right|^2
 \left| \mathcal{M}(N \to \barell \barphi) \right|^2. 
\end{align}
At finite temperature with quasiparticle dispersion relations, we can
not derive equation~\eqref{eq:b43} accurately, but in the narrow-width
approximation\cite{Giudice:2003jh}, one assumes that the influence of
the angle between the external particles is negligible and
equation~\eqref{eq:b43} holds.

Using the relations in equation~\eqref{eq:b27}, we derive
\begin{align}
  \label{eq:b45}
  &\left| \mathcal{M}^{\rm os}(\ell_{h_i} \phi_i \to \barell_{h_f}
    \barphi_f) \right|^2 f_{\ell h_i} f_{\phi,i} (1- f_{\barell h_f})
  (1+ f_{\barphi,f}) - \left| \mathcal{M}^{\rm os}(\barell_{h_i}
    \barphi_i \to \ell_{h_f} \phi_f) \right|^2
  f_{\barell h_i} f_{\barphi,i} (1- f_{\ell h_f}) (1+ f_{\phi,f}) \nonumber \\
  = &\left| D_N^{\rm os} \right|^2 \frac{1}{4} \left|
    \mathcal{M}_{h_i}^0 \right|^2 \left| \mathcal{M}_{h_f}^0 \right|^2
  \left[ f_{\mathcal{L}h_i} (1- f_{\ell h_f}^\rmeq) +f_{\ell
      h_i}^\rmeq f_{\mathcal{L} h_f}-4 \e^N_h f_{\ell h_i}^\rmeq (1-f_{\ell
      h_f}^\rmeq) \right] f_{\phi,i}^\rmeq (1-f_{\phi,f}^\rmeq),
\end{align}
where we have neglected terms of order $\e^2$ and $x^2_{\mathcal{L}}$
and added the subscripts $i$ and $f$ in the Higgs boson
distributions to clarify which momentum to use,
\begin{align}
  \label{eq:b70}
  f_{\phi,i}=f_{\phi}(\omega_{\phi,i})=f_{\phi}(\omega_N-\omega_{\ell h_i})
\end{align}
and likewise for $f_{\phi,f}$.

For the tree-level, $CP$-conserving amplitude, we have
\begin{align}
  \label{eq:b46}
  \left| \mathcal{M}^{\rm tree}(\ell_{h_i} \phi \to \barell_{h_f} \barphi)
  \right|^2
=   \left| \mathcal{M}^{\rm tree}(\barell_{h_i} \barphi \to \ell_{h_f} \phi)
  \right|^2 \equiv   \left| \mathcal{M}_{\D L=2} \right|^2_{h_i h_f}.
\end{align}
For the full amplitude $\left| \mathcal{M}_{\D L=2} \right|^2$, the
on-shell part is also dominant. Since it is $CP$-conserving, we write
\begin{align}
  \label{eq:b48}
  \left| \mathcal{M}_{\D L=2} \right|^2_{h_i h_f} \approx 
  \left| \mathcal{M}^{\rm os}_{\D L=2} \right|^2_{h_i h_f} = 
\left| D_N^{\rm os}
  \right|^2 \frac{1}{4} \left| \mathcal{M}_{h_i}^0 \right|^2
 \left| \mathcal{M}_{h_f}^0 \right|^2
\end{align}
and we get
\begin{align}
  \label{eq:b47}
  \left| \mathcal{M}^{\rm tree}(\ell_{h_i} \phi \right. 
& \left. 
 \to \barell_{h_f} \barphi)
  \right|^2 f_{\ell h_i} f_\phi (1- f_{\barell h_f}) (1+ f_{\barphi})
  \nonumber \\
&-  \left|
    \mathcal{M}^{\rm tree}(\barell_{h_i} \barphi \to \ell_{h_f} \phi) \right|^2
  f_{\barell h_i} f_{\barphi} (1- f_{\ell h_f}) (1+ f_{\phi}) \nonumber \\
&  = \left| \mathcal{M}_{\D L=2} \right|^2_{h_i h_f} \left[
    f_{\mathcal{L}, h_i} (1 - f_{\ell h_f}^\rmeq) + f_{\ell h_i}^\rmeq
    f_{\mathcal{L} h_f} \right].
\end{align}
Subtracting equations~\eqref{eq:b45} and~\eqref{eq:b47}, we derive
\begin{align}
  \label{eq:b49}
  \gamma^{\rm sub} (\ell_{h_i}\phi\to \barell_{h_f} \barphi) -
  \gamma^{\rm sub}(\barell_{h_i} \barphi \to \ell_{h_f} \phi) = \int &
  \rmd \tilde{p}_{\ell h_i} \rmd \tilde{p}_\phi \rmd
  \tilde{p}_{\barell h_f} \tilde{p}_{\barphi} (2 \pi)^4 \delta^4
  (p_{\ell h_i} + p_\phi- p_{\barell h_f} -p_{\barphi})
  \nonumber \\
  & \times \e_h^N \left| D_N^{\rm os} \right|^2 \left|
    \mathcal{M}_{h_i}^0 \right|^2 \left| \mathcal{M}_{h_f}^0 \right|^2
  f^\rmeq_{\ell h_i} f^\rmeq_\phi (1- f^\rmeq_{\ell h_f}) (1+
  f^\rmeq_{\phi}) \nonumber \\
  \equiv & \e^N_h \gamma^{\rm os}_\rmeq(L_{h_i} H \to L_{h_f} H)
\end{align}
Using the relations
\begin{align}
  \label{eq:b71}
  (1-f_{\ell h}^\rmeq)(1+f_\phi^\rmeq) & = (1-f_N^\rmeq)
  (1-f_{\ell h}^\rmeq+f_\phi^\rmeq),
  \\
  f_{\ell h}^\rmeq f_\phi^\rmeq & = f_N^\rmeq
  (1-f_{\ell h}^\rmeq+f_\phi^\rmeq)
  \\
  \text{and } f_\phi^\rmeq f_{\ell h}^\rmeq (1-f_N^\rmeq) & =
  (1+f_\phi^\rmeq) (1-f_{\ell h}^\rmeq) f_N^\rmeq,
\end{align}
which hold for $\omega_N=\omega_{\ell h}+\omega_\phi$, it is
straightforward to derive
\begin{align}
  \label{eq:b72}
    \gamma^{\rm sub} (\ell_{h_i}\phi\to \barell_{h_f} \barphi) - \gamma^{\rm
    sub}(\barell_{h_i} \barphi \to \ell_{h_f} \phi) 
&=   \gamma^{\rm sub} (\ell_{h_f}\phi\to \barell_{h_i} \barphi) - \gamma^{\rm
    sub}(\barell_{h_f} \barphi \to \ell_{h_i} \phi), \nonumber \\
\gamma^{\rm os}_\rmeq(L_{h_i} H \to L_{h_f} H)
& = \gamma^{\rm os}_\rmeq(L_{h_f} H \to L_{h_i} H)
\end{align}

Inserting $1=\int \rmd^4 p_N/(2 \pi)^4 \delta^4(p_N-p_{\ell h_i}
-p_\phi)$ into equation~\eqref{eq:b72}, again using the first relation
from equations~\eqref{eq:b49} and the expression for the total neutrino
width, as derived in chapter~\ref{decayrate},
\begin{align}
  \label{eq:b51}
  \Gamma_N(p_N^0)=\frac{1}{2 p_N^0} \sum_{h_f=\pm 1} \int \rmd \tilde{p}_{L h_f}
  \tilde{p}_H (2 \pi)^4 \delta^4 (p_N-p_{L h_f}-p_H)
  \left| \mathcal{M}_{h_f}^0 \right|^2 (1-f^\rmeq_{L h_f} +f^\rmeq_\phi),
\end{align}
we arrive at equation~\eqref{eq:b31},
\begin{align}
  \sum_{h_f} \left[ \gamma^{\rm sub} (\ell_{h_i}\phi \right.
&
\left. \to \barell_{h_f}
    \barphi) - \gamma^{\rm sub}(\barell_{h_i} \barphi \to \ell_{h_f} \phi)
  \right] \nonumber \\
& =   \sum_{h_f} \left[ \gamma^{\rm sub} (\ell_{h_f}\phi\to \barell_{h_i}
    \barphi) - \gamma^{\rm sub}(\barell_{h_f} \barphi \to \ell_{h_i} \phi)
  \right] \nonumber \\
& = \int \rmd \tilde{p}_N \rmd \tilde{p}_{\ell h_i} \rmd
  \tilde{p}_\phi (2 \pi)^4 \delta^4(p_N-p_{\ell h_i}-p_\phi)
  \e_h^N \left| \mathcal{M}_{h_i}^0 \right|^2 f_{\ell h_i}^\rmeq f_\phi^\rmeq
  (1-f_N^\rmeq) \nonumber \\
& \equiv \e_h^N \gamma_\rmeq(L_{h_i} H \to N).
\end{align}

\section{High Temperature}
\label{sec:high-temperature-2}

For the $u$-channel resonance at high temperature when Higgs bosons
decay into neutrinos and leptons while the neutrinos are stable, we
can derive a relation similar to equation~\eqref{eq:b31}. The width in
the on-shell neutrino propagator is then not the decay rate but an
interaction rate which accounts for the processes where the neutrino
interacts with the medium, that is, $H \to NL$ and $NL \to H$. This
width acts as a regulator of the $u$-channel resonance.

In the narrow-width approximation, the on-shell amplitude reads
\begin{align}
  \label{eq:55}
  \sum_{s_{\ell},s_{\barell}} \left| \mathcal{M}^{\rm os}(\ell_{h_i}
    \phi \to \barell_{h_f} \barphi) \right|^2 = \sum_{s_\ell,s_{\barell}}
  \left| \mathcal{M}(\phi \to N \barell_{h_f}) \right|^2 \left| D_N^{\rm os}
  \right|^2 \left| \mathcal{M}(N \ell_{h_i}\to \barphi) \right|^2,
\end{align}
where the on-shell propagator is the same as in
equation~\eqref{eq:b42}, but the width $\Gamma_N$ is given by the kinematically
allowed processes, $H \to NL$ and $NL \to H$. 

Using the relations in equation~\eqref{eq:b66}, we derive
\begin{align}
  \label{eq:56}
    &\left| \mathcal{M}^{\rm os}(\ell_{h_i} \phi_i \to \barell_{h_f}
    \barphi_f) \right|^2 f_{\ell h_i} f_{\phi,i} (1- f_{\barell h_f})
  (1+ f_{\barphi,f}) - \left| \mathcal{M}^{\rm os}(\barell_{h_i}
    \barphi_i \to \ell_{h_f} \phi_f) \right|^2
  f_{\barell h_i} f_{\barphi,i} (1- f_{\ell h_f}) (1+ f_{\phi,f}) \nonumber \\
  = &\left| D_N^{\rm os} \right|^2 \frac{1}{4} \left|
    \mathcal{M}_{h_i}^0 \right|^2 \left| \mathcal{M}_{h_f}^0 \right|^2
  \left[ f_{\mathcal{L}h_i} (1- f_{\ell h_f}^\rmeq) +f_{\ell
      h_i}^\rmeq f_{\mathcal{L} h_f}+4 \e_h^\phi f_{\ell h_i}^\rmeq (1-f_{\ell
      h_f}^\rmeq) \right] f_{\phi,i}^\rmeq (1-f_{\phi,f}^\rmeq),
\end{align}
Analogous to equation~\eqref{eq:b47}, we derive
\begin{align}
  \label{eq:57}
    \left| \mathcal{M}^{\rm tree}(\ell_{h_i} \phi \right. 
& \left. 
 \to \barell_{h_f} \barphi)
  \right|^2 f_{\ell h_i} f_\phi (1- f_{\barell h_f}) (1+ f_{\barphi})
  \nonumber \\
&-  \left|
    \mathcal{M}^{\rm tree}(\barell_{h_i} \barphi \to \ell_{h_f} \phi) \right|^2
  f_{\barell h_i} f_{\barphi} (1- f_{\ell h_f}) (1+ f_{\phi}) \nonumber \\
&  = \left| \mathcal{M}_{\D L=2} \right|^2_{h_i h_f} \left[
    f_{\mathcal{L}, h_i} (1 - f_{\ell h_f}^\rmeq) + f_{\ell h_i}^\rmeq
    f_{\mathcal{L} h_f} \right],
\end{align}
so that
\begin{align}
  \label{eq:58}
    \gamma^{\rm sub} (\ell_{h_i}\phi\to \barell_{h_f} \barphi) - \gamma^{\rm
    sub}(\barell_{h_i} \barphi \to \ell_{h_f} \phi) =
- \e_h^\phi \gamma^{\rm os}_\rmeq(L_{h_i} H \to L_{h_f} H)
\end{align}

Using the relations
\begin{align}
\label{eq:59}
  (1-f_{\ell h}^\rmeq)f_\phi^\rmeq & = f_N^\rmeq
  (f_{\ell h}^\rmeq+f_\phi^\rmeq),
  \\
  f_{\ell h}^\rmeq (1+f_\phi^\rmeq) & = (1-f_N^\rmeq)
  (f_{\ell h}^\rmeq+f_\phi^\rmeq)
  \\
  \text{and } f_\phi^\rmeq (1-f_{\ell h}^\rmeq) (1-f_n^\rmeq) & =
  (1+f_\phi^\rmeq) f_{\ell h}^\rmeq f_N^\rmeq,
\end{align}
which hold for $\omega_\phi=\omega_{\ell h}+\omega_N$, it is
straightforward to derive
\begin{align}
\label{eq:61}
    \gamma^{\rm sub} (\ell_{h_i}\phi\to \barell_{h_f} \barphi) - \gamma^{\rm
    sub}(\barell_{h_i} \barphi \to \ell_{h_f} \phi) 
&=   \gamma^{\rm sub} (\ell_{h_f}\phi\to \barell_{h_i} \barphi) - \gamma^{\rm
    sub}(\barell_{h_f} \barphi \to \ell_{h_i} \phi), \nonumber \\
\gamma^{\rm os}_\rmeq(L_{h_i} H \to L_{h_f} H)
& = \gamma^{\rm os}_\rmeq(L_{h_f} H \to L_{h_i} H).
\end{align}

Inserting $1=\int \rmd^4 p_N/(2 \pi)^4 \delta^4(p_\phi-p_{\ell h_i}
-p_N)$ into equation~\eqref{eq:61}, again using the first relation
from equations~\eqref{eq:59} and the expression for the total neutrino
width at high temperature, as derived in chapter~\ref{decayrate},
\begin{align}
\label{eq:60}
  \Gamma_N(p_N^0)=\frac{1}{2 p_N^0} \sum_{h_f=\pm 1} \int \rmd \tilde{p}_{L h_f}
  \tilde{p}_H (2 \pi)^4 \delta^4 (p_H-p_{L h_f}-p_N)
  \left| \mathcal{M}_{h_f}^0 \right|^2 (f^\rmeq_{L h_f} +f^\rmeq_H),
\end{align}
we arrive at equation~\eqref{eq:b67} ,
\begin{align}
  \sum_{h_f} \left[ \gamma^{\rm sub} (\ell_{h_i}\phi \right.
&
\left. \to \barell_{h_f}
    \barphi) - \gamma^{\rm sub}(\barell_{h_i} \barphi \to \ell_{h_f} \phi)
  \right] \nonumber \\
& =   \sum_{h_f} \left[ \gamma^{\rm sub} (\ell_{h_f}\phi\to \barell_{h_i}
    \barphi) - \gamma^{\rm sub}(\barell_{h_f} \barphi \to \ell_{h_i} \phi)
  \right] \nonumber \\
& = - \int \rmd \tilde{p}_N \rmd \tilde{p}_{\ell h_i} \rmd
  \tilde{p}_\phi (2 \pi)^4 \delta^4(p_N-p_{\ell h_i}-p_\phi)
  \e_h^\phi \left| \mathcal{M}_{h_i}^0 \right|^2 f_{\ell h_i}^\rmeq (1+f_\phi^\rmeq)
  f_N^\rmeq \, .
\end{align}

    \newpage

  \backmatter
\addcontentsline{toc}{chapter}{{Bibliography}}
\renewcommand*{\bibname}{B\MakeLowercase{ibliography}} 
\markboth{Bibliography}{Bibliography}
  \bibliographystyle{JHEPMS}
  \bibliography{main}
  \markboth{}{}

  \addcontentsline{toc}{chapter}{\protect Acknowledgments}
\chapter*{Acknowledgements}

First of all, I would like to thank Dr.~Michael Pl\"umacher for his
supervision of the Ph.D.~project and our collaborator Prof.~Dr.~Markus H.~Thoma
for his support. I also thank my thesis referees PD Dr.~Georg Raffelt
and Prof.~Dr.~Gerhard Buchalla for agreeing on reading this
thesis. Thanks as well to Mathias Garny for fruitful and inspiring
discussions about leptogenesis.

Moverover, I thank all colleagues from the MPI, especially my office mates
Martin Spinrath and Philipp Kostka for an enjoyable work atmosphere,
but all the others as well. Special thanks goes to the people who
proofread this thesis, Valerie Domcke, Mathias Garny, Jan Germer,
Peter Graf, Daniel Greenwald, Philipp Kostka, Ananda Landwehr, Michael
Schmidt, Martin Spinrath and Maike Trenkel.

Last but not least, I thank my family for their support during this
thesis.



\end{document}